\documentclass[twoside,12pt]{article}
\usepackage{epsfig}
\usepackage{german}
\usepackage{epsfig}
\usepackage{here}
\usepackage{amsmath}
\usepackage{amssymb}
\def\Journal#1#2#3#4{{#1} {#2} (#4) #3 }

\def\NPA{{\em Nucl. Phys.} A}

\def\PRO{{\em Prog. Theor. Phys.}}
\def\NPB{{\em Nucl. Phys.} B}

\def\PLB{{\em Phys. Lett.} B}

\def\PL{{\em Phys. Lett.}}
\def\PRL{\em Phys. Rev. Lett.}
\def\PREV{\em Phys. Rev.}
\def\PREP{\em Phys. Rep.}

\def\PRD{{\em Phys. Rev.} D}
\def\PRC{{\em Phys. Rev.} C}

\def\ZPC{{\em Z. Phys.} C}
\def\ZPA{{\em Z. Phys.} A}

\def\RMP{{\em Rev. Mod. Phys.}}

\def\INT{{\em Int. J. Mod. Phys.} E}
\def\NIM{{\em Nucl. Instr. and Meth.} A}
\def\EPJD{{\em Eur. Phys. J.} D}
\def\EPJC{{\em Eur. Phys. J.} C}
\def\EPJA{{\em Eur. Phys. J.} A}
\def\CJP{\em Can. J. Phys. }
\def\JP{{\em J. Phys.} G}
\def\ARNS{\em Ann. Rev. Nucl. Sci.}
\def\PS{\em Phys. Scripta}
\def\APPB{{\em Acta Phys. Pol.} B}
\def\ANP{\em Adv. Nucl. Phys.}
\def\PPNP{\em Prog. Part. Nucl. Phys.}
\def\FBSS{\em Few Body Systems Suppl.}
\def\RPP{\em Rep. Prog. Phys.}
\def\NC{\em Nuovo Cimento}
\def\MPLA{{\em Mod. Phys. Lett.} A}

\newcommand{\be}{\begin{equation}}
\newcommand{\ee}{\end{equation}}
\newcommand{\bea}{\begin{eqnarray}}
\newcommand{\eea}{\end{eqnarray}}
\newcommand{\nn}{\nonumber}
\usepackage{cite}
\begin{document}
\title{Close--to--threshold Meson Production in Hadronic Interactions$^{*}$}
\author{P.\ Moskal,$^{1,2}$ M.\ Wolke,$^{1}$ A.\ Khoukaz,$^3$
W.\ Oelert$^{1}$ \\ 
\\
$^1$IKP, Forschungszentrum J\"ulich, D -- 52425 J\"ulich, Germany\\
$^2$IP, Jagellonian University, PL -- 30-056 Cracow, Poland \\
$^3$IKP, Westf\"alische Wilhelms--Universit\"at M\"unster, D -- 48149 M\"unster,
Germany}
\date{}
\maketitle
\begin{abstract}
Studies of meson production at threshold in the hadron--hadron 
interaction began in the fifties when sufficient energies of accelerated 
protons were available. 
A strong interdependence between developments in accelerator physics, detector 
performance and theoretical understanding led to a unique vivid field of 
physics. 
Early experiments performed with bubble chambers revealed already typical 
ingredients of threshold studies, which were superseded by more complete meson 
production investigations at the nucleon beam facilities TRIUMF, LAMPF, PSI, 
LEAR and SATURNE. 
Currently, with the advent of the new cooler rings as IUCF, CELSIUS and COSY 
the field is entering a new domain of precision and the next step of further 
progress.

The analysis of this new data in the short range limit permits a more 
fundamental consideration and a quantitative comparison of the production 
processes for different mesons in the few--body final states. 
The interpretation of the data take advantage of the fact that 
production reactions close--to--threshold are characterized by only a few 
degrees of freedom between a well defined combination of 
initial and exit channels.
Deviations from predictions of phase--space controlled one--meson--exchange 
models are indications of new and exciting physics. 
Precision data on differential cross sections, isospin and spin observables 
--- partly but by no means adequately available --- are presently turning up 
on the horizon. 
There is work for the next years and excitement of the physics expected. 
Here we try to give a brief and at the same time comprehensive overview of 
this field of hadronic threshold production studies.
\end{abstract}
\vspace{1.7cm}
$~~~~~~~~~~~~~$14.40.Cs, 25.40.-h, 25.40.Ve, 29.20.Dh 

\vspace{3cm}
{\small{$^{*}$This review article will be published 
in ``Prog. Part. Nucl. Phys.'' Vol. 49, issue 1 (2002).}}

\newpage
\tableofcontents
\newpage

\section{Introduction}  
\label{Intro}    
The complexity of the hadronic structure is one of the actual and topical 
fields of physics concerning the microscopic scale. 
Different, complementary as well as competing experimental approaches are used 
to attack the challenge of the hadronic spectroscopy. 

A detailed understanding of the strong interaction dynamics on the symmetries 
of QCD is a fundamental aim of hadron physics. 
At high energies the QCD serves as a reliable theory whereas at low energies 
the non--perturbative QCD phenomena must be investigated. 
Here calculations are performed in an effective field theory of hadrons 
represented by the chiral effective Lagrangian, a theory of Goldstone bosons 
coupled to the octet of baryons and vector mesons. 
The QCD Lagrange function is written as:
\begin{equation}
L_{QCD} = \bar q\;(i\gamma^{\mu}\partial_{\mu} - M)\;q\;+\;L_{gluon}\;+\; 
L_{quark-gluon}.
\end{equation}
In principle, $q$ describes the column vector of the quarks $u, d, s, c, b, t$ 
with each of them represented by a spinor, identical to the Dirac equation for 
spin--$\frac{1}{2}$ electrons. 
At energies in the few GeV range only the first three quarks are of any 
relevance, $M$ represents the diagonal of the matrix for the quark current 
masses. 

Data of high quality and precision on hadronic processes at low and 
intermediate momenta are necessary in order to verify the systematic low 
energy expansion of the Chiral Perturbation Theory ($\chi PT$), which has 
already enforced an important insight into the structure and dynamics of 
nucleons and mesons~\cite{V.Bernard, A.M.Bernstein}.
It is known that the current masses of the lightest three quarks are 
significantly smaller than the typical hadronic energy scale, represented by 
the proton mass of $\mbox{m}_p \approx 1\,\mbox{GeV}/\mbox{c}^2$.  
This is the reason why in first order the quark masses of the light quarks are 
neglected in theoretical considerations and the spin--$\frac{1}{2}$ fields can 
be separated into two independent left-- and right--handed parts, reflected by 
the chiral symmetry.
The QCD Lagrangian separates into two identical images:

\begin{equation}
L_{QCD}[\bar q , q] = L_{QCD}\;[\bar q_{left}, q_{left}] + 
 L_{QCD}\;[\bar q_{right}, q_{right}].
\end{equation}

Left--handed quarks do not communicate with right--handed quarks and {\em vice 
versa\/}. 
There is no Lorentz transformation which can change the handedness of a 
mass--less quark or particle. 
This symmetry of the Lagrange function is not the symmetry of the spectrum of 
the particles since otherwise each hadron should have a partner of equal mass 
but inverted parity.
Consequently there would be a second proton with negative parity which does not 
exist. 
There are no parity--doublets in the spectrum of the strong interacting 
particles and therefore the symmetry is spontaneously broken.
Due to this breaking the mass--less so--called Goldstone bosons, the 
pseudoscalar $\pi$, $K$ and $\eta$ mesons appear.
The restriction of the interaction of these particles among each other and with 
other hadrons can be used to analyze 
consequences of the chiral symmetry and its breaking in the framework of an 
effective field theory.

In reality the lightest hadrons --- the Goldstone bosons --- are by no means 
mass--less, which can be understood by an explicit breaking of the chiral 
symmetry including a term for the quark masses. 
For each massive Fermion there exists a Lorentz transformation which 
transforms a left--handed to a right--handed field. 
The effect of this explicit breaking now can be treated perturbatively since 
the masses of the light quarks are much smaller than the typical hadronic mass.

Predicting the low--energy properties of nuclear and particle physics, 
$\chi PT$ plays an important role in understanding hadron physics in the 
non--perturbative regime. 
Therefore it is very important to know the properties of the mesons, their 
structure and interaction in the hadronic environment.\vspace{1ex}

The physics program at the medium energy hadron accelerators was and is 
focusing on studies of the production and the decay of light mesons and baryon 
resonances and the conservation or violation of symmetries.\vspace{1ex}

\begin{figure}[H]
\begin{center}
\epsfig{file=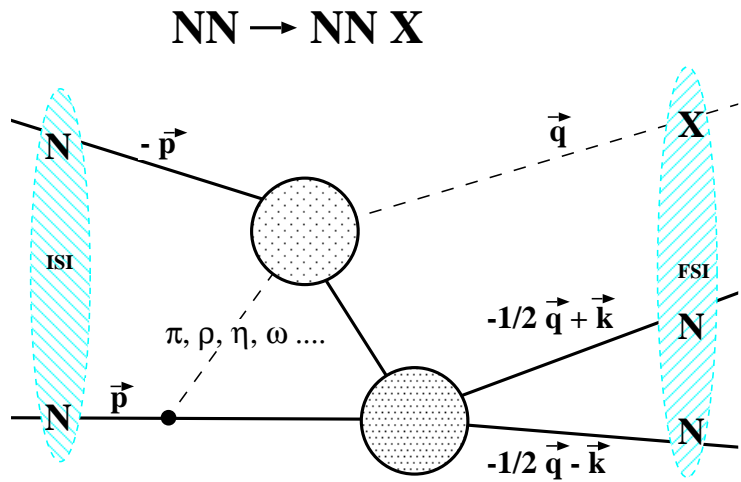,scale=1.05}
\caption{\label{MESON_EXCHANGE} Diagrammatic view of the one--boson--exchange 
meson production process. Produced meson = $X$, nucleon momentum = 
$\vec{\mbox{p}}$, meson momentum = $\vec{\mbox{q}}$, relative $NN$ momentum = 
$2\,\vec{\mbox{k}}$~\cite{wil99}.} 
\end{center}
\end{figure}

Following this sequence we will present and discuss production of mesons and 
meson pairs at threshold with the questions depicted by the interaction view 
of figure~\ref{MESON_EXCHANGE}~\cite{wil99} where in the nucleon--nucleon 
($NN$) scattering a meson $X$ is created in a one--boson--exchange model.
For the particular case the questions have to be answered:
how is the distortion of the incident $NN$ waves (ISI) included, which mesons 
contribute to the exchange process, is there an intermediate baryon resonance, 
how significant are rescattering contributions of the exchange mesons and what 
is a reasonable treatment of the $NN$ and $NX$ final state interactions (FSI)? 
Especially for the $NN$--interaction the FSI is crucial because of the nearby 
poles in the S--wave amplitudes corresponding to the deuteron bound state in the 
$^3\mbox{S}_1$ channel or the $^1\mbox{S}_0$ virtual state~\cite{wil99}.  
These poles and the phase--space factors tend to determine much of the energy 
dependence of the total cross section for meson production.
Furthermore, in any region where these poles dominate, it is possible to link 
quantitatively meson production in cases where the two nucleons emerge 
separately or as a bound deuteron state.
Even at threshold the reaction mechanism of the basic process for the 
interrelation between real pions and virtual exchange meson currents as for 
instance in the pion production ($NN \rightarrow NN \pi$) is still not fully 
understood. 
Though first data suggested that s--wave pions were produced in a 
heavy--meson--exchange process, later measurements resulted in an 
interpretation of an interference among transition amplitudes as Ss, Ps, Pp, 
Sd and Ds, where the capital letter indicates the $NN$ final state wave and 
the small letter the angular momentum between the two nucleons and the meson 
produced. 
In addition, it has been concluded that for higher partial waves $\pi$ 
exchange rather than a heavier meson exchange is more significant.

Especially the pion--nucleon interaction went through several stages of 
increasing sensitivity for tests of nuclear theory. 
As indicated above more than a decade after the initial 
work~\cite{GMW,Rosenfeld} only in the sixties explicit calculations on cross 
sections of the $pp \rightarrow pp\pi^0,\;pn \pi^+$ transitions were 
possible~\cite{KuR,SSY} which, however, failed the total cross 
section~\cite{meyer633,meyer2846,WWD_95} by a factor of five. 
After further theoretical developments, see e.g.\ reference~\cite{LEE_RISKA}, 
only recently detailed calculations were provided within the J\"ulich meson 
exchange model~\cite{hanhart21,hanhart25,hanhart064008} including transitions 
in the threshold region beyond $l_{\pi} = 0$ and predicting analyzing powers 
and spin correlation coefficients making use of i) the basic diagrams, ii) 
final state interactions, iii) off--shell effects, iv) the exchange of heavier 
mesons and v) influences of intermediate excitation of the $\Delta$ resonance. 
On the other hand, the J\"ulich model does not account for quark degrees of 
freedom i.e.\ calculations within the framework of the $\chi PT$ which, in 
fact, predict cross sections too small~\cite{BuK_2000} compared to 
experiments.\newline
The heavier the meson produced in nucleon--nucleon scattering the larger the
momentum transfers and thus in such processes the short--range parts of the 
production operators are tested.\vspace{1ex}

Double meson production processes with both mesons being either identical 
(e.g.\ $\pi^0 \pi^0$ or $K^0 \overline{K^0}$) or different (e.g.\ $\pi \eta$) 
are in principle similar to the single ones, however, the possibly associated 
baryon resonances as intermediate states differ significantly. 
For instance, the two--pion production very likely is dominated by either the 
$\mbox{P}_{11}$ $N^*(1440)$ resonance --- via an effective $\sigma$ exchange 
giving information of a scalar meson excitation of the $\mbox{P}_{11}$ 
$N^*(1440)$ resonance --- or the simultaneously excited $\mbox{P}_{33}$ 
$\Delta(1232)$ resonances. 
Due to selection rules here the choice of definite quantum numbers in the 
initial and final state might help to sort out the reaction mechanism.

In any case, a large amount of the knowledge about the $\pi$--$\pi$ interaction
has been obtained by the $\pi N \rightarrow \pi \pi N$ reaction, whereas data 
employing proton beams just start to come.\vspace{1ex}

As long as only S--wave processes are involved, which is the privilege of 
threshold production studies, scattering length and effective range approaches 
are used to describe the interaction sign and strength. 
In case of strong attractive interactions a distinction between the 
final state scattering 
and the formation of a baryonic resonance leading to bound or 
quasi--bound states can not be made uniquely. 
Such investigation will be presented.\newline
Isospin violation or charge symmetry breaking processes are a topical and 
interesting field of threshold production physics. 
The essential contribution to isospin breaking is the possible $\pi^0 - \eta$ 
mixing. 
Therefore experiments around the $\eta$ threshold as e.g.\ $dd \rightarrow 
\alpha\,\pi^0 (\eta)$ or $pd \rightarrow {^3He}\,\pi^0 (^3H\,\pi^+)$ should 
identify the $\pi^0 - \eta$ mixing angle which in turn might give hints to the mass 
difference of the up and down quarks.\vspace{1ex}

\subsection{Aspects of threshold production}       
\label{Aotp}          
Threshold production experiments are characterized by excess energies which 
are small compared to the produced masses~\footnote{With meson production up to 
masses of $\approx 1\,\mbox{GeV}/\mbox{c}^2$ discussed in this article, we 
will principally refer to experimental results and theoretical investigations 
covering excess energies up to $140\,\mbox{MeV}$ (the $\pi$--mass) in excess 
of the respective thresholds.}.
Consequently, in fixed target experiments the momenta of the final state 
particles transverse with respect to the direction of the incoming beam are 
small compared to the longitudinal components.
Thus, ejectiles are confined to a narrow forward cone in the laboratory system 
around the centre--of--mass movement and --- close--to--threshold --- an 
experimental acceptance covering the full solid angle is feasible with 
comparatively small dedicated detector arrangements.

Small relative momenta in the final state effectively limit the number of 
partial waves contributing, simplifying the theoretical interpretation of 
experimental results.
It should be noted, that already three--body final states require --- in 
principle --- a three--body Faddeev like approach, which has not been 
accomplished so far~\cite{wyc00,fix02}.
However, as first described by Watson~\cite{wat52}, with two strongly 
interacting particles in the final state, the energy dependence of the total 
cross section close--to--threshold is essentially determined by the 
(three--body) phase--space and the energy dependence of the on--shell final 
state interaction (FSI).
Due to small relative velocities FSI effects are inherent to the experimental 
observables.
Thus, the interpretation of data in terms of reaction dynamics requires a 
correct treatment of both initial~\footnote{At high relative momenta of the 
colliding nucleons necessary for meson production the on--shell 
nucleon--nucleon interaction exhibits a rather weak energy dependence and 
might well be approximated by an overall reduction of the cross section in 
magnitude~\cite{Batinic,nak01}. However, for produced masses in the 
$1\,\mbox{GeV}/\mbox{c}^2$ regime, even for proton--proton scattering no 
reliable model exists to allow a consistent evaluation of the initial state 
interaction~\cite{nak01}.} (ISI) and final state 
interactions~\cite{hanhart176} (for a review see~\cite{kle01}).
On the other hand, FSI effects might provide access to low--energy scattering 
parameters, which are difficult to obtain otherwise in case of unstable 
particles.

Meson production at threshold implies high momentum transfers 
$\Delta\mbox{p}$, given by
\begin{equation} 
\label{momtrans_eq}
\Delta\mbox{p} = \frac{1}{2} 
  \sqrt{\left(\mbox{m}_N + \mbox{m}_B + \mbox{m}_X\right)^2 - 
        4\,\mbox{m}_N^2}
\end{equation}
for the reaction type $N N \rightarrow N B X$ with $N$, $B$ and $X$ denoting 
a nucleon, baryon or meson in the initial and final state and $\mbox{m}_N$, 
$\mbox{m}_B$ and $\mbox{m}_X$ as the respective masses.
With momentum transfers in the range of $0.37\,\mbox{GeV/c}$ to 
$1.10\,\mbox{GeV/c}$ for $\pi^0$ and $\phi$ meson production, threshold 
production probes with corresponding distances between $0.53\,\mbox{fm}$ and 
$0.18\,\mbox{fm}$ the short range part of the nucleon--nucleon interaction.
Consequently, the energy dependence of the primary production amplitude is 
expected to be weak, motivating Watson's approach~\cite{wat52}.\vspace{1ex}

Theoretical studies have been carried out mainly within the framework of 
hadronic meson exchange models, i.e.\ with baryons and mesons as effective 
degrees of freedom (for a recent review see~\cite{nak01}).
Chiral perturbation theory has been applied for the description of data on 
$\pi$ production close--to--threshold~\cite{han00}.
However, in view of the characteristic distances mentioned above, QCD inspired 
models~\cite{koc00,dil02} with constituent quarks and gluons or instantons as 
relevant degrees of freedom, might turn out to be appropriate.
High quality exclusive data on close--to--threshold meson production in the 
energy range of non--perturbative strong interaction physics will be crucial 
for exploring the boundary between effective meson exchange models and (so far 
phenomenological) approaches based on quark--gluon degrees of freedom.
\subsection{Hadronic and electromagnetic probes} 
\label{Haemp}            
It seems that the  $\chi PT$ is of limited value once strangeness is involved 
in the hadronic systems and non--perturbative coupled--channel considerations 
beyond $\chi PT$ are on the market~\cite{N.Kaiser-I, N.Kaiser-II}. 
High quality data on the meson ($\pi, \eta, \eta^{\prime}, \omega, K^+, K^-$) 
production at threshold have been and are being produced at the hadronic beam 
cooler rings using the baryon number $\mbox{B} = 2$ systems (essentially $pp$ 
and $pn$). 
At least equally important are studies with hadronic interactions in the 
$\mbox{B} = 1$ sector. 
However, most measurements with $\pi$-- or $K$--beams used bubble chamber 
techniques which naturally suffer from very poor statistics.

Accurate investigations of differential cross sections for meson production 
from the initial $\pi$ plus proton or $K$ plus proton systems are needed at 
several momenta to study the s-- and p--wave dynamics and their interplay. 
There are several indications that as intermediate states both the 
$\mbox{S}_{11}$ resonance and the $K\Sigma$ quasi--bound state --- followed by 
a final state interaction leading to nucleon and meson --- are essential and 
could be investigated best via the $\mbox{B} = 1$ sector.

Until now, most of the data for threshold meson production were observed with 
proton beams. 
The use of antiprotons has contributed considerably to the knowledge of the 
meson spectroscopy up to serious candidates for glueballs and exotic hybrids 
but is still  
suffering from the lack of unique interpretations.

A comparison of the meson production in the fundamental process to productions 
on heavier nucleons would give constraints to eliminate uncertainties in the 
basic interactions. 
As outlined by W.~Weise and U.--G.~Mei\3ner~\cite{WpluM}, precise measurements 
of such processes set the necessary constraints for the effective Lagrangian of 
low--energy QCD with strange quarks. 
Its detailed knowledge has impacts on several other important issues, such as 
the strange quark content of the nucleon.\vspace{1ex}
 
\begin{figure}[H]
\vspace{0.8cm}
\begin{center} 
\epsfig{file=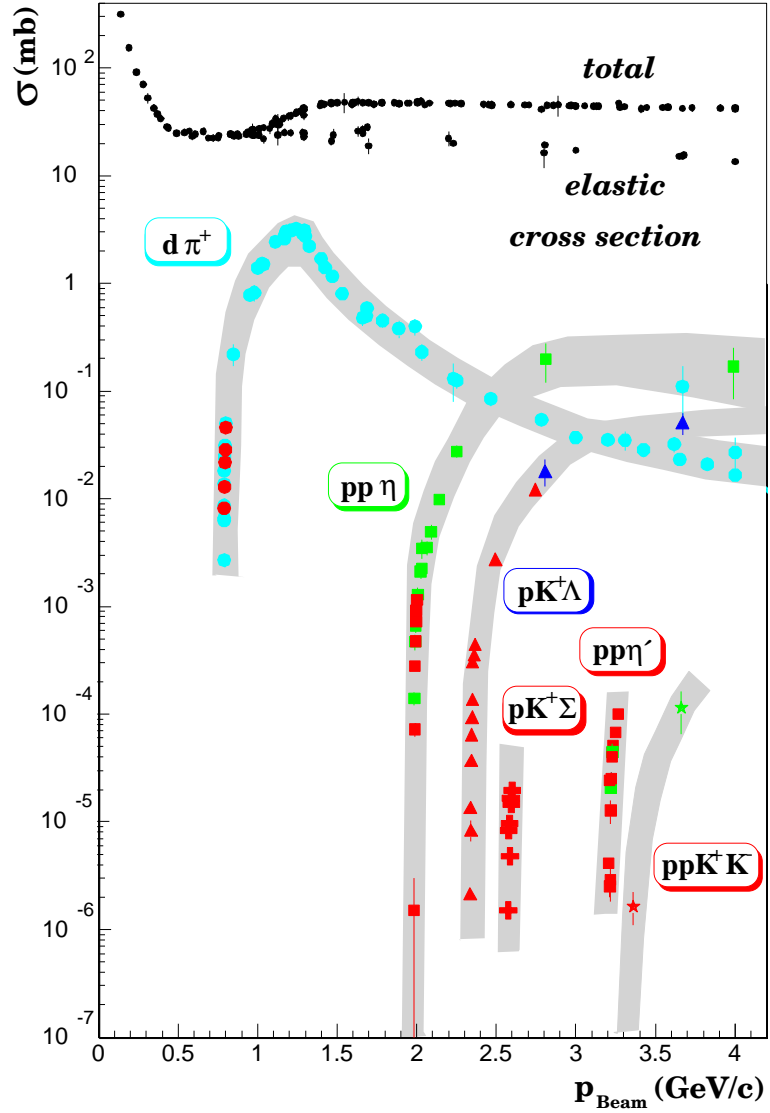,scale=0.59}
\vspace{-0.5cm}
\caption{\label{Dieter_Wasserfall} Proton--proton cross sections for different 
reactions of the light meson production in the threshold region. For 
comparison also the total and elastic cross sections of proton--proton 
scattering are shown. Relevant references to the data will be given throughout 
the present article at the appropriate sections.} 
\end{center}
\end{figure}

Though in the present article we will restrict ourselves to hadron induced 
reactions, in general induced with $p$--, $d$--, $\pi$-- and 
$\overline p$--beams, investigations on the structure of the nucleon as seen 
with leptons should at least be mentioned~\cite{HARRACH}. 
The use of leptonic interactions via photon or electroweak boson exchanges has 
the attraction of combining a weakly interacting particle of well controlled 
energy momentum transfer with the complex multi particle system containing 
charged constituents.
Photoproduction provides a crucial complementary information compared to 
hadronic processes, since they are characterized by the electromagnetic probe 
as a virtual vector meson or an $s \bar s$ fluctuation and as such are a 
powerful source of spin--1 states. 
The well proven perturbative methods allow us to define a hierarchy of 
operators which, at least in principle, can be extracted separately. 
Perturbative methods finally allow to overcome one of the limitations of the 
electroweak interaction, that is the access to gluon distributions. 
In fact the strong interaction in the limit of hard scales is becoming a probe 
for the soft region. 
All this can be complemented by the addition of the spin and flavour degrees 
of freedom~\cite{HARRACH}.

Cross sections are very small in the threshold region down to eight orders of 
magnitude compared to the total yield. 
Due to the rapid growth of the phase--space volume the total cross section of 
the meson production reactions increases by orders of magnitudes in a few MeV 
range of excess energy. 
Such studies have been made possible only due to the low emittance and small 
momentum spread proton beams available at the storage ring facilities.
Figure~\ref{Dieter_Wasserfall} demonstrates threshold production measurements 
of light mesons in the proton--proton interaction.

Especially hadron accelerators offer the possibility to excite nuleonic 
resonances as $N^*$'s and $\Delta^{(*)}$'s with the same quantum numbers as 
those of the proton via the $p \alpha \rightarrow N^* \alpha$ reaction, where 
the $\alpha$--channel --- with its spin and isospin zero --- serves as a 
spin--isospin filter.

This scalar--isoscalar tool is complementary to the vector excitation of 
nuclei with electromagnetic probes. Recent results from different electron beam
facilities as ELSA and MAMI in Germany and NIKHEF in The Netherlands certainly
deserve a detailed and circumstantial presentation. Due to the present
limitation to hadronic interactions we can not give appropriate credit to this
field of physics and refer to~\cite{HARRACH,Workman}. 
At the Continuous Electron Beam Accelerator (CEBAF) of the Thomas Jefferson 
National Accelerator Facility (TJNAF) an electron beam is used for exciting 
particular nuclear resonances and observing their individual decay modes.

\subsection{Hadronic interactions} 
\label{Hinertact}                          
The interaction of hadrons --- being the reflection of the strong force acting 
between their constituents --- delivers indirect information about their 
structure and the strong interaction itself. 
In the frame of the optical potential model the hadronic interaction can be 
expressed in terms of phase--shifts, which in turn are described by the 
scattering length and effective range parameters. 
These are quite well known for the low--energy nucleon--nucleon 
interaction~\cite{machleidtR69,machleidt024001} yet they are still poorly 
established in case of meson--nucleon or even meson--meson interactions. 
This is partly due to the absorption of mesons when scattering on 
a baryon. To account for this effect the scattering length becomes a complex 
quantity where the imaginary part --- e.g.\ in case of the nucleon--$\eta$ 
interaction --- describes the $\eta N \rightarrow \pi N$ and $\eta N \rightarrow 
multi\mbox{--}\pi\,N$ processes. 
Moreover, the short life--time of all {\em neutral} ground state mesons 
prohibits their utilization as secondary beams and therefore the study of 
their interaction with hadrons is accessible only via observations of their 
influence on the cross section of the reactions in which they were produced 
($NN \rightarrow NN\, Meson$).  
When created close to the kinematical threshold with the relative kinetic 
energy being in the order of a few MeV, the final state particles remain much 
longer in the range of the strong interaction than the typical life--time of 
$N^{*}$ or $\Delta$ baryon resonances with $10^{-23}\,\mbox{s}$ and hence 
they can easily experience a mutual interaction before escaping the area of an 
influence of the hadronic force. 
This --- as introduced by Watson~\cite{wat52} --- final state interaction 
(FSI) may significantly modify both the original distributions of relative 
momenta of the outgoing reaction products and the magnitude of the production 
cross section. 
Though it is easy to understand that the FSI changes the distributions of the 
differential cross sections it is rather difficult to cope for the influences 
on the magnitude of the total reaction rate, since one tends to separate the 
primary production from the final state interaction in space as well as in 
time~\cite{wat52}.
Considering the primary production as a separate process it is well worth 
trying to understand the phenomenon qualitatively. 
If there were no final state interactions the total cross section would be 
fully determined by the kinematically available phase--space volume, $V_{ps}$, 
where each interval is populated with a probability governed by the primary 
production amplitude only:
\begin{equation} 
\sigma = \frac{1}{\mbox{F}} \int dV_{ps} \, |M|^2 \ \approx \ 
  const. \cdot V_{ps}.
\end{equation}
The approximation in the equation results from the assumption that 
$|M|^2 = \mbox{constant}$ in a few MeV range above the production 
threshold~\cite{moalem445,bernard259,gedalin471}. 
F denotes the flux factor of the colliding particles.\vspace{1ex}

\begin{figure}[H]
\vspace{-0.65cm}
\parbox{0.50\textwidth}
  {\epsfig{figure=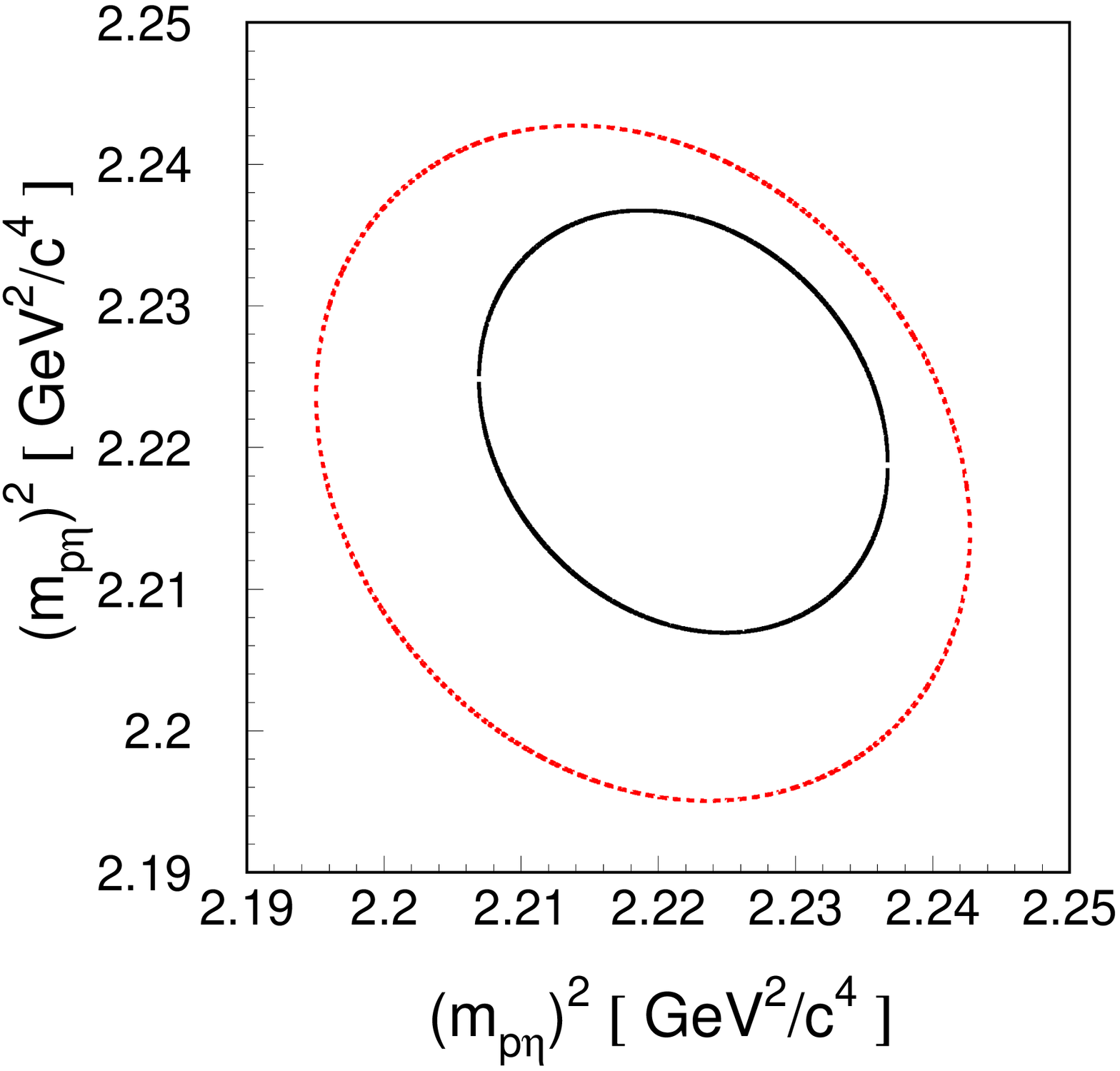,width=0.50\textwidth}} \hfill
\parbox{0.50\textwidth}
  {\epsfig{figure=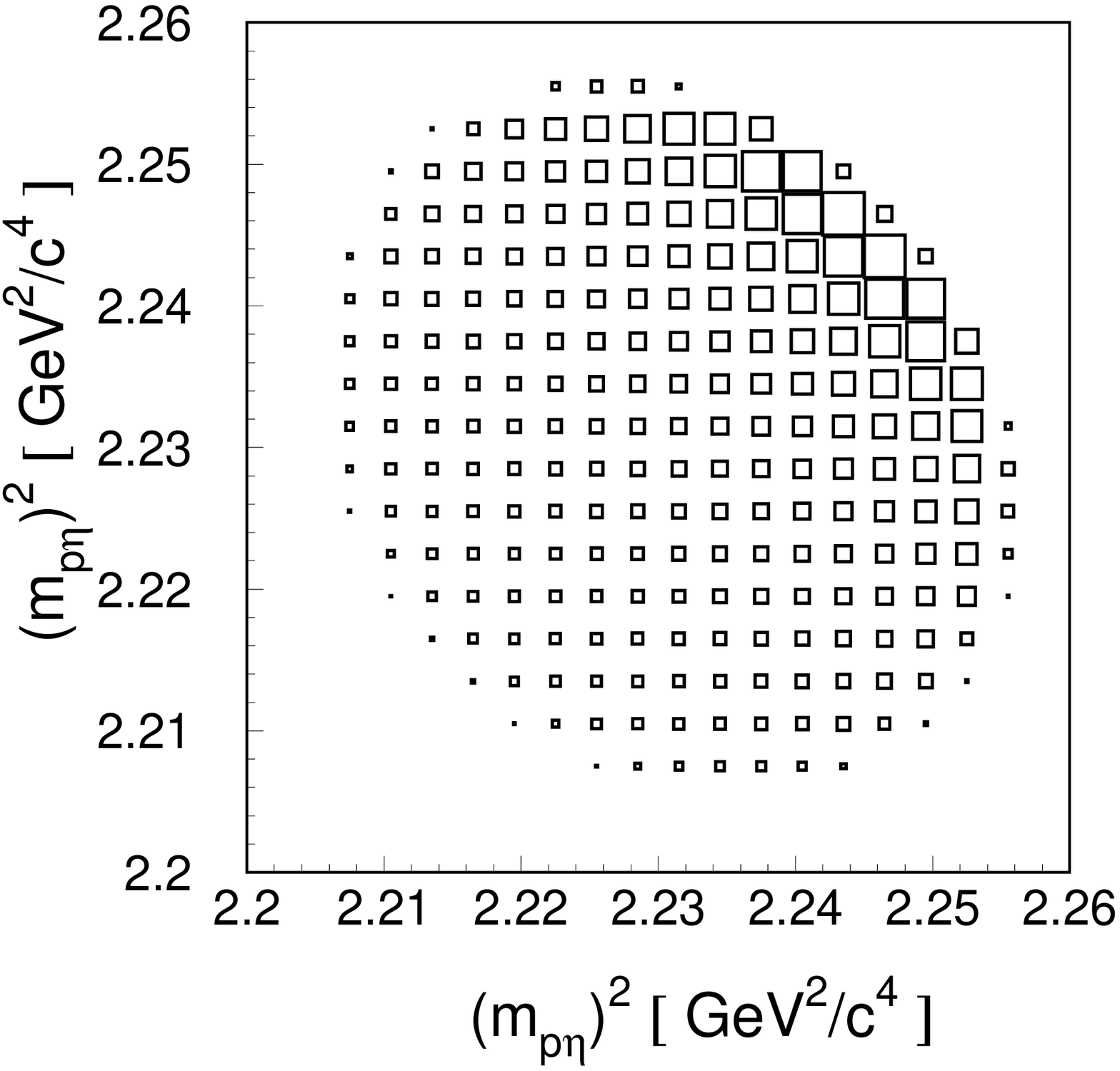,width=0.51\textwidth}}

\caption{\label{dalitz_phasespace} (left) The solid line indicates the 
kinematically available area for the $pp \eta$ system with a total 
centre--of--mass energy $\sqrt{\mbox{s}} = 2433.8\,\mbox{MeV}$. The dotted 
line depicts the range assuming a reduction of the proton and $\eta$--meson 
masses by $2\,\mbox{MeV}$. The phase--space volume results from an integral 
over the closed area: $V_{ps} = \frac{\pi^{2}}{4\,\mbox{\scriptsize s}}
 \int\int d\,\mbox{m}^2_{p_{1}\eta}\,d\,\mbox{m}^2_{p_{2}\eta}$. (The powers 
of $2\pi$ are skipped here and will be included into the flux factor F 
according to the convention introduced by Byckling and 
Kajantie~\cite{bycklingkajantie}).\\
(right) Distribution of the phase--space for the $pp\eta$ system modified by 
the proton--proton interaction and calculated for an excess energy of 
$\mbox{Q} = 16\,\mbox{MeV}$.The area of squares are proportional to the number 
of entries. The largest square of the figure corresponds to 260 events. }
\end{figure}

In the classical picture we might imagine that the reaction particles are 
created together with their appropriate force field and when escaping the 
interaction region they acquire a potential energy increasing or decreasing 
their kinetic energy depending whether the interaction is repulsive or 
attractive. 
For an attractive interaction they could be created also in those phase--space 
partitions which are not available for non--interacting particles and 
subsequently be ``pulled down'' to the energetically allowed regions by final 
state interaction. 
The temporary growth of the primary production phase--space would than 
increase the reaction rate. 
Contrary, in case of a repulsive interaction the particles must be produced in 
the lower phase--space volume since leaving the interaction area they will 
acquire additional kinetic energy.
The reduction of the total cross section, for example in case of the repulsive 
Coulomb force, can easily be understood when considering the production in a 
coordinate space of point--like objects. 
Here --- in contrast to non--interactive particles --- a strongly repulsive 
object can not be produced at appropriately small distances since their later 
acceleration would lead to the breaking of the energy conservation and thus 
the space available to primary production is reduced.
One can also argue relativistically that the primary mechanism creates the 
particles off the mass shell and subsequently they are lifted onto the mass 
shell by the final state interaction. 
The solid line in figure~\ref{dalitz_phasespace} (left) depicts the boundary 
of the Dalitz plot in case of the $pp \rightarrow pp \eta$ reaction calculated 
at the total centre--of--mass energy $\sqrt{\mbox{s}} = 2433.8\,\mbox{MeV}$ 
exceeding by $10\,\mbox{MeV}$ the threshold energy. 
The area surrounded by that curve is a direct measure of the kinematically 
available phase--space volume. 
The dotted line shows the 
\hspace{-0.7cm}corresponding plot at the moment of the primary 
creation if the mass of each particle was reduced by $2\,\mbox{MeV}$, 
demonstrating that now the available phase--space grows significantly. 
Indeed, as shall be 
inferred from the experimental results presented in 
subsequent sections, at excess energies of a few MeV above threshold, the 
mutual interaction among the outgoing particles enhances drastically --- by 
more than an order of magnitude --- the total cross section and modifies 
appreciably the occupation of the phase--space. 
Figure~\ref{dalitz_phasespace} (right) indicates the phase--space distribution 
expected for the $pp \eta$ system at an excess energy of $\mbox{Q} = 
\sqrt{\mbox{s}} - 2\,\mbox{m}_{p} - \mbox{m}_{\eta} = 16\,\mbox{MeV}$, 
assuming a homogeneous primary production and taking into account the S--wave 
interaction between the protons. 
The proton--proton FSI modifies the homogeneous Dalitz plot distribution of 
``non--interacting particles'', enhancing its population at a region where the 
protons have small relative momenta. 
The interaction of the proton--$\eta$ system would manifest itself at low 
invariant masses $\mbox{m}^2_{p\eta}$ corresponding to small relative momenta 
between the proton and the $\eta$ meson.
Such effects observed in the experiments are presented in section~\ref{SmpiNNs}.

At the vicinity of the threshold, relative angular momenta larger than 
$l = 0\,\hbar$ will not play any role due to the short range of the strong 
interaction and small relative momenta of the produced particles. 
It can be inferred from parity and angular momentum conservation laws that 
the production of an $N$--$N$--$Meson$ system (for pseudoscalar or vector
mesons) in relative S--waves may only occur if the nucleons collide in P--wave. 
Thus at threshold the transition $\mbox{P} \rightarrow \mbox{S}\mbox{s}$ is 
the only possible one, with capital letters denoting the angular momentum 
between nucleons and the small letter for the meson angular momentum with 
respect to the pair of nucleons. 
An interaction of the nucleons in the entrance channel influences the 
production process appreciably~\cite{hanhart176}, similarly as that described 
above between the outgoing particles. 
For example, for $\eta$ production in the $pp \to pp \eta$ reaction the 
initial state interaction (ISI) reduces the total cross section by about a 
factor of 3--5~\cite{Batinic,hanhart176} due to the repulsive proton--proton 
$^3\mbox{P}_{0}$--wave potential. 
This factor keeps constant in the range of a few tens of MeV~\cite{Batinic} 
and hence, does not influence the energy dependence of the total cross section 
of the meson production, which remains predominantly determined by the final 
state interaction. 
In particular, the dominant S--wave nucleon--nucleon final state interaction 
is by far stronger than any of the low--energy meson--nucleon ones, with the 
only exception of the proton--$K^-$ FSI~(see section~\ref{Mwhs}). 
The effect of the nucleon--nucleon FSI diminishes with increasing excess 
energy since it significantly influences only that partition of the 
phase--space at which the nucleons have small relative momenta. 
Whereas this fraction keeps constant, the full volume of the phase--space 
grows rapidly: 
An increase of the excess energy from $\mbox{Q} = 0.5\,\mbox{MeV}$ to 
$\mbox{Q} = 30\,\mbox{MeV}$ corresponds to a growth of $V_{ps}$ by more than 
three orders of magnitude~\cite{swave}. 
As a result the S--wave nucleon--nucleon FSI is of less importance for higher 
excess energies where it affects a small fraction of the available 
phase--space volume only.
A more quantitative discussion about the influence of the ISI and FSI effects 
at close--to--threshold production cross sections will  be presented in 
section~\ref{Mwhs}.

\section{Basics of free and quasi--free production}
\label{Bofaqfp}

\subsection{Free NN scattering}
\label{Siv}
Investigations of the production of mesons and their interactions with 
nucleons are based on measurements determining the total and differential 
production cross sections and their dependence on the energy of the 
interacting nucleons.
Therefore, to enable a quantitative discussion on the mechanisms leading to 
the transformation of the energy--of--motion of nucleons into matter in the 
form of mesons let us recall the formula of the reaction cross section. 
In case of the $NN \rightarrow NN \,Meson$ process --- with the four--momenta 
of the colliding nucleons denoted by $\mathbb{P}_a$ and $\mathbb{P}_b$ and 
with $n = 3$ particles in the exit channel --- this reads:
\begin{equation} 
\label{phasespacegeneral}
\sigma = \frac{1}{\mbox{F}} \int dV_{ps} |M_{ab\,\rightarrow\,123}|^2 = 
  \frac{1}{\mbox{F}} \int \prod_{i=1}^n d^{4}\mathbb{P}_{i} 
    \cdot \delta(\mathbb{P}_{i}^2 = \mbox{m}_{i}^2)
    \cdot \Theta(\mathbb{P}_{i}^2) 
    \cdot \delta^{4}(\mathbb{P}_a + \mathbb{P}_b - \sum_{j=1}^{n}\mathbb{P}_{j})
    \cdot |M_{ab\,\rightarrow\,123}|^2,
\end{equation}
where $|M_{ab\,\rightarrow\,123}|^2$ denotes the square of the 
Lorentz--invariant spin averaged matrix element describing the probability to 
create two nucleons and a meson with four--momenta of 
$\mathbb{P}_{i} = (\mbox{E}_i,\vec{\mbox{p}}_i)$ and $i = 1..n$, respectively.
The energy and momentum conservation as well as the on--shellness of the 
created particles is ensured by the Dirac--$\delta$ and the 
Heaviside--$\Theta$ functions.
The formula holds also for $n \neq 3$.\vspace{1ex}

The total cross section is then defined as an integral of the probabilities to 
populate a given phase--space interval over the whole kinematically available 
range --- determined by energy and momentum conservation --- normalized to the 
flux factor F of the colliding nucleons. 
In the case of non--interacting final state particles the matrix element close 
to threshold $|M_{ab\,\rightarrow\,123}|^2$, is nearly 
constant~\cite{moalem445,bernard259,gedalin471} and hence the allowed 
phase--space volume, $V_{ps}$, is the decisive quantity which governs the 
growth of the cross section with increasing excess energy Q.
The latter --- defined as the total kinetic energy --- is shared among the 
outgoing particles in the reaction centre--of--mass 
frame~\footnote{Traditionally by centre--of--mass system we understand a frame 
in which the momenta of all particles add to zero, called sometimes more 
explicitly centre--of--momentum frame.}:
\begin{equation} 
\mbox{Q} = \sqrt{\mbox{s}} - \sum_{i=1}^{n} \mbox{m}_{i}, 
\end{equation}
where $\mbox{s} = |\mathbb{P}_a + \mathbb{P}_b|^2 = 
|\sum_{i=1}^{n} \mathbb{P}_{i}|^2$ denotes the square of total 
centre--of--mass energy.  
Exactly at threshold, where the particles' kinetic energy in centre--of--mass 
system is equal to zero ($\mbox{Q} = 0\,\mbox{MeV}$), the total reaction 
energy $\sqrt{\mbox{s}}$ amounts to: 
$\sqrt{\mbox{s}_{th}} = \sum_{i=1}^{n} \mbox{m}_{i}$.
Before writing explicitly the formula for $V_{ps}$ let us
introduce the kinematical triangle function $\lambda$
defined as~\cite{bycklingkajantie}:
\begin{equation} 
\label{kaellen}
\lambda(x, y, z) = x^2 + y^2 + z^2 - 2xy - 2yz - 2zx,
\end{equation}
which enables the expression of many useful kinematical variables in a very 
compact and Lorentz--invariant form.
In particular, the momenta of particle $i$ and $j$ in their rest system equal:
\begin{equation} 
\label{momentum_kaellen}
\mbox{p}_{i}^{*} = \mbox{p}_{j}^{*} = 
 \frac{\sqrt{\lambda(\mbox{s}_{ij},\mbox{m}_{i}^2,\mbox{m}_{j}^{2})}}
      {2\sqrt{\mbox{s}_{ij}}},
\end{equation}
where $\mbox{s}_{ij} = |\mathbb{P}_i + \mathbb{P}_j|^2$ stands for the square 
of the invariant mass of the $ij$ system considered as one quasi--particle. 
The above relation gives an expression for the flux factor F from 
equation~\eqref{phasespacegeneral} in terms of the colliding masses of the 
nucleons and the total energy s only, namely:
\begin{equation} 
\label{fluxfactor}
\mbox{F} = 4\,\sqrt{\mbox{s}} \,(2\pi)^{3n-4}\;\mbox{p}_{a}^{*} = 
  2\,(2\pi)^{3n-4}\;\sqrt{\lambda(\mbox{s},\mbox{m}_a^2,\mbox{m}_b^2)},
\end{equation}
where we have chosen the convention introduced by Byckling and 
Kajantie~\cite{bycklingkajantie} and included the $(2\pi)$ factors for the 
phase--space $(2\pi)^{3n}$ and for the matrix element $(2\pi)^{-4}$ into the 
definition of F. 
It is important to note that at threshold, for an excess energy range of a few 
tens of MeV, the small fractional changes of total energy 
$(\frac{\mbox{\scriptsize Q}}{\sqrt{\mbox{\scriptsize s}}} = 
\frac{\mbox{\scriptsize Q}}{\mbox{\scriptsize m}_1\,+\,
\mbox{\scriptsize m}_2\,+\,\mbox{\scriptsize m}_3\,+\,\mbox{\scriptsize Q}})$ 
causes weak variations of the flux factor and influences only slightly the 
shape of the energy dependence of the total cross section.\vspace{1ex}

For a two--particle final state $ab \rightarrow 12$ (for instance for the 
reaction $pn \rightarrow d \eta$) the phase--space integral defined in 
equation~\eqref{phasespacegeneral} reduces to $V_{ps} := \int dV_{ps} = 
\frac{\pi}{\sqrt{\mbox{\scriptsize s}}}\,\mbox{p}_1^* = 
\frac{\pi}{2\,\mbox{\scriptsize s}}
\sqrt{\lambda(\mbox{s},\mbox{m}_1^2,\mbox{m}_2^2)}$ and the total cross 
section for such reactions (when neglecting variations due to the dynamical 
effects $(|M_{ab\,\rightarrow\,12}| = const.)$) should increase linearly with 
the centre--of--mass momentum of the produced meson in the vicinity of the 
threshold. 
The total cross section can be expressed analytically as a function of the 
masses of the particles participating in the reaction and the square of total 
reaction energy s:
\begin{equation} 
\label{Vps_two_body}
\sigma_{ab\,\rightarrow\,12} = const \cdot \frac{V_{ps}}{\mbox{F}} = 
 \frac{const}{16 \pi\,\mbox{s}} \frac{\mbox{p}_1^*}{\mbox{p}_a^*} = 
 \frac{1}{16 \pi\,\mbox{s}} 
 \frac{\sqrt{\lambda(\mbox{s},\mbox{m}_1^2,\mbox{m}_2^2)}}
      {\sqrt{\lambda(\mbox{s},\mbox{m}_a^2,\mbox{m}_b^2)}}.
\end{equation}
Near threshold, at a given excess energy $\mbox{Q} = \sqrt{\mbox{s}} - 
\mbox{m}_1 - \mbox{m}_2$, the emission of the reaction products in the 
centre--of--mass frame is isotropic and the whole dynamics of the process 
manifests itself in the absolute value of the transition matrix element 
$|M_{ab\,\rightarrow\,12}|$.
The underlying production mechanisms can also be extracted from the deviations 
of the total cross section energy dependence following the prediction of 
relation~\eqref{Vps_two_body}. 
A visualization of possible differences in the dynamics for the production of 
various mesons can be realized by comparing the total cross section of the 
studied reactions at the same value of the phase--space volume normalized to 
the flux factor ($V_{ps}/\mbox{F}$). 
It is important to recognize that the comparison at the same centre--of--mass 
meson momentum ($\mbox{p}_2^*$) gives direct information about the amplitude 
differences only if the production of mesons with the same masses is concerned 
(for example $pp \rightarrow d \pi^+$ and $nn \rightarrow d \pi^-$), yet the 
reactions $pn \rightarrow d \eta$ and $pn \rightarrow d \pi^0$ have by far 
different $V_{ps}$ at the corresponding $\mbox{p}_2^*$.

However, if one is interested in the decomposition of the production amplitude 
according to the angular momenta of the final state particles then indeed the 
appropriate variable for the comparison even for particles with different 
masses is the meson momentum in the reaction centre--of--mass frame.
Correspondingly for the more than two--body final state, the adequate variable 
is the maximum meson momentum, since it is directly connected to the maximum 
angular momentum by the interaction range. 
Considering for a three--body exit channel ($ab \rightarrow 123$) that the 
meson ($\mbox{m}_3$) possesses maximum momentum when the remaining two 
particles are at rest relative to each other and employing 
definition~\eqref{momentum_kaellen} one obtains:

\begin{equation} 
\mbox{q}_{max} = 
  \frac{\sqrt{\lambda(\mbox{s},(\mbox{m}_1 + \mbox{m}_2)^2,\mbox{m}_3^2)}}
       {2\sqrt{\mbox{s}}}.
\end{equation}

Contrary to the two--body final state, at a fixed excess energy the dynamics 
of the meson production associated with two or more particles reflects itself 
not only in the absolute value of the square of the matrix element but also in 
distributions of variables determining the final state kinematics.
Usually, in non--relativistic calculations of the total cross section, one 
takes the Jacobi momenta, choosing as independent variables the 'q'--meson 
momentum in the reaction centre--of--mass frame and 'k'--momentum of either 
nucleon in the rest frame of the nucleon--nucleon subsystem. 
By means of the $\lambda$ function they can be expressed as:
\begin{equation} 
\mbox{q} = 
 \frac{\sqrt{\lambda(\mbox{s},\mbox{s}_{12},\mbox{m}_3^2)}}{2\,\sqrt{\mbox{s}}},
 \;\;\;\mbox{and}\;\;\;
\mbox{k} = 
 \frac{\sqrt{\lambda(\mbox{s}_{12},\mbox{m}_1^2,\mbox{m}_2^2)}}
      {2\,\sqrt{\mbox{s}_{12}}},
\end{equation}
with $\mbox{s}_{12} = |\mathbb{P}_1+\mathbb{P}_2|^2$ denoting the square of 
the invariant mass of the nucleon--nucleon subsystem.
In a non--relativistic approximation the expression of the total cross section 
defined by formula~\eqref{phasespacegeneral} for a meson production reaction 
in nucleon--nucleon interactions of the type $NN \rightarrow NN\,Meson$ 
simplifies to:
\begin{equation} 
\label{eq:sigma}
\sigma \propto \int_0^{\mbox{\scriptsize q}_{max}} 
  \mbox{k}\,\mbox{q}^2 \,|M_{ab\,\rightarrow\,123}|^2\,d\mbox{q}. 
\end{equation}
Denoting by $L$ and $l$ the relative angular momentum of the nucleon--nucleon 
pair and of the meson relative to the $NN$ system, respectively and 
approximating the final state particles by non--distorted plane waves (case of 
non--interacting objects), whose radial parts $\psi_l(\mbox{q},\mbox{r})$ are 
given by the spherical Bessel functions:
\begin{equation} 
\psi_l(\mbox{q},\mbox{r}) \propto j_l(\mbox{qr}) 
  \stackrel{\mbox{\scriptsize qr }\rightarrow 0}{\longrightarrow} 
  \frac{(\mbox{qr})^l}{(2l+1)!},
\end{equation} 
an expansion of the transition amplitude $|M_{Ll}|^2$ for the $Ll$ partial 
wave around $\mbox{qr} = 0$ leads to
\begin{equation} 
|M_{Ll}|^2 \propto \mbox{q}^{2l}\,\mbox{k}^{2L}.
\end{equation}
Applying the above proportionality in equation~\eqref{eq:sigma} and solving 
the integral, yields the partial cross section
\begin{equation} 
\label{sigmaLl}
\sigma_{Ll} \propto \mbox{q}_{\,max}^{\,2L+2l+4} \propto \eta_{\,M}^{\,2L+2l+4},
\end{equation}
where $\eta_{M} = \mbox{q}_{max}/\mbox{m}_M$ with $\mbox{m}_M$ denoting the 
mass of the created meson~\footnote{In former works~\cite{Rosenfeld,GMW} 
dealing only with pions this parameter is denoted by $\eta$, here in order to 
avoid ambiguities with the abbreviation for the eta-meson, we introduce an 
additional suffix $M$.}.
Thus --- at threshold --- for the Ss partial wave the cross section for the 
$NN \rightarrow NN \,Meson$ reaction should increase with energy of the fourth 
power of $\eta_{M}$.
The dimensionless parameter $\eta_{M}$ was introduced by 
Rosenfeld~\cite{Rosenfeld} as a variable allowing for the qualitative 
estimations of the final state partial waves involved in pion production. 
He argued that if the phenomenon of pion production takes place at a 
characteristic distance R from the centre of the collision, with R in the 
order of $\hbar/\mbox{Mc}$, then the angular momentum of the produced meson is 
equal to $l =\mbox{R}\,\mbox{q} = \hbar\,\mbox{q}/\mbox{Mc}$. 
Hence, $\eta_M$  denotes the classically calculated angular maximum momentum of the 
meson relative to the centre--of--reaction. 
The same arguments one finds in the work of Gell--Mann and Watson~\cite{GMW}, 
where the authors do not expect the range of interaction to be greater than 
the Compton wave-length of the produced meson $\left(\hbar/\mbox{Mc}\right)$. 
However, it is rather a momentum transfer $\Delta\mbox{p}$ between the 
colliding nucleons which determines the distance to the centre of the collision 
$\mbox{R} \approx \hbar/\Delta\mbox{p}$ at which the production occurs. 
Based on the data indications, it is emphasized in the original article of 
Rosenfeld~\cite{Rosenfeld} that R is slightly less than $\hbar/(2\,\mbox{Mc})$, 
which is numerically close to the value of $\hbar/\Delta\mbox{p}$. 
Directly at threshold, where all ejectiles are at rest in the centre--of--mass 
frame, $\Delta\mbox{p}$ is equal to the centre--of--mass momentum of the 
interacting nucleons and hence, exploring equation~\eqref{momentum_kaellen} it 
can be expressed as:
\begin{equation} 
\label{momtranseq}
\Delta\mbox{p}_{th} = 
  \frac{\sqrt{\lambda(\mbox{s}_{th},\mbox{m}_{a}^2,\mbox{m}_{b}^{2})}}
       {2\,\sqrt{\mbox{s}_{th}}} 
  \stackrel{\mbox{\scriptsize for}\,pp \to ppX}{=\!=\!=\!=\!=\!=\!=\!=} 
  \sqrt{\mbox{m}_p \mbox{m}_X + \frac{\mbox{m}_X^2}{4}}.
\end{equation}
Though the present considerations are constrained to the spin averaged 
production only, it is worth noting that very close--to--threshold --- due to 
the conservation laws and the Pauli excluding principle~\footnote{The Pauli 
principle for the nucleon--nucleon system implies that 
$(-1)^{\mbox{\scriptsize L}+\mbox{\scriptsize S}+\mbox{\scriptsize T}} = -1$, 
where L, S and T denote angular momentum, spin and isospin of the nucleon 
pair, respectively. For example if the nucleon--nucleon wave function is 
symmetric in space ($\mbox{L} = 0$) and spin ($\mbox{S} = 1$) then it must be 
antisymmetric in isospin--space ($\mbox{T} = 0$) to be totally antisymmetric.} 
--- for many reactions there is only one possible angular momentum and spin 
orientation for the incoming and outgoing particles.

\begin{table}[H]
\caption{\label{partialtransitions}Partial wave transitions for the $pp 
\rightarrow pp\,Meson$ and $nn \rightarrow nn\,Meson$ reactions at threshold}
\vskip -0.25cm
\tabskip=1em plus2em minus.5em
\halign to \hsize{\hfil#&\hfil#\hfil&\hfill#\hfill&\hfill#\hfill \cr
\noalign{\hrulefill}
type & meson  & spin and parity & transition  \cr
\noalign{\vskip -0.2cm}
\noalign{\hrulefill}
\noalign{\vskip -0.5cm}
\noalign{\hrulefill}
pseudoscalar & $\pi,\eta,\eta^{\prime}$ & $0^-$ & 
 $^3\mbox{P}_0 \,\rightarrow\, ^1\mbox{S}_0\,\mbox{s}$   \cr
vector       & $\rho,\omega,\phi $      & $1^-$ & 
 $^3\mbox{P}_1 \,\rightarrow\, ^1\mbox{S}_0\,\mbox{s}$   \cr
scalar       & $a_0, f_0$               & $0^+$ & 
 $^1\mbox{S}_0 \,\rightarrow\, ^1\mbox{S}_0\,\mbox{s}$   \cr
\noalign{\hrulefill} }
\end{table}

For example the production of neutral mesons with negative parity --- as 
pseudoscalar or vector mesons --- may proceed in the proton--proton collision 
near threshold only via the transition between $^3\mbox{P}$ and 
$^1\mbox{S}_0\mbox{s}$ partial waves~(see table~\ref{partialtransitions}). 
Therefore only the collision of protons with relative angular momentum equal 
to $1\,\hbar$ and a parallel spin orientation --- which flips during the 
reaction --- may lead to the production of negative parity neutral mesons. 
In the case of the production of neutral scalar mesons, the protons or 
neutrons must collide with anti--parallel spin orientations which remains 
unchanged after the reaction.
These simple considerations imply that in the close to threshold measurements
with the polarized beam and target one should see a drastic effect in the 
reaction yield depending whether the polarization of reacting protons is 
parallel or anti-parallel.
In the ideal case, having the product of beam and target polarization equal to 
unity, the close to threshold $\pi^0$, $\eta$ or $\eta^{\prime}$ production
can be realized only in case of the parallel spin orientation
of the reacting protons and for the anti-parallel polarizations
zero yield should be observed. Indeed, in reality, strong differences in the 
yield for various combinations of beam and target polarization have been 
determined in the pioneering measurements of the reaction
$\vec{p}\vec{p} \to pp\pi^{0}$ at the IUCF facility~\cite{HOM}.

Due to both the strong nucleon--nucleon low--energy interaction in the 
$^1\mbox{S}_0$ state and the meson--nucleon forces in the exit channel, the 
assumption of the non--interactive plane waves leading to 
equation~\eqref{sigmaLl} failed when confronted with experimental data. 
However, these deviations give the possibility for determining the still 
poorly known nucleon--meson interactions.
\vspace{-0.1cm}
\begin{figure}[H]
\hspace{1cm}
\parbox{0.49\textwidth}{\epsfig{file=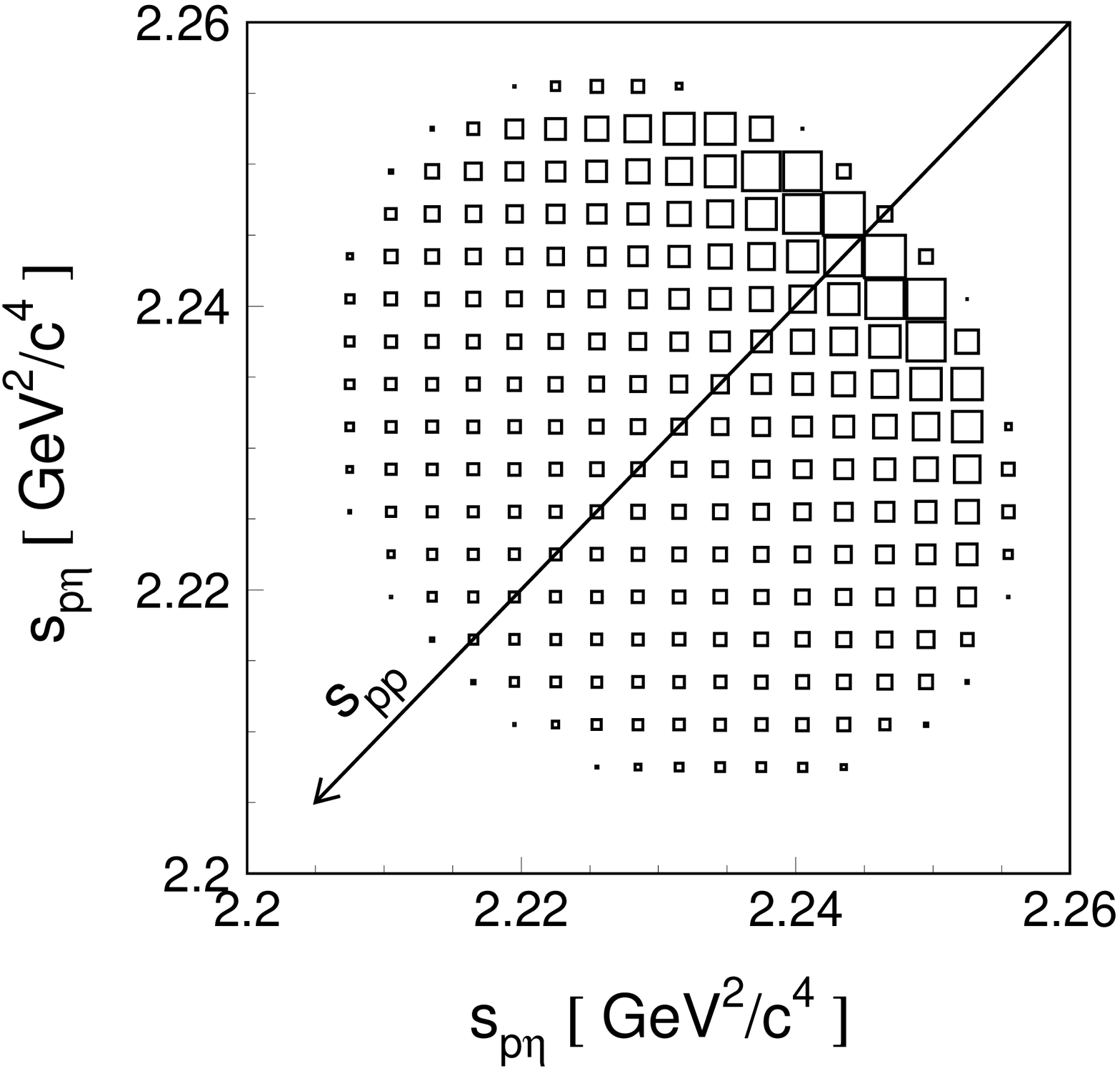,width=0.36\textwidth}}
\hfill
\parbox{0.49\textwidth}{\epsfig{file=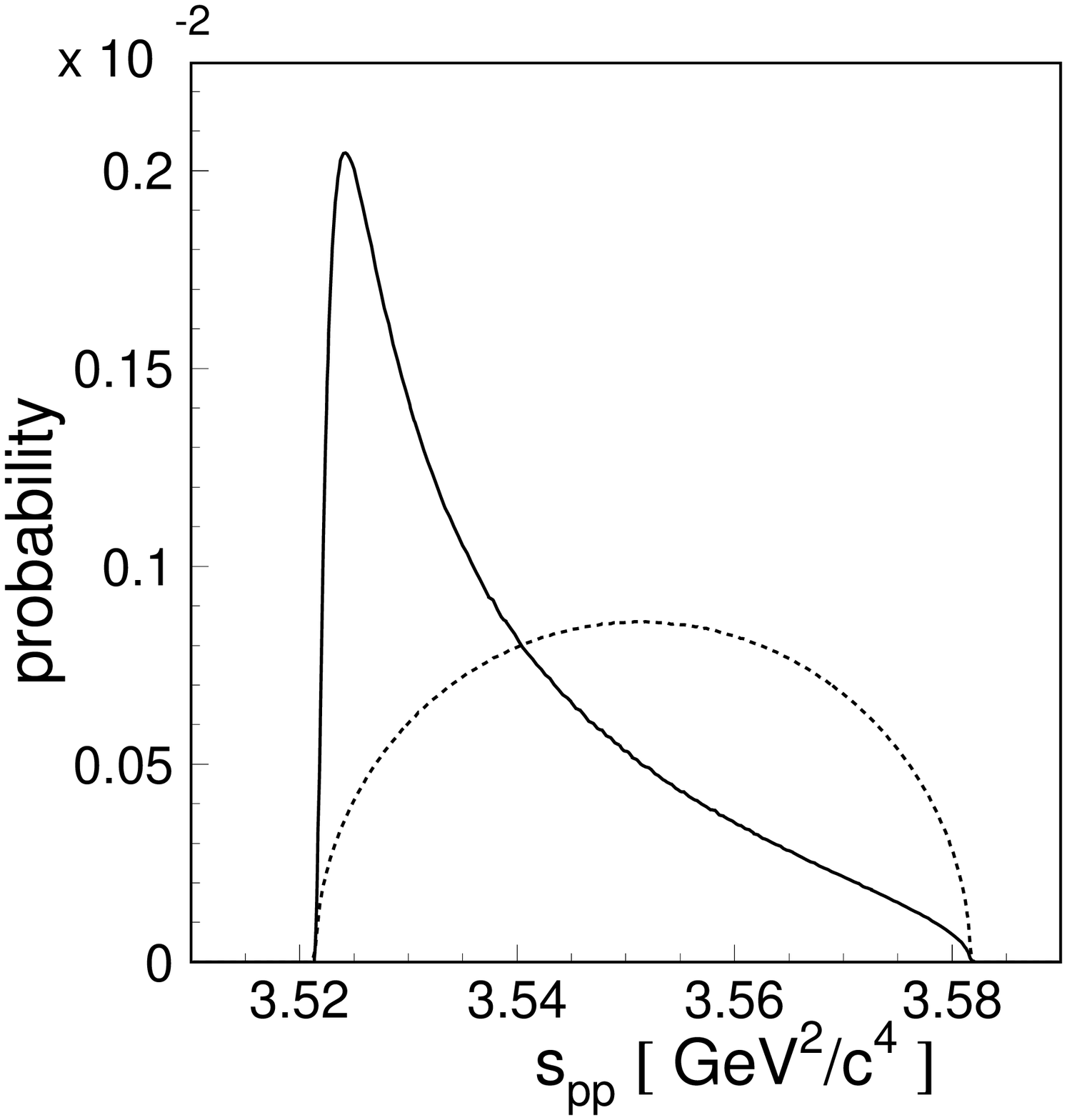,width=0.36\textwidth}}

\parbox{0.39\textwidth}{\raisebox{1ex}[0ex][0ex]{\mbox{}}}\hfill
\parbox{0.50\textwidth}{\raisebox{1ex}[0ex][0ex]{\large a)}}\hfill
\parbox{0.09\textwidth}{\raisebox{1ex}[0ex][0ex]{\large b)}}

\hspace{1cm}
\parbox{0.49\textwidth}{\epsfig{file=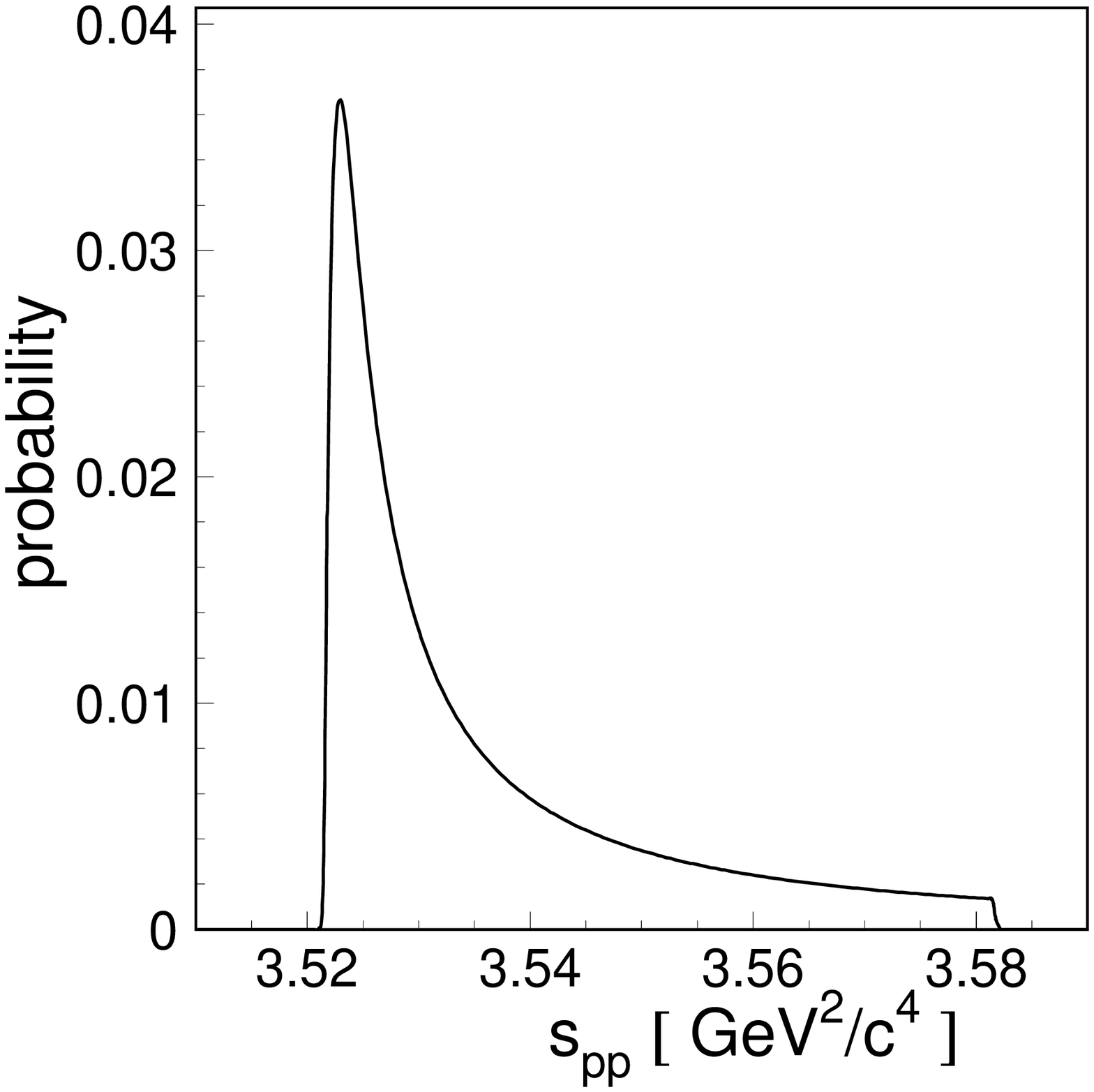,width=0.36\textwidth}}
\hfill
\parbox{0.49\textwidth}{\epsfig{file=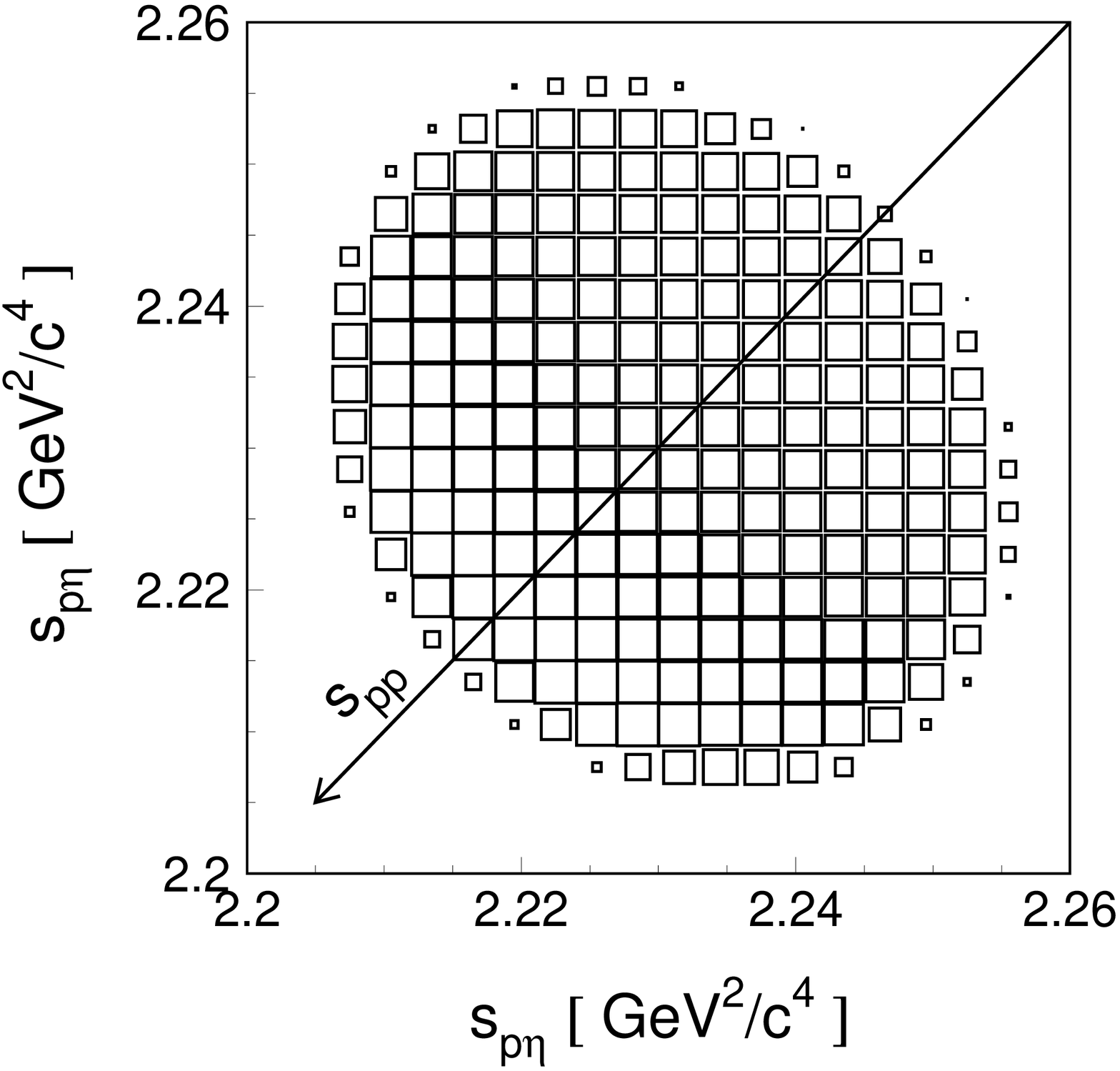,width=0.36\textwidth}}

\parbox{0.39\textwidth}{\raisebox{1ex}[0ex][0ex]{\mbox{}}}\hfill
\parbox{0.50\textwidth}{\raisebox{1ex}[0ex][0ex]{\large c)}}\hfill
\parbox{0.09\textwidth}{\raisebox{1ex}[0ex][0ex]{\large d)}}
\caption{\label{dalitz_examples} 
    Monte--Carlo Simulations: 
(a) Phase--space distribution for the $pp \rightarrow pp \eta$ reaction at 
    $\mbox{Q} = 16\,\mbox{MeV}$ modified by the proton--proton final state 
    interaction. \
(b) The dotted line shows the projection of the pure phase--space density 
    distribution onto the $\mbox{s}_{pp}$ axis and the solid curve presents 
    its modification by the proton--proton FSI. \
(c) The square of the scattering amplitude for the $pp \rightarrow pp$ elastic 
    scattering as a function of the proton--proton invariant mass in the range 
    $2\,\mbox{m}_p < \sqrt{\mbox{s}_{pp}} < 2\,\mbox{m}_p + \mbox{Q}$, 
    calculated according to formula~\eqref{Mpppp} of section~\ref{Mwhs}. \
(d) Phase--space density distribution modified by the proton--$\eta$ 
    interaction, with scattering length equal to $\mbox{a}_{p\eta} = 
    0.7\,\mbox{fm} + i\,0.3\,\mbox{fm}$. 
     The proton--$\eta$ scattering amplitude has been calculated according
     to the equation~\eqref{Mpeta}.  
     A detailed descussion of the nucleon--nucleon and nucleon--meson
     interaction will be presented in section~3.1.}
\end{figure}
\vspace{-0.2cm}
The interaction between particles depends on their relative momenta. 
Consequently, for investigations of final state interactions more instructive 
coordinates, than the q and k momenta, are the squared invariant masses of the 
two--body subsystems~\cite{grzonkakilian}.
These are the coordinates of the Dalitz plot. In the original 
paper~\cite{dalitz} Dalitz has proposed a representation for the energy 
partitions of three bodies in an equilateral triangle whose sides are the axes 
of the centre--of--mass energies. He took advantage of the fact that the sum of 
distances from a point within the triangle to its sides is a constant equal to 
the height. Therefore, the height of the triangle measures the total energy 
$\sqrt{\mbox{s}} = \mbox{E}_1^* + \mbox{E}_2^* + \mbox{E}_3^*$ and interior 
points --- fulfilling four--momentum conservation --- represent energy 
partitions. For a constant $\sqrt{\mbox{s}}$ due to the energy conservation, 
without loosing any information, it is enough to consider the projection on any 
of the $\mbox{E}_i\,\mbox{E}_j$ planes. The linear relation between 
$\mbox{E}_i^*$ and $\mbox{s}_{jk}$ ($\mbox{s}_{jk} = \mbox{s} + \mbox{m}_i^2 - 
2\,\sqrt{\mbox{s}} \mbox{E}_i^*$) allows to use $\mbox{s}_{jk}\,\mbox{s}_{ik}$ 
or $\mbox{E}_i\,\mbox{E}_j$ coordinates equivalently, with the following 
relation between the phase--space intervals: $d\mbox{E}_i^*\,d\mbox{E}_j^* = 
\frac{1}{4\mbox{\tiny s}}\;d\mbox{s}_{jk}\,d\mbox{s}_{ki}$. 
According to the Kilian's geometrical representation~\cite{grzonkakilian} 
a Dalitz lies on a plane in the three dimensional space ($\mbox{s}_{12},\mbox{s}_{13},
\mbox{s}_{23}$) orthogonal to the space diagonal. 
The plane including a Dalitz plot corresponding to a given total energy s is 
then given by the following scalar product~\cite{grzonkakilian}:
\begin{equation} 
\label{scalarproduct}
(1,1,1)(\mbox{s}_{12},\mbox{s}_{13},\mbox{s}_{23}) = 
 \mbox{s}_{12} + \mbox{s}_{13} + \mbox{s}_{23} = 
 \mbox{s} + \mbox{m}_1^2 + \mbox{m}_2^2 + \mbox{m}_3^2.
\end{equation}
The second equality of equation~\eqref{scalarproduct} means that there are only 
two independent invariant masses of the three subsystems and therefore a 
projection onto any of the $\mbox{s}_{ij}\,\mbox{s}_{jk}$ planes still 
comprises the whole principally accessible information about the final state 
interaction of the three--particle system.
In the case of no dynamics whatsoever and the absence of any final state 
interaction the occupation of the Dalitz plot would be fully homogeneous since 
each phase--space interval would be equally probable. 
The final state interaction would then appear as a structure on that area.
Figure~\ref{dalitz_examples}a shows --- for the example of the $pp \rightarrow 
pp \eta$ reaction --- how the uniformly populated phase--space density is 
modified by the S--wave ($^1\mbox{S}_0$) interaction between outgoing protons. 
An enhancement in the range corresponding to the low relative momentum between 
protons is clearly visible. 
A steep decrease of the occupation density with increasing invariant mass of 
the proton--proton subsystem is even better seen in 
figure~\ref{dalitz_examples}b presenting the projection of the phase--space 
distribution onto the $\mbox{s}_{pp}$ axis indicated by an arrow in 
figure~\ref{dalitz_examples}a.
This is a direct reflection of the shape of the proton--proton ($^1\mbox{S}_0$) 
partial wave amplitude shown in figure~\ref{dalitz_examples}c.
Figure~\ref{dalitz_examples}d shows the Dalitz plot distribution simulated when 
switching off the proton--proton interaction but accounting for the interaction 
between the $\eta$--meson and the proton. 
Due to the lower strength of this interaction the expected deviations from the 
uniform distributions are by about two orders of magnitude smaller, but one 
recognizes a slight enhancement of the density in the range of low invariant 
masses of proton--$\eta$ subsystems.  
However, due to weak variations of the proton--$\eta$ scattering amplitude the 
enhancement originating from the $\eta$--meson interaction with one proton is 
not separated from the $\eta$--meson interaction with the second proton. 
Therefore an overlapping of broad structures occurs. 
It is observed that the occupation density grows slowly with increasing 
$\mbox{s}_{pp}$ opposite to the effects caused by the S--wave proton--proton 
interaction, yet similar to the modifications expected for the P--wave 
one~\cite{dyringPHD}.
From the above example it is obvious that only in high statistics experiments 
signals from the meson--nucleon interaction can appear over the overwhelming 
nucleon--nucleon final state interaction. 
It is worth noting, however, that the Dalitz plot does not reflect any possible 
correlations between the entrance and exit channel~\cite{grzonkakilian}. 

The Dalitz plot representation allows also for a simple interpretation of the 
kinematically available phase--space volume as an area of that plot. 
Namely, equation~\eqref{phasespacegeneral} becomes:
\begin{equation} 
\label{Vpsdalitz}
\sigma = \frac{1}{\mbox{F}} \frac{\pi^{2}}{4\,\mbox{s}}
 \int\limits_{(\mbox{\scriptsize m}_1+\mbox{\scriptsize m}_2)^2}^{
   (\sqrt{\mbox{\scriptsize s}}-\mbox{\scriptsize m}_3)^2} d\,\mbox{s}_{12}
 \int\limits_{\mbox{\scriptsize s}_{23}^{min}(\mbox{\scriptsize s}_{12})}^{
   \mbox{\scriptsize s}_{23}^{max}(\mbox{\scriptsize s}_{12})} 
   d\,\mbox{s}_{23}\; |M_{ab\,\rightarrow\,123}|^2,
\end{equation}
where the limits of integrations defining the boundaries of the Dalitz plot can 
be expressed as~\cite{bycklingkajantie}:
\begin{eqnarray}
\mbox{s}_{23}^{max}(\mbox{s}_{12}) = 
  \mbox{m}_2^2 + \mbox{m}_3^2 - \frac{1}{2\,\mbox{s}_{12}} 
  \left\{(\mbox{s}_{12} - \mbox{s} + \mbox{m}_3^2) 
         (\mbox{s}_{12} + \mbox{m}_2^2 - \mbox{m}_1^2) - 
         \sqrt{\lambda(\mbox{s}_{12},\mbox{s},\mbox{m}_3^2) 
               \lambda(\mbox{s}_{12},\mbox{m}_2^2,\mbox{m}_1^2)}\right\},\nn \\
\mbox{s}_{23}^{min}(\mbox{s}_{12}) = 
  \mbox{m}_2^2 + \mbox{m}_3^2 - \frac{1}{2\,\mbox{s}_{12}}
  \left\{(\mbox{s}_{12} - \mbox{s} + \mbox{m}_3^2) 
         (\mbox{s}_{12} + \mbox{m}_2^2 - \mbox{m}_1^2) + 
         \sqrt{\lambda(\mbox{s}_{12},\mbox{s},\mbox{m}_3^2) 
               \lambda(\mbox{s}_{12},\mbox{m}_2^2,\mbox{m}_1^2)}\right\}.
\end{eqnarray}
Thus, the phase--space volume kinematically available for the three--body final 
state can be written by means of only one integral:
\begin{equation}
\label{Vps_relativistic}
V_{ps} = \int dV_{ps} = 
 \frac{\pi^{2}}{4\,\mbox{s}}
 \int\limits_{(\mbox{\scriptsize m}_1+\mbox{\scriptsize m}_2)^2}^{
   (\sqrt{\mbox{\scriptsize s}}-\mbox{\scriptsize m}_3)^2} d\,\mbox{s}_{12}
 \int\limits_{\mbox{\scriptsize s}_{23}^{min}(\mbox{\scriptsize s}_{12})}^{
   \mbox{\scriptsize s}_{23}^{max}(\mbox{\scriptsize s}_{12})} 
   d\,\mbox{s}_{23}\; = \;\frac{\pi^{2}}{4\,\mbox{s}}
 \int\limits_{(\mbox{\scriptsize m}_1+\mbox{\scriptsize m}_2)^2}^{
  (\sqrt{\mbox{\scriptsize s}}-\mbox{\scriptsize m}_3)^2}\;
 \frac{d\,\mbox{s}_{12}}{\mbox{s}_{12}}
 \sqrt{\lambda(\mbox{s}_{12},\mbox{s},\mbox{m}_3^2) 
       \lambda(\mbox{s}_{12},\mbox{m}_2^2,\mbox{m}_1^2)}, 
\end{equation}
whose solution leads, in general, to elliptic functions~\cite{bycklingkajantie}.
However, in the nonrelativistic approximation it has the following closed form:
\begin{equation} 
\label{Vps_nonrelativistic}
V_{ps} = \frac{\pi^{3}}{2} 
  \frac{\sqrt{\mbox{m}_1\,\mbox{m}_2\,\mbox{m}_3}}
       {(\mbox{m}_1 + \mbox{m}_2 + \mbox{m}_3)^\frac{3}{2}} \; \mbox{Q}^2,
\end{equation}
where the substitution of the non--relativistic relation between $\eta_{m_3}$ 
and Q
\begin{equation*}
\mbox{Q} = 
  \frac{\mbox{m}_3^2 + 2\,\mbox{m}_3^2\,(\mbox{m}_1 + \mbox{m}_2)}
       {2\,(\mbox{m}_1 + \mbox{m}_2)} \; \eta_{m_3}^2 
\end{equation*}
gives the $\mbox{S}\,\mbox{s}$ partial cross section of 
equation~\eqref{sigmaLl}.
On the basis of the formula~\eqref{Vps_nonrelativistic} the kinematically 
available phase--space volume ($V_{ps}$) can be as easily calculated as the 
excess energy Q. 
Close--to--threshold --- in the range of a few tens of MeV --- the 
non--relativistic approximation differs only by a few per cent from the full 
solution given in equation~\eqref{Vps_relativistic}, which in fact with an 
up--to--date computer can be solved numerically with little effort. 
Therefore, in the following chapters we will describe the data as a function of 
$V_{ps}$ as well as of Q or $\eta_{M}$, if it is found to be appropriate.

The choice of the proper observables for picking up the dynamical effects in 
the case of a four--particle exit channel is by far more complicated.
It was for example discussed by Chodrow~\cite{chodrow} who has given the 
extension of the Dalitz representation to the four--body final states.
He has proven that the two--body correlation in the case where there are two 
identical particles in the four--body final state --- say $ij$ --- can be 
directly visible in the density distribution of events on the 
$\mbox{F}_j\,\mbox{E}_i$--plane, where $\mbox{E}_i$ denotes the energy of 
particle $i$ and $\mbox{F}_j$ is defined as:

\begin{equation} 
\mbox{F}_{j} = \frac{1}{2} 
  \left\{\mbox{E}_j\,\sqrt{(\mbox{E}_j^2 - \mbox{m}_j^2)} - \mbox{m}_j^2 \, 
         \cosh^{-1}\left(\frac{\mbox{E}_j}{\mbox{m}_j}\right)\right\}.
\end{equation} 
 
\subsection{Scattering inside the deuteron}
\label{Sitd}
\begin{figure}[H]
\begin{center}
\epsfig{file=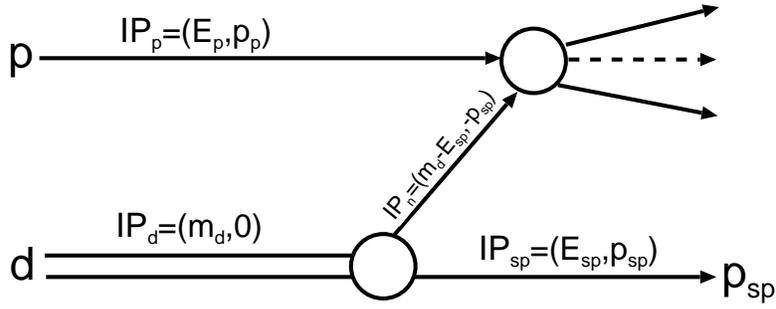,scale=0.9}
\end{center}
\caption{\label{qfree} Spectator model for a particle production reaction via 
$p d \rightarrow p_{sp} X$.}
\end{figure}

Extending meson production experiments from the free nucleon--nucleon 
interaction to interactions of light nuclei is interesting for many reasons.
First of all it offers the possibility to study more complex production 
processes with more than only one target nucleon involved. 
Furthermore, in the absence of free neutron targets, it gives the possibility 
to study the meson creation in the proton--neutron interaction at proton--beam 
facilities, complementary to a neutron--induced $\pi$ meson production with use 
of high--resolution (energy spread less than $1\,\mbox{MeV}$) secondary neutron 
beams~\cite{hutcheon176,hutcheon618,bachman495} investigated at 
TRIUMF~\cite{helmer588}. 
The technique based on quasi--free proton--nuclear reactions, with a beam of 
protons reacting with a nucleus in the target is utilized at the CELSIUS and 
COSY accelerators~\cite{bilger64,calen2069,calen2667,calen2642,moskal_c11proc}.
First experiments of $\pi^{0}$ meson production in the proton--neutron reaction 
with simultaneous tagging of the spectator proton have recently been carried 
out at CELSIUS and resulted in a resolution of the excess energy equal to 
$\sigma = 1.8\,\mbox{MeV}$~\cite{bilger64}.
In the quasi-free approximation (fig.~\ref{qfree}) it is assumed that the 
bombarding proton interacts exclusively with one nucleon in the target nucleus 
and that the other nucleons affect the reaction by providing a momentum 
distribution to the struck constituent only.
This assumption is justified if the kinetic energy of a projectile is large 
compared to the binding energy of the hit nucleus.
In fact, as noticed by Slobodrian~\cite{slobodrian175}, also the scattering of 
protons on a hydrogen target, where the protons are bound by molecular forces, 
may serve as an extreme example of the quasi--free reaction. 
In that case, although the hydrogen atoms rotate or vibrate in the molecule, 
their velocities and binding forces are totally negligible with respect to the 
velocity and nuclear forces operating the scattering of the relativistic 
proton~\cite{slobodrian175}.

\begin{figure}[H]
\parbox{0.49\textwidth}
  {\epsfig{file=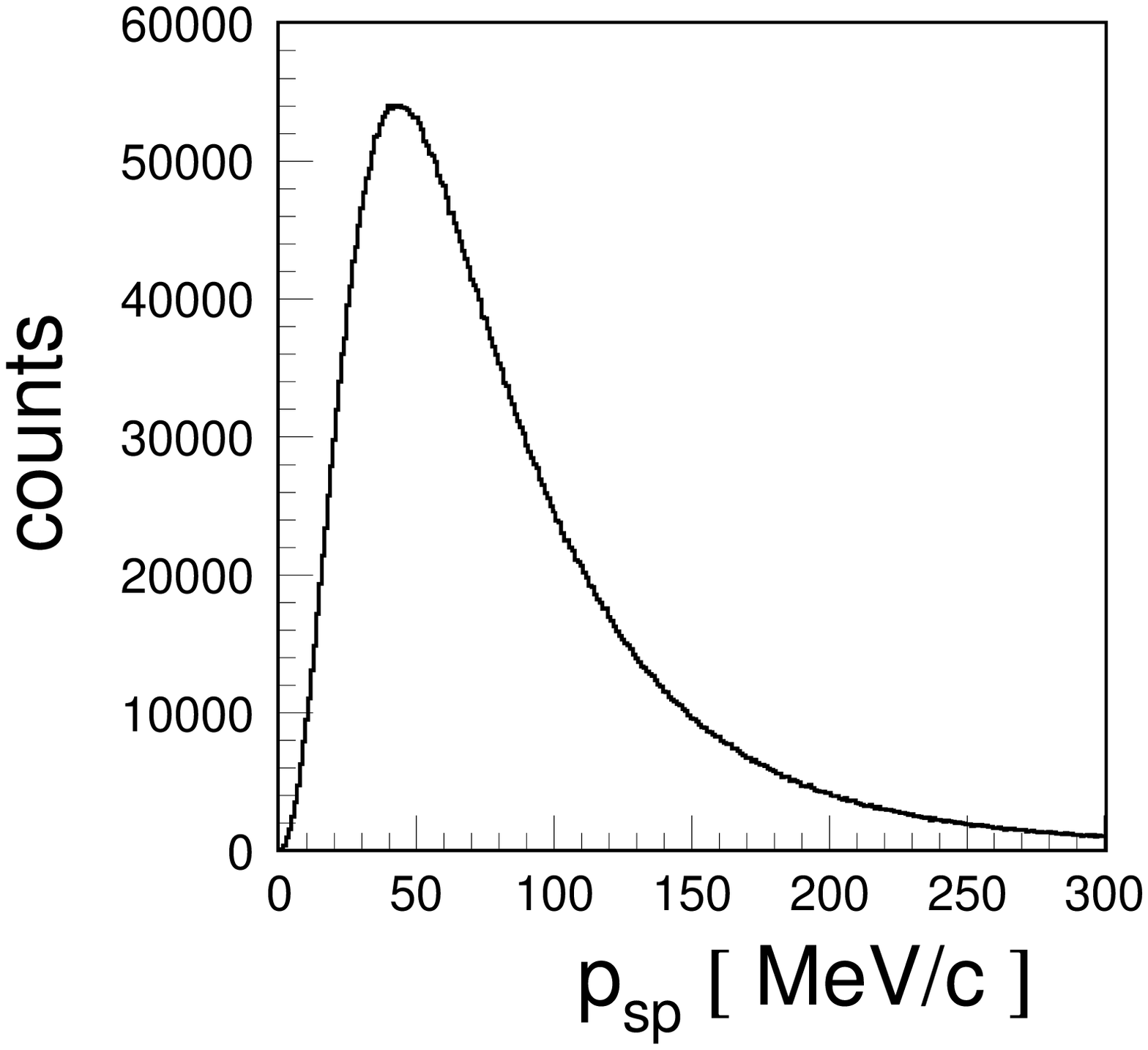,width=0.49\textwidth}}\hfill
\parbox{0.49\textwidth}
  {\epsfig{file=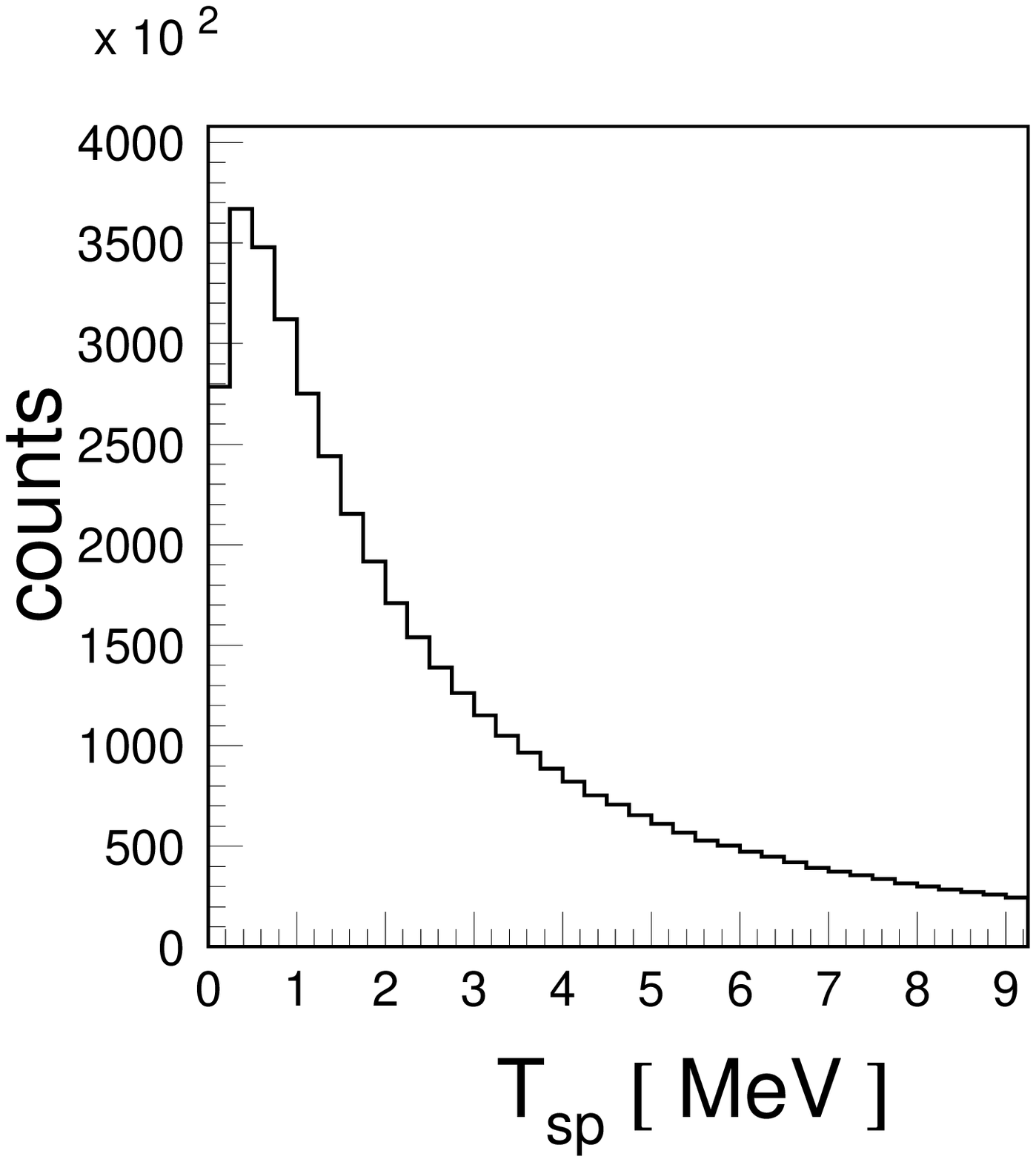,width=0.49\textwidth}}

\parbox{.44\textwidth}{\raisebox{4ex}[0ex][0ex]{\mbox{}}} \hfill
\parbox{.50\textwidth}{\raisebox{4ex}[0ex][0ex]{\large a)}} \hfill
\parbox{.04\textwidth}{\raisebox{4ex}[0ex][0ex]{\large b)}}

\caption{\label{fermi_mom_and_kin} (a) Momentum and (b) kinetic energy 
distribution of the nucleons in the deuteron, generated according to an 
analytical parametrization of the deuteron wave 
function~\cite{lacombe139,stina} calculated from the Paris 
potential~\cite{lacombe861}.}
\end{figure}

The deuteron is also relatively weakly bound with a binding energy of 
$\mbox{E}_B \approx 2.2\,\mbox{MeV}$, which is by far smaller --- more than two 
orders of magnitude in the case of pion and already more than three orders of 
magnitude in the case of $\phi$ meson production --- than the kinetic energy of 
the bombarding protons needed for the creation of mesons in the proton--neutron 
interaction.
However, it has to be considered that even the low binding energy of 
$\mbox{E}_B \approx 2.2\,\mbox{MeV}$ results in large Fermi momenta of the 
nucleons which can not be neglected. 
The momentum and kinetic energy spectra of the nucleons in the deuteron are 
shown in figure~\ref{fermi_mom_and_kin}.

\begin{figure}[H]
\hspace{0.3cm}
\parbox{0.47\textwidth}{\epsfig{file=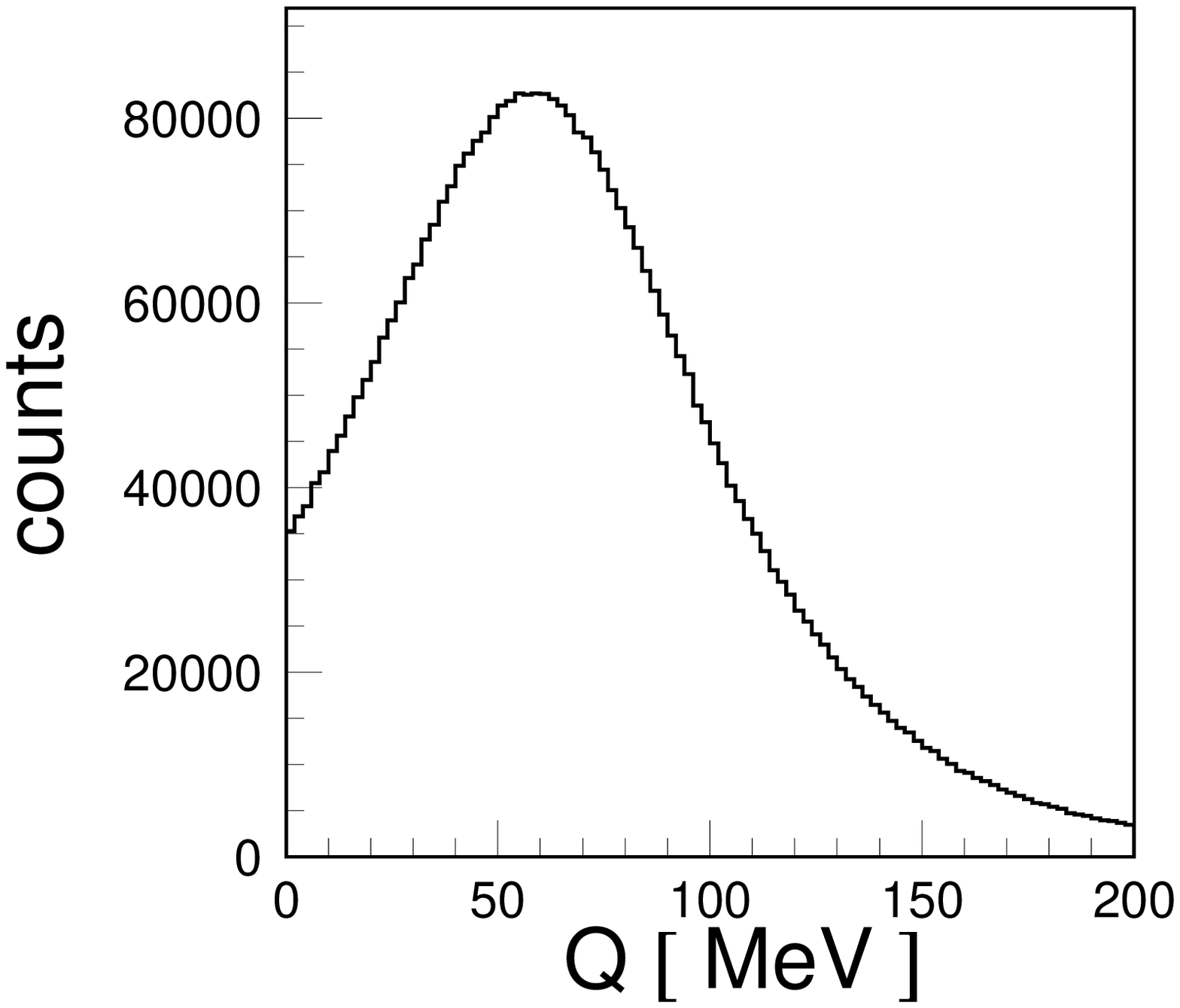,width=0.53\textwidth}}
\hfill
\parbox{0.53\textwidth}{\epsfig{file=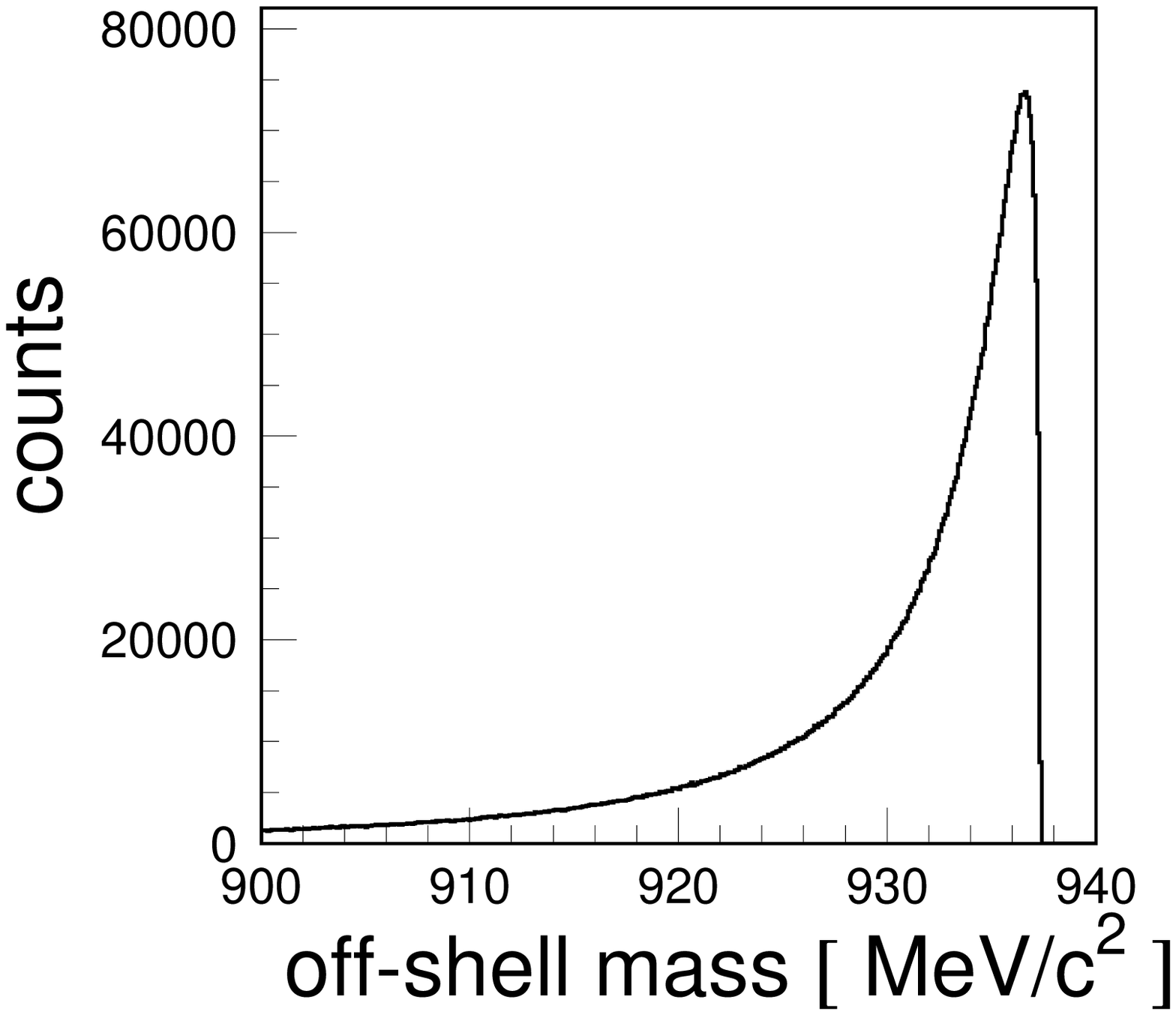,width=0.53\textwidth}}

\parbox{.44\textwidth}{\raisebox{5ex}[0ex][0ex]{\mbox{}}}\hfill
\parbox{.50\textwidth}{\raisebox{5ex}[0ex][0ex]{\large a)}}\hfill
\parbox{.04\textwidth}{\raisebox{5ex}[0ex][0ex]{\large b)}}\hfill
\vspace{-1.0cm}
\caption{\label{Q_and_mass_off} (a) Distribution of the excess energy Q for the 
$pn \eta^{\prime}$ system originating from the reaction $pd \rightarrow 
p_{sp} pn \eta^{\prime}$ calculated with a proton beam momentum of 
$3.350\,\mbox{GeV/c}$ and the neutron momentum smeared out according to the 
Fermi distribution shown in figure~\ref{fermi_mom_and_kin}a. 
(b) Spectrum of the off--shell mass of the interacting neutron, as calculated 
under the assumption of the impulse approximation.}
\end{figure}

In the considered approximation the Fermi motion of the nucleons is the only 
influence of the internucleon forces on the proton--neutron reaction and hence 
the struck neutron is treated as a free particle 
\begin{equation} 
\begin{array}{c}
  p \; \\
  \\
\end{array}
\left(\begin{array}{c}
  n \\
  p \\
\end{array}\right) 
\begin{array}{c}
  \longrightarrow \\
  \\
\end{array}
\begin{array}{c}
  p \; n \; Meson \\
  p_{sp} \\
\end{array}
\end{equation}
in the sense that the matrix element for quasi--free meson production off a 
bound neutron is identical to that for the free $pn \rightarrow p n\,Meson$ 
reaction. 
The proton from the deuteron is considered as a spectator which does not 
interact with the bombarding particle, but rather escapes untouched and hits 
the detectors carrying the Fermi momentum possessed at the moment of the 
collision. 
From the measurement of the momentum vector of the spectator proton 
$\vec{\mbox{p}}_{sp}$ one can infer the momentum vector of the neutron 
$\mbox{p}_n = - \vec{\mbox{p}}_{sp}$ at the time of the reaction and hence 
calculate the excess energy Q for each event, provided that the beam momentum 
is known. 
As an example, a distribution of the excess energy in a quasi--free $pn 
\rightarrow pn \eta^{\prime}$ reaction is presented in 
figure~\ref{Q_and_mass_off}a~\cite{moskal_c11proc}. 
Due to the large centre--of--mass velocity ($\beta \approx 0.75$) with respect 
to the colliding nucleons, a few MeV wide spectrum of the neutron kinetic 
energy inside a deuteron (fig.~\ref{fermi_mom_and_kin}b) is spread by more than 
a factor of 30.\\[-1.8cm]
Therefore, especially in the case of near--threshold measurements, where the 
cross section grows rapidly with increasing excess energy~(see e.g.\ 
fig.~\ref{Dieter_Wasserfall}) the total centre--of--mass energy 
$\sqrt{\mbox{s}}$ has to be determined on an event--by--event level. 
For this purpose, in experiments, the spectator protons are usually registered 
by silicon pad-- or $\mu$--strip 
detectors~\cite{bilger64,moskal_c11proc,ankespec} which determine their kinetic 
energy ($\mbox{T}_{sp}$) and polar emission angle ($\theta$).
Thus, it is useful to express s as function of these variables:
\begin{equation} 
\label{eq:sqf}
\mbox{s} = |\mathbb{P}_p + \mathbb{P}_n|^2 =
 \mbox{s}_0 - 2\,\mbox{T}_{sp}\,(\mbox{m}_d + \mbox{E}_{p} ) + 
 2\,\mbox{p}_p\,\sqrt{\mbox{T}_{sp}^2 + 2\mbox{m}_p\mbox{T}_{sp}}\,\cos(\theta) 
\end{equation}
with $\mbox{s}_0$ denoting the centre--of--mass energy, assuming a vanishing 
Fermi motion. 
Measuring both the energy and the emission angle of the spectator protons it is 
possible to study the energy dependence of a meson production cross section 
from data taken at only one fixed beam momentum. 

\begin{figure}[H]
\parbox{0.49\textwidth}
  {\epsfig{file=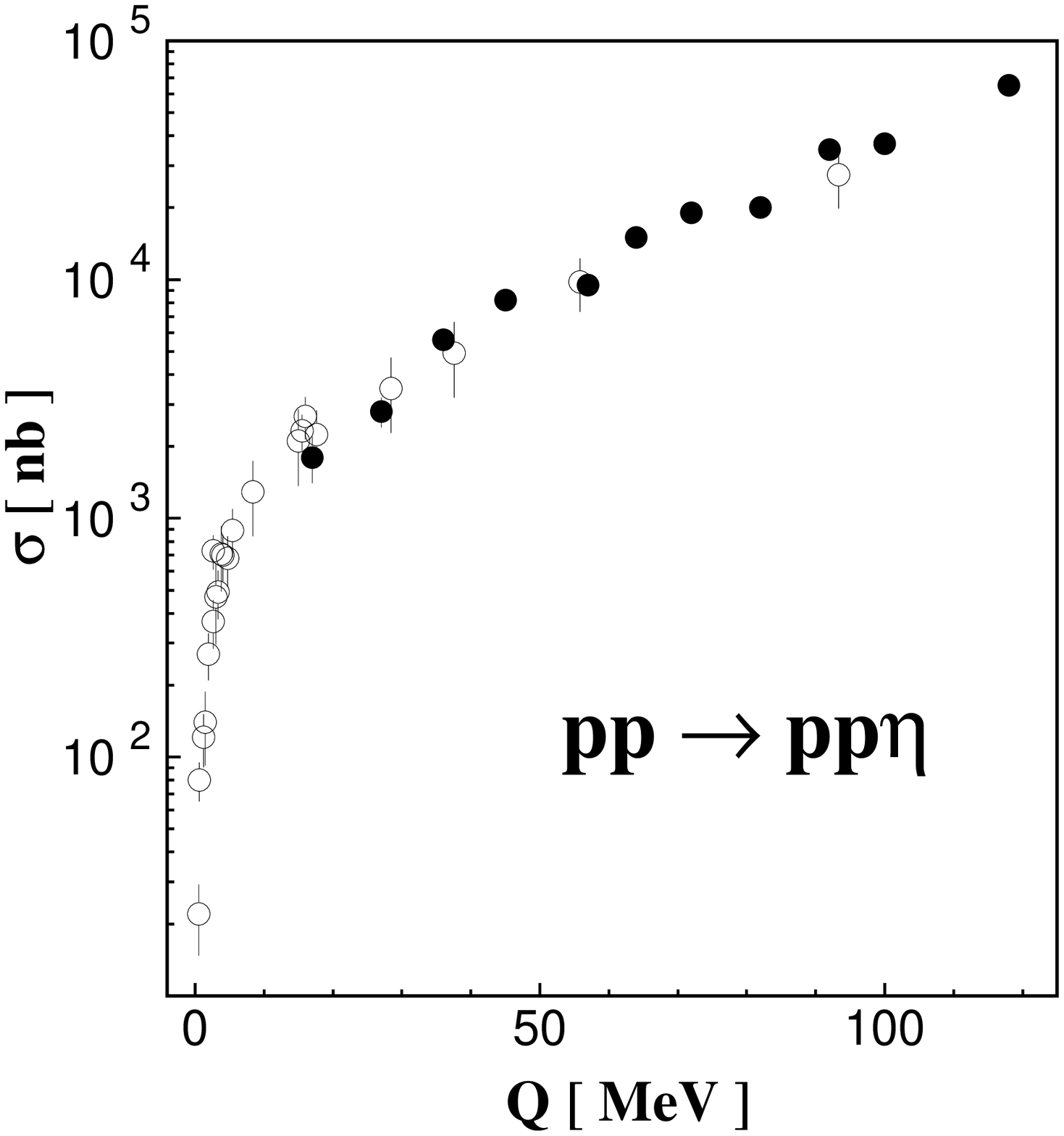,width=0.40\textwidth}}\hfill
\parbox{0.49\textwidth}
  {\epsfig{file=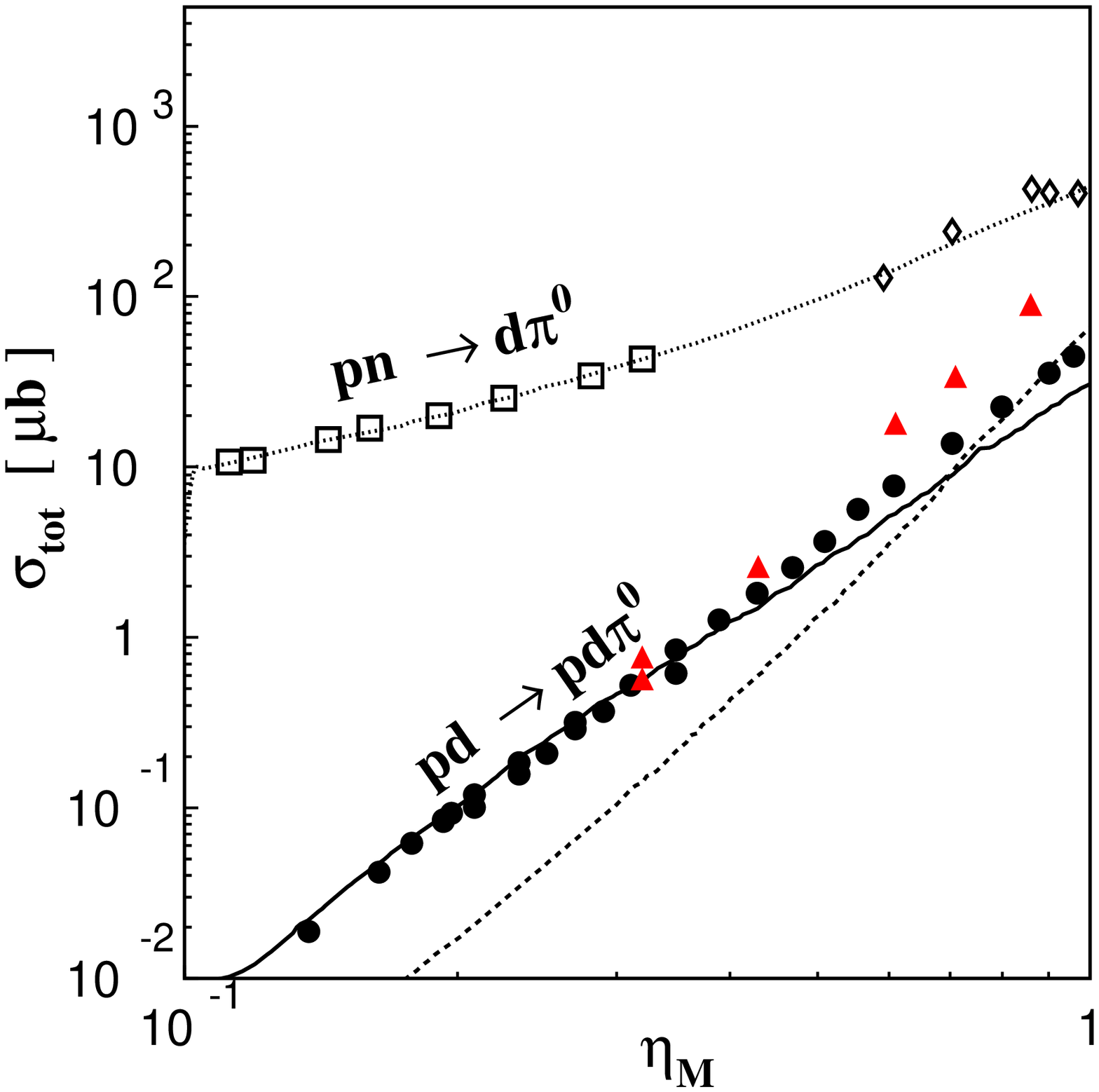,width=0.40\textwidth}}

\parbox{.44\textwidth}{\raisebox{3ex}[0ex][0ex]{\mbox{}}}\hfill
\parbox{.50\textwidth}{\raisebox{3ex}[0ex][0ex]{\large a)}}\hfill
\parbox{.04\textwidth}{\raisebox{3ex}[0ex][0ex]{\large b)}}\hfill

\caption{\label{free_vs_quasifree} (a) Total cross sections for the $pp 
\rightarrow pp \eta$ reaction as a function of excess energy measured for the 
scattering of protons in vacuum (open 
symbols)~\cite{smyrski182,hibou41,calen39,chiavassa270,bergdoltR2969,eta_menu}
and inside a deuteron (filled symbols)~\cite{calen2642}.\\
(b) Close--to--threshold total cross sections for the reactions $pn 
\rightarrow d \pi^{0}$ (squares~\cite{hutcheon618}, 
diamonds~\cite{bystricky98}) and $pd \rightarrow pd \pi^0$ 
(circles~\cite{rohdjess2864}, triangles~\cite{greiff064002}). 
The solid curve illustrates the predictions for the total 
cross section of the $pd \to pd\pi^0$ reaction assuming 
that the process is utterly due to the quasi-free 
$pn \to d\pi^0$ reaction and taking into account
the final state interaction between proton and deuteron.
Dashed lines shows the result when the final state interaction 
was neglected. In the calculations the cross section of the 
$pn \to d\pi^0$ was parametrized as presented by the dotted line. }
\end{figure}

It must be noted, however, that in the framework of the impulse approximation, 
illustrated in figure~\ref{qfree}, the measured spectator proton is a physical 
particle, yet the reacting neutron is off its mass shell, where the explicit 
expression for its four--momentum vector $\mathbb{P}_n$, in the rest frame of 
the deuteron, reads:
\begin{equation} 
\label{neutronfourvector}
\mathbb{P}_n = (\mbox{m}_d - \mbox{m}_p - \mbox{T}_{sp},\,-\vec{\mbox{p}}_{sp}),
\end{equation}
with $\mbox{T}_{sp}$ and $\vec{\mbox{p}}_{sp}$ denoting the kinetic energy and 
the momentum vector of a spectator proton, respectively.
The  mass spectrum of the interacting neutron ($\mbox{m}_n^2 = 
|\mathbb{P}_n|^2$) resulting from the distribution of Fermi momentum is shown 
in figure~\ref{Q_and_mass_off}b.
It can be seen that the maximum of this spectrum differs only by about 
$3\,\mbox{MeV}/\mbox{c}^2$ from the free neutron mass ($\mbox{m}_n = 
939.57\,\mbox{MeV}/\mbox{c}^2$), however on the average it is off by about 
$9\,\mbox{MeV}/\mbox{c}^2$. 
In the frame of the discussed approximation, the struck neutron is never on its 
mass shell and the minimum deviation from the real mass occurs for vanishing 
Fermi--momentum and --- as can be inferred from 
equation~\eqref{neutronfourvector} --- is equal to the binding energy 
$\mbox{E}_B = \mbox{m}_d - \mbox{m}_n - \mbox{m}_p$. 
Measurements performed at the CELSIUS and TRIUMF accelerators for the 
$pp \rightarrow pp \eta$~\cite{calen2642} and $pp \rightarrow 
d \pi^+$~\cite{duncan4390} reactions, respectively, have shown that within the 
statistical errors there is no difference between the total cross section of 
the free and quasi--free processes. 
In figure~\ref{free_vs_quasifree}a the production of the $\eta$ meson in free
proton--proton collisions is compared to the production inside a deuteron and 
in the overlapping regions the data agree within the statistical errors. 
These observations allow to anticipate that indeed the assumption of the 
identity for the transition matrix element for the meson production off free 
and quasi--free nucleons bound in the deuteron is correct. 
In case of the meson production off the deuteron, one can also geometrically 
justify the assumption of the quasi--free scattering since the average 
distance between the proton and the neutron is 
in the order of~\footnote{The matter radius of the deuteron amounts to 
$\approx 2\,\mbox{fm}$~\cite{garcon049}.} $3\,\mbox{fm}$.
Of course, the other nucleon may scatter the incoming proton and the outgoing 
meson.
However, these nuclear phenomena are rather of minor importance in case of the 
production on the neutron bound in the deuteron, but should be taken into 
account for derivations of total cross sections from experimental data.

The reduction of the beam flux on a neutron, due to the presence of the proton, 
referred to as a shadow effect, decreases for example the total cross section 
by about $4.5\,\%$~\cite{chiavassa192} for the $\eta$--meson production.
Similarly, the reduction of the total cross section due to the reabsorption of 
the outgoing $\eta$ meson on the spectator proton was found to be only about 
$3\,\%$~\cite{chiavassa192}. 
The appraisals were performed according to a formula derived in 
reference~\cite{smith647} which shows that the cross section for the deuteron 
reduces by a factor of:
\begin{equation} 
R = 1 - \sigma_{\eta N}^{inel} <\mbox{r}^{-2}> / 4\pi
\end{equation}
compared to the free nucleon cross sections. 
Here $\sigma_{\eta N}^{inel}$ denotes the $\eta N$ inelastic cross section and 
$<\mbox{r}^{-2}>$ stands for the average of the inverse square nucleon 
separation in the deuteron taking the nucleon size into 
account~\cite{chiavassa192}.
The latter effect for the production of mesons like $(\pi,\omega,\eta^{\prime},
\phi)$ is expected to be much smaller, since (as shall be discussed in 
section~\ref{Mwhs}) the s--wave interaction of the $\eta$--meson with nucleons is by far 
stronger than for any of the mentioned ones.

Due to the relatively large distance between the nucleons in the deuteron one 
could argue that the quasi--free reactions will dominate the production of 
mesons even with a deuteron in the final state.
The conviction that the production of mesons in nucleon--deuteron collisions 
proceeds predominantly via an elementary scattering of nucleons is also based 
on the comparative investigations of the $\pi^0$--meson production in 
proton--neutron and proton--deuteron collisions~\cite{meyer2474}. 
The total cross section for the $pd \rightarrow pd \pi^0$ reaction close to 
the kinematical threshold is by more than two orders of magnitude lower compared to 
the elementary $pn \rightarrow d \pi^0$ process (see 
figure~\ref{free_vs_quasifree}b). 
This observation has been successfully explained assuming that the production 
of the $\pi^0$ meson in the $pd \rightarrow pd \pi^0$ reaction is proceeding 
entirely due to the quasi--free $pn \rightarrow d \pi^0$ 
reaction~\cite{meyer2474} for $\eta_{M}$ larger than 0.7~(dashed line).
In the threshold region the result of the calculation (solid line) describes 
exactly the experimental data when the proton--deuteron FSI is additionally 
taken into account.
The large difference in magnitude between the threshold cross section for the 
$pn \rightarrow d \pi^0$ and the $pd \rightarrow pd \pi^0$ reaction can be 
explained since the incident proton momentum needed to create a $\pi^0$ meson 
via the $pd \rightarrow pd \pi^0$ process is much below the kinematical 
threshold for the production in elementary proton--neutron collisions. 
Therefore only nucleons possessing large Fermi momenta can contribute to the 
$\pi^0$ production and this appears with a small probability (compare 
fig.~\ref{fermi_mom_and_kin}a).
The other possible elementary processes $pp \rightarrow pp \pi^0$ and $pn 
\rightarrow pn \pi^0$ can be neglected since they are expected to lead 
predominantly to a four--body exit channel.
This explanation could serve as a heuristic example, however recent 
data~\cite{greiff064002} on the cross sections of $dp \rightarrow dp \pi^0$ 
show that the supposed spectator protons to the elementary $np \rightarrow 
d \pi^0$ reaction are observed in phase--space regions corresponding to zero 
Fermi momenta and hence the quasi--free production could be significant only 
at energies close to the $NN \rightarrow NN \pi$ threshold ($\eta_M\,(pd 
\rightarrow pd \pi^0) \approx 0.83$).
In this experiment the deuteron beam was impinging onto the hydrogen target, 
and hence the spectator protons were boosted in the laboratory permitting 
registration also of the low--energy part of the spectators' spectral 
distribution.

\begin{figure}[H]
\parbox{0.58\textwidth}{\epsfig{file=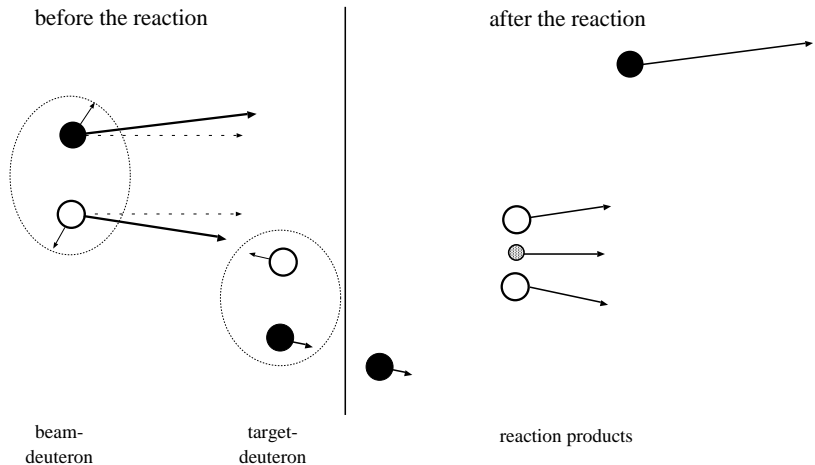,width=0.58\textwidth}}
\hfill
\parbox{0.4\textwidth}
{\caption{\label{dd_ppnneta} Schematic depiction of the double quasi--free 
$nn \rightarrow nnX$ reaction. During the collision of deuterons (left hand 
side of the figure, with the total momentum (solid arrow) resulting from the 
sum of the beam momentum (dotted arrow) plus the Fermi momentum (short arrow)) 
a double quasi--free neutron--neutron reaction may lead to the creation of 
mesons (small gray circle). The spectator protons (black circles) leave the 
reaction region with their initial momentum plus the Fermi momentum, which 
they possessed at the moment of the reaction. Neutrons are plotted as open 
circles. Due to the large relative momenta between spectators and the outgoing 
neutrons ($\sim 1\,\mbox{GeV/c}$ close to the threshold for the $\eta$ meson 
production) a distortion of the $nnX$ system by the accompanying protons can 
be neglected.}}
\end{figure}

A natural extension of the meson production in quasi--free proton--neutron 
collisions with either a deuteron used as the target or as the beam is the 
combination of both methods into a double quasi--free production occurring 
during the collisions of deuterons as depicted in figure~\ref{dd_ppnneta}.

Utilizing this technique~\cite{nn} and registering both the slow and fast 
spectator protons could allow for the study of meson production in 
neutron--neutron collisions with a very precise determination of the excess 
energy which depends on the accuracy for the registration of the momentum 
vectors of the spectators.

\section{Phenomenology of the initial and final state interaction}  
\label{Mwhs} 
\begin{figure}[H]
\parbox{0.32\textwidth}{\epsfig{file=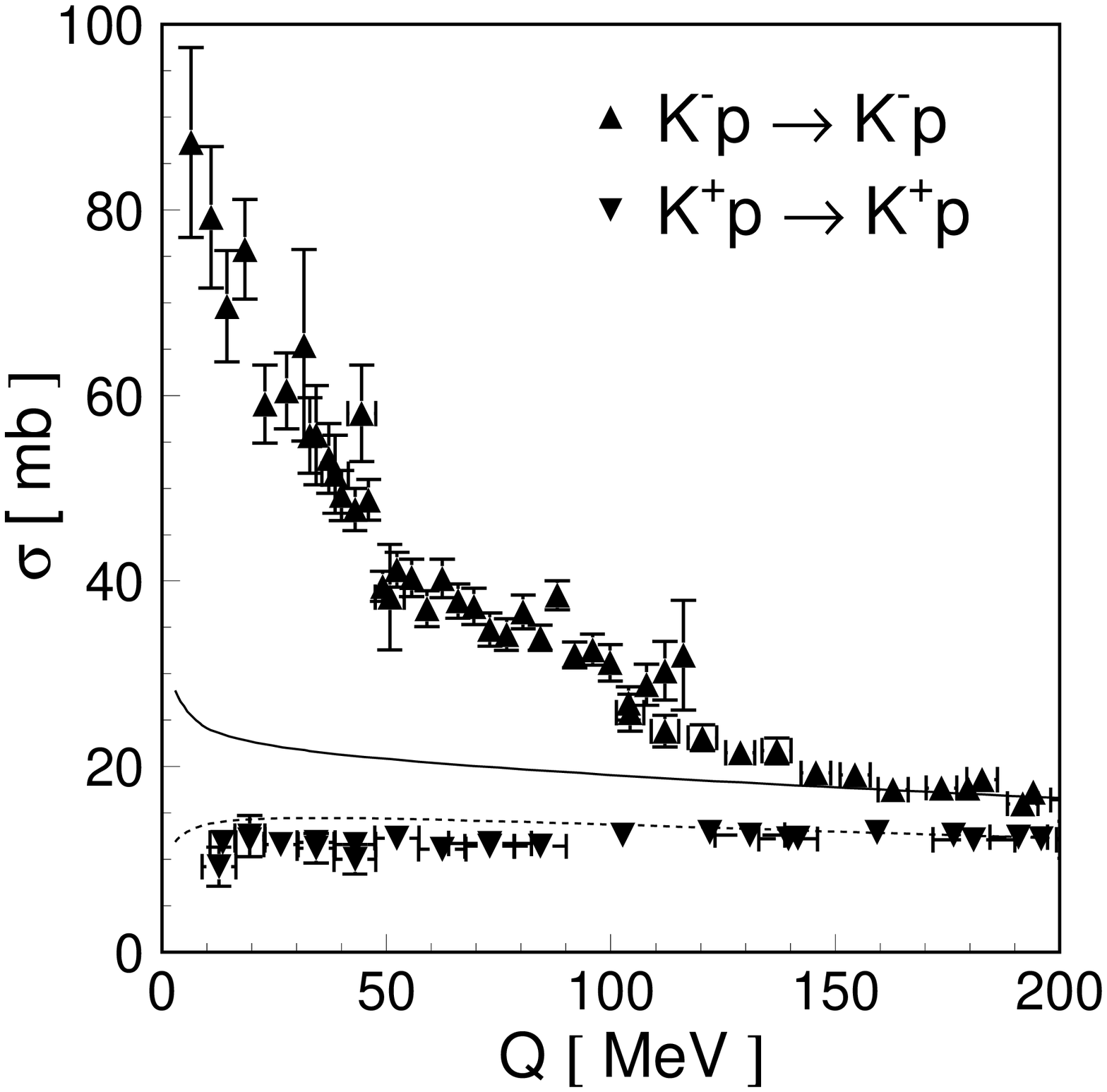,width=0.34\textwidth}}\hfill
\parbox{0.32\textwidth}{\epsfig{file=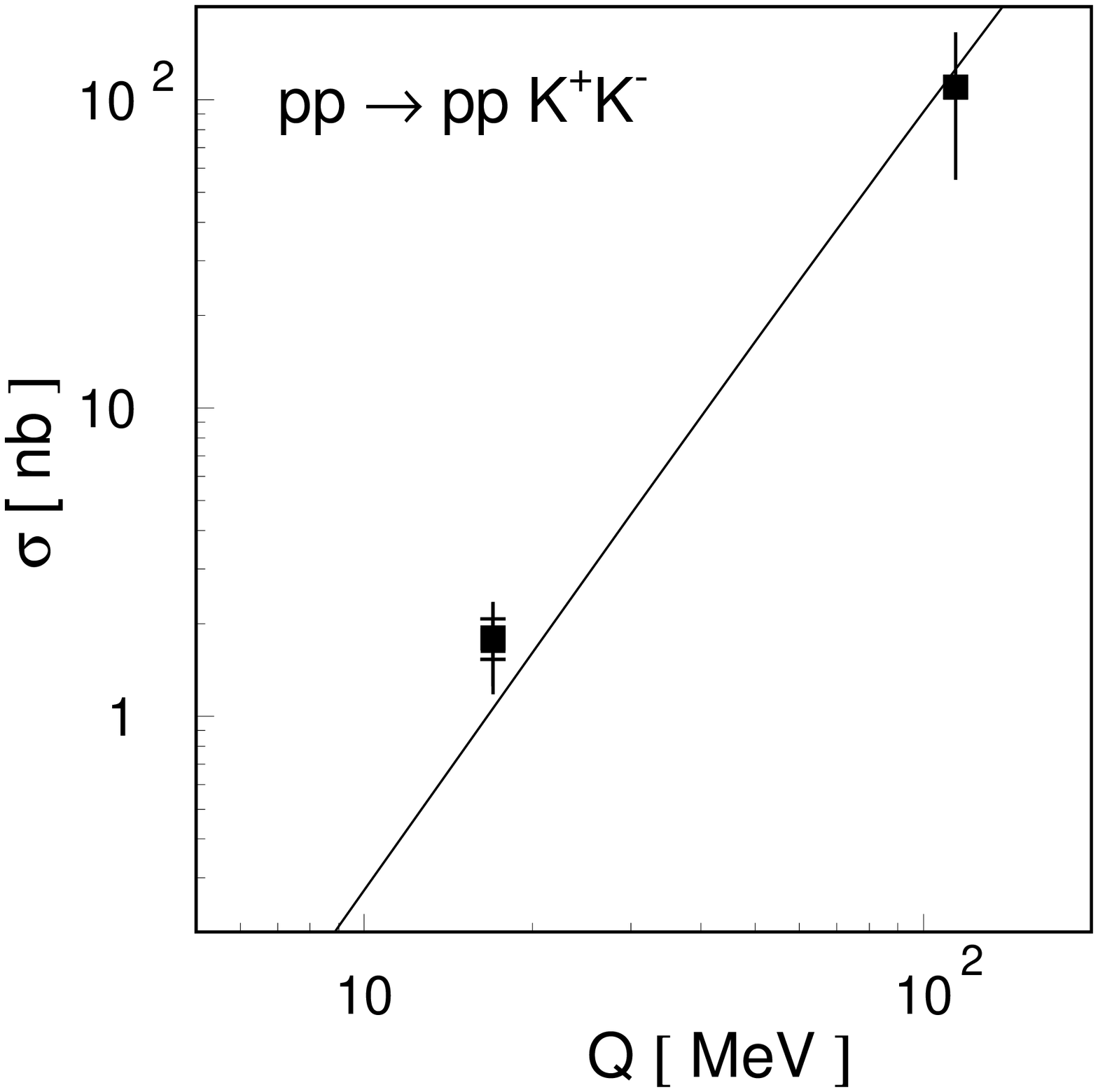,width=0.34\textwidth}}\hfill
\parbox{0.32\textwidth}{\epsfig{file=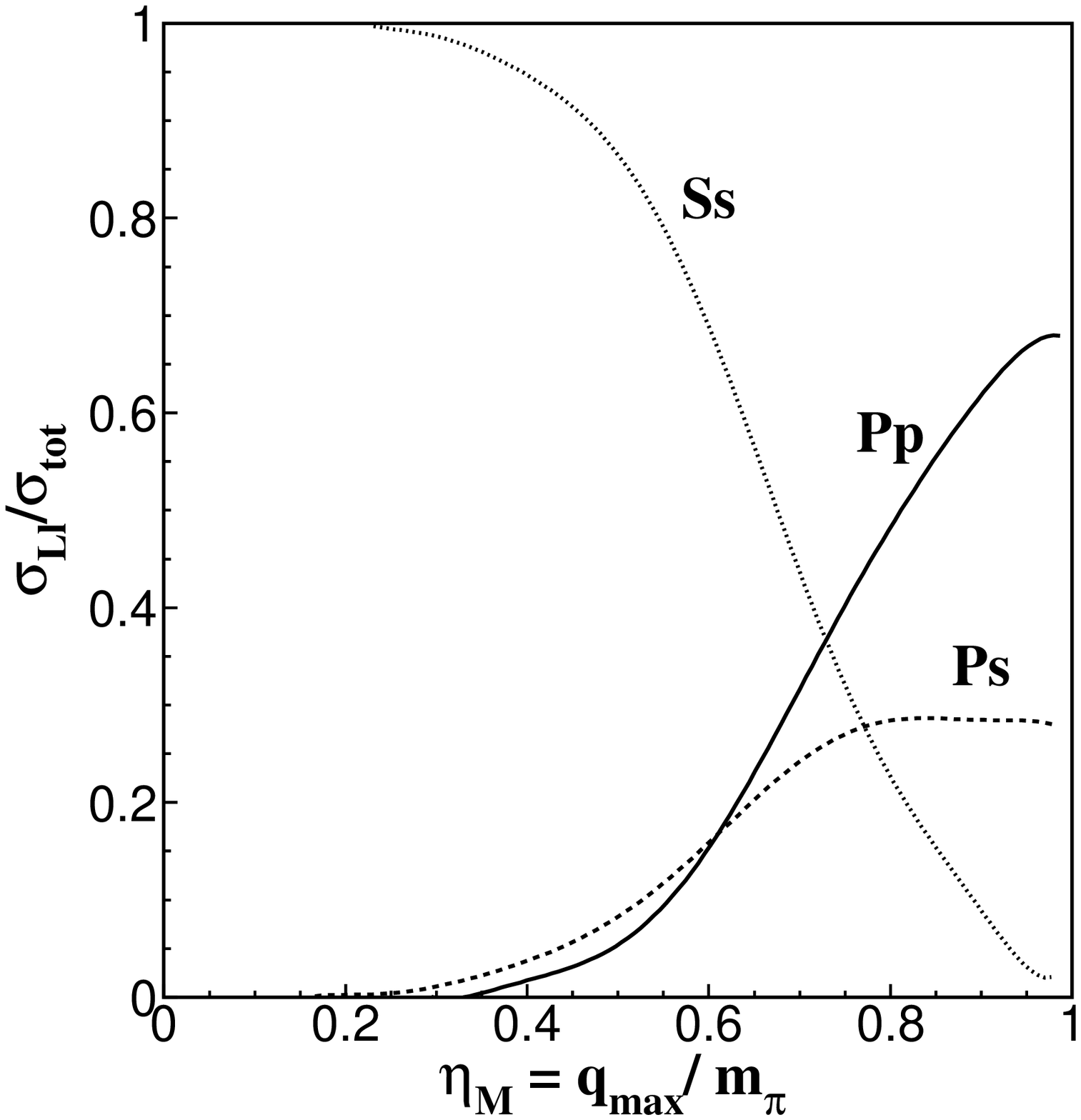,width=0.32\textwidth}}

\parbox{.28\textwidth}{\mbox{}}\hfill
\parbox{.34\textwidth}{\raisebox{0ex}[0ex][0ex]{\large a)}}\hfill
\parbox{.30\textwidth}{\raisebox{0ex}[0ex][0ex]{\large b)}}\hfill
\parbox{.04\textwidth}{\raisebox{0ex}[0ex][0ex]{\large c)}}

\caption{\label{kpkp_kk_SsPsPp} (a) Cross section for the 
$K^+$--proton~\cite{goldhaber135} and $K^-$--proton~\cite{adams54} elastic 
scattering. Solid and dashed lines represent the changes of the phase--space 
integral (eq.~\eqref{Vps_two_body}) modified by the initial and final state 
Coulomb interaction $\left(\sigma_{Kp \rightarrow Kp} = 
const \cdot C^4 /\left(16 \pi\,\mbox{s}\right)\right)$. Both curves 
are normalized to points of large excess energies. 
(b) Total cross section of the $pp \rightarrow pp K^+ K^-$ 
reaction~\cite{quentmeier276,balestra7}. Statistical and systematical errors 
are separated by dashes. The solid line indicating calculations of 
reference~\cite{sibirtsev101} is described in the text.
(c) Decomposition of the total cross section of the $pp \rightarrow pp \pi^0$ 
reaction into Ss, Ps and Pp final state angular momenta. The dashed and solid 
lines represent the $\eta_M^6$ and $\eta_M^8$ dependence of Ps and Pp partial 
cross sections, respectively. The remainder is indicated as the dotted line. 
The Sp partial wave is forbidden by the conservation laws and the Pauli 
excluding principle. Note that at $\eta_M = 1$ the Pp and Ps partial waves 
seem to dominate. However, the analysis of the differential cross section 
measured at CELSIUS at $\eta_M = 0.449$~\cite{zlomanczuk251,bilger633} showed 
that also a d--wave pion production --- due to the interference between Ss and 
Sd states --- constitutes $7\,\%$ of the total cross section, when a 
meson--exchange model is assumed. The figure has been adapted 
from~\cite{meyer064002}.}
\end{figure}

The simplest method to study the interactions between meson and nucleon is to 
scatter a beam of mesons off hydrogen or deuteron targets and to measure the 
behaviour of the total and differential cross section in the elastic 
scattering of the considered particles.
Figure~\ref{kpkp_kk_SsPsPp}a  presents the total cross section for $K^+ p$ and 
$K^- p$ elastic scattering as a function of the centre--of--mass excess 
energy. 
When comparing the data to calculations~\cite{moskalsqm2001} --- including the 
changes of phase--space integral~(eq.~\eqref{Vps_two_body}) and Coulomb 
interaction in the initial and final states --- one observes a huge 
enhancement for the $K^- p$ cross section with decreasing excess energy and a 
slight suppression in the case of $K^+ p$ scattering.
This observation may be attributed to the slight repulsion due to the 
kaon--proton hadronic interaction and the significantly larger scattering  
caused by the strong interaction between the $K^-$ and proton due to the 
vicinity of the $\Lambda(1405)$ hyperon resonance. 
The effect has a direct influence on the effective mass splitting of kaons 
immersed in dense nuclear matter~\cite{senger209,schaffner325}.

A Coulomb repulsion was taken into account by multiplying the 
expression~\eqref{Vps_two_body} by the Coulomb penetration factor $C^2$ for 
both exit and entrance channels.
$C^2$ determines the ratio of the probability of finding two particles close 
together to the probability of finding two uncharged particles together, all 
other things being equal~\cite{jackson77} and can be expressed 
as~\cite{bethe38}:
\begin{equation} 
\label{penetrationfactor}
  C^{2} = \frac{2\pi\eta_c}{e^{2\pi\eta_c} - 1},
\end{equation}
where
$\eta_c$ is the relativistic Coulomb parameter, which for the collision of 
particles $i,j$ reads:
\begin{equation*}
\eta_c = \frac{\mbox{q}_i\,\mbox{q}_j \,\alpha}{\mbox{v}} = 
  \mbox{q}_i\,\mbox{q}_j \,\alpha 
  \frac{\mbox{s}_{ij} - \mbox{m}_i^2 - \mbox{m}_j^2}
       {\sqrt{\lambda(\mbox{s}_{ij},\mbox{m}_i^2,\mbox{m}_j^2)}},
\end{equation*}
with the fine structure constant $\alpha$, the relative velocity v of the 
colliding particles and with $\mbox{q}_i$, $\mbox{q}_j$ denoting their 
charges~\footnote{For collisions at an angular momentum $l$ larger than 
$0\,\hbar$ the $C^2$ of equation~\eqref{penetrationfactor} needs to be 
multiplied by a factor of $\prod_{n=1}^{l} 
\left(1 + (\eta_c/n)^2 \right)$~\cite{wong1866,arndt1002}.}.
The comparison of the $K^{-}$p scattering data with the phase space
integral presented in figure~\ref{kpkp_kk_SsPsPp}a aims only to illustrate
the qualitative influence of the meson-nucleon dynamics on the total cross 
section energy dependence. It is worth noting, however, that at present
the data can be qualitatively well described in the frame of the effective 
chiral Lagrangian with a coupled channel 
potential~\cite{N.Kaiser-I,oset99,ramon249}. 
 
Another method permitting such investigations is the production of a meson in 
the nucleon--nucleon interaction close to the kinematical threshold where the 
outgoing particles possess small relative velocities. 
The experimental observables are the excitation functions and the double 
differential cross section.
A possible deviation from the expectation based on the assumption of 
homogeneous phase--space abundance corrected for the known nucleon--nucleon 
interaction delivers information about the meson--nucleon forces.
For example, figure~\ref{kpkp_kk_SsPsPp}b presents the first data points on 
the close--to--threshold  production of the $K^-$ meson via the $pp 
\rightarrow pp K^+ K^-$ reaction. 
The superimposed solid line depicts the results of 
calculations~\cite{sibirtsev101} taking into account the changes of the 
production amplitude as deduced from the $K^+ p$ and $K^- p$ elastic 
scattering shown in figure~\ref{kpkp_kk_SsPsPp}a, but neglecting the influence 
of the dominant proton--proton interaction!
On the other hand the proton--proton FSI can not be neglected in case of 
three--body final states, e.g.\ $pp \rightarrow 
pp \pi^0$~\cite{meyer633,bondar8}, $pp \rightarrow 
pp \eta$~\cite{calen39,smyrski182}, or $pp \rightarrow 
pp \eta^{\prime}$~\cite{swave,moskal416}, since it has strong influence on the 
total cross section energy dependence enhancing it by more than an order of 
magnitude for excess energies below $\mbox{Q} \approx 15\,\mbox{MeV}$ as shall 
be shown below.
Thus it is surprising that in spite of its neglection one can describe the 
data of the $pp \rightarrow pp K^+ K^-$ reaction. 
The origin of that effect will be investigated experimentally in the near 
future~\cite{magnusproposal}.
At present one can only speculate whether it is due to the partial 
compensation of the $pp$ and $K^- p$ hadronic interaction or maybe due to the 
additional degree of freedom given by the four--body final 
state~\cite{magnus_sibi}.

\begin{figure}[H]
       \parbox{0.33\textwidth}{\epsfig{file=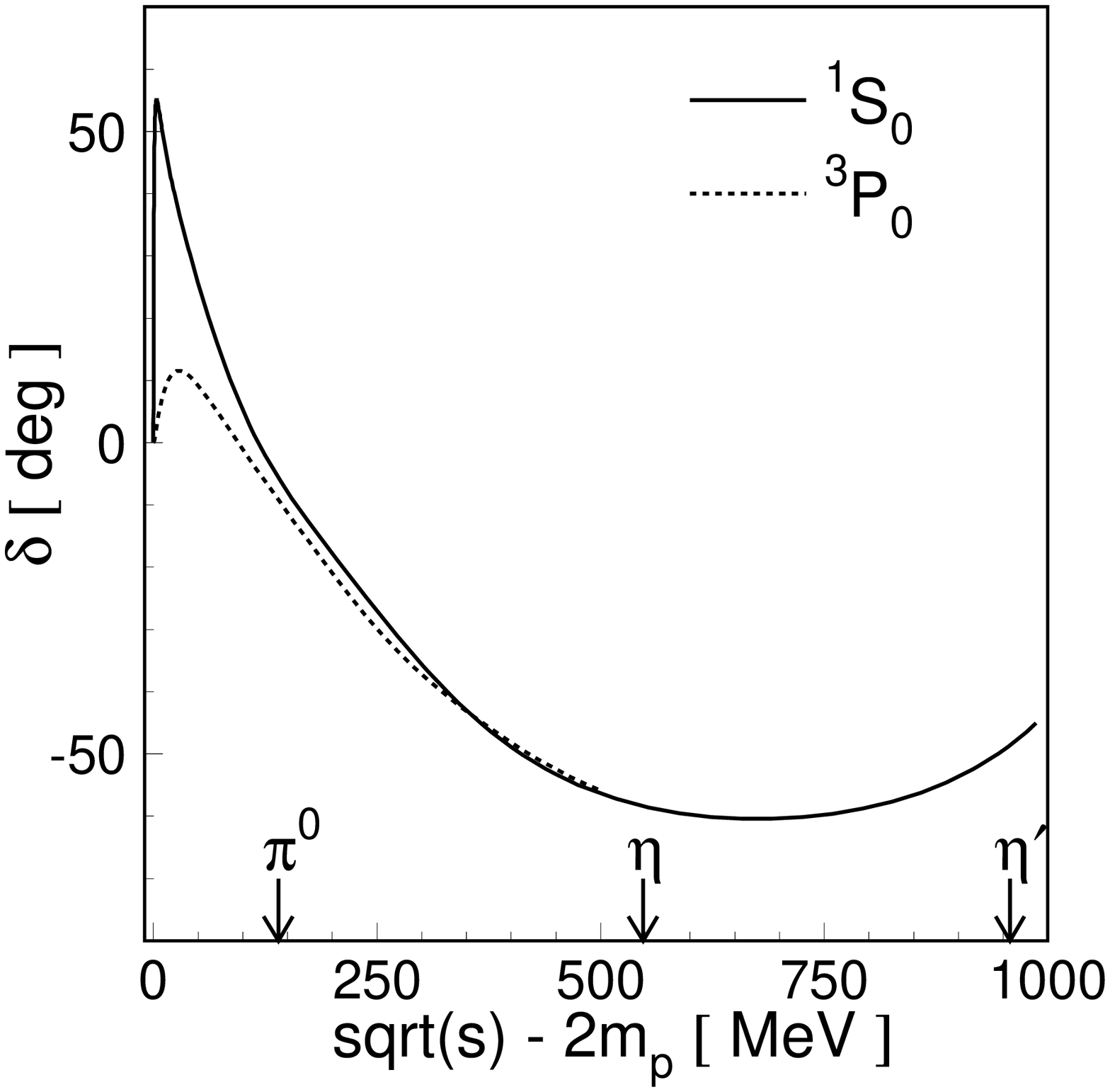,width=0.36\textwidth}}\hfill
       \parbox{0.32\textwidth}{\epsfig{file=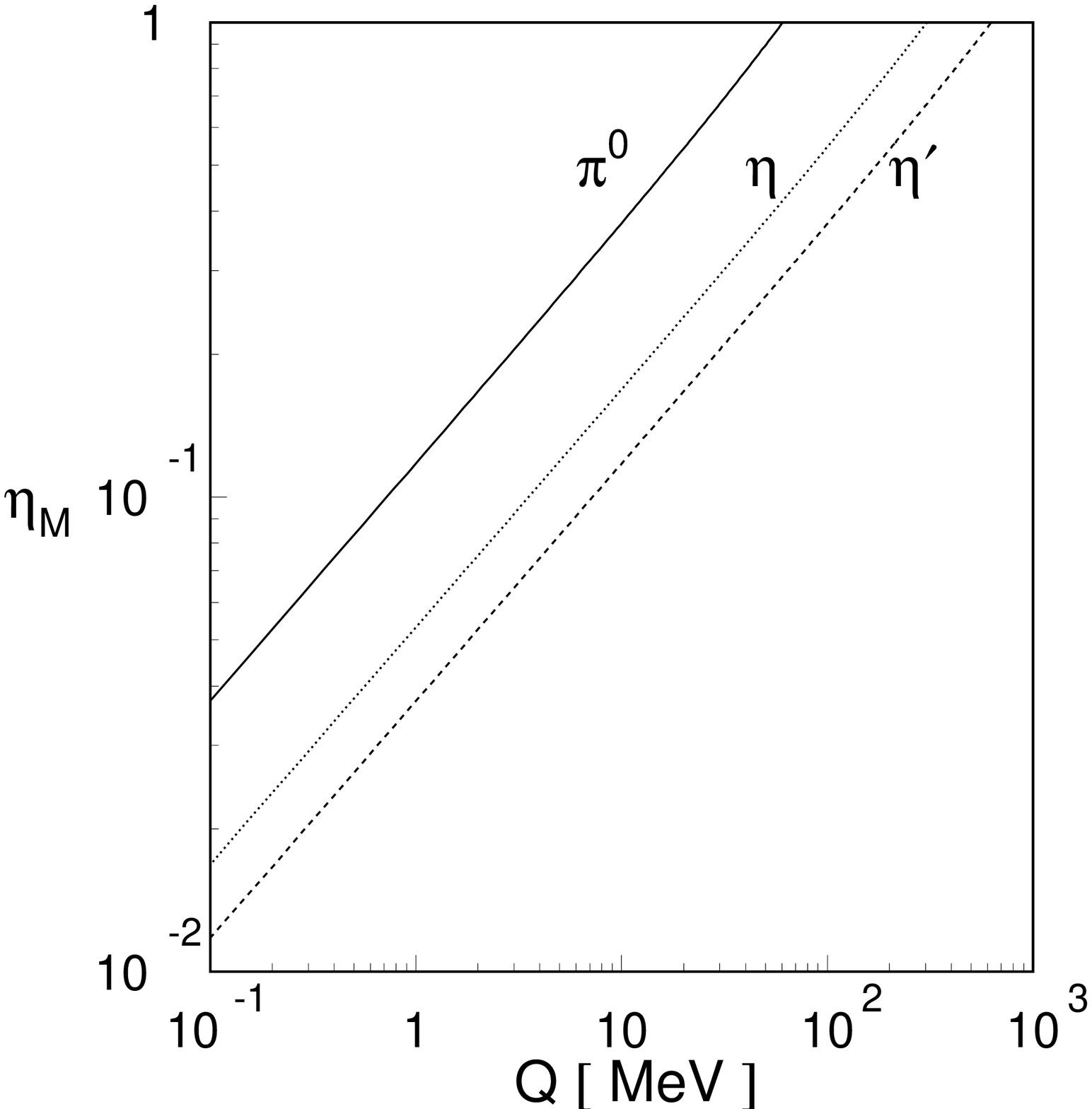,width=0.36\textwidth}}\hfill
       \parbox{0.33\textwidth}{\epsfig{file=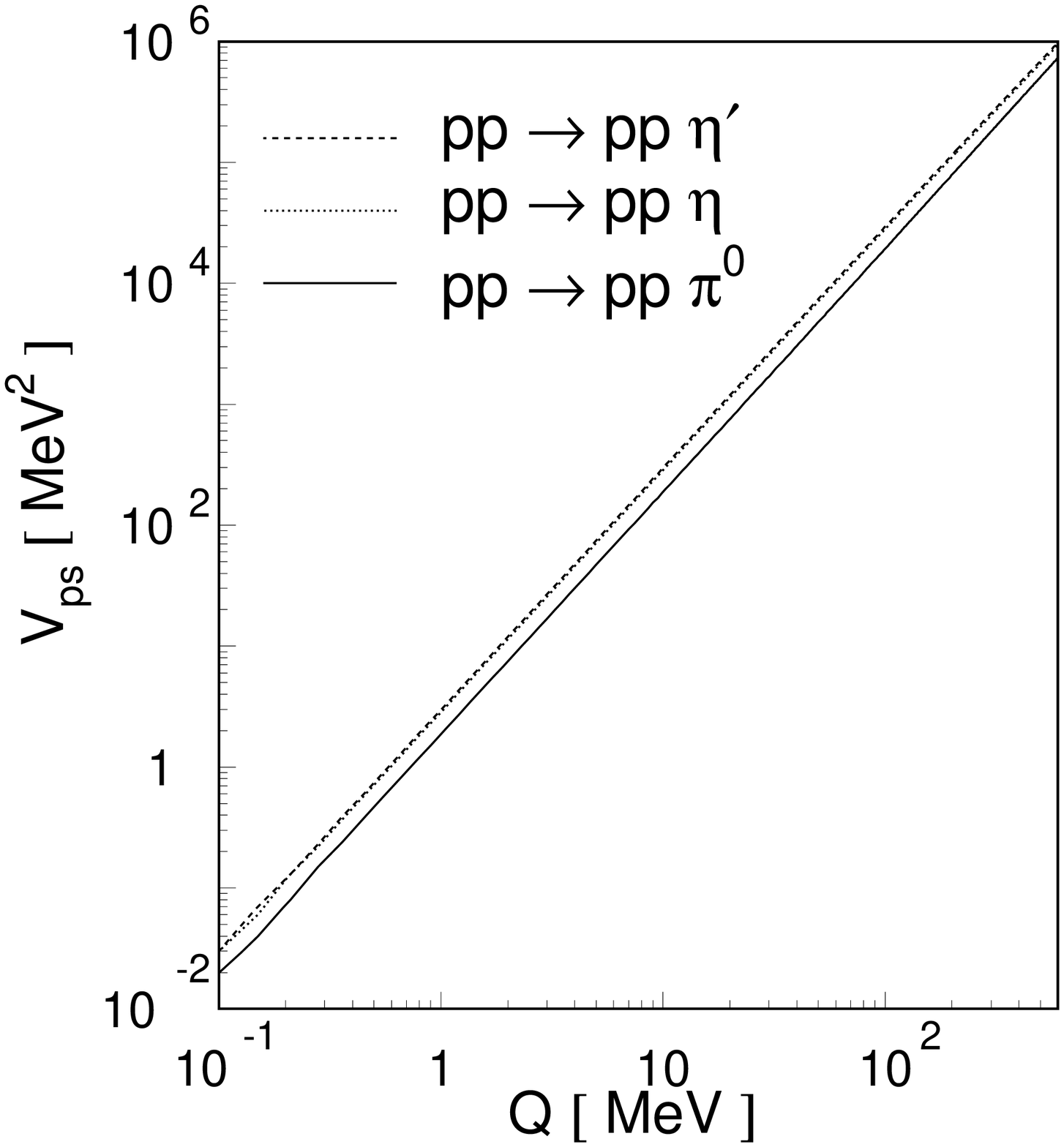,width=0.36\textwidth}}
 
 \parbox{0.30\textwidth}{\raisebox{0ex}[0ex][0ex]{\mbox{}}} \hfill
 \parbox{0.32\textwidth}{\raisebox{0ex}[0ex][0ex]{\large a)}} \hfill
 \parbox{0.33\textwidth}{\raisebox{0ex}[0ex][0ex]{\large b)}} \hfill
 \parbox{0.02\textwidth}{\raisebox{0ex}[0ex][0ex]{\large c)}}
 \vspace{-0.2cm}
\caption{\label{1s03p0_eta_Q_Vps} (a) The $^1\mbox{S}_0$ and $^3\mbox{P}_0$ 
phase--shifts of the nucleon--nucleon potential shown versus the 
centre--of--mass kinetic energy available in the proton--proton system. The 
values have been extracted from the SAID data base~\cite{arndt3005} (solution 
SM97). For higher energies the S-- and P--wave phase--shifts are nearly the 
same. This is because the collision parameter required to yield the angular 
momentum of $1\,\hbar$ diminishes significantly below $1\,\mbox{fm}$ with 
increasing energy and consequently the interaction of nucleons --- objects of 
about $1\,\mbox{fm}$ size --- becomes almost central. 
(b) The variable $\eta_M$ as a function of the excess energy for $\pi^0 $, 
$\eta$ and $\eta^{\prime}$ mesons produced via $pp \to pp\,Meson$ 
reactions. 
(c) The phase--space volume $V_{ps}$ (defined by 
equation~\eqref{Vps_relativistic}) versus the excess energy Q. The picture 
indicates that for the production of ``heavy mesons'' in the nucleon--nucleon 
interaction at a given Q value there is only a slight difference of the 
available phase--space volume on the produced meson mass, which is larger 
than that of $\pi^0$ production by about $30\,\%$ only. Therefore, for the 
comparative studies of the production dynamics of different mesons, Q is as 
much a suitable variable as $V_{ps}$.}
\end{figure}

In order to demonstrate the influence of the nucleon--nucleon interaction on 
the cross section dependence on the excess energy let us consider the 
production of neutral pseudoscalar mesons ($\pi^0$, $\eta$, $\eta^{\prime}$) 
in the collision of protons. Due to the short life--time of these 
mesons the study of their low--energy interaction with nucleons in direct 
scattering experiments is not feasible.  
In general for a three--body exit channel one expects a dependence of the 
total cross section on the energy which can be described by the linear 
combination of partial cross sections of equation~\eqref{sigmaLl}. 
Therefore, to be able to extract the information about the final state 
interaction of the outgoing particles from the energy dependence it is 
necessary to know precisely the contribution originating from different 
partial waves. 
Yet appropriately close--to--threshold there is only one important combination 
of the angular momenta of emitted particles (Ss) and in this region the energy 
dependence of the total cross section is uniquely determined. 
However, the range of excess energies for the S--wave dominance changes 
strongly with the mass of the produced meson.

For the production of the $\pi^0$ meson in proton--proton collisions the 
investigations with polarized beam and targets~\cite{meyer064002,meyer5439} 
allowed to deduce that the Ss partial--wave accounts for more than $95\,\%$ of 
the total cross section up to $\eta_M \approx 0.4$, as can be seen in 
figure~\ref{kpkp_kk_SsPsPp}c, where the Ss contribution is indicated by the 
dotted line. 
The Ss contribution was inferred assuming the $\eta_M^6$ and $\eta_M^8$ 
dependence for Ps and Pp partial waves, respectively.
These are power--laws taken from proportionality~\eqref{sigmaLl}, which was 
derived under the assumption of non--interacting particles.
Relatively small values of $^3\mbox{P}_0$--wave nucleon--nucleon phase--shifts 
at low energies (compared to $^1\mbox{S}_0$ phase--shifts in 
figure~\ref{1s03p0_eta_Q_Vps}a) and similarly weak low--energy interactions of 
P--wave protons in other spin combinations~\cite{bergervoet1435} justify this 
assumption. 

The measurements with the polarized beam and target allow a model independent 
determination of the contribution from individual partial waves. In case of 
the $\vec{p}\vec{p} \to pp\pi^{0}$ reaction, due to the identical particles in 
the initial state and the rotational equivalences~\cite{meyer064002,knutson177}
there are only seven independent polarization observables, e.g. two beam 
analyzing powers and five linear combination of spin correlation coefficients.
   
In particular, as shown in references~\cite{meyer064002,meyer5439} the close 
to threshold contributions of the $Ps$ and $Pp$ partial waves can be
determined in the model-free way from the measurements of the spin dependent 
total cross sections only. For example the strength of the $Ps$ final state 
can be expressed as~\cite{meyer064002}:
      \begin{equation}
        \sigma(Ps) = \frac{1}{4}
                     \left( \sigma_{tot} +
                            \Delta{\sigma_T} +
                            \frac{1}{2}\Delta{\sigma_L},
                     \right)
      \end{equation}
where $\sigma_{tot}$ denotes the total unpolarized cross section and
$\Delta{\sigma_T}$ and $\Delta{\sigma_L}$ stand for differences between the 
total cross sections measured with anti-parallel and parallel beam and target 
polarizations. Subscripts $T$ and $L$ associate the measurements with
the transverse and longitudinal polarizations, respectively.

\begin{figure}[H]
\parbox{0.32\textwidth}
  {\epsfig{file=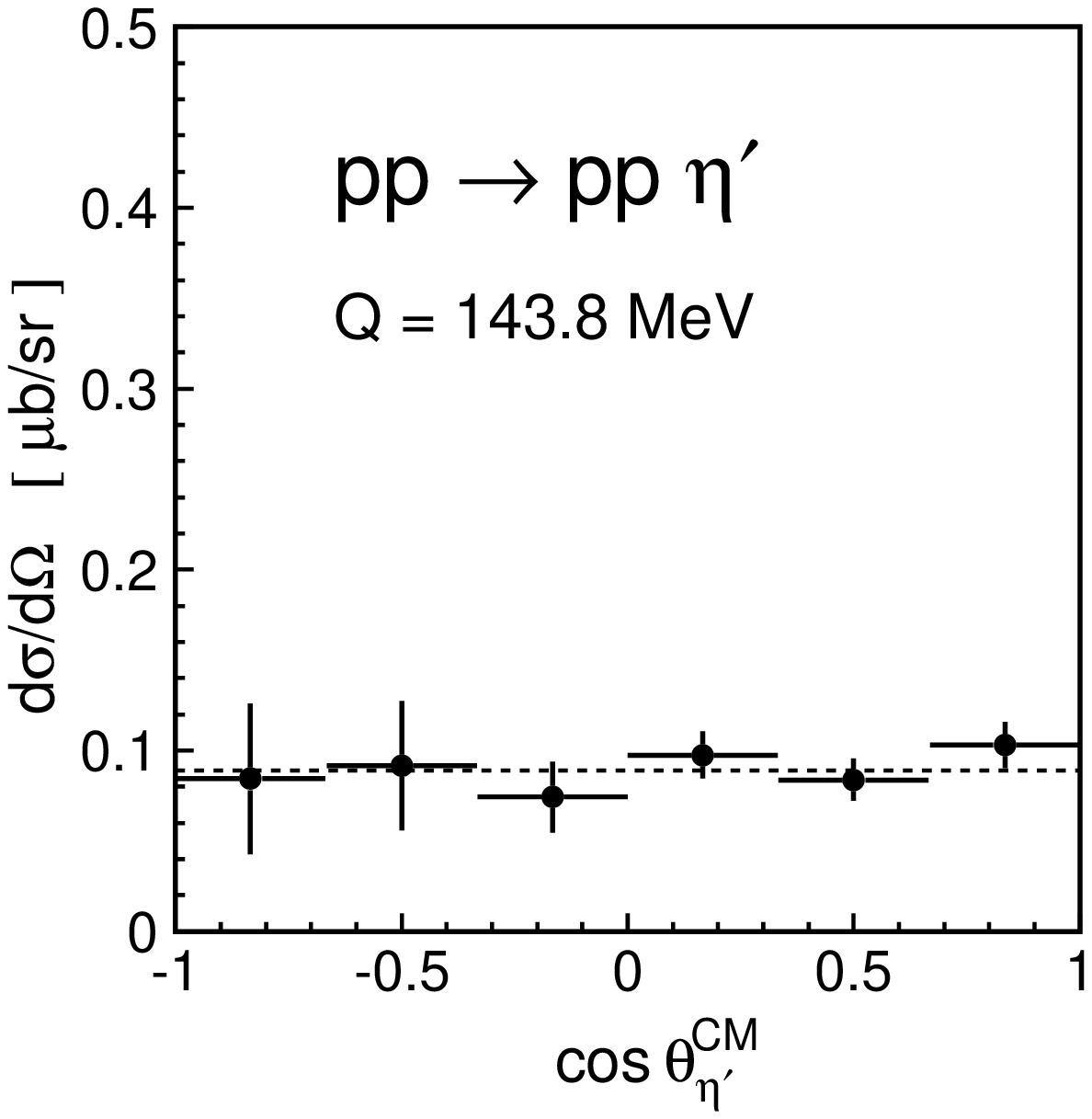,width=0.33\textwidth}}\hfill
\parbox{0.32\textwidth}
  {\epsfig{file=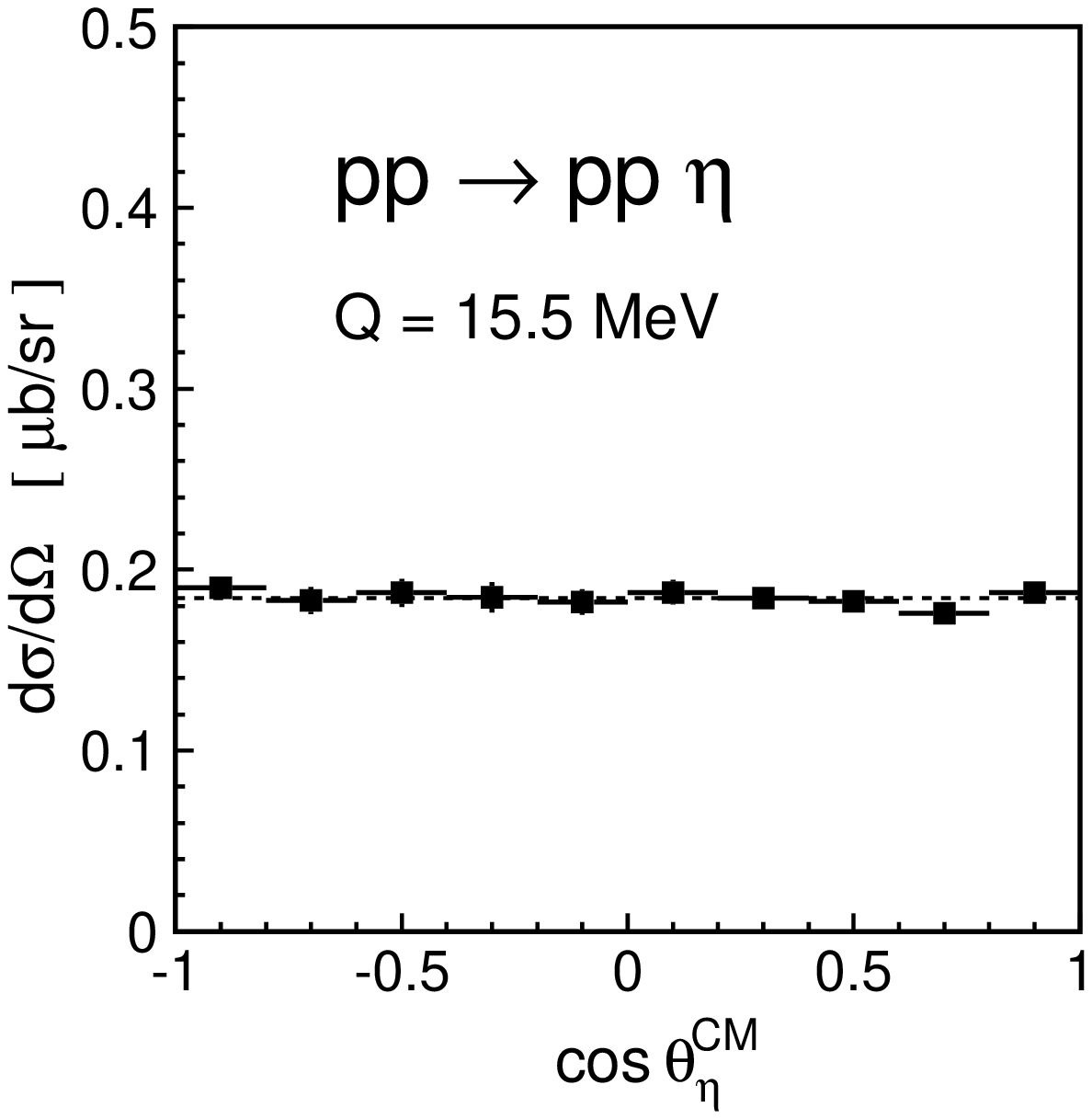,width=0.33\textwidth}}\hfill
\parbox{0.32\textwidth}
  {\epsfig{file=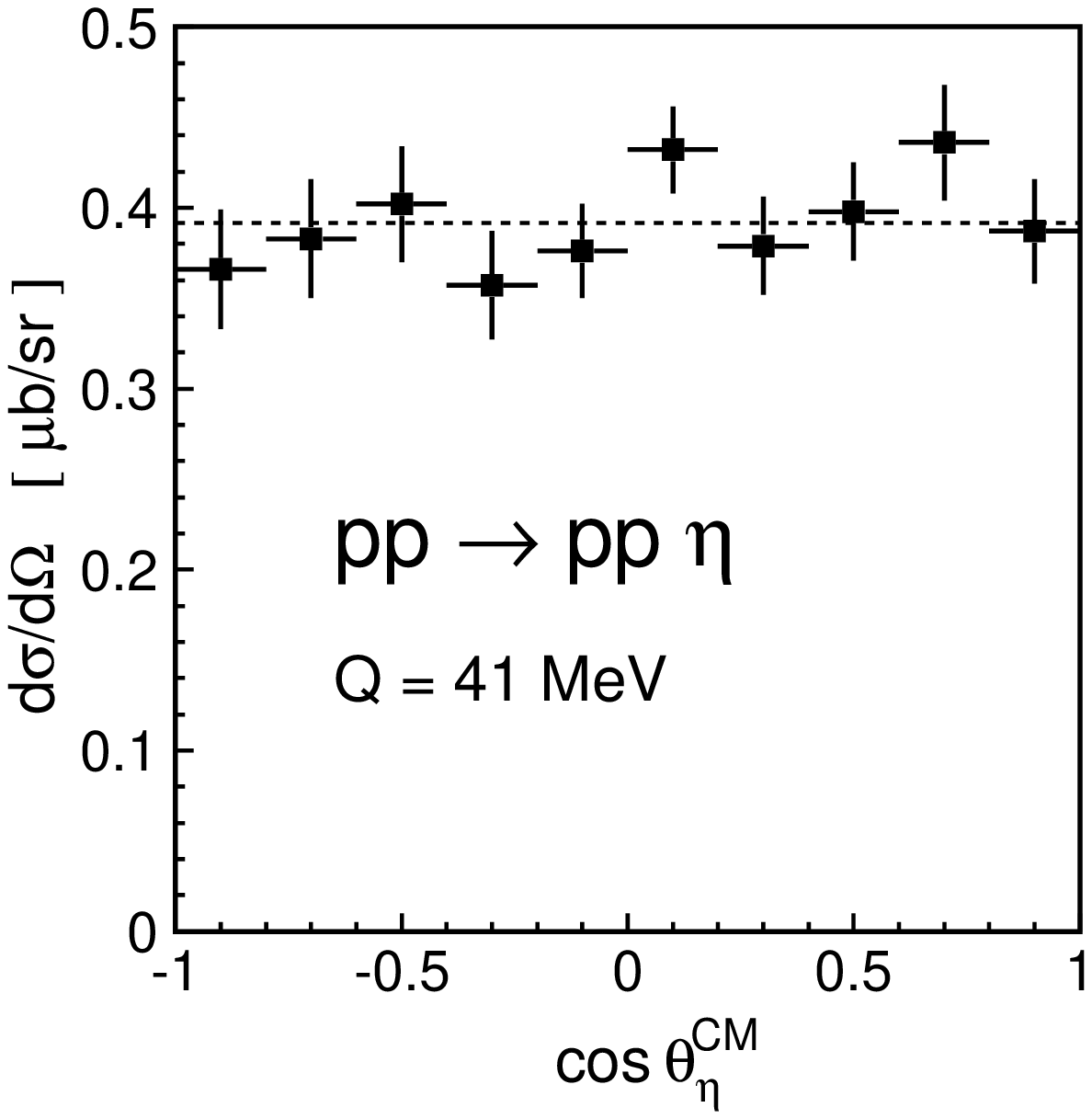,width=0.33\textwidth}}
 \parbox{0.27\textwidth}{\raisebox{1ex}[0ex][0ex]{\mbox{}}} \hfill
\parbox{0.31\textwidth}{\raisebox{1ex}[0ex][0ex]{\large a)}} \hfill
\parbox{0.32\textwidth}{\raisebox{1ex}[0ex][0ex]{\large b)}} \hfill
\parbox{0.03\textwidth}{\raisebox{1ex}[0ex][0ex]{\large c)}}
\caption{\label{differential} Differential cross section of the $pp 
\rightarrow pp\,Meson$ reaction as a function of the meson centre--of--mass 
polar angle. Dashed lines indicate the isotropic distribution. Shown are 
results of measurements for the $pp \rightarrow pp \eta^{\prime}$ reaction 
taken at $\mbox{Q} = 143.8\,\mbox{MeV}$~\cite{balestra29} (a) and for the 
$pp \rightarrow pp \eta$ reaction at $\mbox{Q}  = 
15.5\,\mbox{MeV}$~\cite{eta_menu} (b) and $\mbox{Q} = 
41\,\mbox{MeV}$~\cite{TOFeta} (c). In all pictures only 
statistical errors are plotted, which in figure (b) are smaller than the size 
of symbols. The distribution presented in picture (b) is consistent with a 
measurement performed at an excess energy of $\mbox{Q} = 16\,\mbox{MeV}$ by 
means of the PROMICE/WASA detector~\cite{calen190}, whereas the data at 
$\mbox{Q} = 37\,\mbox{MeV}$ also from reference~\cite{calen190} deviate 
significantly from isotropy. However, data shown in picture (c) have been 
taken with a detector of much higher angular acceptance.}
\end{figure}

In accordance with the phenomenology of Gell--Mann and Watson~\cite{GMW} 
described in the previous section one expects that also in the case of heavier 
mesons the Ss partial wave combination will constitute the overwhelming 
fraction of the total production cross section for $\eta_M$ smaller than 0.4. 
This implies --- as can be deduced from the relation between $\eta_M$ and Q 
illustrated in figure~\ref{1s03p0_eta_Q_Vps}b --- that mesons heavier than the 
pion are produced exclusively via the Ss state in a much larger excess energy 
range and hence larger phase--space volume (see figure~\ref{1s03p0_eta_Q_Vps}c).
In particular in case of $\eta^{\prime}$ this is larger by more than one order 
of magnitude. 
Thus, whereas in case of $\pi^0$ the onset of higher partial waves is observed 
at Q around $10\,\mbox{MeV}$ it is expected only above $100\,\mbox{MeV}$ and 
above  $\approx 40\,\mbox{MeV}$ for $\eta^{\prime}$ and $\eta$ mesons, 
respectively. 
Figures~\ref{differential}b and~\ref{differential}c present the angular 
distributions of the created $\eta$ meson in proton collisions.
It is evident that at $\mbox{Q} = 15.5\,\mbox{MeV}$ and still at $\mbox{Q} = 
41\,\mbox{MeV}$ the production of the $\eta$ meson is completely isotropic 
within the shown statistical errors. 
Although at $\mbox{Q} = 41\,\mbox{MeV}$ the accuracy of the data does not 
exclude a few per cent of the contribution originating from higher partial 
waves, the dominance of the s--wave creation is evident.
Similarly, the measurements of the differential cross section 
(fig.~\ref{differential}a) for the $pp \rightarrow pp \eta^{\prime}$ reaction 
performed at SATURNE~\cite{balestra29} at $\mbox{Q} = 143.8\,\mbox{MeV}$ are 
still consistent with pure Ss--wave production, though the relatively large 
error bars would allow for other contributions on a few per cent level 
($\approx 10\,\%$).

\begin{table}[H]
\caption{\label{momtranstable} Momentum transfer $\Delta\mbox{p}$ calculated 
according to equation~\ref{momtranseq} and the corresponding distance 
$\mbox{R} \approx \hbar/\Delta\mbox{p}$ probed by the $NN \rightarrow 
NN\,Boson$ reaction at the kinematical threshold for different particles 
produced. The table has been adapted from~\cite{nak01}.}
\vskip -0.25cm
\tabskip=1em plus2em minus.5em
\halign to \hsize{\hfil#&\hfil#\hfil&\hfil#\hfil&\hfil#\hfil\cr
\noalign{\hrulefill}
particle  & mass [MeV] & $\Delta\mbox{p}$ [$\mbox{fm}^{-1}$] & R [fm] \cr
\noalign{\vskip -0.20cm}
\noalign{\hrulefill}
\noalign{\vskip -0.5cm}
\noalign{\hrulefill}
$\gamma$        &   0   &  0.0  &  $\infty$ \cr
$\pi$           & 140   &  1.9  &  0.53 \cr
$\eta$          & 550   &  3.9  &  0.26 \cr
$\rho , \omega$ & 780   &  4.8  &  0.21 \cr
$\eta^\prime$   & 960   &  5.4  &  0.19 \cr
$\phi$          &1020   &  5.6  &  0.18 \cr
\noalign{\hrulefill} }
\end{table}

Let us now consider to what extent the energy dependence of the total cross 
section in the estimated range of the dominance of the Ss partial waves can be 
understood in terms of the phase--space variation and the interaction between 
the particles participating in the reaction.
Watson~\cite{wat52} and Migdal~\cite{migdal2} have proposed the factorization 
of the amplitude in the case where the production is of short range and the 
interaction among the outgoing particles is of long range.
This requirement is well fulfilled in the case of the close--to--threshold 
meson production due to the large momentum transfer ($\Delta\mbox{p}$) between 
the interacting nucleons needed to create the considered mesons ($\pi, \eta, 
\ldots, \phi$). 
According to the Heisenberg uncertainty relation the large momentum transfer 
brings about a small space in which the primary creation of the meson takes 
place.
In table~\ref{momtranstable} the distance probed by the $NN \rightarrow 
NN\,Meson$ reaction at threshold is listed for particular mesons. 
It ranges from $0.53\,\mbox{fm}$ for pion production to $0.18\,\mbox{fm}$ for 
the $\phi$ meson, whereas the typical range of the strong nucleon--nucleon 
interaction at low energies determined by the pion exchange may exceed a 
distance of few Fermi and hence is by one order of magnitude larger than the 
values listed in table~\ref{momtranstable}.
Thus in analogy to the Watson--Migdal approximation for two--body 
processes~\cite{wat52} the complete transition matrix element of 
equation~\eqref{phasespacegeneral} may be factorized approximately 
as~\footnote{For a comprehensive discussion of the FSI and ISI issue including 
a historical overview and a criticism of various approaches the reader is 
referred to~\cite{kle01}.}
\begin{equation} 
\label{M0FSIISI}
|M_{pp \rightarrow pp X}|^2 \approx |M_{FSI}|^2 \cdot |M_0|^2 \cdot F_{ISI},
\end{equation}
where $M_{0}$ represents the total short range production amplitude, 
$M_{FSI}$ describes the elastic interaction among particles in the exit 
channel and $F_{ISI}$ denotes the reduction factor accounting for the 
interaction of the colliding protons.
Further, in the first order approximation one assumes that the particles are 
produced on their mass shell and that the created meson does not interact with 
nucleons.
This assumption implies that the $|M_{FSI}|^2$ term can be substituted by the 
square of the on--shell amplitude of the nucleon--nucleon elastic scattering:
\begin{equation} 
\label{FSI_elastic}
|M_{FSI}|^2 = |M_{NN \rightarrow NN}|^2.
\end{equation}
Effects of this rather strong assumption will be considered later, while 
comparing the estimation with the experimental data.

In the frame of the optical potential model the scattering amplitude is 
determined by phase--shifts.
Particularly, the $^{1}\mbox{S}_0$ proton--proton partial wave --- relevant 
for further consideration --- can be expressed explicitly as 
follows~\cite{morton825}:
\begin{equation} 
\label{amppp}
M_{pp \rightarrow pp} = 
  \frac{e^{-i\delta_{pp}({^{1}\mbox{\scriptsize S}_{0}})} \cdot 
        \sin{\delta_{pp}({^{1}\mbox{S}_0})}}
       {C \cdot \mbox{k}},
\end{equation}
where $C$ denotes the square root of the Coulomb penetration factor defined by 
equation~\eqref{penetrationfactor}, k stands for either proton momentum in the 
proton--proton rest frame and the phase--shift is indicated by $\delta_{pp}$. 
The phase--shifts $\delta_{pp}({^{1}\mbox{S}_0})$ can be extracted from the 
SAID data base (see fig.~\ref{1s03p0_eta_Q_Vps}a) or, alternatively, can be 
calculated according to the modified Cini--Fubini--Stanghellini formula 
including the Wong--Noyes Coulomb 
correction~\cite{naisse506,noyes995,noyes465}\\[-0.3cm]
\begin{equation} 
\label{CFS}
C^2 \;\mbox{k}\; ctg(\delta_{pp}) \;+\; 2\,\mbox{k}\,\eta_c\,h(\eta_c) = 
  - \frac{1}{a_{pp}} + \frac{b_{pp}\,\mbox{k}^2}{2} 
  - \frac{P_{pp}\,\mbox{k}^4}{1 + Q_{pp}\,\mbox{k}^2},
\end{equation}
where $h(\eta_c) = -ln(\eta_c) - 0.57721 + \eta_c^2\,
\sum_{n=1}^{\infty}\frac{1}{n \cdot (n^2 + \eta_c^2)}$~\cite{jackson77}, with 
$\eta_c$ the same as in eq.~\eqref{penetrationfactor}.\\[-0.1cm]

\begin{figure}[H]
\parbox{0.45\textwidth}{\epsfig{file=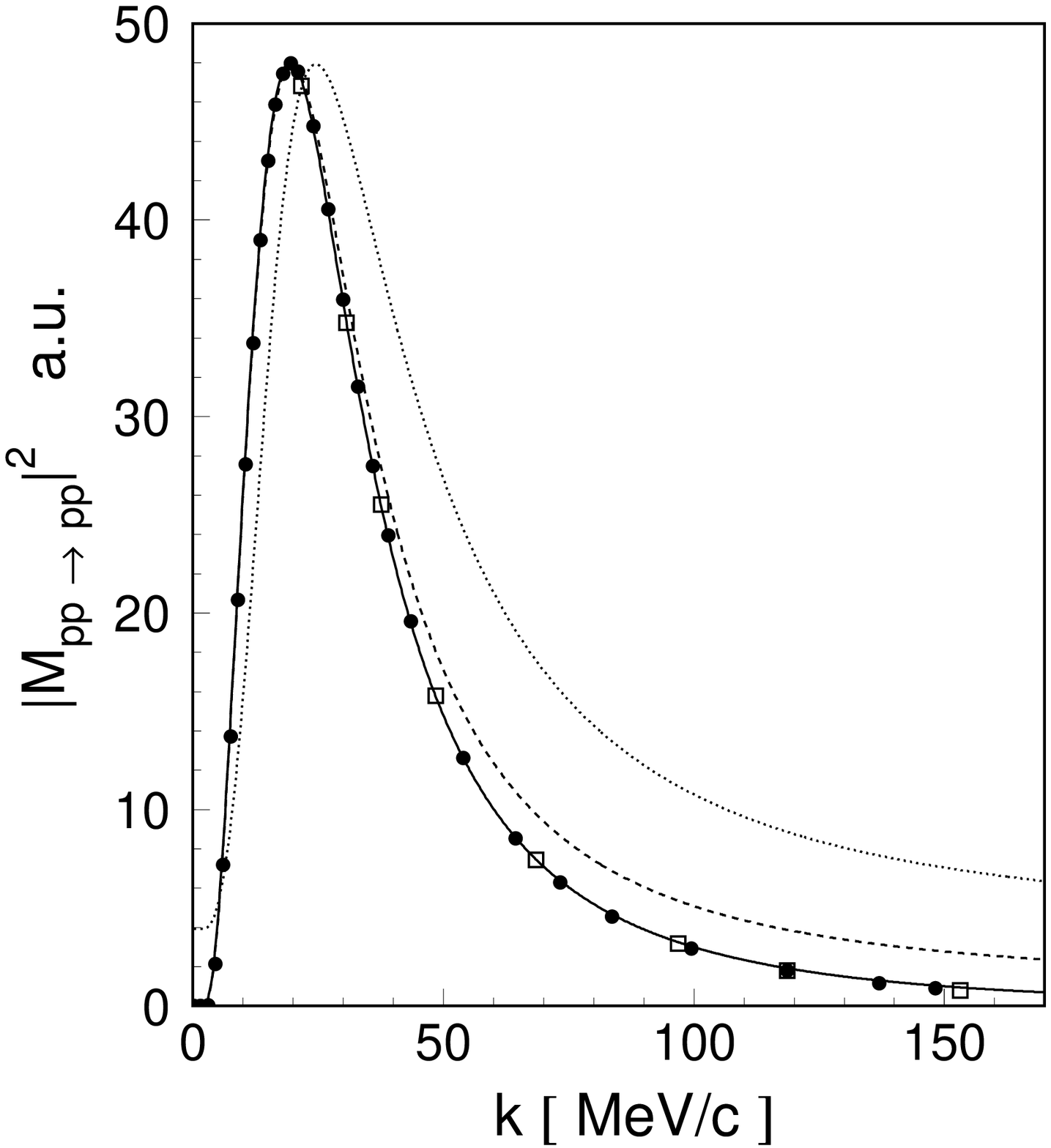,width=0.37\textwidth}} \hfill
\parbox{0.49\textwidth}
  {\epsfig{file=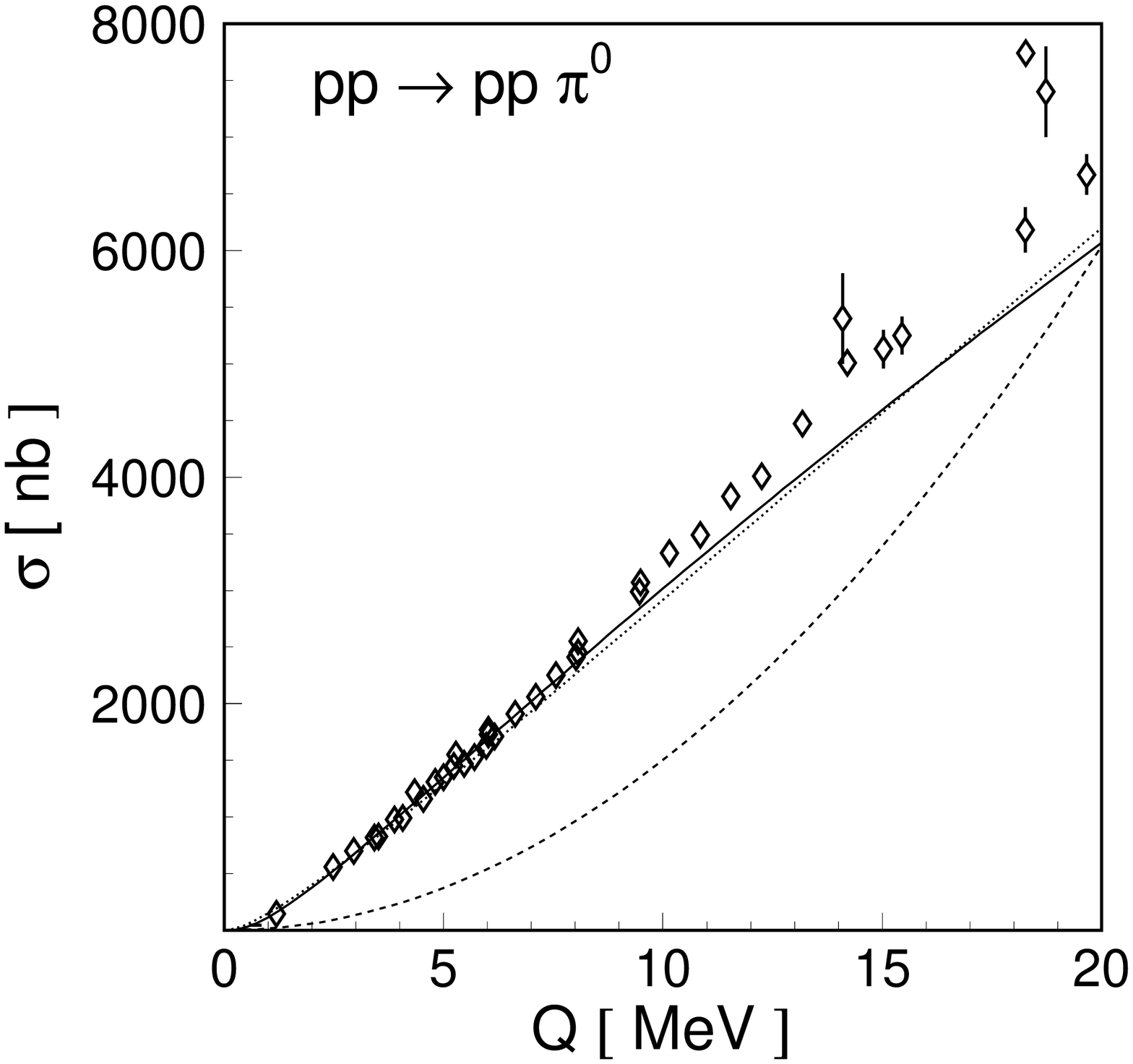,width=0.40\textwidth}}

\vspace{-0.3cm}
\parbox{0.37\textwidth}{\raisebox{1ex}[0ex][0ex]{\mbox{}}} \hfill
\parbox{0.54\textwidth}{\raisebox{1ex}[0ex][0ex]{\large a)}} \hfill
\parbox{0.07\textwidth}{\raisebox{1ex}[0ex][0ex]{\large b)}}

\vspace{-0.5cm}
\caption{\label{Mpppp_cross_pi} (a) Square of the proton--proton scattering 
amplitude versus k, the proton momentum in the proton--proton subsystem 
from~\cite{morton825,naisse506}~(solid line), \cite{druzhinin}~(dashed line), 
and \cite{shyammosel,goldbergerwatson}~(dotted line). The filled circles are 
extracted from~\cite{arndt3005} and the opened squares from~\cite{nijmpsa}.
The curves and symbols are arbitrarily normalized to be equal at maximum to 
the result from reference~\cite{druzhinin}, shown as the dashed line. 
(b) Total cross section for the $pp \rightarrow pp \pi^0$ reaction as a 
function of the centre--of--mass excess energy Q. Data are from 
refs.~\cite{bondar8,meyer633,stanislausR1913,bilger633}. The dashed line 
indicates a phase--space integral normalized arbitrarily. The phase--space 
distribution with inclusion of proton--proton strong and Coulomb interactions 
fitted to the data at low excess energies is shown as the solid line. The 
dotted line indicates the parametrization of reference~\cite{faldt209} written 
explicitly in equation~\eqref{faldtwilkin}, with $\epsilon = 0.3$.}
\end{figure}

The phenomenological quantities $a_{pp} = -7.83\,\mbox{fm}$ and $b_{pp} = 
2.8\,\mbox{fm}$ denote the scattering length and effective 
range~\cite{naisse506}, respectively. 
The parameters $P_{pp} = 0.73\,\mbox{fm}^3$ and $Q_{pp} = 3.35\,\mbox{fm}^2$ 
are related to the detailed shape of the nuclear potential and derived from a 
one--pion--exchange model~\cite{naisse506}.
Substituting equation~\eqref{CFS} into equation~\eqref{amppp} allows to 
calculate the low--energy amplitude for the proton--proton elastic 
scattering~\footnote{In principle the formula is valid for
$k \leq 133~MeV/c$~\cite{noyes995}.  
$|M_{pp \rightarrow pp}|^2$ is taken to be constant for larger 
values of $k$.}: \\[-0.4cm]
\begin{equation} 
\label{Mpppp}
|M_{pp \rightarrow pp}|^2 \;=\; 
 \frac{C^2}
  {C^4\,\mbox{k}^2 \;+\; 
  \left(- \frac{1}{a_{pp}}\,+\,\frac{b_{pp}\,\mbox{\scriptsize k}^2}{2}\, 
 -\,\frac{P_{pp}\,\mbox{\scriptsize k}^4}{1 + Q_{pp}\,\mbox{\scriptsize k}^2}\, 
 -\,2\,\mbox{k}\,\eta_c\,h(\eta_c)\right)^2}.
\end{equation}
The result is presented as a solid line in figure~\ref{Mpppp_cross_pi}a and is 
in good agreement with the values obtained from the phase--shifts of the VPI 
partial wave analysis~\cite{arndt3005}, shown as solid circles and with the 
phase--shifts of the Nijmegen analysis~\cite{nijmpsa}, shown as open squares.
The factor $C^2$ is always less than unity due to the Coulomb repulsion 
between protons. 
At higher energies, where $C^2$ is close to unity, the nuclear scattering will 
be predominant and for the very low energies the Coulomb and nuclear 
interactions are competing. 
The Coulomb scattering dominates approximately up to about $0.8\,\mbox{MeV}$ 
of the proton energy in the rest frame of the other proton, where $C^2$ equals 
to one--half~\cite{jackson77}. 
In the case of the $pp \rightarrow pp\,Meson$ reaction the maximum possible 
energy of a proton seen from another proton is equal to $0.8\,\mbox{MeV}$ 
already at an excess energy of about $\mbox{Q} = 0.4\,\mbox{MeV}$. 
\begin{figure}[H]
\vspace{-0.3cm}
\hspace{1.3cm}
\parbox{0.35\textwidth}
  {\epsfig{file=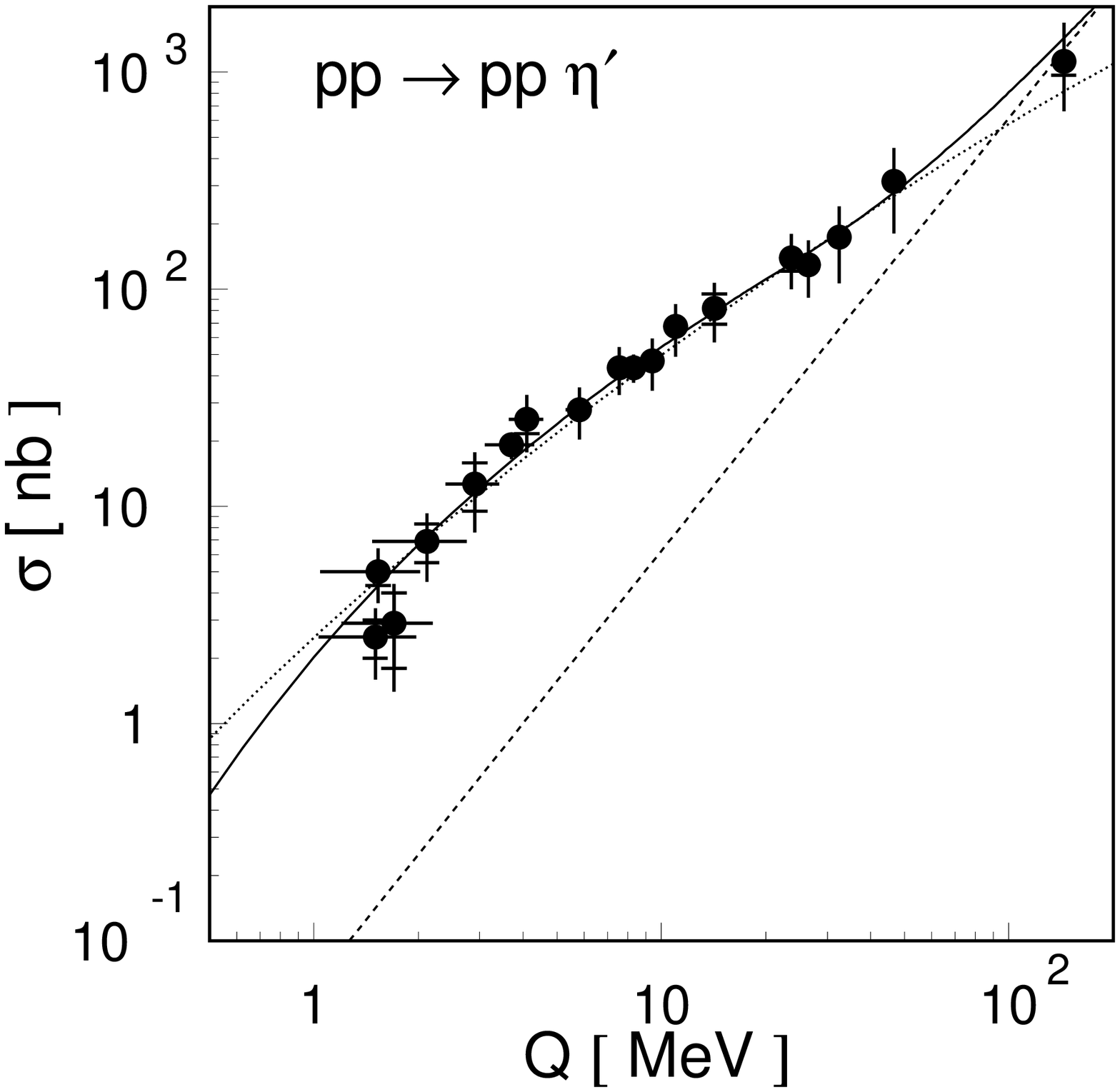,width=0.35\textwidth}} \hfill
\parbox{0.49\textwidth}
  {\epsfig{file=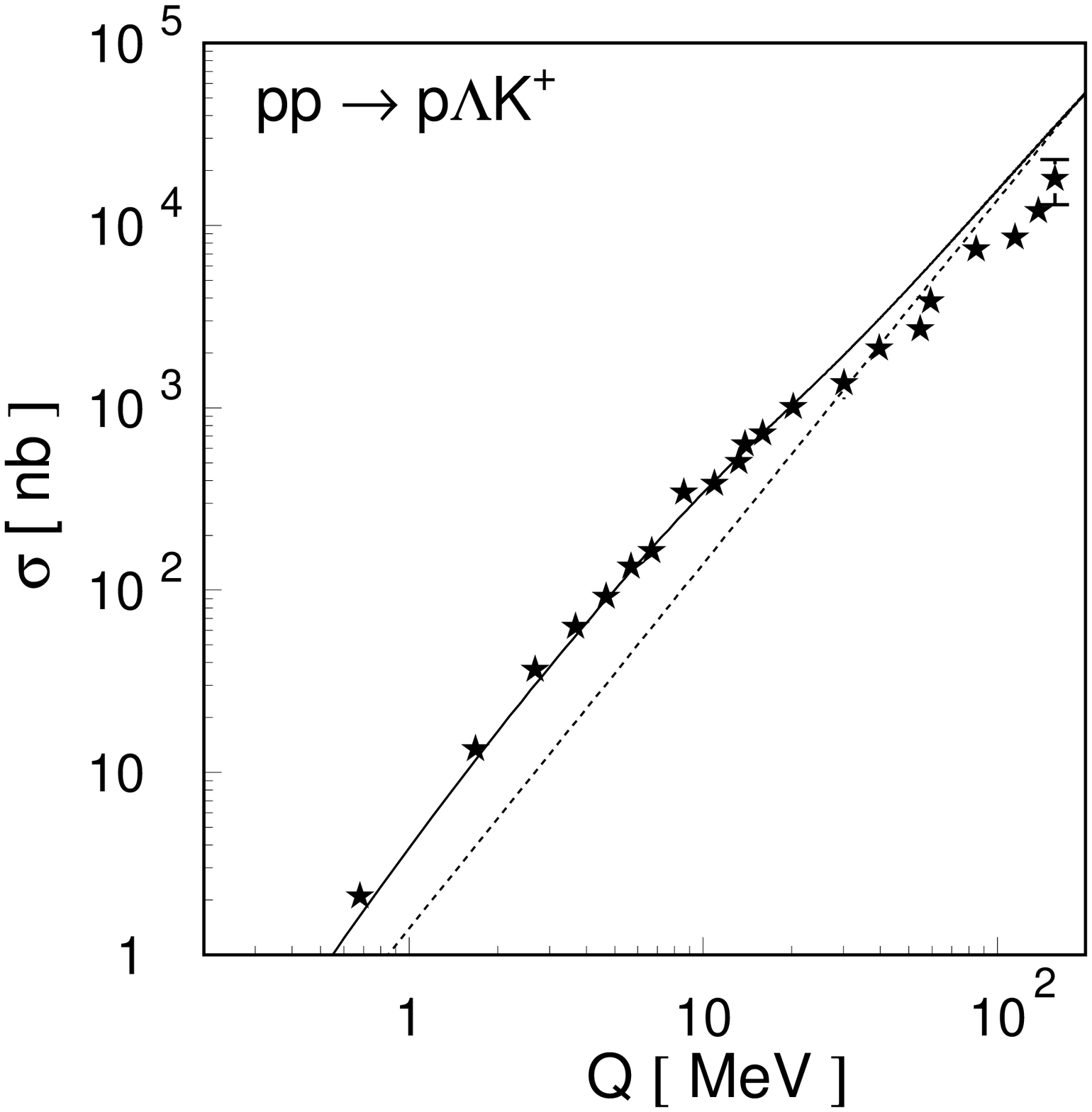,width=0.35\textwidth}}

\vspace{-0.5cm}
\parbox{0.41\textwidth}{\raisebox{1ex}[0ex][0ex]{\mbox{}}} \hfill
\parbox{0.51\textwidth}{\raisebox{1ex}[0ex][0ex]{\large a)}} \hfill
\parbox{0.06\textwidth}{\raisebox{1ex}[0ex][0ex]{\large b)}}

\vspace{-0.3cm}
\caption{\label{cross_etap_pkl} (a) Total cross section for the $pp 
\rightarrow pp \eta^{\prime}$ reaction as a function of the centre--of--mass 
excess energy Q. Data are from 
refs.~\cite{hibou41,balestra29,moskal3202,moskal416,khoukazetap}. 
Statistical and systematical errors are separated by dashes. The solid line 
shows the phase--space distribution with inclusion of proton--proton strong 
and Coulomb interactions. The dotted line indicates the parametrization of 
reference~\cite{faldt209} written explicitely in equation~\eqref{faldtwilkin}, 
with $\epsilon = 0.3$ and the dashed line
indicates a phase--space integral normalized arbitrarily.
(b) Stars represent the data of the $pp \rightarrow p K^+ \Lambda$ 
reaction~\cite{bal98,sewerin682,kow02,bilger217,mar01,hes01,fickinger2082}. The dashed line 
presents the energy dependence defined by the phase--space. The solid line 
shows the calculation with proton--$K^+$ Coulomb repulsion and a 
proton--$\Lambda$ strong interaction taken into account as described in the 
text.}
\end{figure}
Therefore, a significant influence of the Coulomb repulsion on the energy 
dependence of the total production cross section is expected only at very low 
excess energies, i.e.\ conservatively for $\mbox{Q} \le 2\,\mbox{MeV}$.

Assuming that the on--shell proton--proton amplitude exclusively determines 
the phase--space population one can obtain the total cross section energy 
dependence substituting equation~\eqref{Mpppp} into the 
formula~(\ref{Vpsdalitz}).
Solid lines in figures~\ref{Mpppp_cross_pi}b and \ref{cross_etap_pkl}a 
represent the determined dependence for the $pp \rightarrow pp \pi^0$ and 
$pp \rightarrow pp \eta^{\prime}$ reactions, respectively.
The absolute scale of the calculations has been fixed by the normalization to 
the data.
One recognizes the good agreement with the experimental points in the excess 
energy range up to $\mbox{Q} \approx 10\,\mbox{MeV}$ for $\pi^0$ meson 
production and up to $\mbox{Q} \approx 150\,\mbox{MeV}$ for $\eta^{\prime}$.
This confirms the earlier considerations that the domain of Ss partial wave 
dominance should widen significantly with the mass of the created meson and 
as expected it extends above $100\,\mbox{MeV}$ in case of the $\eta^{\prime}$ 
meson.
Comparing the data to the arbitrarily normalized phase--space integral of 
equation~\eqref{Vps_relativistic} reveals that the proton--proton FSI enhanced 
the total cross section by more than an order of magnitude for low excess 
energies.
This is particularly clear in figure~\ref{cross_etap_pkl}a where the 
logarithmic scale is utilized. 
However, the much weaker nucleon--hyperon interaction increases the total 
cross section for the close--to--threshold $K^+$ meson production only by about 
a factor of two. 
Again in that case --- as shown in figure~\ref{cross_etap_pkl}b --- the 
factorization of the matrix element into constant on--shell primary production 
amplitude and the elastic scattering between outgoing baryons leads to the 
good description of the close--to--threshold data. 
In this case the Coulomb interaction between $K^+$ meson and proton was taken 
into account multiplying the scattering amplitude by the Coulomb penetration 
factor. 

The dotted line in figures~\ref{Mpppp_cross_pi}b and~\ref{cross_etap_pkl}a, 
which for $\mbox{Q} < 50\,\mbox{MeV}$ is practically indistinguishable from 
the solid line, presents the excess energy dependence of the total cross 
section taking into account the proton--proton FSI effects according to the 
model developed by F{\"a}ldt and Wilkin~\cite{faldt209,faldt2067}.
Representing the scattering wave function in terms of bound state wave 
function the authors derived a closed formula which describes the effects of 
the nucleon--nucleon FSI as a function of the excess energy Q only. 
This approach is specifically useful for the description of the spin--triplet 
proton--neutron FSI, due to the existence of a bound state (deuteron) with the 
same quantum numbers.
Though the bound state of the proton--proton system does not exist, the model 
allows also to express the total cross section energy dependence for a $pp 
\rightarrow pp\,Meson$ reaction by a simple and easily utilizable formula: 
\begin{equation} 
\label{faldtwilkin}
\sigma \;=\; 
  const. \cdot \frac{V_{ps}}{\mbox{F}} \cdot 
  \frac{1}
    {\left(1\;+\;\sqrt{1\,+\,\frac{\mbox{\scriptsize Q}}{\epsilon}}\right)^2}\;
=\; const.^{\prime} \cdot 
  \frac{\mbox{Q}^{2}}{ \sqrt{\lambda(\mbox{s},\mbox{m}_p^2,\mbox{m}_p^2)}} 
  \cdot 
  \frac{1}
    {\left(1\;+\;\sqrt{1\,+\,\frac{\mbox{\scriptsize Q}}{\epsilon}}\right)^2},
\end{equation}
where the parameter $\epsilon$ has to be settled from the data.
The flux factor F and the phase--space volume $V_{ps}$ are given by 
equations~\eqref{fluxfactor} and~\eqref{Vps_nonrelativistic}, respectively.
The normalization can be determined from the fit of the data which must be 
performed for each reaction separately.

\begin{figure}[H]
\parbox{0.32\textwidth}
  {\epsfig{file=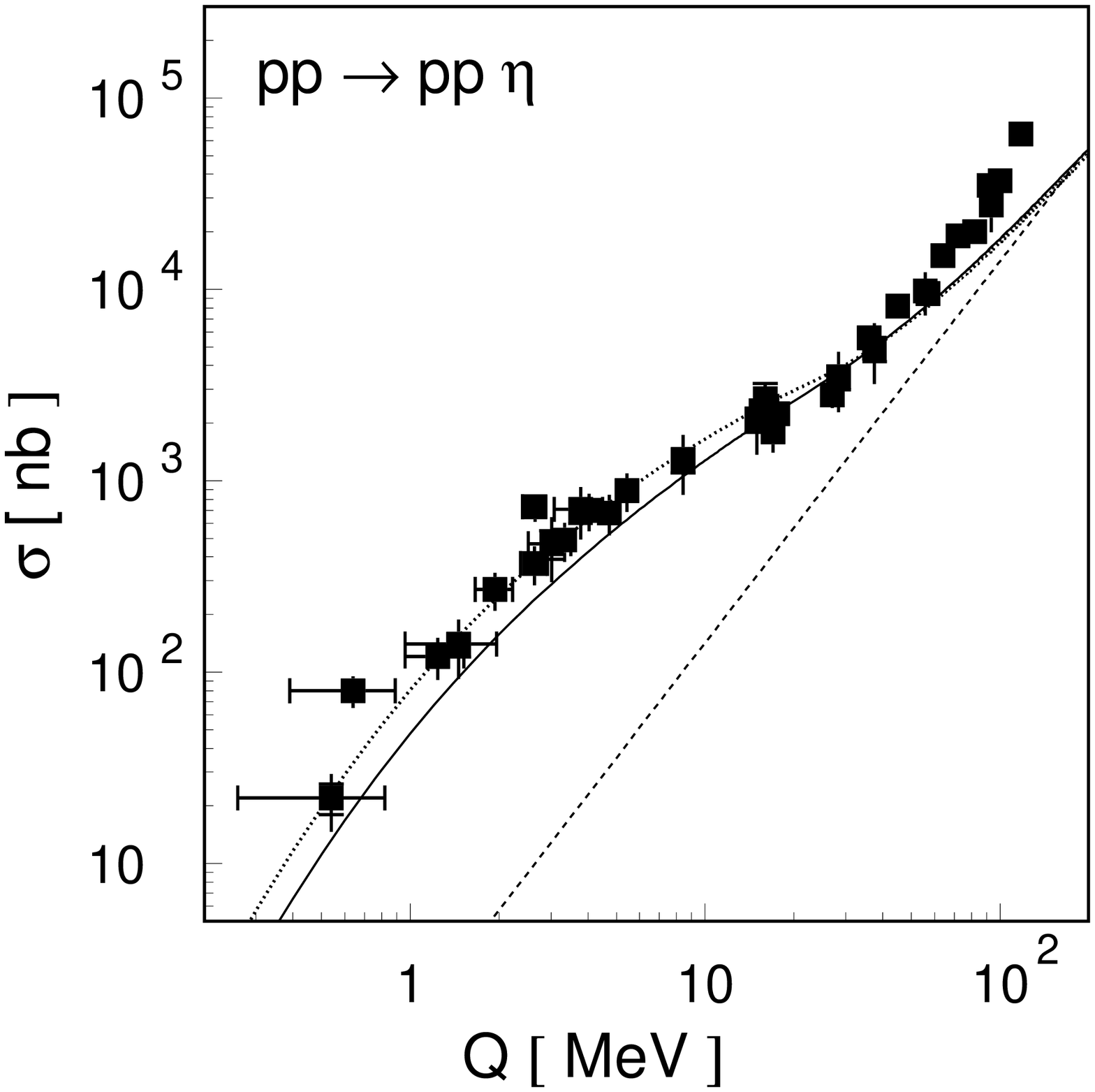,width=0.31\textwidth}} \hfill
\parbox{0.32\textwidth}
  {\epsfig{file=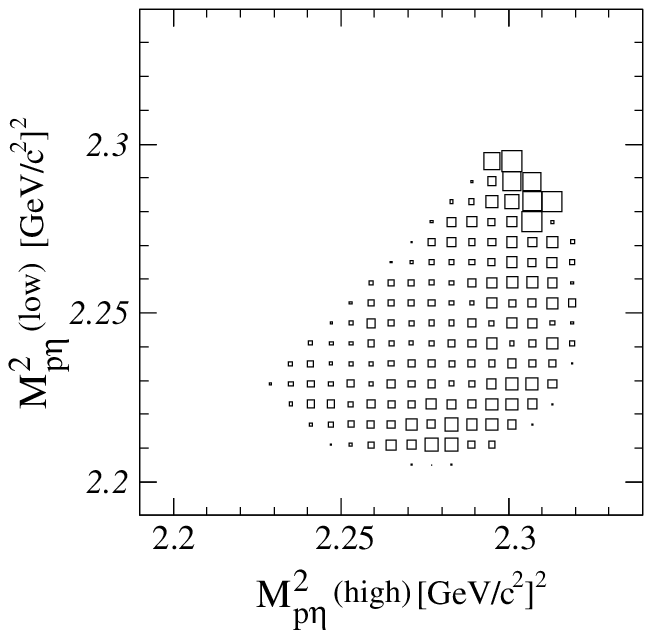,width=0.32\textwidth}} \hfill
\parbox{0.33\textwidth}
  {\epsfig{file=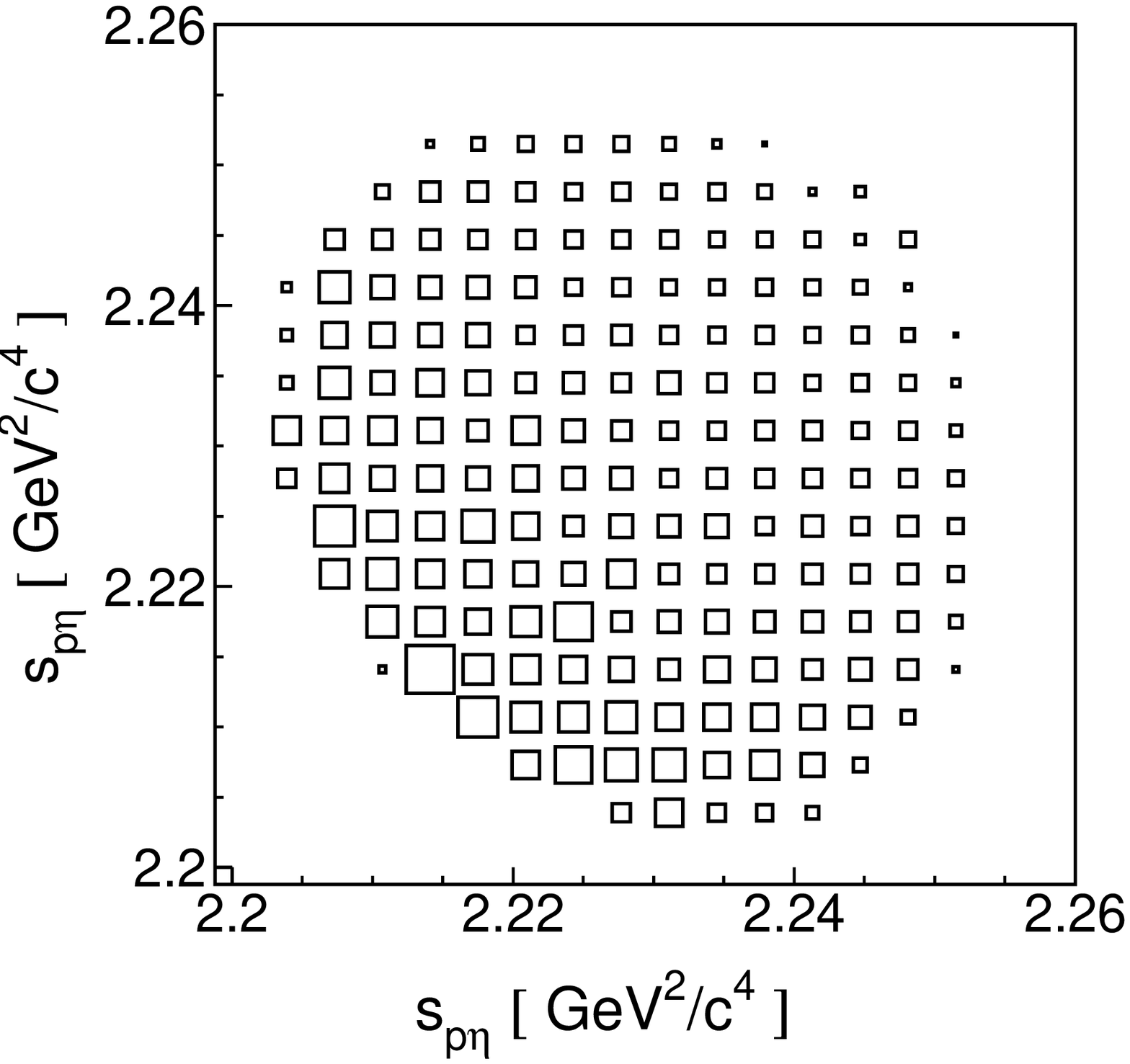,width=0.33\textwidth}}
 
\parbox{0.28\textwidth}{\raisebox{2ex}[0ex][0ex]{\mbox{}}} \hfill
\parbox{0.33\textwidth}{\raisebox{2ex}[0ex][0ex]{\large a)}} \hfill
\parbox{0.32\textwidth}{\raisebox{2ex}[0ex][0ex]{\large b)}} \hfill
\parbox{0.04\textwidth}{\raisebox{2ex}[0ex][0ex]{\large c)}}

\caption{\label{cross_eta_fig} (a) Total cross section for the $pp \rightarrow 
pp \eta$ reaction as a function of the centre--of--mass excess energy Q. Data 
are from refs.~\cite{hibou41,bergdoltR2969,chiavassa270,calen39,smyrski182,
eta_menu,calen2642}. Statistical and systematical errors are separated by 
dashes. The dashed line indicates a phase--space integral normalized 
arbitrarily. The phase--space distribution with inclusion of proton--proton 
strong and Coulomb interactions fitted to the data in the excess energy range 
between 15 and $40\,\mbox{MeV}$ is shown as the solid line. Additional 
inclusion of the proton--$\eta$ interaction is indicated by the dotted line. 
The scattering length $a_{p\eta} = 0.7\,\mbox{fm} + i\,0.4\,\mbox{fm}$ and the 
effective range parameter $b_{p\eta} = -1.50\,\mbox{fm} - 
i\,0.24\,\mbox{fm}$~\cite{greenR2167} have been arbitrarily chosen for the 
calculations described in the text.
(b) Dalitz plot distribution of the $pp \rightarrow pp\eta$ reaction at an 
excess energy of $\mbox{Q} = 37.6\,\mbox{MeV}$, corrected for the detection 
acceptance. Out of the two invariant masses corresponding to two $p-\eta$ 
pairs the one being larger is plotted along the x--axis. The figure is taken 
from~\cite{calen39} (A similar spectrum for the $pp \pi^0$ 
system~\cite{bondar8} reveals the influence of the proton--proton interaction, 
yet again, as for the total cross section energy dependence 
(fig.~\ref{Mpppp_cross_pi}b), the proton--$\pi^0$ interaction is too weak to 
affect observably the density distribution of the Dalitz plot.).
(c) Dalitz plot distribution of the $pp \rightarrow pp \eta$ reaction at 
$\mbox{Q} = 15.5\,\mbox{MeV}$, corrected for the detection acceptance and the 
proton--proton FSI~\cite{eta_menu}. The proton--proton FSI enhancement factor 
was calculated as square of the on--shell proton--proton scattering amplitude 
as written in equation~\eqref{Mpppp}.}
\end{figure}

In the case of $\eta$ meson production the interaction between nucleons is 
evidently not sufficient to describe the increase of the total cross section 
for very low and very high excess energies, as can be seen in 
figure~\ref{cross_eta_fig}a.
The solid line in the figure was normalized to the data at an excess energy 
range between $15\,\mbox{MeV}$ and $40\,\mbox{MeV}$. 
The increase of the total cross section for higher energies can be assigned to 
the outset of higher partial waves.
As expected from the previous considerations, this is indeed seen at 
$\mbox{Q} \approx 40\,\mbox{MeV}$ where the energy dependence of the total 
cross section starts to change its shape.
On the contrary, the close--to--threshold enhancement --- being by about 
factor of two larger than in the case of the $\pi^0$ and $\eta^{\prime}$ 
mesons --- can be assigned neither to the contribution from other than Ss 
partial waves nor to the variation of the primary production amplitude 
$|M_{0}|$. 
The latter is expected to change at most by a few per cent for excess energies 
below $20\,\mbox{MeV}$~\cite{moalem649}.
Instead, this discrepancy can be plausibly explained by the influence of the 
attractive interaction between the $\eta$ meson and the proton.
Note, that the real part of the scattering length of the $\eta$--proton 
potential --- depending on the analysis method and the studied reaction --- is 
3 to 10 times~\cite{green053,sibirtsev086} larger than the scattering length 
for $\pi^0$--proton scattering ($a_{p\pi} \approx 
0.13\,\mbox{fm}$)~\cite{sigg269}.
Hence, the modifications of the total cross section energy dependence due to 
$\pi^0$--proton interaction are too weak to be observed within the 
up--to--date accuracy of measurements and calculations. 
However, the influence of the $\eta$--proton interaction presented in 
figure~\ref{cross_eta_fig} is evident and hereafter it will be considered 
whether it may serve for the estimation of the $\eta$--proton scattering 
parameters.

The strict quantitative calculation requires, however, the evaluation of the 
three--body Faddeev equation~\footnote{An exact derivation of the Faddeev 
equation can be found for example in~\cite{gloeckle107}.}.
Here, we will rather present a simple phenomenological treatment which shall 
lead to the qualitative understanding how the mutual interaction among three 
outgoing particles affects the total cross section dependence on excess energy.
One of the simplest possibilities based on the naive probabilistic 
interpretation of the incoherent pairwise interaction would be to factorize 
the overall enhancement factor into corresponding pair 
interactions~\cite{schuberthPHD,bernard259}:
\begin{equation} 
\label{pairinteraction} 
|M_{FSI}|^2 \;=\; |M_{12 \rightarrow 12}|^2 \cdot |M_{13 \rightarrow 13}|^2 
  \cdot |M_{23 \rightarrow 23}|^2,
\end{equation}
where $|M_{ij \rightarrow ij}|^2$ denotes the square of the elastic scattering 
amplitude of particles $i$ and $j$. 
The $|M_{pp\to pp}|^2$ can be evaluated according to the formula~\eqref{Mpppp}, 
which for the s--wave $\eta$--proton scattering, after substitution of $C^2 
= 1$ and $\eta_c = P_{pp} = 0$, reduces to~\footnote{Note that the sign of the 
term $-1/a$ from equation~\eqref{Mpppp} was changed because the imaginary part 
of the proton--$\eta$ scattering length is positive~\cite{greenR2167}, as we 
also adopted here, whereas in the majority of works concerning 
nucleon--nucleon interaction, the scattering length is 
negative~\cite{machleidtR69}. We are grateful for this remark to A. Gasparyan.}:
\begin{equation} 
\label{Mpeta}
|M_{p\eta \rightarrow p\eta}|^2 \;=\; 
  \left| \frac{1}
    {\frac{1}{a_{p\eta}}\,+\,
     \frac{b_{p\eta}\,\mbox{\scriptsize k}_{p\eta}^2}{2} - i\,\mbox{k}_{p\eta}}
  \right|^2,
\end{equation}
where
\begin{equation*}
\mbox{k}_{p\eta} \;=\;
  \frac{\sqrt{\lambda(\mbox{s}_{p\eta},\mbox{m}_{\eta}^2,\mbox{m}_p^2)}}
       {2\,\sqrt{\mbox{s}_{p\eta}}} 
\end{equation*}
denotes the $\eta$ momentum in the proton--$\eta$ rest frame. 
The scattering length $a_{p\eta}$ and effective range $b_{p\eta}$ are complex 
variables with the imaginary part responsible e.g.\ for the $p\eta 
\rightarrow p \pi^0$ conversion.\vspace{1ex}

\begin{figure}[H]
\vspace{-0.7cm}
\begin{center}
\epsfig{file=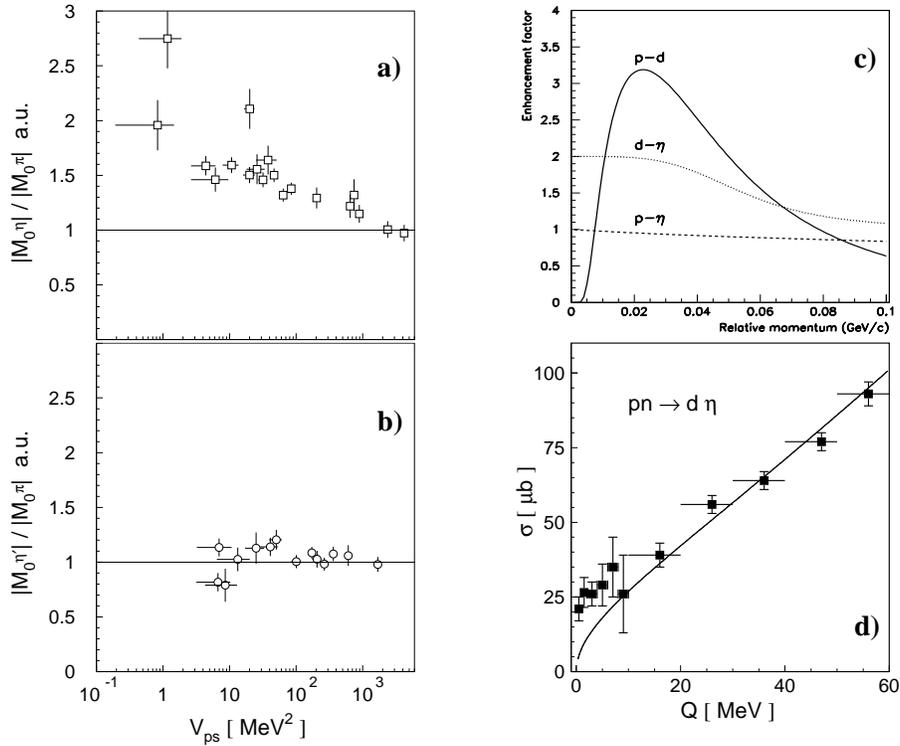,scale=0.71}
\end{center}
\caption{\label{mmnisk}\label{ratio_and_factors} The ratios of \hspace{1ex}
a) $|M^{\eta}_0|/|M^{\pi^0}_0|$ and \hspace{1ex} 
b) $|M^{\eta^{\prime}}_0|/|M^{\pi^0}_0|$ extracted from the data by means of 
equation~\eqref{cross_with_FSI}, assuming a $pp$--FSI enhancement factor as 
depicted by the dotted line in figure~\ref{Mpppp_cross_pi}a and neglecting 
the proton--meson interaction~\cite{swave}.
c) Arbitrarily normalized enhancement factors for $pd$--, $d \eta$-- and 
$p \eta$--FSI. The $\eta$--proton factor is calculated according to 
equation~\eqref{Mpeta}, with $a_{p \eta} = 0.717\,\mbox{fm} + 
i\,0.263\,\mbox{fm}$~\cite{Batinic023} and $b_{p \eta} = -1.50\,\mbox{fm} - 
i\,0.24\,\mbox{fm}$~\cite{greenR2167}. The enhancement factor for 
$\eta$--deuteron has been extracted from the data of panel~d) parametrizing 
the ratio of the cross section to the phase--space volume by the 
expression~\cite{smyrskic11proc}: $F_{d \eta} = 1 + 0.5/(0.5 + (\mbox{Q}/5)^2)$.
The proton--deuteron FSI factor is calculated according to 
ref.~\cite{meyer2474}.
The figure is taken from~\cite{smyrskic11proc}.
d) Total cross section of the quasi--free $pn \rightarrow d\eta$ reaction as a 
function of the excess energy~\cite{calen2069,calen2642}. The curve indicates 
the energy dependence proportional to the function $\sqrt{\mbox{Q}} \cdot 
\left(1 + \mbox{Q}/83.5 \right)$. The relative s-- and p--wave contribution 
was taken as determined for the $np \rightarrow d \pi^0$ 
reaction~\cite{bilger64,hutcheon176,hutcheon618}. It was assumed that the 
p--wave to s--wave ratio is the same for the $pn \rightarrow d \eta$ and $pn 
\rightarrow d \pi^0$ reactions at a corresponding value of $\eta_M$.} 
\end{figure}

The factorization of both \hspace{1ex}i) the overall production matrix element 
(eq.~\eqref{M0FSIISI}) and \hspace{1ex} ii) the three particle final state 
interactions (eq.~\eqref{pairinteraction}) applied to the 
formula~\eqref{Vpsdalitz} gives the following expression for the total cross 
section of the $pp \rightarrow pp \eta$ reaction:
\begin{equation} 
\label{cross_with_FSI}
\sigma \;=\; \frac{F_{ISI}\,|M_0|^2}{\mbox{F}}\,\frac{\pi^2}{4\,\mbox{s}} 
  \int\limits_{(\mbox{\scriptsize m}_p+\mbox{\scriptsize m}_p)^2}^{
               (\sqrt{\mbox{\scriptsize s}}-\mbox{\scriptsize m}_{\eta})^2} 
    d\,\mbox{s}_{pp}\;
  \int\limits_{\mbox{\scriptsize s}_{p_2 \eta}^{min} 
                 (\mbox{\scriptsize s}_{pp})}^{
               \mbox{\scriptsize s}_{p_2 \eta}^{max}
                 (\mbox{\scriptsize s}_{pp})} d\,\mbox{s}_{p_2 \eta} \;
  |M_{pp \rightarrow pp}(\mbox{s}_{pp})|^2 \cdot 
  |M_{p_1 \eta \rightarrow p_1 \eta} (\mbox{s}_{p_1 \eta})|^2 \cdot 
  |M_{p_2 \eta \rightarrow p_2 \eta} (\mbox{s}_{p_2 \eta})|^2,
\end{equation}
where the protons are distinguished by subscripts.
Exploring formulas~\eqref{Mpppp} and~\eqref{Mpeta} gives the results shown as 
the dotted line in figure~\ref{cross_eta_fig}a. 
Evidently, the inclusion of the proton--$\eta$ interaction enhances the total 
cross section close--to--threshold by about a factor of 1.5 and leads to a 
better description of the data.
The effect of the proton--$\eta$ interaction is also seen in the experimental 
distribution of the differential cross sections 
$d^2\,\sigma/(d\,\mbox{s}_{p_1 \eta}\:d\,\mbox{s}_{p_2 \eta})$ shown in 
figures~\ref{cross_eta_fig}b and~\ref{cross_eta_fig}c.
These distributions originating from kinematically complete measurements 
comprise the whole experimentally available information about the interactions 
of the $pp\eta$--system. 
In figure~\ref{cross_eta_fig}b one recognizes the increase of the distribution 
density at regions of small invariant masses of the proton--proton and 
proton--$\eta$ subsystems. 
At this excess energy ($\mbox{Q} = 37.6\,\mbox{MeV}$) these regions are quite 
well separated. 
However, since this is close to the energy where the advent of higher partial 
waves is awaited, the possible contribution from the P--wave proton--proton 
interaction can not be a priori excluded. 
Specifically, the latter leads to the enhancement at large invariant masses of 
the proton--proton pair~\cite{dyringPHD} and hence affects the phase--space 
region where the modification from the proton--$\eta$ interaction 
is expected.
Figure~\ref{cross_eta_fig}c presents the high statistics data taken at an 
excess energy of $\mbox{Q} = 15.5\,\mbox{MeV}$, where the assumption of the 
total dominance of the Ss partial wave is rather save. 
At this excess energy, corresponding to the small relative momentum range
$\mbox{k}_{p \eta}^{max} \approx 105\,\mbox{MeV/c}$, the variations of the 
proton--$\eta$ scattering amplitude are quite moderate 
(fig.~\ref{ratio_and_factors}c) and as expected from simulations presented in 
the previous section, the increase of the phase--space population due to the 
$\eta$--meson interaction with one proton is not separated from the other one. 
Interestingly, the density growth with increasing invariant mass of the 
proton--proton system is much faster than expected from the simulations 
presented in figure~\ref{dalitz_examples}d, which have been performed with a 
scattering length equal to $a_{p \eta} = 0.7\,\mbox{fm} +  i\,0.3\,\mbox{fm}$.
However, when reducing the proton--proton FSI effect to a multiplicative 
factor, one finds that it depends on the assumed nucleon--nucleon potential 
and on the produced meson mass~\cite{baru579}. 

This issue was recently vigorously investigated e.g.\ by authors of 
references~\cite{baru579,kle01,hanhart176,niskanen107,nak01} and we shall 
briefly report this here as well.
Up to now we factorized the transition matrix element into a primary 
production of particles and its on--shell rescattering in the exit channel 
(eqs.~\eqref{M0FSIISI}\eqref{FSI_elastic}\eqref{pairinteraction}).
Though it is a crude approximation, neglecting the off--shell effects of the 
production process completely, it astoundingly leads to a good description of 
the energy dependence of the total cross section, as already demonstrated in 
figures~\ref{Mpppp_cross_pi}b,~\ref{cross_etap_pkl}a,b and~\ref{cross_eta_fig}a.
The off--shell effects, as pointed out by Kleefeld~\cite{kle01}, could have 
been safely neglected in case of the electromagnetic transitions in atoms or 
$\beta$ decays, where the excitation energy of the involved nucleons is by 
many orders of magnitude smaller than their masses and the initial and final 
states go hardly off--shell~\cite{kle01}.
However, in case of the $NN \rightarrow NN\,Meson$ process the large 
excitation energy of the colliding nucleons is comparable with the nucleon 
masses and the primary interaction may create the particles significantly far 
from their physical masses, so that the off--shell effects can not be a priori 
disregarded.
\vspace{0.1cm}
\begin{figure}[H]
\begin{center}
\parbox{0.9\textwidth}
  {\epsfig{file=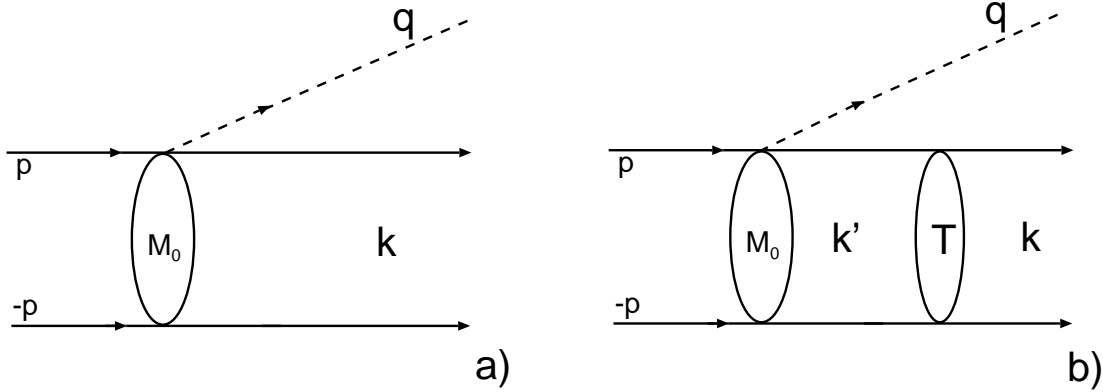,width=0.8\textwidth}}
\end{center}
\caption{\label{loop} Diagrammatic representation of the DWBA expressed by 
equation~\eqref{eqbaru1}. (a) The primary production term. (b) The loop 
diagram including the nucleon--nucleon FSI. $T(\mbox{k}^{\prime},\mbox{k})$ 
stands for the half--off--shell $T$ matrix with k and $\mbox{k}^{\prime}$ 
denoting the on-- and off--shell centre--of--mass momentum in the 
nucleon--nucleon system, respectively. q indicates the momentum of the created 
meson in the reaction centre--of--mass and p the momentum of the colliding 
nucleons.}
\end{figure}

\begin{figure}[H]
\parbox{0.32\textwidth}
  {\epsfig{file=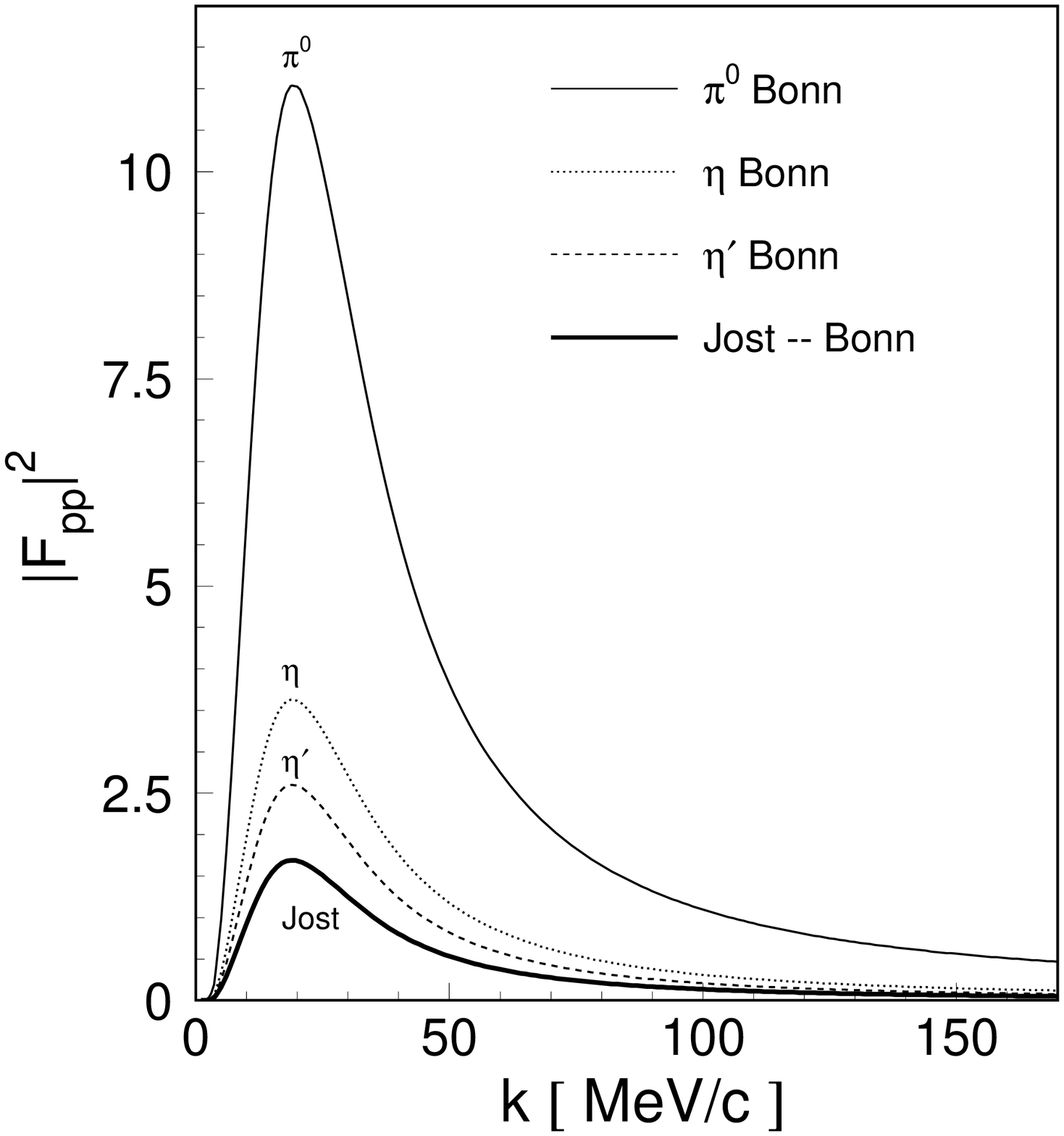,width=0.32\textwidth}} \hfill
\parbox{0.32\textwidth}
  {\epsfig{file=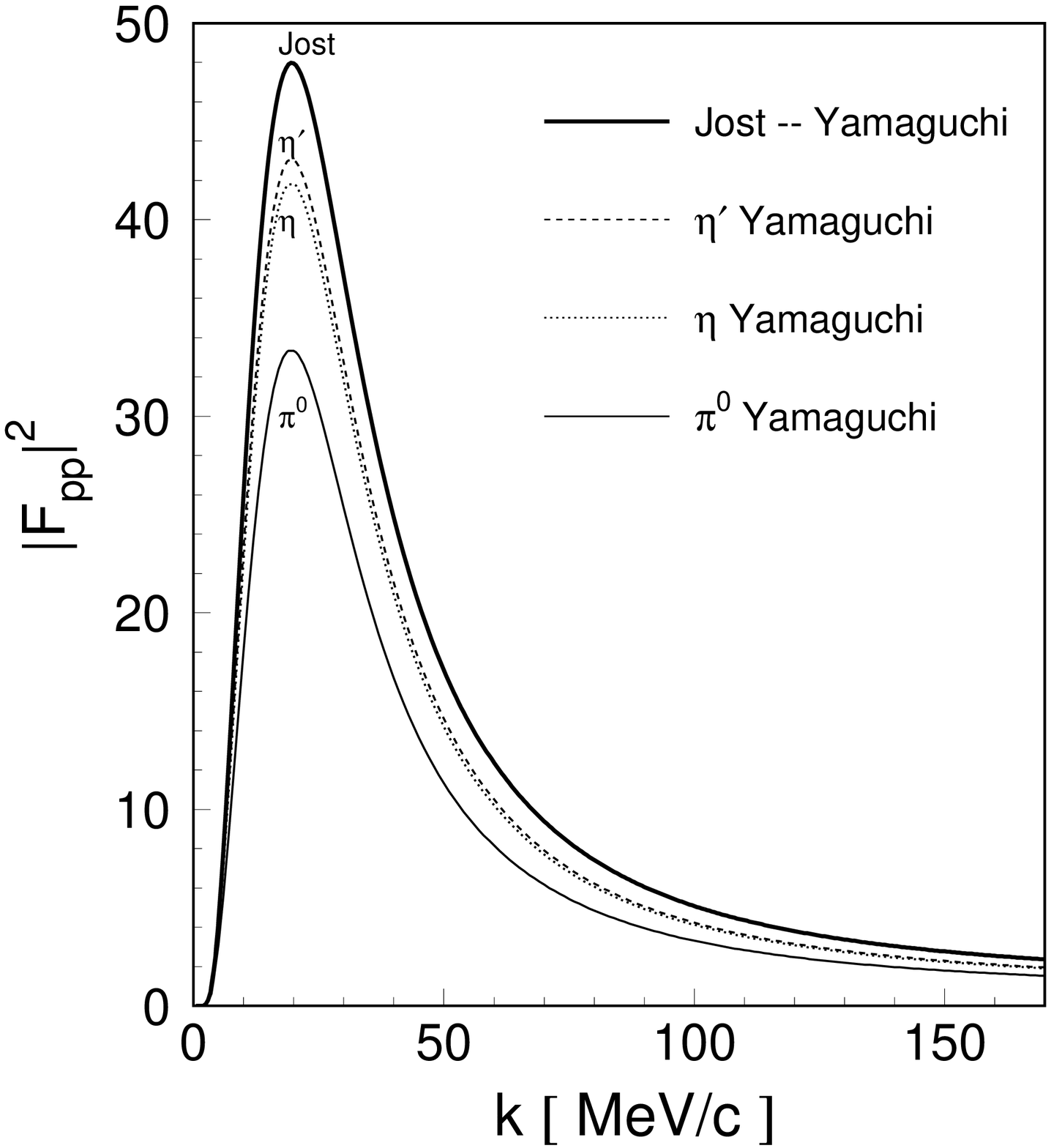,width=0.32\textwidth}} \hfill
\parbox{0.32\textwidth}
  {\epsfig{file=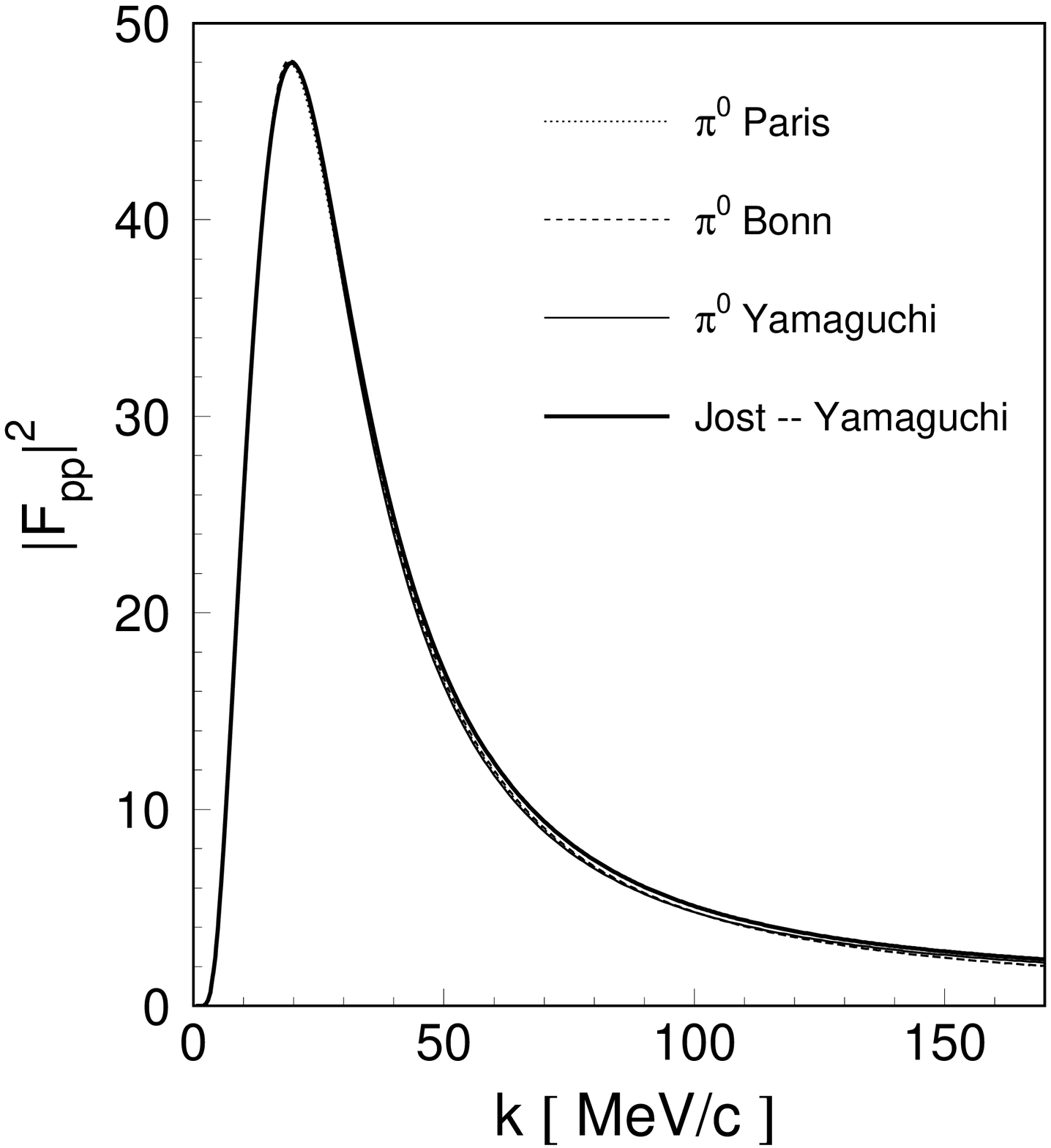,width=0.32\textwidth}}
\parbox{0.28\textwidth}{\raisebox{0ex}[0ex][0ex]{\mbox{}}} \hfill
\parbox{0.33\textwidth}{\raisebox{0ex}[0ex][0ex]{\large a)}} \hfill
\parbox{0.335\textwidth}{\raisebox{0ex}[0ex][0ex]{\large b)}} \hfill
\parbox{0.025\textwidth}{\raisebox{0ex}[0ex][0ex]{\large c)}}
\caption{\label{baruFSI} The FSI factor for Bonn~\cite{haidenbauer2190} a) and 
Yamaguchi~\cite{haeringen355} b) potentials. The solid, dotted and dashed 
lines correspond to $\pi^0$, $\eta$ and $\eta^{\prime}$ meson production, 
respectively. Thick solid lines indicate the inverse of the squared Jost 
function ($|F_{pp}(\mbox{k})|^2 = |J(-\mbox{k})|^{-2}$). 
c) The FSI factors for $\pi^0$ production of Paris (dotted 
curve)~\cite{lacombe861}, Bonn (dashed curve) and Yamaguchi (solid curve) 
potentials normalized to be equal at maximum to the inverse of the squared 
Jost function of the Yamaguchi potential (thick solid line). The shapes 
stemming from different potentials are almost indistinguishable.
Note that the thick solid line corresponds to the dashed line in 
figure~\ref{Mpppp_cross_pi}a. The figures are adapted from 
reference~\cite{baru579}.}
\end{figure}

Generally, the decomposition of the total production amplitude into the 
primary production and the subsequent nucleon--nucleon interaction visualized 
in figure~\ref{loop} is expressed by the formula:
\begin{equation} 
\label{eqbaru1}
M \;=\; M_0^{on} \:+\:  M_0^{off}\,G\,T_{NN},
\end{equation}
where the second term of the equation represents the integration over the 
intermediate ($\mbox{k}^{\prime}$) momenta of the off--shell production 
amplitude and the half--off--shell nucleon--nucleon $T$ matrix~\cite{baru579}.
Assuming that the primary production occurs in such a way that one nucleon 
emits the meson which then re--scatters on the other nucleon and appears as a 
real particle the authors of reference~\cite{baru579} found that the 
enhancement of the cross section due to the nucleon--nucleon interaction 
depends strongly on the mass of the created meson. 
This is because with the increasing mass of the produced meson the distance 
probed by the nucleon--nucleon interaction decreases (see 
table~\ref{momtranstable}) and hence the relevant range of the off--shell 
momenta becomes larger.
The effect for the $pp \rightarrow pp\,Meson$ reactions is presented in 
figure~\ref{baruFSI}a, where one can see that, when utilizing the Bonn 
potential model for the nucleon--nucleon $T$ matrix, the enhancement in case 
of the $\pi^0$ production is by about a factor of four larger than for the 
$\eta$ or $\eta^{\prime}$ mesons.
A similar conclusion, but with the absolute values larger by about $40\,\%$, 
was drawn for the Paris $NN$ potential~\cite{baru579}.
On the  contrary, when applying the Yamaguchi potential into calculations the 
enhancement grows with the increasing mass of the meson, as shown in 
figure~\ref{baruFSI}b. 
Thick solid curves in figures~\ref{baruFSI}a and~\ref{baruFSI}b show the 
results of the frequently applied approximation of the nucleon--nucleon FSI 
effects:
\begin{equation} 
\label{eqbaru2}
M \;=\; M_0^{on} \:+\:  M_0^{off}\,G\,T_{NN} \;\approx\; 
        M_0^{on} \cdot (1 \:+\: G\,T_{NN}) \;=\; M_0^{on}\,J^{-1}(-\mbox{k})
  \;\equiv\; M_0^{on}\,F_{NN}(k) ,
\end{equation}
where the overall transition matrix element $M$ is factorized to the primary 
on--shell production and the $NN$ FSI  expressed as the inverse of the Jost 
function $J^{-1}(-\mbox{k})$~\cite{goldbergerwatson}.
As can be seen in figures~\ref{baruFSI}a and~\ref{baruFSI}b, the variation of 
the absolute values --- of such obtained enhancement factors --- with the 
applied potential is severe. \\

Since the physical value of the total cross section can not depend on the 
off--shell features of the potential used in calculations, which are in 
principal not measurable~\cite{fearing758}, the differences in the magnitude 
of the $|F_{pp}|$ factor must reflect itself in a corresponding dependence of 
the primary production amplitude on the potential used. 
Therefore, it is of crucial importance to realize that the values of the 
threshold amplitudes $|M_0|$ are significant only in the context of the 
potential they were extracted from.
Figure~\ref{baruFSI}c demonstrates, however, that the shapes of the 
enhancement factors, with the meson exchange mechanism assumed for the primary 
production, are pretty much the same, independently of the applied $NN$ 
potential and correspond to the form of the Jost function inferred from the 
Yamaguchi potential.
This indicates that the energy dependence of the $NN$ FSI factors is 
predominantly determined by the on--shell $NN$ $T$ matrix.

In references~\cite{hanhart176,nak01,kle01} the formula for the transition 
matrix element with explicit dependence on the considered off--shell features 
for the initial and final state interaction is derived:
\begin{equation} 
M \:=\: \left\{1 \,+\, \frac{1}{2} 
  \left[\eta(\mbox{k}) e^{2i\delta(\mbox{\scriptsize k})} \,-\, 1 \right] 
  \cdot \left[ 1 \,+\, P_f(\mbox{p},\mbox{k})\right] \right\}  
  \: M_0 \: \left\{1 \,+\, \frac{1}{2}
  \left[\eta(\mbox{p}) e^{2i\delta(\mbox{\scriptsize p})} \,-\, 1 \right]
  \cdot \left[ 1 \,+\, P_i(\mbox{p},\mbox{k})\right] \right\},
\end{equation}
where subscripts $i$ and $f$ indicate the initial and final state, 
respectively. 
The functions $P(\mbox{p},\mbox{k})$ exhibit all the off--shell effects of the 
$NN$ interaction and the primary production current~\cite{nak01} and $\delta$ 
and $\eta$ denote the phase--shift and inelasticity, correspondingly.
At threshold, the inelasticity in the exit channel is equal to unity 
($\eta(\mbox{k}) = 1$) due to the small relative momentum of the outgoing 
nucleons.
The last term of the formula expresses the influence of the initial state 
interaction on the production process. 
Due to the large relative momenta of the colliding protons needed to create a 
meson it is characterized by a weak energy dependence in the excess energy 
range of a few tens of MeV.
For example in figure~\ref{1s03p0_eta_Q_Vps}a one can see that the phase--shift 
variation of the $^3\mbox{P}_0$ partial wave (having predominant contribution 
to the threshold production of pseudoscalar mesons) in the vicinity of 
threshold for mesons heavier than $\pi^0$ is indeed very weak.
Taking additionally into account that the initial state off--shell function 
$P_i(\mbox{p},\mbox{k})$ is small (at least it is the case for meson exchange 
models~\cite{hanhart176}) one reduces the influence of the $NN$ initial state 
interaction to the reduction factor $F_{ISI}$ which can be estimated from the 
phase--shifts and inelasticities only:
\begin{equation} 
\label{F_ISI}
F_{ISI} \;=\; \frac{1}{4} \:\left| 
  \,\eta(\mbox{p}) \,e^{2i\delta(\mbox{\scriptsize p})} \,+\, 1 \:\right|^2. 
\end{equation}

At the threshold for $\pi$ meson production this is close to unity since at 
this energy the inelasticity is still nearly 1 and the $^3\mbox{P}_0$ 
phase--shift is close to zero (see figure~\ref{1s03p0_eta_Q_Vps}a). 
However, at the $\eta$ threshold, where the phase--shift approaches its 
minimum, the proton--proton ISI diminishes the total cross section already by 
a factor of 0.2~\cite{hanhart176}. 
A similar result was obtained using a meson exchange model for $\eta$ 
production in the $pp \rightarrow pp \eta$ reaction and calculating the 
proton--proton distortion from the coupled--channel $\pi NN$ 
model~\cite{Batinic}.
The authors of reference~\cite{Batinic} concluded that the initial 
proton--proton distortion reduces the total cross section by about a factor of 
$\approx 0.26$, which keeps constant at least in the studied range of 
$100\,\mbox{MeV}$ in kinetic beam energy.
Hence, the closed formula~\eqref{F_ISI} disregarding the off--shell effects 
($P_i(\mbox{p},\mbox{k})$) permits to estimate the cross section reduction due 
to the initial state distortion with an accuracy of about $25\,\%$. 
The shape of the $^3\mbox{P}_0$ phase--shift shown in 
figure~\ref{1s03p0_eta_Q_Vps}a indicates that the effect is at most pronounced 
close to the $\eta$ production threshold, yet for the $\eta^{\prime}$ meson 
the formula~\eqref{F_ISI} leads to a factor $F_{ISI} = 
0.33$~\cite{nakayama024001}.
The primary production amplitude as well as the off--shell effects of the 
nucleon--nucleon FSI ($P_{f}(\mbox{p},\mbox{k})$) are also weakly energy 
dependent~\cite{nak01}, since they account for the short range creation 
mechanism, which shall be considered in the next section.
The accordance of the experimental data with the simple factorization 
represented by solid lines in figures~\ref{Mpppp_cross_pi}b, 
\ref{cross_etap_pkl} and~\ref{cross_eta_fig}a fully confirms the above 
considerations which imply that the total cross section energy dependence and 
the occupation density on a Dalitz plot is in the first order determined by 
the on--shell scattering of the outgoing particles. 
However, since the distortion caused by the nucleons is by some orders of 
magnitude larger than that due to the meson--nucleon interaction, even small 
fractional inaccuracies in the description of nucleon--nucleon effects may 
obscure the inference on the meson--nucleon interaction.\vspace{1ex}

The differences between the square of the on--shell proton--proton scattering 
amplitude and the Jost function prescription are presented in 
figure~\ref{Mpppp_cross_pi}a.
To minimize the ambiguities that may result from these discrepancies at least 
for the quantitative estimation of the effects of the unknown meson--nucleon 
interaction one can compare the spectra from the production of one meson to 
the spectra determined for the production of a meson whose interaction with 
nucleons is established.
For instance, to estimate the strength of the $\eta pp$ and $\eta^{\prime} pp$ 
FSI one can compare the appropriate observables to those of the $\pi^0 pp$ 
system.

Figures~\ref{ratio_and_factors}a and~\ref{ratio_and_factors}b show the 
dependence of $|M_0|$ on the phase--space volume for $\eta$ and 
$\eta^{\prime}$ production normalized to $|M_0^{\pi^0}|$.
The values of $|M_0|$ were extracted from the experimental data by means of 
equation~\eqref{cross_with_FSI} disregarding the proton--meson interaction 
($|M_{p \eta (\eta^{\prime}) \rightarrow p \eta (\eta^{\prime})}|$ was set 
to 1).
If the influence of the neglected interactions were the same in case of 
$\eta (\eta^{\prime})$ and $\pi^0$ the points would be consistent with the 
solid line. 
This is the case for the $pp \rightarrow pp \eta^{\prime}$ reaction 
visualizing the weakness of the proton--$\eta^{\prime}$ interaction 
independently of the prescription used for the proton--proton FSI~\cite{swave}.
In case of the $\eta^{\prime}$ meson its low--energy interaction with the 
nucleons was expected to be very weak since there exists no baryonic resonance 
which would decay into $N \eta^{\prime}$~\cite{groom1}. 
Figure~\ref{ratio_and_factors}a shows --- independently of the model used for 
the correction of the proton--proton FSI --- the strong effects of the 
$\eta pp$ FSI at low $V_{ps}$.

With the up--to--date experimental accuracy, from all $Meson\,NN$--systems the 
$\eta NN$ one reveals by far the most interesting features.
The dynamics of the $\eta NN$ system has become recently a subject of 
theoretical investigations in view of the possible existence of quasi--bound 
or resonant states~\cite{fix119}.
According to reference~\cite{rakityanskyR2043} --- within the present 
inaccuracy of a$_{\eta N}$ --- the existence of quasi--bound $\eta$--mesic 
light nuclei could be possible.
A direct measure of the formation --- or non--formation --- of an 
$\eta$--nuclear quasi--bound state is the real part of the $\eta$--nucleon 
scattering length~\cite{svarc024}.
The determined values of $\mbox{Re}\,(a_{\eta N}$) range between
$0.20\,\mbox{fm}$~\cite{N.Kaiser-II} and $1.05\,\mbox{fm}$~\cite{green035208}
depending on the analysis method and the reaction studied~\cite{green053},
and at present an univocal answer whether the attractive interaction between 
the $\eta$ meson and nucleons is strong enough to form a quasi--bound state is 
not possible. 

The shape of the energy dependence of the $pd \rightarrow {^3He}\,\eta$ cross 
section implies that either the real or imaginary part of the $\eta\;{^3He}$ 
scattering length has to be very large~\cite{wilkinR938}, which may be 
associated with a bound $\eta\;{^3He}$ system.
Similarly encouraging are results of reference~\cite{shevchenko143}, where it 
is argued that a three--body $\eta NN$ resonant state, which may be formed 
close to the $\eta d$ threshold, may evolve into a quasi--bound state for 
$\mbox{Re}\,(a_{\eta N}) \ge 0.733\,\mbox{fm}$.
Also the close--to--threshold enhancement of the total cross section of the 
$pp \rightarrow pp \eta$ reaction was interpreted as being either a Borromean 
(quasi--bound) or a resonant $\eta pp$ state~\cite{wycech2981} provided that 
$\mbox{Re}\,(a_{\eta N}) \ge 0.7\,\mbox{fm}$.
Contrary, recent calculations performed within a three--body 
formalism~\cite{fix119} indicate that a formation of a three--body $\eta NN$ 
resonance state is not possible, independently of the $\eta N$ scattering 
parameters.
Moreover, the authors of reference~\cite{garcilazo021001} exclude the 
possibility of the existence of an $\eta NN$ quasi-bound state.
Results of both calculations~\cite{fix119,garcilazo021001}, although performed 
within a three--body formalism, have been based on the assumption of a 
separability of the two--body $\eta N$ and $NN$ interactions.
However, in the three--body system characterized by the pairwise attractive 
interactions, the particles can be pulled together so that their two--body 
potentials overlap, which may cause the appearance of qualitatively new 
features in the $\eta NN$ system~\cite{fix119}.
Specifically interesting is the $\eta d$ final state where the pair of 
nucleons alone is bound by the strong interaction.
Figure~\ref{ratio_and_factors}d shows the total cross section for the $pn 
\rightarrow d \eta$ reaction measured close to the production threshold. 
For excess energies below $10\,\mbox{MeV}$ the data are enhanced over the 
energy dependence determined for the $pn \rightarrow d \pi^0$ reaction 
indicated by the solid curve.
This is in qualitative agreement with the calculations of Ueda~\cite{ueda297}
for the three--body $\eta NN$--$\pi NN$ coupled system, which predict the 
existence of an $\eta NN$ quasi--bound state with a width of $20\,\mbox{MeV}$. 
Ueda pointed out that the binding of the $\eta NN$ system is due to the 
$\mbox{S}_{11}$ $\eta N$ and $^{3}\mbox{S}_1$ $NN(d)$ interaction which is 
characterized by no centrifugal repulsion. 
Such repulsion makes the $\pi NN$ system, in spite of the strong 
$\mbox{P}_{33}$ $\pi N$ attraction, hard to be bound~\cite{ueda68}.
Whether the observed cusp at the $pn \rightarrow d \eta$ threshold is large 
enough to confirm the existence of the $\eta NN$ bound state is recently 
vigorously discussed~\cite{wycech045206,deloff024004}.
The considerations concerning the possible existence of the light 
$\eta$--mesic nuclei will be continued in section~\ref{Qbs}.

The enhancement factor for the deuteron--$\eta$ interaction inferred from the 
data in figure~\ref{ratio_and_factors}d varies much stronger in comparison to 
the proton--$\eta$ one, as demonstrated in figure~\ref{ratio_and_factors}c. 
This suggests that the effects of this interaction should be even more 
pronounced in the differential distributions of the cross section for the $pd 
\rightarrow pd \eta$ reaction as those observed in case of the $pp \rightarrow 
pp \eta$ process, especially because the ``screening'' from the 
proton--deuteron interaction is by more than an order of magnitude smaller 
compared to the proton--proton interaction as can be deduced from the 
comparison of the solid lines in figures~\ref{ratio_and_factors}c 
and~\ref{Mpppp_cross_pi}a.
The experimental investigations on that issue~\cite{hibou537,smyrskic11proc}
as well as the search for the $\eta$--mesic nuclei~\cite{gillitzerTOF} by 
measuring proton--deuteron induced reactions in the vicinity of the $\eta$ 
production threshold are on the way.
 
\section{Single meson production in NN scattering}
\label{SmpiNNs}
\subsection{Dynamics of pseudoscalar meson production} 
\label{Dopsmp}    
The considerations from the previous section led to the conclusion, that close 
to the kinematical threshold, the energy dependence of the total cross section 
is in the first approximation determined via the interaction among the 
outgoing particles and that the entire production dynamics manifests itself 
in a single constant,  which determines the absolute scale of the total cross 
section. 
As a first step towards the understanding of the creation mechanism underlying 
the production let us compare the total cross sections of neutral pseudoscalar 
mesons $\pi^0$, $\eta$ and $\eta^{\prime}$.
Since the masses of these mesons are significantly different~\footnote{The 
$\pi^0$, $\eta$ and $\eta^{\prime}$ masses amount to 
$134.98\,\mbox{MeV}/\mbox{c}^2$, $547.30\,\mbox{MeV}/\mbox{c}^2$ and 
$957.78\,\mbox{MeV}/\mbox{c}^2$, respectively~\cite{groom1}.} the influence  
of the kinematical flux factor F on the total cross section and the 
suppression due to the initial state interaction $F_{ISI}$ depend 
substantially on the created meson. 
Therefore, for the comparison of the primary dynamics we will correct for 
these factors and instead of comparing the total cross section we will 
introduce --- according to reference~\cite{swave} --- a dimensionless quantity 
$\sigma \cdot \mbox{F}/F_{ISI}$, which depends only on the primary production 
amplitude $M_0$ and on the final state interaction among the produced 
particles.

\begin{figure}[H]
\parbox{0.49\textwidth}
  {\epsfig{figure=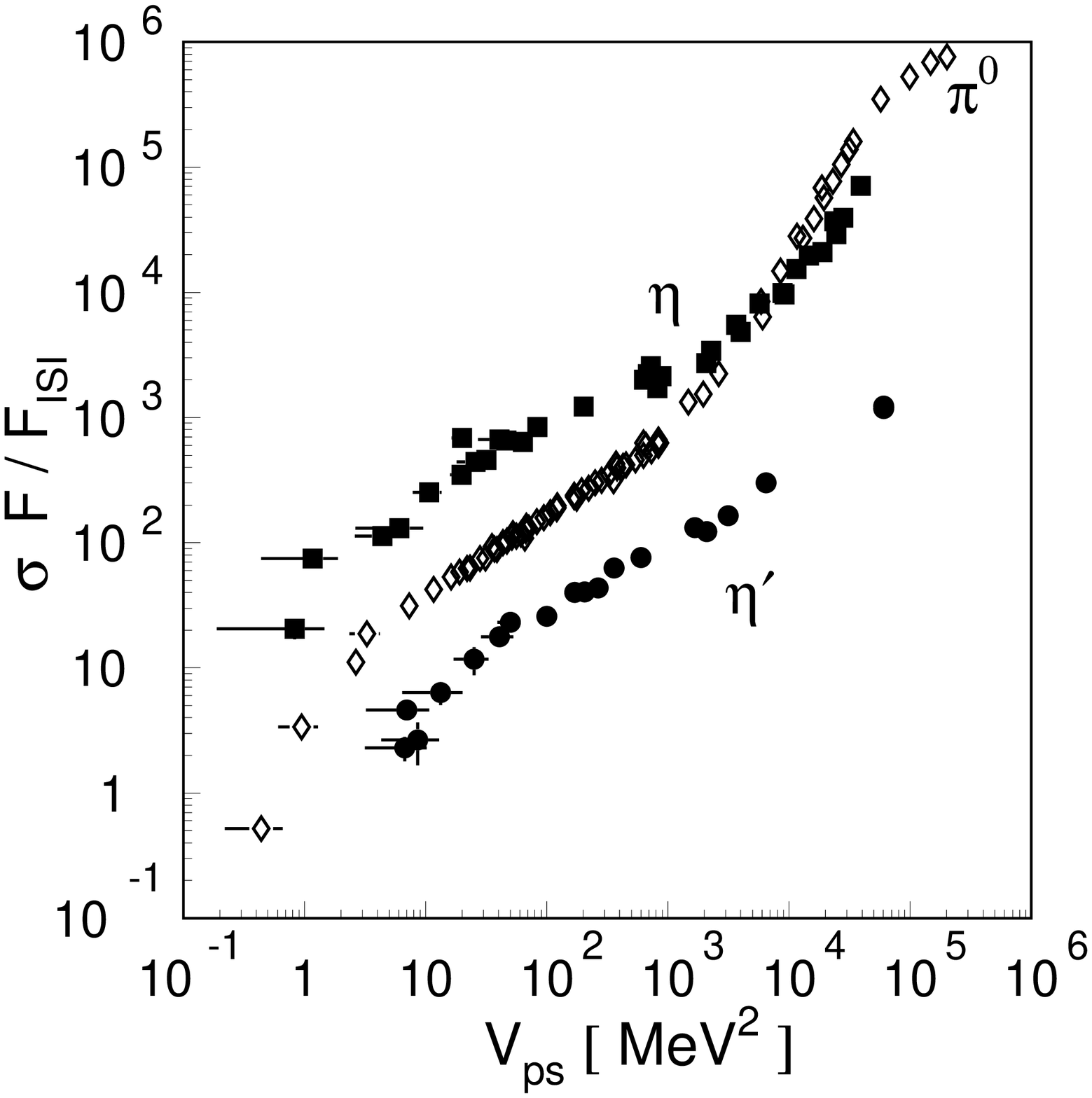,width=0.49\textwidth}} \hfill
\parbox{0.49\textwidth}
  {\epsfig{figure=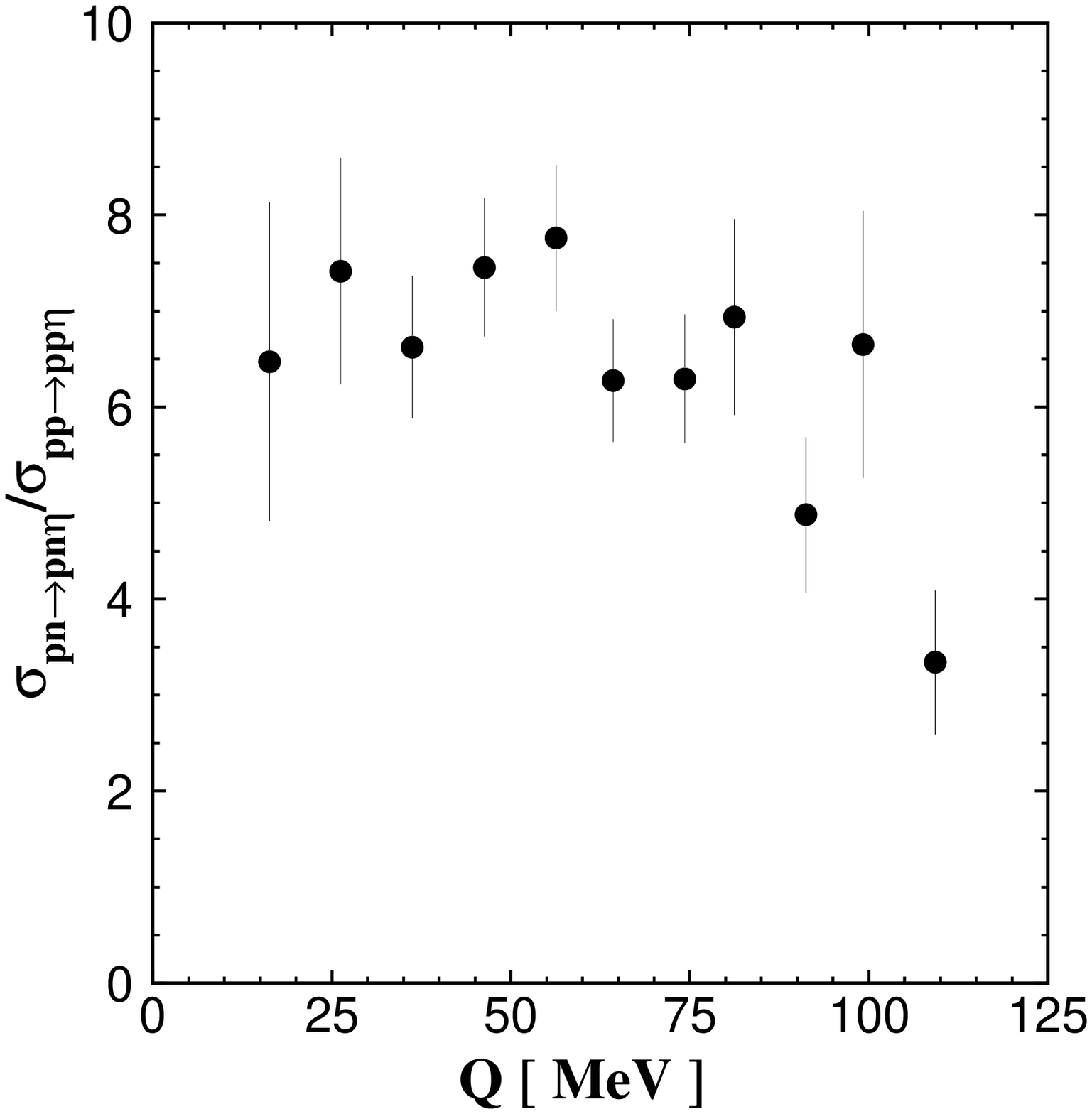,width=0.49\textwidth}}

\parbox{.43\textwidth}{\raisebox{2ex}[0ex][0ex]{\mbox{}}} \hfill
\parbox{.49\textwidth}{\raisebox{2ex}[0ex][0ex]{\large a)}} \hfill
\parbox{.06\textwidth}{\raisebox{2ex}[0ex][0ex]{\large b)}}

\caption{\label{cfisi_Vps} (a) Total cross section ($\sigma$) multiplied by 
the flux factor F and divided by the initial state interaction reduction factor 
$F_{ISI}$ versus the available phase--space volume for the reactions $pp 
\rightarrow pp \eta$ (squares~\cite{hibou41,bergdoltR2969,chiavassa270,calen39,
smyrski182,eta_menu,calen2642}), $pp \rightarrow pp \pi^0$ 
(diamonds~\cite{bondar8,meyer633,stanislausR1913,bilger633,rappenecker763}), 
and $pp \rightarrow pp\eta^{\prime}$ (circles~\cite{hibou41,balestra29,
moskal3202,moskal416,khoukazetap}).\hspace{1ex} (b) Experimentally 
determined ratio between $pn \rightarrow pn \eta$ and $pp \rightarrow pp \eta$ 
cross sections as a function of excess energy Q. The figure is adapted from 
reference~\cite{calen2667}.}
\end{figure}

Close--to--threshold the initial state interaction, which reduces the total 
cross section, is dominated by proton--proton scattering in the 
$^{3}\mbox{P}_0$ state which may be estimated in terms of phase--shifts and 
inelasticities by employing equation~\eqref{F_ISI}. 
The $F_{ISI}$ factor is close to unity for pion production and amounts to 
$\sim 0.2$~\cite{hanhart176} and $\sim 0.33$~\cite{nakayama024001} for the 
$\eta$ and $\eta^{\prime}$ meson, respectively, at threshold.  

A comparative study of the production of mesons with significantly different \
masses encounters the difficulty of finding a proper variable at which the 
observed yield can be compared. 
The total cross section is defined as the integral over the available 
phase--space volume of transition probabilities --- reflecting the dynamics of 
the process --- from the initial to the final state as written explicitly in 
equation~\ref{phasespacegeneral}.
Thus, if the dynamics of the production process of two different mesons were 
exactly the same then the above introduced yield would also be strictly the 
same for both mesons provided it was extracted at the same value of $V_{ps}$. 
This inference would however not be valid if the production yields were 
compared at the $\eta_M$ or Q variables.
Therefore the volume of the available phase--space for the produced particles 
is the most suited quantity for the regarded comparison~\cite{swave}. 
This could be also the best choice for the investigation of isospin breaking 
where the cross sections for the production of particles with different masses 
need to be compared (e.g.\ $\pi^+ d \rightarrow pp \eta$ and $\pi^- d 
\rightarrow nn \eta$~\cite{tippens052001}).

Figure~\ref{cfisi_Vps}a shows the yield of $\pi^0$, $\eta$ and 
$\eta^{\prime}$ mesons in the proton--proton interaction as a function of the 
available phase--space volume. 
The onset of higher partial waves is seen for $\pi^0$ and $\eta$ mesons in the 
$V_{ps}$ range between $10^3$ and $10^4\,\mbox{MeV}^2$, whereas the whole 
range covered by the $\eta^{\prime}$ data seems to be consistent with the pure 
Ss production.
One can also recognize that the data for the Ss final state are grouped on 
parallel lines indicating a dependence according to the power 
law $\sigma \cdot \mbox{F}/F_{ISI} \approx 
\alpha \cdot V_{ps}^{0.61}$~\cite{raport48} and that over the relevant range 
of $V_{ps}$ the dynamics for $\eta^{\prime}$ meson production is about six 
times weaker than for the $\pi^0$ meson, which again is a further factor of 
six weaker than that of the $\eta$ meson.
This is an interesting observation, since the quark wave functions of $\eta$ 
and $\eta^{\prime}$ comprise a similar amount of strangeness 
($\approx 70\,\%$~\cite{moskalphd}) and hence, in the nucleon--nucleon 
collision one would expect both these mesons to be produced much less 
copiously than the meson $\pi^0$ being predominantly built out of {\em up} and 
{\em down} quarks. 
On the hadronic level, however, one can qualitatively argue that the $\eta$ 
meson owes its rich creation in the threshold nucleon--nucleon collisions to 
the existence of the baryonic resonance $N^*(1535)$ whose branching ratio into 
the $N \eta$ system amounts to 30--$55\,\%$~\cite{groom1}.
There is no such established resonance, which may decay into an s--wave 
$\eta^{\prime} N$ system~\cite{groom1} and the $\pi^0$ meson production with 
the formation of the intermediate $\Delta(1232)$ state is strongly suppressed 
close--to--threshold, because of conservation laws.
 
Prior to the more comprehensive discussion of the production mechanism for 
different mesons let us present one more interesting observation visualizing 
that the production dynamics depends strongly not only on the structure of the 
created meson but also on the isospin configuration of the colliding nucleons.

\begin{figure}[H]
\begin{center}
\epsfig{file=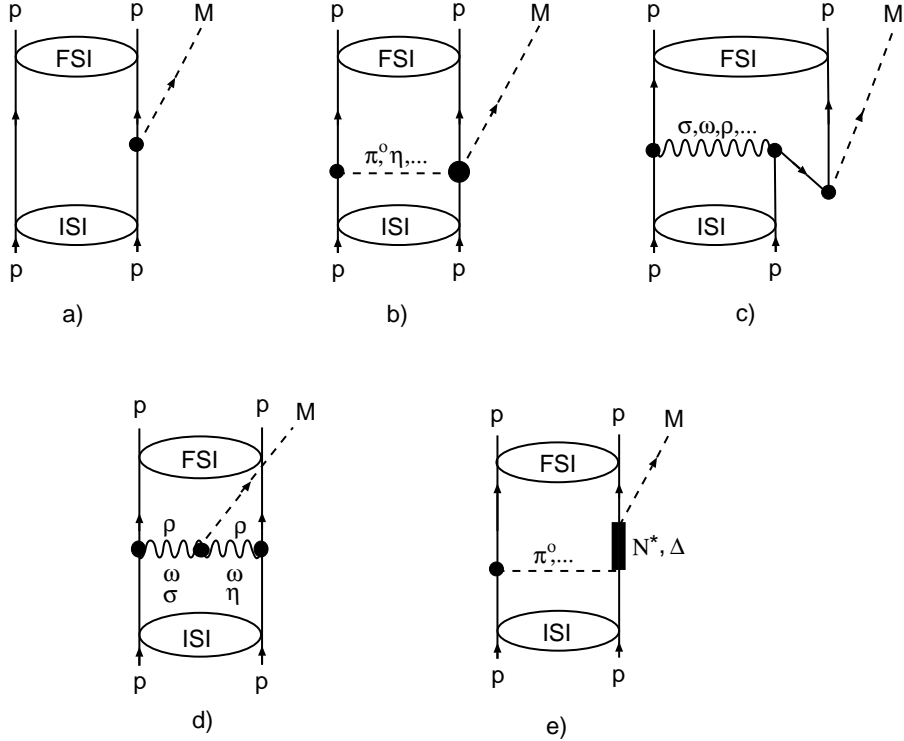,width=0.65\textwidth}
\end{center}

\vspace{-0.9cm}
\caption{\label{graph_mesonexchange} Diagrams for the $pp \rightarrow 
pp\,Meson$ reaction near threshold: 
(a) --- $Meson$--bremsstrahlung (nucleonic current) \hspace{1ex}
(b) --- ``rescattering'' term (nucleonic current) \hspace{1ex}
(c) --- production via heavy--meson--exchange \hspace{1ex}
(d) --- emission from virtual meson (mesonic current) \hspace{1ex}
(e) --- excitation of an intermediate resonance (nucleon resonance current).}
\end{figure}

\begin{figure}[H]
\begin{center}
\epsfig{file=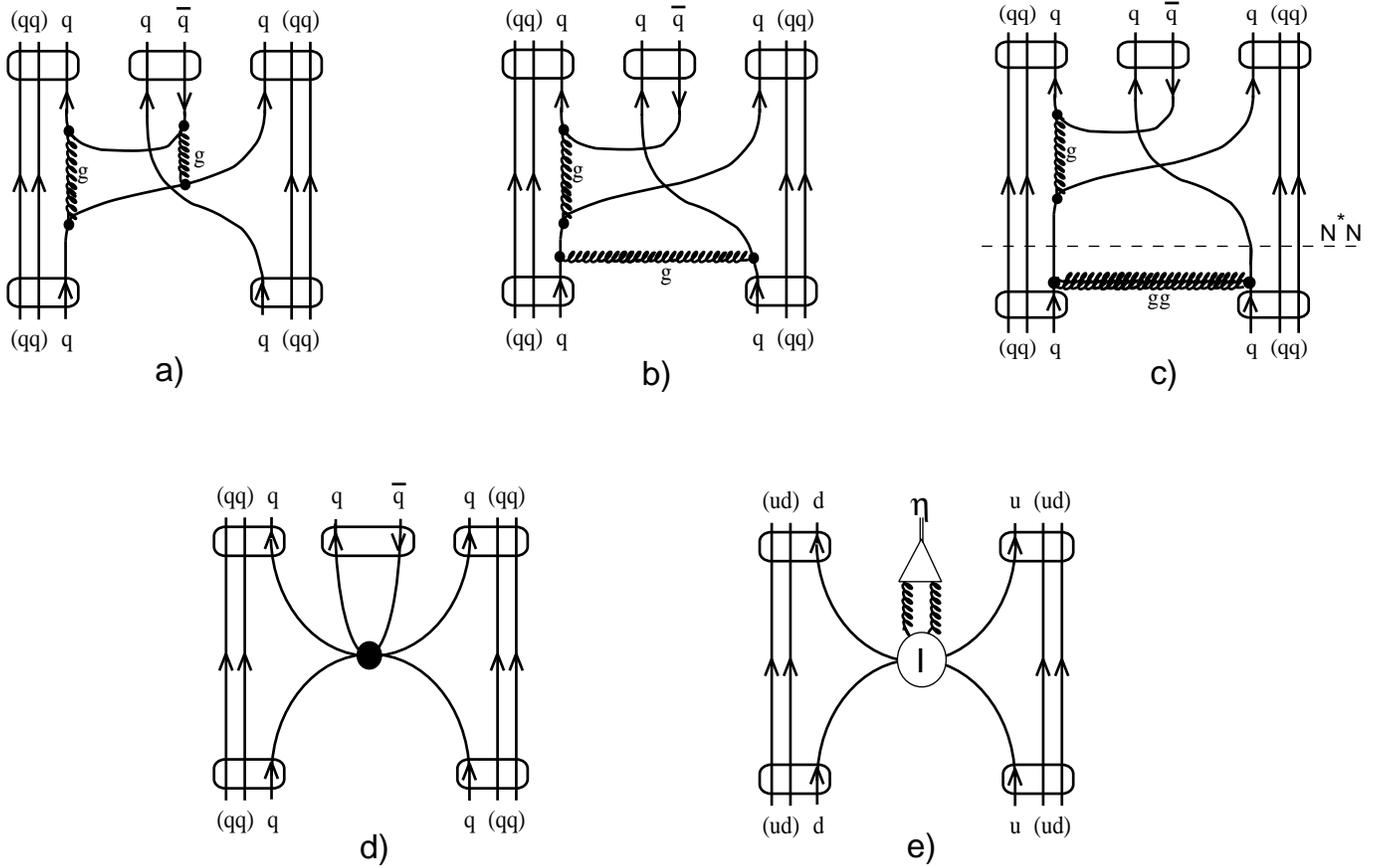,height=0.98\textwidth,angle=-90}
\end{center}
\caption{\label{dillig_gluons} Diagrams for the $qq \rightarrow qq (q\bar{q})$ 
production operator: \hspace{1ex}
(a), (b) Two--gluon exchange and rescattering mechanism \hspace{1ex}
(c) Correlated, colourless two-gluon exchange. The dashed line indicates the 
excitation of an intermediate $N N^*$ system. \hspace{1ex}
(d) Instanton induced 6--quark interaction. Figures (a--d) according 
to~\cite{dil02}. \hspace{1ex}
(e) The instanton contribution to the $\eta$ meson production in the 
proton--neutron interaction. Figure copied from~\cite{koc00}.}
\end{figure}

For example, in the case of the $\eta$ meson the comparison of the cross 
section from the production in proton--proton and proton--neutron collisions 
revealed that the yield of the $\eta$ meson in the isosinglet nucleon 
configuration exceeds the one in the isospin triplet state by more than an 
order of magnitude.
The ratio 
\begin{equation*}
R_{\,\eta} \;=\; \frac{\sigma(pn \rightarrow pn \eta)}
                      {\sigma(pp \rightarrow pp\eta)}\;,
\end{equation*}
presented in figure~\ref{cfisi_Vps}b, was determined to be about 6.5 in the 
excess energy range between $16\,\mbox{MeV}$ and 
$109\,\mbox{MeV}$~\cite{calen2667}. 
This implies that the production of the $\eta$ meson with the total isospin 
equal to zero exceeds the production with isospin $I = 1$ by a factor of 12, 
since $\sigma(pn \rightarrow pn \eta) = (\sigma_{I=0} + \sigma_{I=1})/2$ and 
$\sigma(pp \rightarrow pp \eta) = \sigma_{I=1}$ and hence $\sigma_{I=0} = 
(2\,R_{\,\eta} - 1)\,\sigma_{I=1}$.
On the mesonic level, this large difference of the total cross section between 
the isospin channels suggests the dominance of isovector meson ($\pi$ and 
$\rho$) exchanges (figure~\ref{graph_mesonexchange}~graph~b) in the creation 
of $\eta$ in nucleon--nucleon collisions~\cite{wil99,calen2667}, but 
alternatively it can also be explained on the quark--gluon level assuming the 
instanton induced flavour--dependent quark--gluon interaction~\cite{koc00} 
(see below figure~\ref{dillig_gluons}~graph~e).
The above examples visualize one of the most interesting problems in the 
investigations of the dynamics of the close--to--threshold meson production: 
Namely the  determination of the relevant degrees of freedom for the 
description of the nucleon--nucleon interaction, especially in case when the 
nucleons are very close together~\footnote{These investigations are listed as 
one of the key issues in hadronic physics~\cite{capstick238}.}. 
As pointed out by Nakayama~\cite{nak01}, the transition region from the 
hadronic to constituent quark degrees of freedom does not have a well defined 
boundary and at present both approaches should be evaluated in order to test 
their relevance in the description of close--to--threshold meson production in 
the collision of nucleons.
\begin{figure}[H]
\hfill
\parbox{0.4\textwidth}
  {\epsfig{file=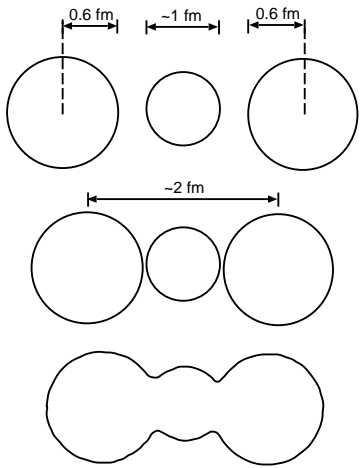,width=0.26\textwidth}} \hfill
\parbox{0.4\textwidth}
  {\caption{\label{cartoon} A cartoon illustrating in naive geometrical terms 
   that for $\mbox{r} < 2\,\mbox{r}_N + 2\,\mbox{r}_M$ meson exchange is 
   unlikely to be appropriate to the description of the internucleon 
   potential. Figure and caption are taken from reference~\cite{maltmanisgur}.}}
\end{figure}
Simple geometrical considerations presented in figure~\ref{cartoon} indicate 
that at distances smaller than $2\,\mbox{fm}$ the internucleon potential 
should begin to be free of meson exchange effects and may be dominated by the 
residual colour forces~\cite{maltmanisgur}. 
In the previous section it was shown, that the close--to--threshold production 
of mesons occurs when the colliding nucleons approach distances of about 
$0.5\,\mbox{fm}$ in case of $\pi^0$ and of about $0.18\,\mbox{fm}$ in case of 
$\phi$ production (see table~\ref{momtranstable}).
This distance is by about an order of magnitude smaller than $2\,\mbox{fm}$ 
and it is rather difficult to imagine --- in coordinate space --- an exchange 
of mesons between nucleons as mechanism of the creation process.
Such small collision parameters imply that the interacting nucleons -- objects 
of about 1~fm -- overlap and their internal degrees of freedom may be of 
importance.
       
Nonetheless, in the last decade, the effective theory based on the meson 
exchanges, which accounts for the size of the participating particles by 
introduction of the momentum transfer dependent form factors, has been 
extensively utilized for the description of the creation process. 
Figure~\ref{graph_mesonexchange} represents the regarded mechanisms. 

\begin{figure}[H]
\hfill
\parbox{0.50\textwidth}
  {\epsfig{file=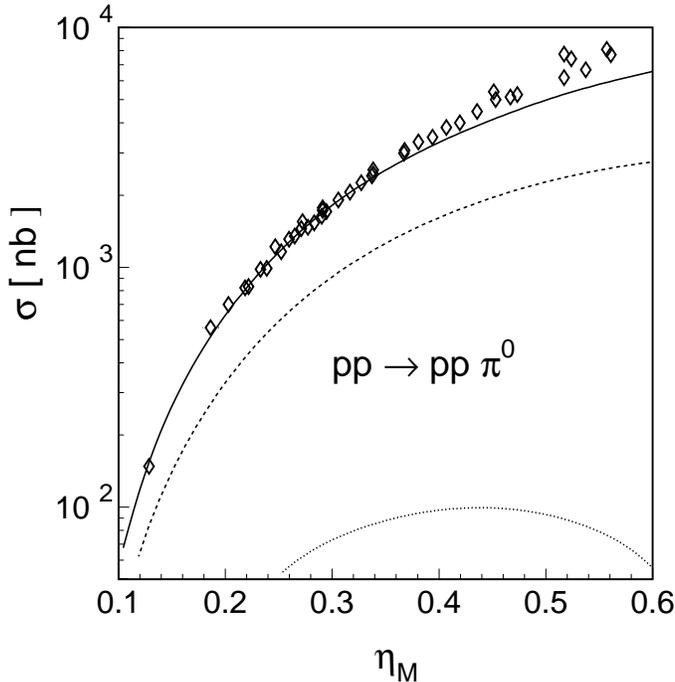,width=0.50\textwidth}} \hfill
\parbox{0.46\textwidth}
  {\vspace{-1cm}\caption{\label{haidenbauer_acta} Total cross section for the reaction $pp 
  \rightarrow pp \pi^0$. The dotted line shows the results for direct 
  production only. Adding rescattering gives the dashed line. Including also 
  the contribution from heavy meson exchange leads to the solid line. The data 
  are from refs.~\cite{meyer633,bondar8}. The figure is adapted from 
  reference~\cite{haidenbauer2893}. Please note, that similarly as shown in 
  figure~\ref{Mpppp_cross_pi}b, the solid line representing the full 
  calculation starts to deviate from the data at $\eta_M \approx 0.4$ 
  ($\mbox{Q} \approx 11\,\mbox{MeV}$), yet here the magnitude reflects the 
  assumed mechanism and in figure~\ref{Mpppp_cross_pi} it was settled to fit 
  the data. Thus indeed, the primary dynamics fixes in practice only the 
  scale.}}
\end{figure}

The question which processes --- on the mesonic level --- are responsible for 
the $\pi^0$ production was considered already in 1966 by Koltun and 
Reitan~\cite{KuR}. 
The authors deemed the direct pion production on one of the protons 
(fig.~\ref{graph_mesonexchange}a) to play the most important role in the $pp 
\rightarrow pp \pi^0$ reaction. 
This supposition was anticipated until the precise experiments performed at 
IUCF~\cite{meyer2846,meyer633} and CELSIUS~\cite{bondar8} revealed that this 
mechanism can not account for more than $5\,\%$ of the total yield, as 
demonstrated in figure~\ref{haidenbauer_acta}. 
Subsequent investigations showed that the inclusion of the processes 
illustrated in figure~\ref{graph_mesonexchange}b, when the pion is produced on 
one of the protons and then scatters on the other one (in the $\pi N$ 
s--wave~\cite{niskanen227} and $\pi N$ p--wave via 
$\Delta(1232)$~\cite{niskanen1285}), increases the cross section essentially 
yet not enough to describe the data, even if the off--shell properties of the 
$\pi N$ amplitude are taken into account~\cite{hanhart21,hernandez158}. 
The contribution of the rescattering mechanism is shown as the dashed line in 
figure~\ref{haidenbauer_acta}. 
The agreement with the data can, however, be achieved --- as shown by the 
solid line --- if one takes additionally into account the mechanisms of 
heavy--meson--exchange associated with an intermediate virtual 
nucleon--antinucleon state~\cite{LEE_RISKA,horowitz1337} illustrated in 
figure~\ref{graph_mesonexchange}c.
A thorough discussion concerning $\pi$ meson production in differently charged 
channels can be found e.g.\ in references~\cite{hanhart25,machnerR231}. 

The dominance of the heavy--meson--exchange and the off--shell $\pi N$ 
rescattering mechanisms, both reflecting the short range 
nucleon--nucleon interactions, confirms the simple estimation of the reaction 
region based on the momentum transfer between the colliding nucleons. 
Thus, in case of the $\pi^0$ meson creation,
this approach succeeded to describe the magnitude 
of the total cross section.
However, the meson-exchange model at the present stage 
does not reproduce the angular distributions of the spin 
observables for all $pp \to NN \pi$ isospin channels, simultaneously.
Specifically, though the only available calculations
of the spin correlation coefficients based on the 
J{\"u}lich model~\cite{hanhart064008}
are in a good agreement with the  experimental data for the
reactions $\vec{p}\vec{p} \to pn\pi^{+}$~\cite{saha175} 
and $\vec{p}\vec{p} \to d\pi^{+}$~\cite{przewoski064604} 
they reveal serious discrepancies when confronted with
the angular dependences of the spin correlation coefficients 
{\mbox{$A_{\Sigma} \equiv A_{xx} + A_{yy}$}} and $A_{zz}$ of the 
$\vec{p}\vec{p} \to pp\pi^{0}$ reaction~\cite{meyer064002}.
Therefore,  further investigations concnerning
the relative contributions of various mechanisms,
as well as the different partial waves are still required.\\

The role of the mechanisms leading to the $\pi$ production 
have been also considered in context of the lowest--order 
chiral perturbation theory~\cite{bernard259} and relativistic 
impulse approximation~\cite{adam97}. In contrast to the 
meson exchange models the chiral perturbation method  
results~\cite{sato1246,cohen2661,park1519}
in the destructive interference
between  the direct and the pion-rescattering terms, 
and consequently implies more significant contributions
originating from the shorter-than-pion-range mechanisms, like
the heavy-meson exchanges or the creation of
pions in the fusion of exchanged mesons~\cite{kolck679}.

In case of the $\eta$ and $\eta^{\prime}$  mesons
the investigations of the mechanisms
underlying the production process are even more difficult.
It is partly due to the fact that in contrary to the pion case
even the coupling constants determining the strength
of the $NN\eta(\eta^{\prime})$ vertex are poorly established.
 
The values derived by various methods differ by more than an order of 
magnitude~\cite{moskalphd,feldmann044}. 
For example, the estimation based on the dispersion relation results in 
$g_{NN \eta^{\prime}}$ smaller than 1~\cite{grein332}, whereas the value 
obtained from the fits to low--energy nucleon--nucleon scattering in 
one--boson--exchange models amounts to 7.3~\cite{nagels189}. 
This uncertainty prevents from the exact estimation of the contribution from 
the nucleonic current. 
However, according to the systematic investigations of Nakayama et 
al.~\cite{nakayama024001,nakayama012,nak01} this mechanism is of minor 
importance. 
The calculations have been performed with $g_{NN \eta^{\prime}} = 
g_{NN \eta} = 6.1$ --- used in the $NN$ scattering 
analysis~\cite{machleidt189} --- which is close to the upper limit of the 
predicted values.\\

\begin{figure}[H]
\hfill
\parbox{0.6\textwidth}
  {\epsfig{file=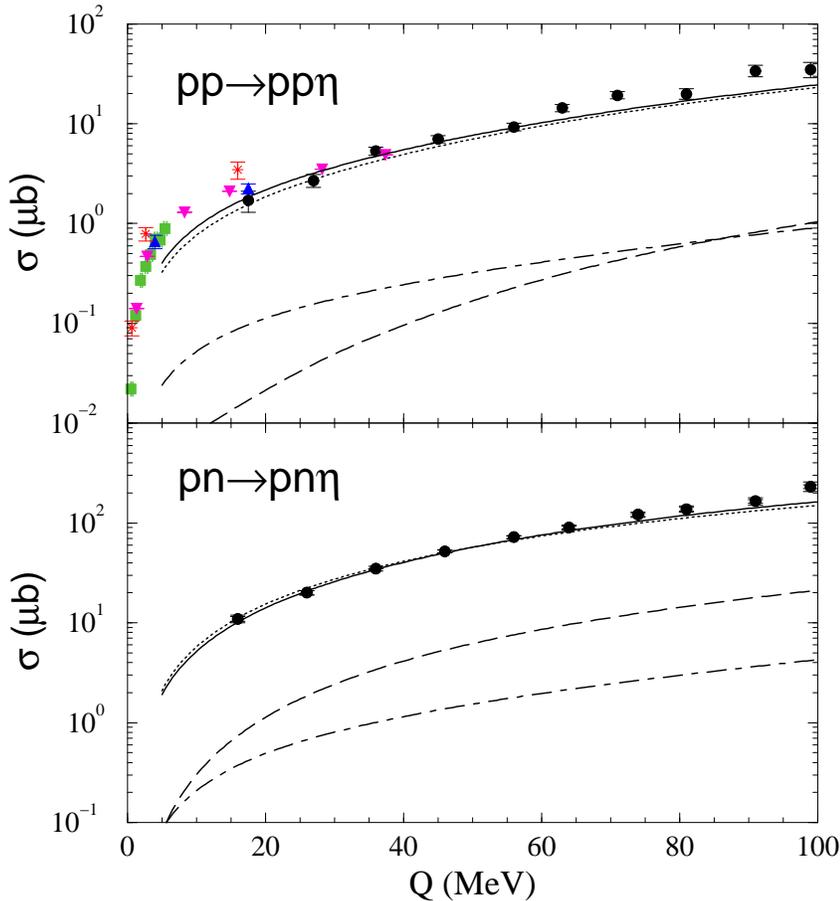,width=0.6\textwidth}} \hfill
\parbox{0.36\textwidth}
  {\caption{\label{nakayama_eta} Total cross sections for the $pp \rightarrow 
  pp\eta$ (upper panel) and $pn \rightarrow pn \eta$ (lower panel) reactions 
  as a function of excess energy. The dashed curves correspond to the 
  nucleonic current contribution and the dash--dotted curves to the mesonic 
  current consisting of $\eta\rho\rho$, $\eta\omega\omega$ and $\eta a_0 \pi$ 
  contributions. The resonance current presented by the dotted line consists 
  of the predominant $\mbox{S}_{11}(1535)$ and of the $\mbox{P}_{11}(1440)$ 
  and $\mbox{D}_{13}(1520)$ resonances excited via exchange of $\pi$, $\eta$, 
  $\rho$ and $\omega$ mesons. The solid curves are the total contribution. The 
  deviation from the data at low Q in the upper panel reflects the $\eta p$ 
  FSI which was not included in the calculations. The data are from 
  refs.~\cite{hibou41,bergdoltR2969,chiavassa270,calen39,smyrski182,calen2667}.
  The figure is taken from~\cite{nak01}.}}
\end{figure}

It is at present rather well established~\cite{germond308,laget254,faldt427,
moalem445,moalem649,vetter153,nakayama012,alvaredo125,Batinic} that the $\eta$ 
meson is produced predominantly via the excitation of the $\mbox{S}_{11}$ 
baryonic resonance $N^*(1535)$ which subsequently decays into $\eta$ and 
nucleon and whose creation is induced through the exchange of $\pi$, $\eta$, 
$\rho$, $\sigma$ and $\omega$ mesons, as shown in 
figure~\ref{graph_mesonexchange}e. 
Though all the quoted groups reproduce the magnitude of the total cross 
section, their models differ significantly as far as the relative contribution 
from the $\pi$, $\eta$ and $\rho$ exchange mechanisms is concerned.
The discrepancies are due to the not well known strength of 
$Meson$--$N$--$\mbox{S}_{11}$ couplings and the $\eta N$ scattering potential.
For example, while the dominance of the $\rho$ meson exchange is anticipated 
by authors of references~\cite{germond308,laget254,faldt427,moalem445,
moalem649}, it is rather the exchange of the $\eta$ meson which dominates the 
production if one takes into account effects of the off--shell $\eta N$ 
scattering~\cite{pena322}, or utilizes the multi--channel multi--resonance 
model~\cite{Batinic}.
In any case, the hitherto performed studies aiming to describe the total cross 
section show that the close--to--threshold production of $\eta$ mesons in 
nucleon--nucleon collisons is dominated by the intermediate virtual 
$\mbox{S}_{11}$ nucleon isobar whose width overlaps with the threshold. 
In order to disentangle between various scenarios of the $\mbox{S}_{11}$ 
excitation a confrontation of the predictions with other observables is needed.
The interference between considered amplitudes causes a different behaviour 
--- depending on the assumed scenario --- e.g.\ of the $\eta$ meson angular 
distributions. 
These differences, however, are too weak in the close--to--threshold region to 
judge between different models. 
Also the ratio of the $\eta$ meson production via the reactions $pp 
\rightarrow pp \eta$ and $pn \rightarrow pn \eta$ can be equally well 
described either by assuming the $\rho$ meson exchange 
dominance~\cite{faldt427} or by taking pseudoscalar and vector mesons for 
exciting the $\mbox{S}_{11}$ resonance~\cite{nakayama012}. 
In the latter case, shown in figure~\ref{nakayama_eta}, the excitation of the 
resonance via the $\rho$ meson exchange was found to be negligible.
Yet, the predictions of the analyzing power depends crucially on the assumed 
mechanism~\cite{faldt427,nakayama012}. 
This fact has triggered already experimental investigations which aim to 
determine the spin observables~\cite{Winter_Dipl}. 

Figure~\ref{nakayama_eta} presents also a contribution from the mesonic 
current where the $\eta$ meson is created in the fusion of virtual e.g.\ 
$\rho$ or $\omega$ mesons emitted from both colliding nucleons. 
In comparison to the overwhelming strength of the resonance current this 
mechanism is however by a factor of 30 too weak. 
In contrast, it is enough to explain the magnitude of the close--to--threshold 
$\eta^{\prime}$ meson production in the proton--proton interaction, which is 
just by about a factor of 30 smaller compared to the $\eta$ meson.
Among the mesonic currents regarded by authors of 
reference~\cite{nakayama024001} the $\rho\rho\eta^{\prime}$ gives the dominant 
contribution, by a factor of five larger than the $\sigma\eta\eta^{\prime}$-- 
and $\omega\omega\eta^{\prime}$--exchange. 
However, the understanding of the $\eta^{\prime}$ production on the hadronic 
level is far from being satisfactory. 
The magnitude of the total cross section was also well reproduced in the frame 
of a one--boson--exchange model (nucleonic current) where the virtual boson 
($B = \pi, \eta, \sigma, \rho, \omega, a_0$) created on one of the colliding 
protons converts to the $\eta^{\prime}$ on the other one~\cite{gedalin471}. 
Taking into account the off--shell effects of the $B p \rightarrow 
\eta^{\prime} p$ amplitude it was found that the short range $\sigma$ and 
$\rho$ meson exchanges dominate the creation process, whereas the $\pi$ 
exchange plays a minor role. 
On the contrary, other authors~\cite{sibirtsev333} reproduced the magnitude of 
the total cross section with $\pi$ exchange only and concluded that the 
$\eta$, $\rho$ and $\omega$ exchange currents either play no role or cancel 
each other. 
However, in both of the above quoted 
calculations~\cite{gedalin471,sibirtsev333} the initial state interaction 
between protons, which reduces the rate by a factor of about 3, was not taken 
into account and hence the obtained results would in any case reproduce only 
$30\,\%$ of the entire magnitude and could be at least qualitatively 
reconciled with the mentioned result of reference~\cite{nakayama024001}, where 
the nucleonic current was found to be small.
Moreover, the choice of the another prescription for the form factors could 
reduce the one--pion--exchange contribution 
substantially~\cite{nakayama024001}. 
However, the picture that the $\eta^{\prime}$ meson is predominantly created 
through the mesonic current, which we would like to advocate, remains at 
present unclear as well. 
This is because the magnitude of the total cross section could have been also 
described assuming that the production of $\eta^{\prime}$ is 
resonant~\cite{nakayama024001}. 
As possible intermediary resonances the recently reported~\cite{ploetzke555} 
$\mbox{S}_{11}(1897)$ and $\mbox{P}_{11}(1986)$ have been considered.
These resonances were deduced from $\eta^{\prime}$ photoproduction data, under 
the assumption that the close--to--threshold enhancement observed for the 
$\gamma p \rightarrow \eta^{\prime} p$ reaction can be utterly assigned to 
resonance production. 
Further, as well strong, assumptions have been made in the derivation of the 
$g_{N N^* \eta^{\prime}}$ and $g_{N N^* \pi}$ coupling 
constants~\cite{nakayama024001}.
Hence, it is rather fair to state that in the case of the close--to--threshold 
$\pi^0$ and $\eta$ meson production in nucleon--nucleon collisons the dynamics 
is roughly understood on the hadronic level, but the mechanisms leading to the 
$\eta^{\prime}$ creation are still relatively unknown.
Unil now it was not possible to estimate satisfactorily the relative 
contributions of the nucleonic, mesonic and resonance current to the 
production process. 
In fact, model uncertainties allow that each one separately could describe the 
absolute values of the $pp \rightarrow pp \eta^{\prime}$ total cross section.
This rather pessimistic conclusion calls for further experimental and 
theoretical research.  
The understanding of the production dynamics of the $\eta^{\prime}$ meson on 
both hadronic and the quark--gluon level is particularly important since its 
wave function comprises a significant gluonic component~\cite{ball367}, 
distinguishing it from other mesons and hence the comprehension of the 
mechanism leading to its creation in collisions of hadrons may help in the 
determination of its quark--gluon structure which is also of importance to 
investigate possible glueball candidates~\cite{ball367}. 
Hereafter we will briefly report about the gluonic mechanisms which may --- 
additionally to the meson exchange processes discussed above --- contribute to 
the $\eta^{\prime}$ production.\vspace{1ex}

As already mentioned, the close--to--threshold production of $\eta$ and 
$\eta^{\prime}$ mesons in the nucleon--nucleon interaction requires a large 
momentum transfer between the nucleons and hence can occur only at distances 
of about $0.3\,\mbox{fm}$~(see table~\ref{momtranstable}). 
This suggests that the quark--gluon degrees of freedom may indeed play a 
significant role in the production dynamics of these mesons. 
A possibly large glue content of the $\eta^{\prime}$ and the dominant 
flavour--singlet combination of its quark wave function may cause that the 
dynamics of its production process in nucleon--nucleon collisions is 
significantly different from that responsible for the production of other 
mesons.
In particular, the $\eta^{\prime}$ meson can be efficiently created via a 
``contact interaction'' from the glue which is excited in the interaction 
region of the colliding nucleons~\cite{bass286}.
A gluon--induced contact interaction contributing to the close--to--threshold 
$pp \rightarrow pp \eta^{\prime}$ reaction derived in the frame of the 
U(1)--anomaly extended chiral Lagrangian is discussed in 
references~\cite{bass286,bass429,bass348}.
The strength of this contact term is related to the amount of spin carried by 
polarized gluons in a polarized proton~\cite{bass348,bass17}, thus making the 
study of the close--to--threshold $\eta^{\prime}$ meson production even more 
interesting.\vspace{1ex}

\begin{figure}[H]
\vspace{0.4cm}
\hfill
\parbox{0.65\textwidth}
  {\epsfig{file=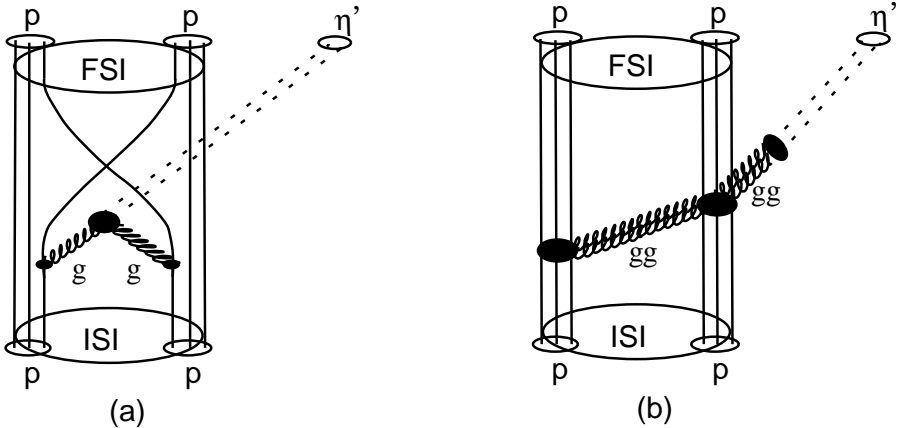,width=0.65\textwidth}} \hfill
\parbox{0.3\textwidth}
  {\caption{\label{graf_glue} Diagrams depicting possible quark--gluon 
  dynamics of the reaction $pp \rightarrow pp \eta^{\prime}$. 
  (a) --- production via a fusion of gluons~\cite{kolacosynews} with 
  rearrangement of quarks.
  (b) --- production via a rescattering of a ``low energy pomeron''.}}
\end{figure}

\begin{figure}[H]
\begin{center}
\epsfig{file=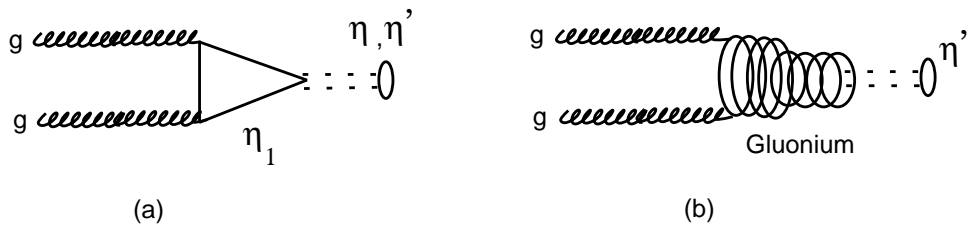,width=0.7\textwidth}
\end{center}
\caption{\label{triangle} Coupling of $\eta$ and $\eta^{\prime}$ to two gluons 
through \hspace{1ex} (a) quark and antiquark triangle loop and \hspace{1ex} \\
(b) gluonic admixture. \hspace{1ex} $\eta_{1}$ denotes the flavour--singlet 
quark--antiquark state. This indicates that gluons may convert into the $\eta$ 
or $\eta^{\prime}$ meson via a triangle quark loop only by coupling through 
their flavour singlet part. The figure is taken from 
reference~\protect\cite{kou054027}.}
\end{figure}

Figure~\ref{graf_glue} depicts possible short--range mechanisms which may lead 
to the creation of the $\eta^{\prime}$ meson via a fusion of gluons emitted 
from the exchanged quarks of the colliding protons~\cite{kolacosynews} or via 
an exchange of a colour--singlet object made up from glue, which then 
re--scatters and converts into $\eta^{\prime}$~\cite{basspriv}.
The hadronization of gluons to the $\eta^{\prime}$ meson may proceed directly 
via its gluonic component or through its overwhelming flavour--singlet 
admixture~$\eta_1$ (see fig.~\ref{triangle}).
Contrary to the significant meson exchange mechanisms and the fusion of gluons 
of figure~\ref{graf_glue}~graph~a), the creation through the colour--singlet 
object proposed by S.D.~Bass (graph~\ref{graf_glue}b) is isospin independent, 
and hence should lead to the same production yield of the $\eta^{\prime}$ 
meson in both reactions ($pp \rightarrow pp \eta^{\prime}$ and $pn \rightarrow 
pn \eta^{\prime}$) because gluons do not distinguish between flavours.
This property should allow to test the relevance of a short range gluonic 
term~\cite{bassproc} by the experimental determination of the cross section 
ratio $R_{\,\eta^{\prime}} = \sigma(pn \rightarrow pn \eta^{\prime})/
\sigma(pp \rightarrow pp \eta^{\prime})$, which in that case should be close 
to unity after correcting for the final and initial state interaction between 
participating baryons.
The other extreme scenario --- assuming the dominance of the isovector meson 
exchange mechanism --- should result in the value of $R_{\,\eta^{\prime}}$ 
close to 6.5 as was already established in the case of the $\eta$ meson (see 
fig.~\ref{cfisi_Vps}b).
The experimental investigations of that issue are in 
preparation~\cite{moskal_c11proc}.\vspace{1ex}

\vspace{1.0cm}

In addition to the QCD motivated investigation presented 
above~\cite{bass286,bass429}, the interesting features of the 
close--to--threshold meson production, in particular the large cross section 
for the $\eta$ meson in proton--proton interactions exceeding the one of the 
pion, or even more surprisingly large cross section of the $\eta$ 
production in proton--neutron collisions, encouraged also other authors to 
seek for the underlying --- OZI rule violating --- creation mechanisms in the 
frame of microscopic models of QCD~\cite{koc00,dil02,kleefeld2867}.
The hitherto regarded processes are presented in figure~\ref{dillig_gluons}.
As indicated in the pictures the structure of participating baryons has been 
modeled as quark--diquark objects with harmonic confinement~\cite{dillig050}.
In the upper graphs the large momentum transfer is shared by the exchanged 
gluons with a subsequent interchange of quarks to provide a colourless object 
in the final state.
The lower graphs depict the two examples of instanton induced interactions 
with a 6--quark--antiquark (fig.~\ref{dillig_gluons}d) and two-gluon vertex 
(fig.~\ref{dillig_gluons}e).

Adjusting the normalization to the cross section of the $pp \rightarrow 
pp \pi^0$ reaction at a single energy point the model~\cite{dil02} accounts 
roughly for close--to--threshold cross sections of other pseudoscalar and 
vector mesons in proton--proton collisions. 
Though this approach is characterized by significantly smaller number of 
parameters than in the meson exchange models, their uncertainties allow for 
the description of the data equally well either by the gluon exchange or by 
the instanton induced interactions, at least in case of the $\pi^0$, 
$\eta^{\prime}$, $\omega$ and $\phi$ mesons. 
Yet for the resonance dominated $\eta$ and $K^+$ meson production in 
proton--proton collisions it was found that the instanton induced interaction 
presented by graph~\ref{dillig_gluons}d is not sufficient.
Similarly, the authors of reference~\cite{koc00} argue that the instanton 
induced interaction with a quark--gluon vertex (graph~\ref{dillig_gluons}e) 
should be of no importance for the $\eta$ production in proton--proton 
collisions. 
This arises from the properties of the vertex which lead to the $\eta$ 
production only in case of an interaction between quarks of different flavours 
($ud \rightarrow udgg$) and in the quark--diquark model of baryons the proton 
consists of $ud$--diquark and $u$--quark and correspondingly the neutron is 
modeled as $ud$--diquark and $d$--quark.
Although negligible in case of proton--proton interactions, this mechanism may 
contribute significantly to the total cross section of the  $pn \rightarrow 
pn \eta$ reaction and indeed as shown in reference~\cite{koc00} it reproduces 
the data. 

However, the magnitude of the cross section is very sensitive to the size of 
the instanton in the QCD vacuum and --- similar to the uncertainty of 
coupling constants in case of the hadronic approach --- at present it makes 
precise predictions impossible.
Thus, despite the partial successes, at the present stage of developments both 
approaches --- on the quark--gluon and on the hadronic level --- do not 
deliver an unambiguous answer to the dynamics of the close--to--threshold 
meson production in the nucleon--nucleon interaction.

\subsection{Vector meson production}
\label{vmp}
Vector meson production studies in the mass range up to 
$1\,\mbox{GeV}/\mbox{c}^2$ in hadronic collisions are especially motivated by 
questions concerning the strangeness content of the nucleon.
Within the quark model the $\phi$ meson is an (almost) ideally mixed 
$s \overline{s}$ state, whereas the nucleon is composed only of $u$-- and 
$d$--quarks.
According to the OZI rule~\cite{ozi66}~\footnote{for a review 
see~\cite{lip92}.}, $\phi$ production in the nucleon--nucleon interaction 
would be forbidden as a process with disconnected quark lines, if the $\phi$ 
was a pure $s \overline{s}$ state.

However, due to a small deviation from ideal mixing ($\delta_V = 
3.7^{\circ}$~\cite{groom1}) the $\phi$ meson is allowed to couple to the 
nucleon through its $\left(u\overline{u} + d\overline{d}\right)$ admixture.
Quantitatively, the naive OZI rule results in a suppression of $\phi$ compared 
to $\omega$ production in hadronic interactions
\begin{equation} 
\label{ozieq}
R = \frac{\sigma\left(A\,B \rightarrow \phi\,X\right)}
         {\sigma\left(A\,B \rightarrow \omega\,X\right)} 
  = \tan^2(\delta_V) \approx 4.2 \times 10^{-3}
\end{equation}
after phase--space corrections, where $A$, $B$ and $X$ denote hadronic systems 
consisting only of light quarks~\cite{lip76}.
Any significant deviation might be interpreted as a hint for a strangeness 
component in the nucleon.
A considerable contribution of strange sea quarks to the nucleon wave function 
is suggested by results on the $\Sigma_{\pi N}$ term in pion nucleon elastic 
scattering~\cite{sigmaterm}, on the nucleon's structure function in deep 
inelastic scattering with polarized muons~\cite{polmuon} and, more recently, 
on charm production in deep inelastic neutrino scattering~\cite{ada99,bar00}.

Large violations of the OZI expectation (eq.~\eqref{ozieq}) have been found in 
$p \overline{p}$ annihilation experiments by the ASTERIX, Crystal Barrel and 
OBELIX collaborations at LEAR (as reviewed in~\cite{ell95}).
However, deviations are predominantly found in $p \overline{p}$ annihilation 
at rest as compared to higher energies and seem to be restricted to S--wave 
processes only.
Furthermore, the results are strongly dependent on the final state:
While the $\phi\,\pi$ and $\phi\,\gamma$ channels exceed the OZI value by a 
factor of 20 and 100, respectively, little or no effect is seen in the 
$\phi\,\pi\,\pi$ and $\phi\,\eta$ final states.

These data have been interpreted as ``shake--out'' and ``re-arrangement'' of a 
negatively polarized $s \overline{s}$ Fock space component of the proton wave 
function~\cite{ell95,gut97}, which also accounts for large double $\phi\,\phi$ 
cross sections~\cite{dov92} in $p \overline{p}$ annihilation measured at the 
JETSET experiment~\cite{jetsetphi}.
However, OZI--allowed two--step processes via intermediate $K \overline{K}$ or 
$\Lambda \overline{\Lambda}$ states have been shown to describe available data 
from $p \overline{p}$ annihilation without any strangeness in the 
nucleon~\cite{loc94a,loc94b,mul94,buz98,gor96,ani96,mei97,mar98}.

In nucleon--nucleon scattering, effects of competing two step processes for 
$\phi$ production via intermediate $\Lambda K$ or $\Sigma K$ states are 
expected to be of minor importance~\cite{gei96,mei97}.
Furthermore, in close--to--threshold $\phi$ production the entrance channel 
must be in a $^3\mbox{P}_1$ state (see table~\ref{partialtransitions}) 
due to parity and angular momentum conservation and it is the spin triplet 
fraction in $p \overline{p}$ annihilation which is strongly correlated to the 
$\phi$ meson yield~\cite{ber96}.
Thus, cross section ratios for $\phi$ and $\omega$ production in 
proton--proton scattering should clearly indicate possible OZI violations and 
probe the $s \overline{s}$ component of the nucleon.

\begin{figure}[H]
\hfill
\parbox{0.52\textwidth}
  {\epsfig{file=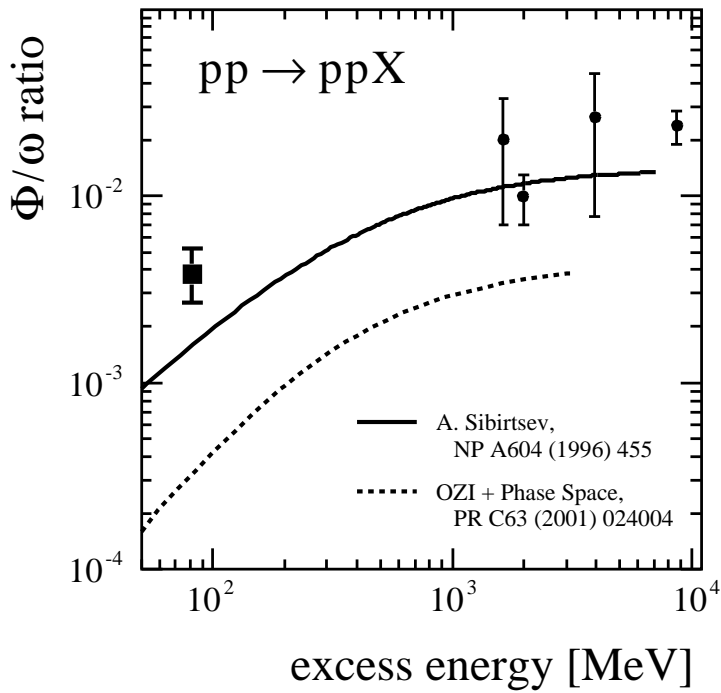,width=0.52\textwidth}}
\hfill
\parbox{0.37\textwidth}
  {\caption{\label{phiomega} $\phi/\omega$ total cross section ratio in 
  proton--proton collisions versus excess energy with respect to the $\phi$ 
  production threshold from~\cite{bal01}. The recent data point from the DISTO 
  collaboration (solid square~\cite{bal01}) is shown together with high energy 
  data (solid circles~\cite{phihigh}) and calculations taking into account the 
  OZI prediction and three--body phase--space~\cite{bal01} and results obtained 
  within a one--boson--exchange model including the proton--proton 
  FSI~\cite{sib96}. Data includes both statistical and systematical errors.}}
\end{figure}

The exclusive production ratio has been determined in the reaction $pp 
\rightarrow pp \phi/\omega$ at a beam momentum of $3.67\,\mbox{GeV/c}$ 
corresponding to excess energies of $82\,\mbox{MeV}$ and $319\,\mbox{MeV}$ to 
the $\phi$ and $\omega$ thresholds, respectively, by the DISTO collaboration 
at SATURNE~\cite{bal01}.
After phase--space corrections, the observed ratio exceeds the OZI expectation 
by an order of magnitude, while data at excess energies larger than 
$1.1\,\mbox{GeV}$~\cite{phihigh} show an enhancement smaller by a factor of 
four (see fig.~\ref{phiomega}).
One--boson--exchange models~\cite{sib96,sib00,chu97,tit99} underestimate the 
absolute value of the $\phi$ production cross section measured at the DISTO 
experiment by at least a factor of three. \\

Figure~\ref{phiomega} includes results obtained within a one--pion--exchange 
model~\cite{sib96}, which provides a reasonable description 
of high energy data.
Model predictions --- taking into account the proton--proton final 
state interaction only --- falls short of the recent close--to--threshold 
value~\cite{bal01}.
Using instead a parametrization of the energy dependence of the $\omega$ production 
cross section and considering both the finite $\omega$ width as well as 
effects of proton--proton FSI~\cite{sib99,sib00}, the $\phi/\omega$ ratio can 
be determined at equal excess energies, thus reducing uncertainties due to the 
available phase--space, partial wave amplitudes~\footnote{The $\omega$ angular 
distribution relative to the $pp$ system measured at an excess energy of 
$319\,\mbox{MeV}$ shows evidence for p-- and d--waves 
contributing~\cite{bal01}.} and effects of the proton--proton FSI. 
In conclusion, the excess over the OZI prediction is decreased to a factor of 
five~\cite{bal01}, i.e.\ to a modest enhancement as compared to the results 
from $p \overline{p}$ annihilation experiments~\footnote{Comparing the DISTO 
data at $\mbox{Q} = 82\,\mbox{MeV}$~\cite{bal01} and the recent COSY--TOF 
result at an excess energy of $92\,\mbox{MeV}$~\cite{abd01} leads to a similar 
enhancement of $\approx 7$ compared to the OZI expectation.}.

\begin{figure}[H]
\hfill
\parbox{0.58\textwidth}
  {\epsfig{file=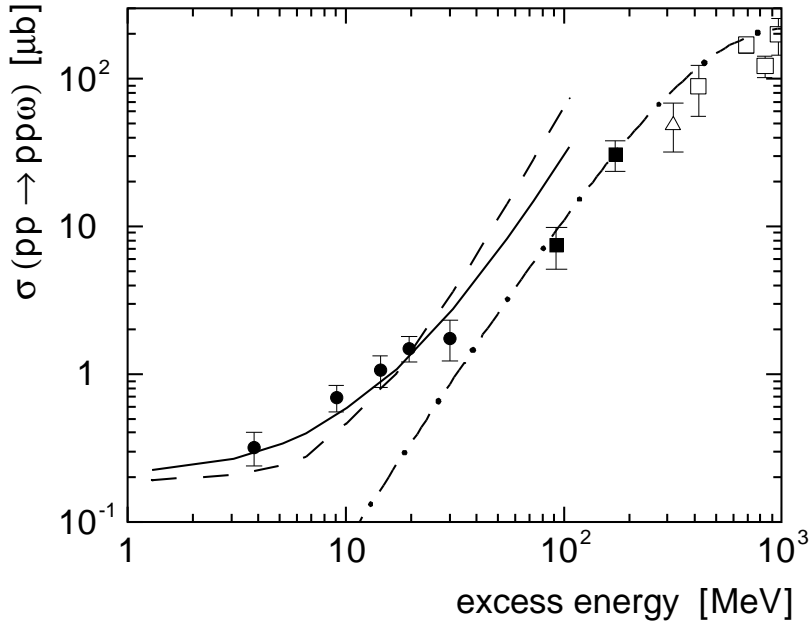,width=0.58\textwidth}} \hfill
\parbox{0.32\textwidth}
  {\caption{\label{omega_cross} Total cross section for $pp \rightarrow 
  pp \omega$ as a function of the excess energy. Data are from refs.\ 
  \cite{bal01} (open triangle), \cite{abd01} (solid squares), \cite{hib99} 
  (solid circles) and \cite{fla84} (open squares). 
  Calculations include a one--pion--exchange based parametrization 
  (dash--dotted curve \cite{sib96}) and results obtained 
  within a relativistic meson--exchange model considering nucleonic and 
  mesonic current contributions (solid and dashed curves \cite{nak98,nak99}, 
  for details see text). The figure is taken from ref.\ \cite{abd01}.}}
\end{figure}

In the DISTO results, $\phi$ production at an excess energy of 
$82\,\mbox{MeV}$ appears to be dominated by s--wave relative to the 
nucleon--nucleon system in agreement with theoretical predictions~\cite{rek97}
and phenomenologically described in section~\ref{Mwhs}.
However, the proton--proton angular distribution relative to the $\phi$ 
direction is found to deviate significantly from isotropic emission at this 
beam energy with evidence for a P--wave contribution in the proton--proton 
subsystem.
Consequently, $\phi$ production at $82\,\mbox{MeV}$ above threshold, although 
dominated by a $^3\mbox{P}_1$ proton--proton entrance channel, partly proceeds 
via the $^1\mbox{S}_0$ and $^1\mbox{D}_2$ spin singlet partial 
waves~\cite{bal01}.
Thus, the influence of the spin--triplet entrance channel, prerequisite to a 
``re--arrangement'' of an $s \overline{s}$ component in the nucleon wave 
function~\cite{ell95}, might be diluted in the presently available data and 
would be expected to be more prominent in data closer to the production 
threshold.\vspace{1ex}
\vspace{-0.5cm}
\begin{figure}[H]
\begin{center}
\epsfig{file=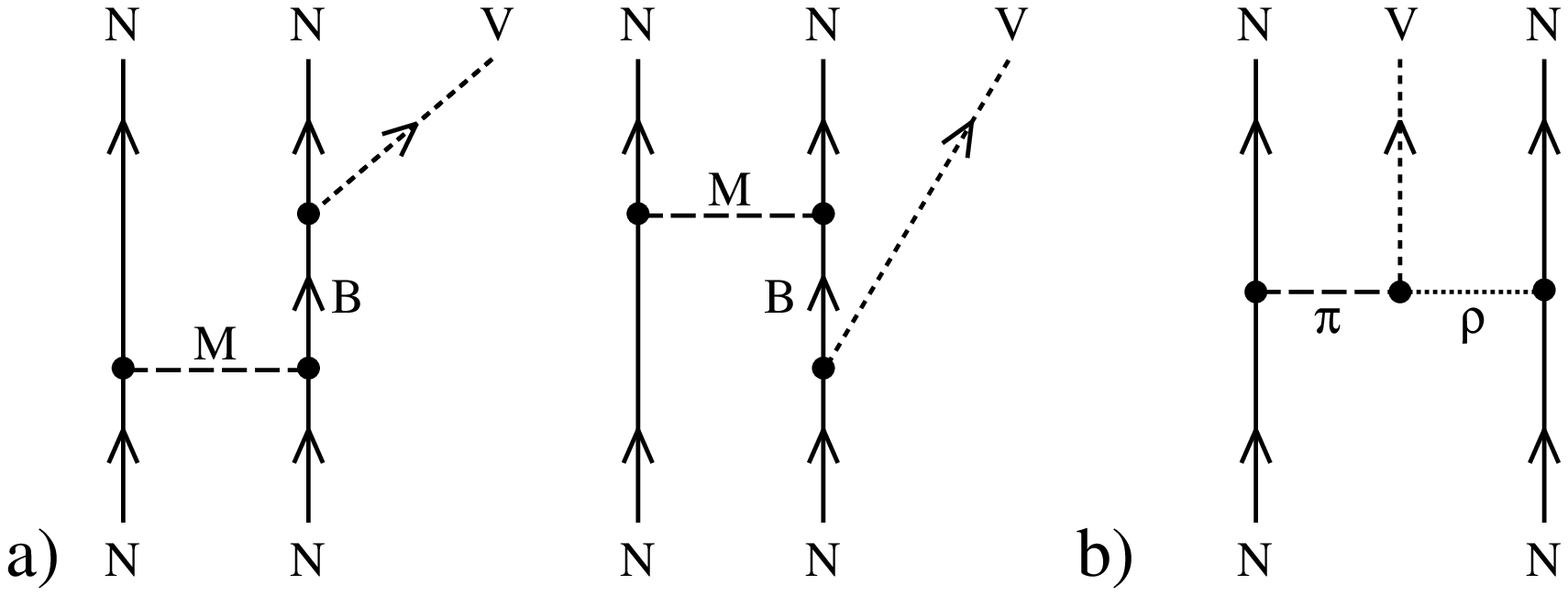,scale=0.67,angle=270}
\caption{\label{kanzo_curr} Vector meson production currents in 
nucleon--nucleon scattering ($NN \rightarrow NNv$) studied within a 
relativistic meson exchange model in~\cite{nak01,nak98,nak99}: \protect \\
(a)~~nucleonic current and (b)~~mesonic current.  \protect \\
$M = \pi, \eta, \rho, \omega, \sigma, a_0$ denotes the exchanged meson and 
$B$ the intermediate off mass-shell baryon.}
\end{center}
\end{figure}
Recently, within the framework of a relativistic meson exchange model, 
nucleonic and meson--exchange currents have been used explicitly to study 
vector meson production in nucleon--nucleon scattering (fig.~\ref{kanzo_curr} 
\cite{nak01,nak98,nak99}).
Uncertainties in the model calculation for the nucleonic current arise mainly 
from a large range of values quoted for vector and tensor $NNv$ coupling 
constants~\cite{hoe76,grein332,fur90,mac87,mer96,spe97,dov84,machleidt189,
mei97}, as well as from the vector and tensor vertex form factors for the 
coupling of an on--mass shell vector meson to off mass--shell nucleons.
Note, that a resonance contribution to the nucleonic current 
(fig.~\ref{kanzo_curr}a) is neglected in the model~\cite{nak98}, since there 
are no isospin--1/2 nucleon resonances experimentally known coupling to the 
$v p$ channel~\cite{cap94}.
Due to the strength of both the $NN \pi$ and $NN \rho$ couplings and the 
$g_{\pi\rho\omega}$ coupling constant as compared to other possible choices, 
only the $\pi \rho v$ vertex is considered in evaluating the meson--exchange 
current.
For $\omega$ production, contributions from the latter and the nucleonic 
current are found to interfere destructively when added coherently, which 
leaves two choices for a given value of the $NN \omega$ tensor--to--vector 
coupling ratio --- when fitting the close--to--threshold data on the reaction 
$pp \rightarrow pp \omega$ taken at SATURNE~\cite{hib99} --- for the only 
remaining free parameter of the model, the cut--off parameter at the 
$NN \omega$ form factor.
Effects from a destructive interference of mesonic exchange current and 
nucleonic current contribution are also concluded by other authors for 
$\omega$ production in~\cite{kai99a}~\footnote{Kaiser's study~\cite{kai99a} is 
based on the calculation of various tree level diagrams in comparison with the 
threshold transition amplitude extracted from data of~\cite{hib99} and follows 
the approach developed in~\cite{bernard259,kai99b} (see section~\ref{Asp}). 
The dominant nucleonic current contribution arises from $\rho^0$ exchange. 
However, the combined result of the studied amplitudes shows similarly 
to~\cite{nak98,nak99} a dependence on the choice of the tensor--to--vector 
coupling ratio.} and for $\phi$ production in~\cite{tit00}. 
In Nakayama's analysis the energy dependence of the total cross section is 
slightly influenced by either choice, as shown by the solid and dashed 
curves~\footnote{The two curves correspond to the extreme cases found for the 
tensor coupling values $f_{NNv} = \pm 0.5\,g_{NNv}$~\cite{nak99}, with 
cut--off masses of $1545\,\mbox{MeV}$ (dashed line) and $1170\,\mbox{MeV}$ 
(solid line), respectively.}
in figure~\ref{omega_cross} together with a parametrization based on 
one--pion--exchange (dash--dotted line~\cite{sib96}).
In comparison, total cross section data taken at the SPES~3 spectrometer up to 
excess energies of $30\,\mbox{MeV}$~\cite{hib99}, the DISTO data point 
discussed above~\cite{bal01}, high energy data from~\cite{fla84} and recent 
results obtained at the TOF facility at COSY at excess energies of $\mbox{Q} = 
92\,\mbox{MeV}$ and $173\,\mbox{MeV}$, respectively~\cite{abd01}, are included.

The one--pion--exchange parametrization~\cite{sib96} deviates strongly from 
the close--to--threshold data, where FSI effects become important, which have 
not been included in the calculation.
Both predictions obtained within the relativistic meson exchange model 
of~\cite{nak01,nak99,nak98} overestimate the recent COSY--TOF 
data~\cite{abd01} by at least a factor of five.
The solid curve in figure~\ref{omega_cross} seems to follow the data slightly 
better, which would correspond to a lower cut--off parameter in the nucleonic 
current and, consequently, a dominating meson exchange current contribution.

Within the model~\cite{nak98,nak99}, angular distributions of the produced 
mesons are expected to differ depending on the dominant production mechanism 
for both $\omega$ and $\phi$ production:
While the mesonic current leads to an isotropic angular distribution, the 
nucleonic current results in a $\cos^2 \Theta_v$ behaviour for the vector 
meson emission angle $\Theta_v$ in the overall centre--of--mass 
system~\footnote{The $\cos^2 \Theta_v$ behaviour arises according 
to~\cite{nak98} from the spin--dependent part of the nucleonic current 
contribution, the ``magnetization current''~\cite{her91}.}.
Thus, angular distributions offer a unique possibility to separate both 
contributions, or to isolate the nucleonic current contribution in order to 
determine the $\phi NN$ coupling directly in a combined analysis of $\omega$ 
and $\phi$ production data in $pp$ induced reactions:
The experimentally observed isotropic $\phi$ angular distribution~\cite{bal01} 
clearly favours a dominant $\phi\rho\pi$ meson exchange over the nucleonic 
current~\cite{mei97,nak01}.
Contrary, the centre--of--mass angular distribution of the $\omega$ meson 
obtained at the COSY--TOF facility at an excess energy of 
$173\,\mbox{MeV}$~\cite{abd01} exhibits a strong anisotropy similar to the 
observation by the DISTO collaboration at an excess energy of 
$319\,\mbox{MeV}$~\cite{bal01}, indicating the presence of higher partial 
waves and rather favours a dominant nucleonic current contribution instead of 
meson exchange currents suggested by the shape of the energy dependence of the 
total cross section in figure~\ref{omega_cross}~\footnote{It should be noted, 
that the excess energy of the COSY--TOF data~\cite{abd01} is slightly out of 
the energy range covered by the predictions from Nakayama et 
al.~\cite{nak98,nak99} due to the onset of inelasticities in the final state.}.

Further constraints on the relevant vector meson production mechanisms in 
nucleon--nucleon scattering might be obtained from comparing with complementary 
meson induced reactions:
An analysis of $\pi N \rightarrow N \phi/\omega$ data concludes $\omega$ 
production to result from strong interferences between the meson exchange 
current and both nucleonic and resonance currents, which contribute with 
similar strengths~\cite{tit01}.
Near threshold, the $\mbox{P}_{11}$ $N^*(1440)$ resonance and the 
$\mbox{S}_{11}$ $N^*(1535)$ and $N^*(1650)$ excitations are found to 
contribute, with the $\mbox{S}_{11}$ contributions canceling to a large 
extent~\footnote{For the resonance contributions in the work of Titov et 
al.~\cite{tit01}, $\omega N N^*$ couplings and relative phases have been taken 
from~\cite{ris01}.}.

In conclusion, at least for the reaction $\pi N \rightarrow N \omega/\phi$, 
the nucleonic current contribution appears to be much larger for $\omega$ 
production relative to the meson exchange current, contrary to $\phi$ 
production~\cite{tit01}.
Thus, the scenario from pion induced vector meson production seems to agree 
--- qualitatively --- with the observations in nucleon--nucleon scattering, 
and is interesting to note especially in view of the $\omega$ angular 
distribution from the COSY--TOF data~\cite{abd01}.
However, angular distributions at lower excess energies would clearly help to 
constrain the interpretation of the presently available nucleon--nucleon data 
in future.\hspace{1ex}

In a combined analysis of $\omega$ and $\phi$ production within the 
relativistic meson exchange framework~\cite{nak01} a value of the $\phi$ 
coupling to the nucleon has been extracted from the DISTO data and is found to 
be consistent with the OZI expectation $g_{NN \Phi} = 
-3\,g_{NN \rho} \sin{\delta_V} \approx 
- \left(0.60 \pm 0.15\right)$~\cite{nak99}.
Yet, as stated by Nakayama, the value still has large uncertainties, 
especially due to the lack of more and precise data in the threshold 
region~\cite{nak01}.
Indeed, on the basis of extracted transition matrix elements~\cite{sib00} 
combining $\pi N$ and $p p$ induced reactions, a $\phi/\omega$ production 
ratio has been derived, which exceeds the OZI estimate of eq.~\eqref{ozieq} by 
a factor of five in both cases.\hspace{1ex}

As suggested in~\cite{bal01}, future experiments on $\phi$ and $\omega$ 
production in close--to--threshold proton--neutron collisions might provide 
further evidence for the intrinsic strangeness content of the nucleon, as 
predictions for $\phi$ production cross sections based on the ``shake--out'' 
and ``rearrangement'' mechanism~\cite{ell00} and on meson exchange 
models~\cite{tit99,rek97} differ by more than an order of magnitude.

\subsection{Scalar sector}                            
\label{Ssec}
\begin{figure}[H]
\hfill
\parbox{0.41\textwidth}
  {\epsfig{file=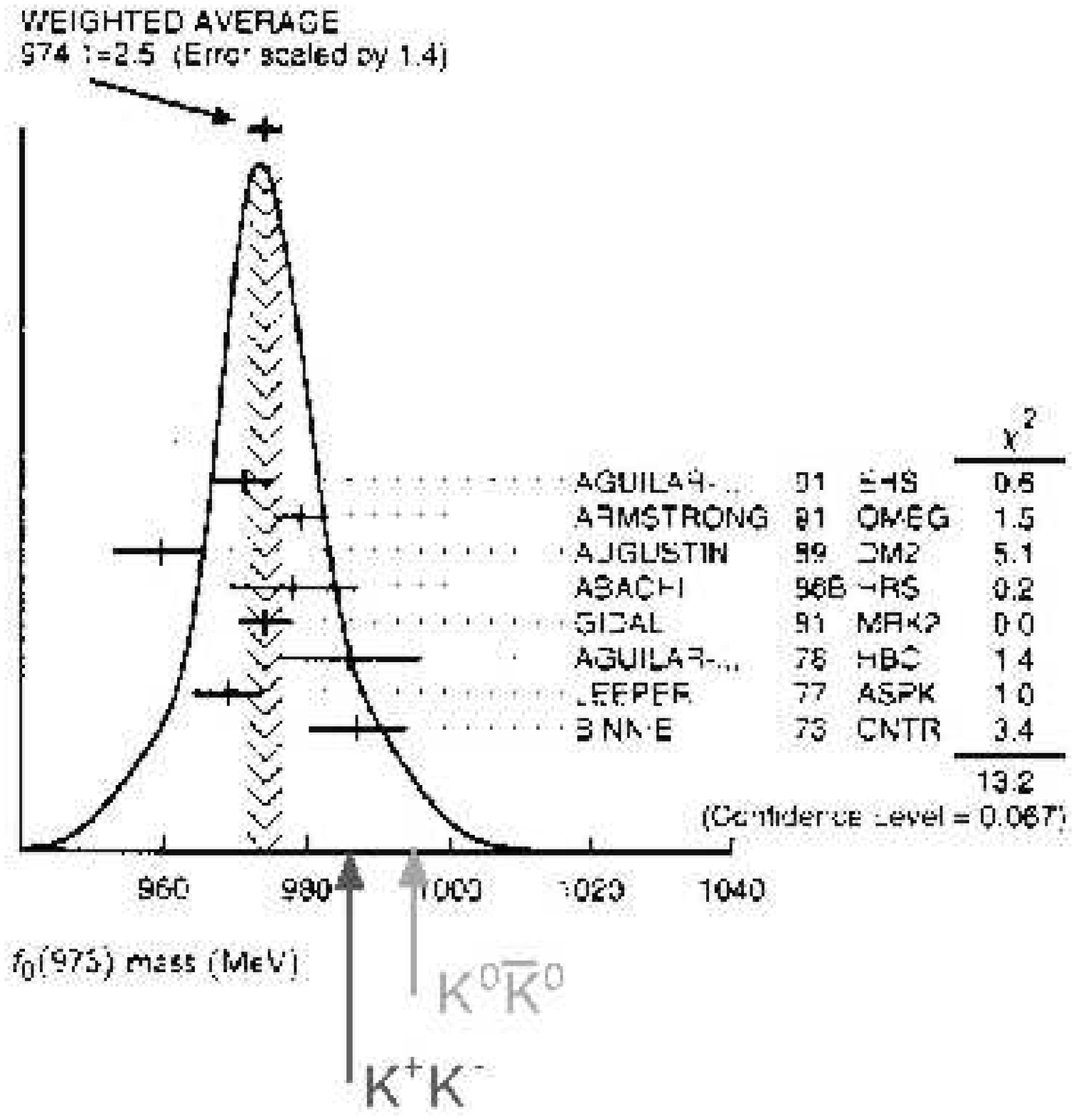,width=0.41\textwidth}} \hfill
\parbox{0.55\textwidth}
  {\epsfig{file=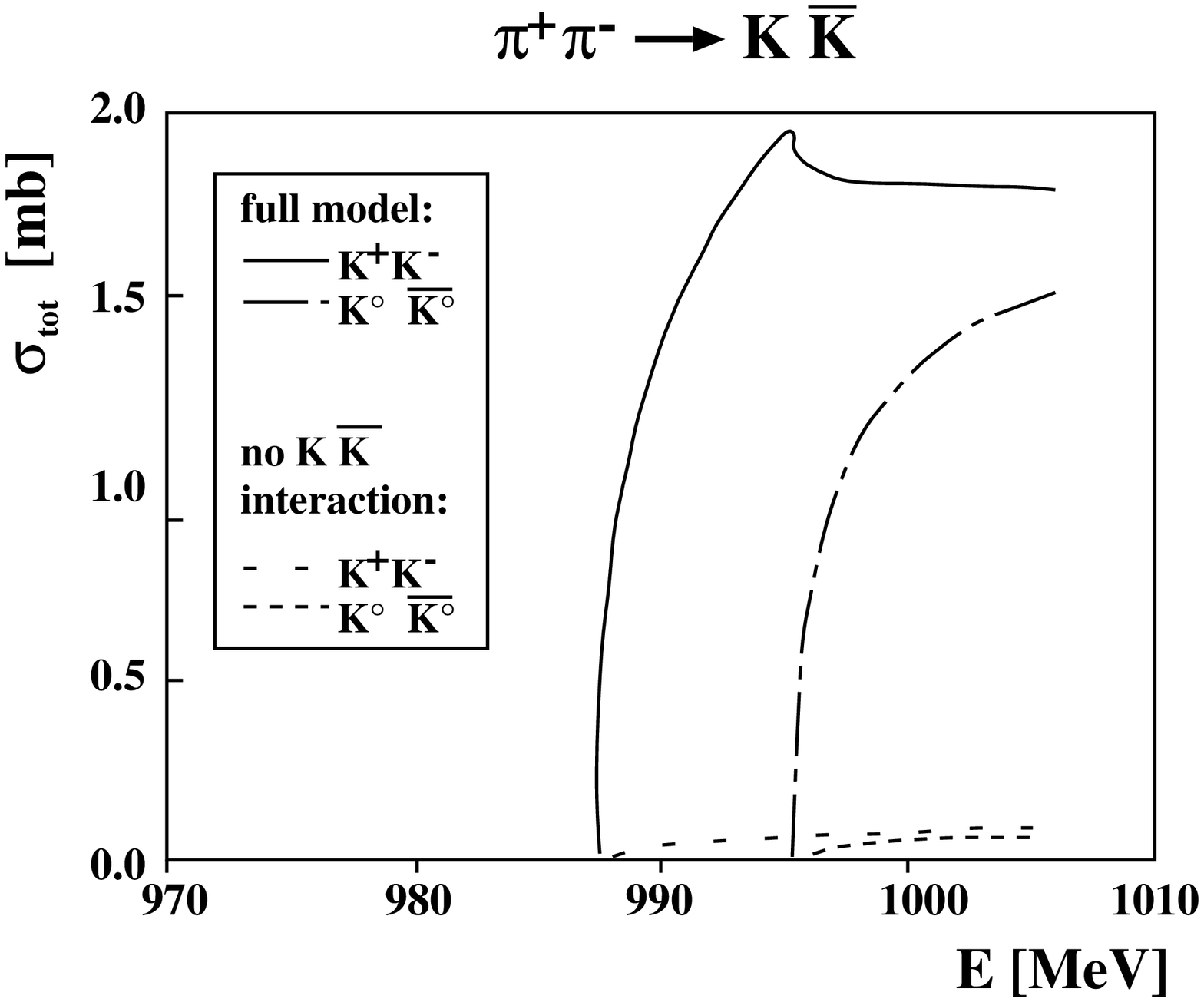,width=0.50\textwidth}} 
\parbox{0.29\textwidth}{\raisebox{1ex}[0ex][0ex]{\mbox{}}}
\parbox{0.40\textwidth}{\raisebox{1ex}[0ex][0ex]{\large a)}}
\parbox{0.09\textwidth}{\raisebox{1ex}[0ex][0ex]{\large b)}}

  {\caption{\label{Mass_PDB_KK_INTERACTION} (a) Weighted mass average for the $f_0(980)$ scalar 
  resonance from~\cite{Mass_PDB}.
  (b) Predictions of the J\"ulich model for 
  $\pi\pi \rightarrow K \overline{K}$ transitions using the full model and 
  neglecting any $K \overline{K}$ interaction, taken from ref.~\cite{KREHL}.}}
\end{figure}

The $1\,\mbox{GeV}/\mbox{c}^2$ meson mass range is continuously under 
discussion regarding the nature of the scalar resonances $f_0(980)$ and 
$a_0(980)$, which have been interpreted as exotic two--quark two--antiquark 
states~\cite{Jaf77}, possible glueball candidates~\cite{Close88}, 
excitations of the light quark vacuum condensate 
({\em minions}~\cite{close1993}), conventional $q \bar q$ 
states~\cite{Mor93,kleefeldf0}, or molecular like $K \overline{K}$ bound 
states~\cite{Wei90,WANG}. 
Just already the similarity of the masses of these objects with twice the kaon 
mass, see figure~\ref{Mass_PDB_KK_INTERACTION}a, suggests a strong correlation between these 
resonances and the $K$--$\overline{K}$ system.

Within the framework of the J\"ulich meson exchange model for 
$\pi\,\pi$~\cite{LOHSE} and $\pi\,\eta$~\cite{KREHL} scattering the 
$K \overline{K}$ interaction dominated by vector meson exchange gives rise to 
a $K \overline{K}$ bound state identified with the $f_0(980)$ in the isoscalar 
sector, while the isovector $a_0(980)$ is concluded to be a dynamically 
generated threshold effect~\cite{KREHL,haidenproc}.

As shown in figure~\ref{Mass_PDB_KK_INTERACTION}b, both shape and absolute scale of 
$\pi\pi \rightarrow K \overline{K}$ transitions turn out to depend crucially 
on the strength of the $K \overline{K}$ interaction, which, in turn is 
prerequisite of a $K \overline{K}$ bound state interpretation of the 
$f_0(980)$. 
Similar effects might be expected for the elementary kaon--antikaon production 
in proton--proton scattering.

Although the $K\overline{K}$ decay mode of $f_0(980)$ and $a_0(980)$ is rather 
weak in comparison to the dominant $\pi\pi$ and $\pi\eta$ decay 
channels~\cite{groom1}, even a new theoretical analysis based on the chiral 
approach can not account for the $f_0(980)$ and $a_0(980)$ if the 
$K \overline{K}$ channel is not introduced additionally to the $\pi\pi$ and 
$\pi\eta$ interaction~\cite{oller}.
An analysis of the $\pi\pi$ and $K\overline{K}$ interaction~\cite{zou} showed 
that the $f_0$ corresponds to two poles on the second and third Riemann sheet, 
respectively, and appears physically as an object with a decay width of about 
$400\,\mbox{MeV}$ and a narrow peak width of about $50\,\mbox{MeV}$. 
The same parameters of the $f_0$ were found by utilizing a unitarized quark 
model, according to which the $f_0$ was interpreted as a $q\bar{q}$ state with 
a large admixture of a $K\overline{K}$ virtual state~\cite{tornqvist}.
The origin of the scalar resonances was also thoroughly studied by means of a 
coupled channel analysis considering $\pi\pi$, $K\overline{K}$ and 
$\sigma\sigma$ meson--meson scattering~\cite{kaminski97,kaminski99}. 
Decreasing gradually the interchannel coupling constants it was inferred that 
for some solutions at the limit of the fully uncoupled case the $f_0$ 
corresponds to a $K\overline{K}$ bound state~\cite{kaminski99,kaminski00}.

Recently notable progress has been made concerning the discussion of this 
topic due to approaches in unitary extensions of chiral perturbation theory 
where the resonances under consideration qualify as dynamically generated 
resonances from the multiple scattering with the lowest order chiral 
Lagrangian, as outlined in the review~\cite{OLLER_NP45}. 
In high energy experiments, the $f_0$ meson is observed as a resonance in the 
system of two pions produced in a variety of hadro--production 
reactions~\cite{alde94,barberis99,barberis,breakstone} and in the hadronic 
decays of either heavier mesons~\cite{aitala,akhmetshin005,akhmetshin006} or 
the $Z^0$ boson~\cite{ackerstaff,abreu} created in $e^+e^-$ collisions. 
These experiments study the invariant masses of the created neutral 
($\pi^0\pi^0$)~\cite{alde94,akhmetshin006,barberis} and charged 
($\pi^+\pi^-$)~\cite{akhmetshin005,aitala,ackerstaff,breakstone,barberis99} 
pion pairs.
Similarly, charged~\cite{ackerstaffC5,gay} and neutral $a_0$~\cite{a0_teige} 
mesons were observed as a clear signal in an invariant mass spectrum of the 
$\eta \pi$ system. 

Complementary to these approaches, studying the interaction of $\pi\pi$, 
$K\overline{K}$ and $\pi\eta$ meson pairs, the COSY--11 collaboration 
presently investigates the possible manifestation of the mesons $f_0$ and/or 
$a_0$ as doorway states leading to meson production in proton--proton 
collisions, namely $pp \rightarrow pp f_0 (a_0) \rightarrow pp\,Mesons$.
By measuring the missing mass of the $pp$--system the overlapping $f_0$--$a_0$ 
resonances are studied as a genuine particle produced directly at the reaction. 
The production of these resonances is carried out at a mass range a few tens of
MeV below the $K\overline{K}$ threshold (where the resonances can only decay 
into non--strange mesons) and above the $K\overline{K}$ thresholds, where the 
equivalent part of these broad resonances can be excited.

\begin{figure}[H]
\hfill
\parbox{0.55\textwidth}
  {\epsfig{file=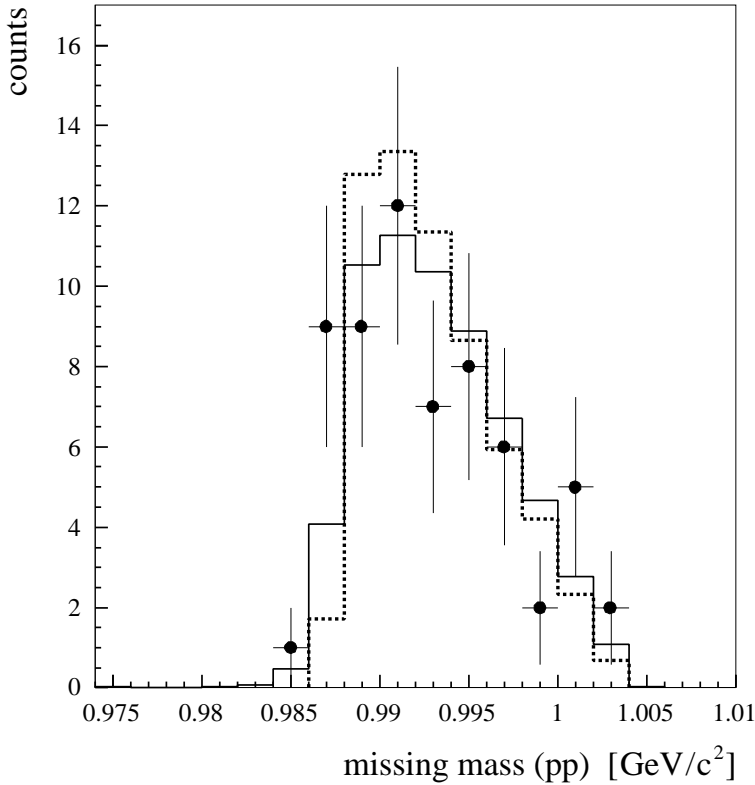,width=0.55\textwidth}}\hfill
\parbox{0.4\textwidth}
  {\caption{\label{mm_kk_f0} Experimental spectrum of the $K^+ K^-$ invariant 
  mass measured for the reaction $pp \rightarrow pp K^+ K^-$ at a beam 
  momentum of $3.356\,\mbox{GeV/c}$ corresponding to an excess energy of 
  $\mbox{Q} = 17\,\mbox{MeV}$ with respect to the $K^+ K^-$ threshold (data 
  points). The width of the bins corresponds to the experimental resolution of 
  the mass determination ($\mbox{FWHM} \approx 2\,\mbox{MeV}$).
  The solid and dashed lines show Monte--Carlo simulations assuming the direct 
  and resonant production, correspondingly\cite{quentmeierphd}.}}
\end{figure}

It is obvious that for the production of a broad resonance the phrase 
``close--to--threshold'' is not well defined and implies here that the beam 
momentum is such that masses just in the range of the resonance can be excited.
Different mass ranges of such objects are produced with different excess 
energies making the interpretation of the experimental data difficult but still
possible~\cite{proceedingscosy11}.
It is worth noting, that recently first measurements of the $f_0$ meson 
production relatively close to its threshold but still above the 
$K\overline{K}$ threshold were performed by observing the $\phi$ meson decays 
into the $f_0 \gamma$ channel~\cite{akhmetshin005,akhmetshin006}, where the 
$\phi$ is only about $40\,\mbox{MeV}/\mbox{c}^2$ heavier than the $f_0$.
In a recent publication~\cite{quentmeier276} data on the close--to--threshold 
$K^+ K^-$ production following the proton--proton interaction at an excess 
energy of $\mbox{Q} = 17\,\mbox{MeV}$ are presented.
The obtained distribution of the missing mass to the $pp$ system is shown in 
figure~\ref{mm_kk_f0} and demonstrates that the non--resonant $K^+ K^-$ 
production (solid line) is hardly distinguishable from the resonant $pp 
\rightarrow pp f_0(980) \rightarrow pp K^+ K^-$ reaction sequence (dashed 
line)~\cite{quentmeierphd}.
Since the statistics of the data were not sufficient to favour one of the two 
processes, the cross section was extracted for both and resulted in nearly 
identical values of: \hspace{1ex} $\sigma_{non-resonant} = 
(1.80 \pm 0.27^{\,+0.28}_{\,-0.35})\,\mbox{nb}$ and $\sigma_{resonant} = 
(1.84 \pm 0.29^{\,+0.25}_{\,-0.33})\,\mbox{nb}$ including statistical and 
systematical errors, respectively.
The issue, whether there is a chance to distinguish between $K\overline{K}$ 
pairs originating from the decay of a genuine $f_0$/$a_0$ resonance and those 
produced by a strong $\pi\pi$--$K\overline{K}$ correlation is at present under 
theoretical investigation~\cite{haidenproc}. Still, calculations based upon the
one--pion exchange and a Breit--Wigner presentation of the $f_0$ resonance
indicate~\cite{Bratkovskaya165} that no $f_0$ signal might be extracted from 
the $K^+ K^-$ invariant mass spectrum in $pp \to pp K^+K^-$ reaction at 
near--threshold energies, due to the large contribution from other reaction 
channels~\cite{sibirtsev101} as arising from pion and kaon exchange.

Considering the $a_0\left(980\right)$ as a conventional $q \overline{q}$ meson,
predictions on the total and differential cross sections for $a_0$ production
via the reaction $p p \rightarrow d a_0^+$ are given for the COSY energy range
in~\cite{gri00}:
Within a two--step model, where the $a_0$ is found to be produced predominantly
from u--channel $\pi$ exchange with a subsequent fusion of the nucleons to a
deuteron bound state, a rather large total cross section is concluded.
An experimental verification at the ANKE facility~\cite{che97} might give an
additional hint on the $q \overline{q}$ content of the isovector scalar
resonance.
The measurements also aim at determining the rather poorly known $a_0$
branching
ratios to the $K\overline{K}$ and $\eta \pi$ final states (see discussion
in~\cite{gri00} and references therein).

The role of final state interactions in the reactions $pp \rightarrow d K^+
\overline{K^0}$ and $pp \rightarrow d \pi^+ \eta$ is explicitly studied
in~\cite{ose01}:
Considering primary production amplitudes within meson--baryon chiral
perturbation theory, $\overline{K^0} d$ final state interactions --- due to the
proximity of the $\Lambda\left(1405\right)$ to the $\overline{K^0} n$ system
---
are expected to change the cross section by up to an order of magnitude and to
significantly influence invariant mass distributions.
The $K^+ \overline{K^0}$ and $\pi^+ \eta$ final state systems are investigated
within a coupled--channel chiral unitary approach, in which the
$a_0\left(980\right)$ is generated dynamically as a meson--meson resonance.
Once the two remaining independent model parameters --- related to the
couplings
of the two pseudoscalar meson systems --- are fixed by sensitive experimental
data, i.e.\ $K^+ \overline{K^0}$ and $\overline{K^0} d$ invariant mass
distributions, the approach will provide absolute predictions for the $\eta
\pi$
system.
Thus, the underlying dynamics of the coupled--channel system giving rise to the
$a_0\left(980\right)$ should be probed directly by the experimental approach
of~\cite{che97}.

Further experimental efforts on studying the iso--singlet $f_0$ and the 
iso--triplet $a_0$ scalar resonances are continuing at the 
COSY--11~\cite{proceedingscosy11,Proposal} and ANKE~\cite{Ankeproposal} 
facilities at COSY using different approaches.

\subsection{Associated strangeness production}                
\label{Asp}
 
In elementary hadronic interactions with no strange valence quark in the 
initial state the associated strangeness production provides a powerful tool 
to study reaction dynamics by introducing an ``impurity'' to hadronic matter.
Thus, quark model descriptions might be related to mesonic or baryonic degrees 
of freedom, with the onset of quark degrees of freedom expected for 
kinematical situations with large enough transverse momentum transfer.

A large fraction of the nucleon spin is carried by the intrinsic
orbital angular momentum of quarks which -- in the condensate -- 
are pairwise in a $^3P_0$ state. The challenge of the strange
quark, with its intermediate mass between light and heavy quarks,
is to explore the effective degrees of freedom via the associated
strangeness production in the low energy structure of QCD.
The $\bar q q$ ($\bar s s$) vacuum condensate is the link between
current quark and gluon of QCD and valence quark and potentials of
the quark model, structures which can give the massless QCD quarks 
their constituent masses.

In close--to--threshold kaon--hyperon production, effects of the 
low energy hyperon--nucleon interaction are inherent to the 
observables, and allow to constrain hyperon nucleon 
interaction models.

As the strange quark is heavy compared to up and down quarks
($\mbox{m}_u,\,\mbox{m}_d << \mbox{m}_s \sim \Lambda_{QCD}$) 
it is not a priori clear, whether the strange quark is appropriately 
treated as a light quark on the hadronic scale in chiral perturbative 
approaches with an expansion parameter of  $\mbox{m}_K/4\pi F_{\pi} = 0.43$ 
(with the kaon mass $m_K$ and the pion decay constant $F_{\pi}$) compared to 
$\mbox{m}_{\pi}/4\pi F_{\pi} = 0.12$. In exploratory calculations, SU(3) 
baryon chiral perturbation theory only has been applied to kaon photo-- and 
electroproduction~\cite{stei97}, for a more detailed discussion  
see~\cite{mei2001}).

With a significant contribution of strange sea quarks to the nucleon wave 
function being suggested by complementary experimental approaches during the 
past years (see section~\ref{vmp} and~\cite{sigmaterm,polmuon,ada99,bar00}) 
the creation of an $s \overline{s}$ pair in the nucleon--nucleon interaction 
should provide a further piece of valuable information on the structure of the 
nucleon~\cite{kleefeld2867}. Especially polarization observables are expected 
to act as very sensitive tools~\cite{alb95}.

At the same time, both the kaon and antikaon (see section~\ref{SdiKKp}) 
production cross section in nucleon--nucleon scattering are important input 
parameters within the framework of transport model calculations used to 
describe the strangeness flow in heavy ion collisions:
Due to strangeness conservation, kaons will not be absorbed in the nuclear 
medium. 
Thus, kaon yields in subthreshold production are expected to probe the early 
stage of the collision, i.e.\ hot, dense matter, the nuclear equation of state 
at high densities~\cite{eos} and the in--medium kaon 
potential~\cite{fan94,Lut98,ram00}, see also section~\ref{SdiKKp}), 
which relates to the partial restoration of chiral symmetry~\cite{kap86,bro94}
(for a review see~\cite{cas99}).
Elementary cross sections are also of interest in view of the recent inclusive 
subthreshold measurements in nucleus--nucleus collisions, resulting in 
comparable $K^+$ and $K^-$ yields at the same energy per nucleon below 
threshold~\cite{lau99,sen01,oes01}.

\begin{figure}[H]
\hfill
\parbox{0.55\textwidth}
  {\epsfig{file=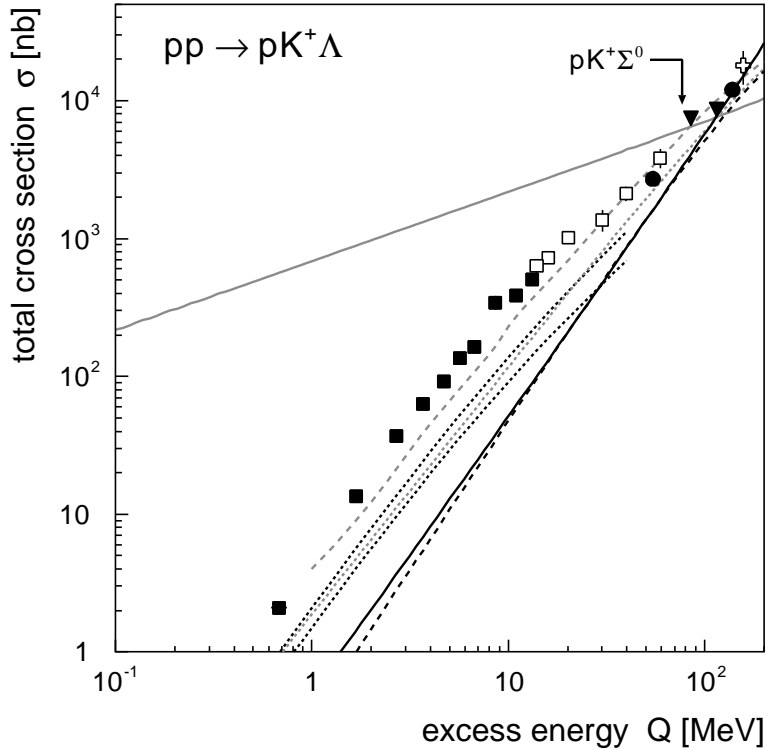,width=0.55\textwidth}} \hfill
\parbox{0.31\textwidth}
  {\caption{\label{pkl_cross} Total cross section of the reaction $pp 
  \rightarrow p K^+ \Lambda$. Data obtained at the COSY--11 facility are 
  marked by filled \cite{bal98,sewerin682} and open squares \cite{kow02}, 
  COSY--TOF data by filled circles \cite{bilger217} and triangles 
  \cite{mar01,hes01} and the BNL result \cite{fickinger2082} by an open cross, 
  respectively. Experimental data are the same as in figure 
  \ref{cross_etap_pkl}b. The arrow denotes the opening of the $p K^+ \Sigma^0$ 
  threshold. For comparison, theoretical descriptions \cite{sch87,zwe88,ran80,
  li95b,sib95,tsu97,fae97} are included (for details see text).}}
\end{figure}

Exclusive data close--to--threshold on the elementary kaon production via $pp 
\rightarrow p K^+ \Lambda$ have only become available recently with the advent 
of dedicated experiments at COSY.
Figure~\ref{pkl_cross} shows the available total cross section data obtained 
at the COSY--11~\cite{bal98,sewerin682,kow02} and COSY--TOF 
facility~\cite{bilger217,mar01,hes01} together with an earlier result 
from the BNL Cosmotron~\cite{fickinger2082} and theoretical calculations and 
parametrizations as a function of the excess energy~Q.\\[-0.2cm]

Assuming the production amplitude to be constant and both the $p \Lambda$ 
system and the $K^+$ relative to the $p \Lambda$ subsystem to be in s--wave 
leads to a production cross section proportional to $\eta^4$ 
(section~\ref{Siv}, eq.~\eqref{sigmaLl}). 
This dependence has been adopted by Sch\"urmann and Zwermann~\cite{sch87} for 
a quartic (in terms of the maximum meson momentum) parametrization of the 
cross section (black solid line in fig.~\ref{pkl_cross}), leading to a 
reasonable fit~\cite{zwe88} of data at excess energies above $150\,\mbox{MeV}$.
Although quite commonly employed in the description of nucleon--nucleus and 
nucleus--nucleus collisions in the past~\cite{li95b}, the quartic 
parametrization underestimates the close--to--threshold data by an order of 
magnitude, as does a linear parametrization once suggested by Randrup and Ko 
(grey solid line~\cite{ran80}), overestimating the presently available data at
small excess energies by more than two orders of magnitude. It should be noted 
that both descriptions have been estimated from high energy data, before any 
of the near--threshold measurements were done.

\begin{figure}[H]
\begin{center}
\epsfig{file=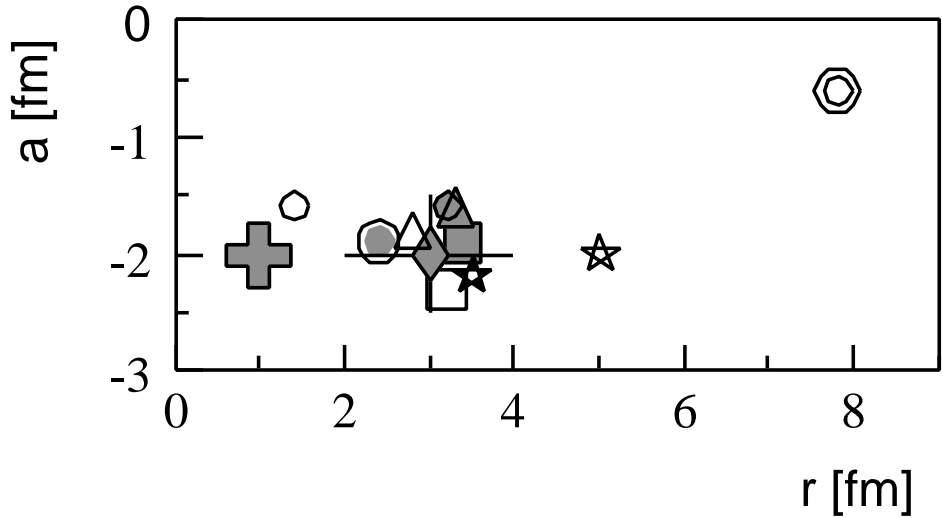,scale=0.8}
\end{center}
\caption{\label{pl_param} $\Lambda p$ S--wave singlet (open symbols) and 
triplet (closed symbols) scattering length $a_{p\Lambda}$ and effective range 
$r_{p\Lambda}$. 
Experimental data are from $\Lambda p$ elastic scattering (stars~\cite{ale68} 
and triangles~\cite{sec68}), $K^-$ capture on a deuteron 
(diamond~\cite{tan69}) and from the FSI approach in threshold production 
(cross~\cite{bal98b}, spin--averaged value). Results of $YN$ interaction 
models are included by squares~\cite{nag79} and circles and double circles 
(solutions A and B of~\cite{hol89,reu94}), respectively. Errors have only been 
determined for the $K^-$ capture experiment~\cite{tan69}.}
\end{figure}

Close--to--threshold, deviations from the $\eta^4$ energy dependence result from 
the strong s--wave final state interactions in the three two--body $p K^+$, 
$p \Lambda$ and $K^+ \Lambda$ subsystems. 
As the strong interaction in the $p K^+$ and $K^+ \Lambda$ systems appears to 
be weaker by an order of magnitude as compared to the proton--$\Lambda$ 
case~\cite{hof95,del89}, an accurate description of the energy dependence is 
expected to be achieved by considering the strong attractive $p \Lambda$--FSI 
and repulsive Coulomb corrections in the proton--kaon subsystem.
Following the discussion in section~\ref{Mwhs}, a reasonable fit of the energy 
dependence is obtained by factorizing the reaction amplitude 
$M_{pp \rightarrow p K^+ \Lambda}$ according to eq.~\eqref{M0FSIISI}:
\begin{eqnarray}
\left|M_{pp \rightarrow p K^+ \Lambda}\right|^2 & \approx & 
  \left| M_{FSI} \right|^2 \cdot \left| M_0 \right|^2 \cdot F_{ISI} \nn \\
 & \approx & F_{FSI}\left(\mbox{p}_{p \Lambda}\right) \, 
     F_{Coul}\left(\mbox{p}_{p K^+}\right) \cdot 
    \left|M_0\right|^2 \cdot F_{ISI} \; 
\end{eqnarray}
Here, $M_0$ is the short--range production amplitude, which is assumed to be 
constant in the energy range close--to--threshold, $M_{FSI}$ and $F_{ISI}$ 
describe the elastic interaction in the final state and the reduction due to 
the $pp$ initial state interaction (see section~\ref{Mwhs}).
According to the above discussion the interaction in the three--body final 
state $M_{FSI}$ is approximated by $F_{Coul}(\mbox{p}_{p K^+})$ and 
$F_{FSI}(\mbox{p}_{p \Lambda})$, which denote the corrections for the Coulomb 
interaction and the $p \Lambda$ FSI, depending on the relative momenta 
$\mbox{p}_{p K^+}$ and $\mbox{p}_{p\Lambda}$ in the proton--kaon and 
proton--$\Lambda$ centre--of--mass systems, respectively.
The Coulomb correction is essentially given by the Coulomb penetration factor 
$C^2$ of eq.~\eqref{penetrationfactor}
\begin{equation}
F_{Coul} \left(\mbox{p}_{p K^+}\right) = C^2 = 
  \frac{2\pi\,\eta_{c,pK^+}}{e^{\,2\pi\,\eta_{c,pK^+}} - 1}\,;\; 
\end{equation}
with the relativistic Coulomb parameter 
\begin{equation*}
\eta_{c,pK^+} \,=\, \alpha\; 
  \frac{\mbox{s}_{p K^+} - m_p^2 - m_{K^+}^2}
       {\sqrt{\lambda(\mbox{s}_{pK^+},m_p^2,m_{K^+}^2)}},
\end{equation*}
where $\alpha$ is the fine structure constant, $\mbox{s}_{p K^+}$ denotes the 
total energy in the $pK^+$ subsystem and the triangle function $\lambda$ is 
defined in eq.~\eqref{kaellen}.\newline
The modification of the energy dependence close--to--threshold due to the 
$p \Lambda$ FSI
\begin{equation}
\label{fsipl}
F_{FSI}\left(\mbox{p}_{p \Lambda}\right) = 
  \frac{1}
    {\mbox{p}_{p\Lambda}^2\left(1 + \cot^2\delta_{p\Lambda}\right)} =
  \frac{1}
    {\mbox{p}_{p\Lambda}^2 + 
     \left(-1/\hat{a}_{p \Lambda} + 
           \hat{r}_{p \Lambda}\,\mbox{p}_{p\Lambda}^2\,/\,2\right)^2}
\end{equation}
is determined by the relative momentum in the $p \Lambda$ centre--of--mass 
system and the $p \Lambda$ S--wave phase--shift $\delta_{p\Lambda}$, which has 
been expressed in terms of an effective range expansion in eq.~\eqref{fsipl}, 
with $\hat{a}_{p \Lambda}$ and $\hat{r}_{p \Lambda}$ as spin--averaged values 
of $p \Lambda$ scattering length and effective range~\footnote{As the 
$p \Lambda$ system can be in a spin singlet ${}^1\mbox{S}_0$ and spin triplet 
${}^3\mbox{S}_1$ state, the amplitudes should be added incoherently. However, 
model calculations~\cite{nag79,hol89,reu94} and experimental 
data~\cite{ale68,sec68} suggest that the scattering parameters are rather 
similar for both spin states (fig.~\ref{pl_param}).}.
Following the result for the $pp \eta$ final state in 
eq.~\eqref{cross_with_FSI} in section~\ref{Mwhs} and using 
relation~\eqref{Vps_relativistic} from section~\ref{Siv}, the total cross 
section for the reaction $pp \rightarrow p K^+ \Lambda$ writes as:
\begin{equation}
\label{pklcross_eq}
\sigma\left(pp \rightarrow p K^+ \Lambda\right) = 
  \frac{F_{ISI} \left|M_0\right|^2}{\mbox{F}} \;
  \int d\,V_{ps}
  \frac{\left(2\pi\eta_{c,pK^+}\right)/\left(e^{2\pi\eta_{c,pK^+}} - 1\right)}
 {\mbox{p}_{p \Lambda}^2 + 
  \left( -1/\hat{a}_{p \Lambda} + 
         \hat{r}_{p \Lambda}\,\mbox{p}_{p \Lambda}^2\,/\,2 \right)^2} 
\end{equation}
where F is the flux factor of the colliding nucleons from defined in 
eq.~\eqref{fluxfactor}.
Using the energy dependence given in eq.~\eqref{pklcross_eq} an accurate 
description close--to--threshold is achieved, as shown in 
figure~\ref{cross_etap_pkl}b in section~\ref{Mwhs} (see 
also~\cite{bal98}).\vspace{1ex}

\begin{figure}[H]
\begin{center}
\epsfig{file=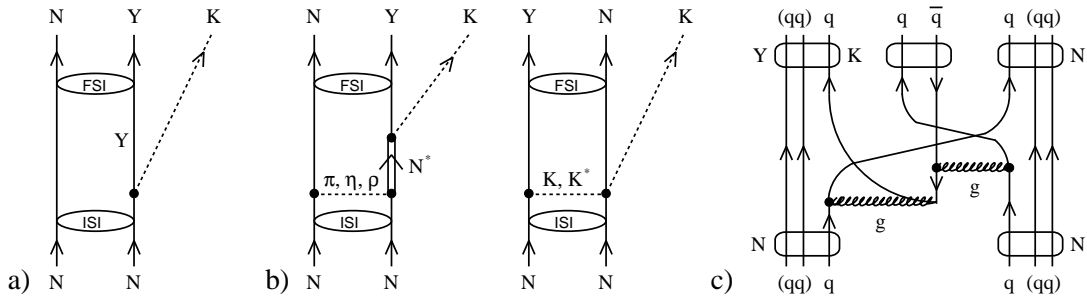,scale=0.5,angle=270}
\end{center}
\caption{\label{pky_graph} Possible production mechanisms for kaon--hyperon 
production in nucleon--nucleon scattering. \hspace{1ex} a) --- Direct kaon 
emission, \hspace{1ex} b) --- Nonstrange and strange meson exchange diagrams, 
\hspace{1ex} c) --- Quark--gluon exchange mechanism. For the latter see the 
discussion in section~\ref{Dopsmp}.}
\end{figure}

Within meson--exchange models, contributions to the reaction dynamics of the 
elementary strange\-ness dissociation are expected from direct production 
(fig.~\ref{pky_graph}a) and from the exchange of nonstrange 
$(\pi,\eta,\rho,\sigma,\omega)$ mesons including $N^*$ resonance excitation 
and $(K,K^*)$--exchange in the strange sector (fig.~\ref{pky_graph}b).\\[-1.0cm]

Pion-- and kaon---exchange were suggested as mechanisms of the associated 
strangeness production in the nucleon--nucleon interaction already by 
Ferrari~\cite{fer60}. As noted by Laget~\cite{lag91}, the contribution from
direct production~(fig.\ref{pky_graph}a) to the total cross section is small but
might be of relevance for differential observables as discussed below.

Modern parameterizations considering $\pi$-- and $K$--exchange reproduce the 
magnitude of the total cross section close--to--threshold within a factor of two 
or three (grey dashed~\cite{li95b} and grey dotted~\cite{sib95} lines in 
fig.~\ref{pkl_cross}).
In refs.~\cite{li95b,sib95} the elementary $\pi N \rightarrow K Y$ and $K N 
\rightarrow K N$ transitions are included phenomenologically~\footnote{For $\pi$ 
induced strangeness production, either experimental values~\cite{pinky} (as in 
the work of Laget~\cite{lag91}) or parametrizations (cross sections determined 
from the resonance model approach~\cite{tsu95} used by Li and Ko~\cite{li95b}, 
and Sibirtsev~\cite{sib95}), respectively, are employed. $KN$ scattering data 
are available experimentally~\cite{gia74} and in terms of phase--shift 
analyses~\cite{mar75}.} and both exchange diagrams are found to be important 
in the case of the reaction $pp \rightarrow p K^+ \Lambda$, while $\Sigma$ 
production appears to be dominated by $\pi$--exchange~\footnote{The relative 
strength of the $K$ exchange mechanism in $\Lambda$ and $\Sigma^0$ production 
reflects the smallness of the $KN \Sigma$ coupling constant 
$\mbox{g}_{KN\Sigma}$ as compared to $\mbox{g}_{KN \Lambda}$ (see 
e.g.~\cite{mae89,reu94})}.
However, interference terms of the amplitudes as well as the effects of 
final state interactions have been neglected, although the latter 
have been shown to be of relevance in earlier studies~\cite{lag91}.

The same holds for a resonance model (dashed line in 
fig.~\ref{pkl_cross}~\cite{tsu97,tsu99}) considering $\pi$--, $\eta$-- and 
$\rho$--exchange exciting baryon resonances up to $2\,\mbox{GeV}/\mbox{c}^2$ 
coupling to the kaon--hyperon channel (fig.~\ref{pky_graph}b left), 
i.e.\ the $\mbox{S}_{11}$ $N^*(1650)$, $\mbox{P}_{11}$ $N^*(1710)$, 
$\mbox{P}_{13}$ $N^*(1720)$ and $\mbox{P}_{33}$ $\Delta(1920)$ resonances.
The fully relativistic model essentially uses parameters fixed by the previous 
study on $\pi$ induced strangeness production~\cite{tsu95} and experimental 
branching ratios, interferences of the amplitudes are neglected.
Non--resonant exchange diagrams, i.e.\ especially kaon exchange,
are not considered from arguments based on the author's studies on
$\pi N \to K Y$ reactions~\cite{tsu95}.  
As a result, $\pi$ exchange and excitation of the $N^*(1650)$, coupling 
strongly to the $\pi N$ channel, is found to give the dominant contribution 
close--to--threshold and is only exceeded by $\rho$ exchange at excess energies 
beyond $1\,\mbox{GeV}$.
However, as FSI effects are neglected --- according to the authors --- the 
approach is neither expected nor intended to describe the close--to--threshold 
regime~\cite{tsu99}.

Within a factor of two, the total cross section data is described by an 
alternative approach~\cite{fae97} considering $\pi$ exchange and excitation of 
the second $\mbox{S}_{11}$ resonance, the $N^*(1650)$ (dotted lines in 
fig.~\ref{pkl_cross}) --- in analogy to $\eta$ production arising from $\pi$ 
exchange and excitation of the first $\mbox{S}_{11}$ resonance ($N^*(1535)$).
Assuming the $N^*(1535)$ and the $N^*(1650)$ to dominate the $\eta N$ and 
$K^+ \Lambda$ systems, respectively, leads to identical forms of the 
production operators and the spin--angular momentum algebra.
Thus, after normalizing the scale of the $p p \rightarrow p K^+ \Lambda$ 
excitation function to low--energy data on the $pp \eta$ final state, the model 
gives an absolute prediction for $\Lambda$ production, while the energy 
dependence close--to--threshold is essentially determined by three--body 
phase--space modified by the $p \Lambda$ final state interaction.
The two corresponding curves shown in figure~\ref{pkl_cross} (dotted lines) 
differ in the choice of low energy $p \Lambda$ scattering 
parameters~\cite{hol89,reu94,ale68} determining the proton--hyperon FSI 
(see eq.~\eqref{fsipl}).

The situation is summarized by Kaiser~\cite{kai99b}:
Separating the on--shell S--wave final state interaction and the invariant 
$T$ matrix at threshold, the information from the close--to--threshold total 
cross section can be ``condensed'' into essentially one number, the threshold 
transition amplitude parametrizing the $T$ matrix, with the energy dependence 
being accurately reproduced by three--body phase--space and FSI (see also our 
discussion in section~\ref{Dopsmp}).
In an analysis based on tree--level meson exchange diagrams pointlike $\omega$ 
exchange alone or the total vector $\rho^0$, $\omega$ and $K^*$ exchange, are 
found to describe the transition amplitude~\footnote{The transition amplitudes 
extracted in~\cite{kai99b} based on the approach developed 
in~\cite{bernard259} are related to the production amplitude $|M_0|$ used in 
this work by simple kinematical factors, i.e.\ $|\,M_0\,|^2 = |\,{\cal{T}}\,|^2 
\cdot (8\,\mbox{m}_p\,\mbox{m}_B \cdot 
\lambda(\mbox{s},\mbox{m}^2_p,\mbox{m}^2_p))\,/\,\mbox{s}$, with 
$\mbox{m}_B$ denoting the mass of the second baryon $B = p,Y$ in the final 
state and $|\,{\cal{T}}\,|$ as threshold transition amplitude, i.e.\ 
$|\,{\cal{A}}\,|$ and $|\,{\cal{C}}\,|$ for $\pi$ and $\eta$ production 
in~\cite{bernard259} and $|\,{\cal{K}}\,|/\sqrt{3}$ in~\cite{kai99b}.}.
However, also pseudoscalar amplitudes, i.e.\ pointlike $K^+$ exchange alone, 
or --- in line with~\cite{fae97,shy99} --- $\pi$ exchange followed by an 
excitation of the $\mbox{S}_{11}$ $N^*(1650)$ resonance when added to the total 
pseudoscalar and vector exchange contributions are in good agreement with the 
experiment.
It is concluded~\cite{kai99a} that the unpolarized $pp \rightarrow 
p K^+ \Lambda$ total cross section close--to--threshold does not provide enough 
information to determine the underlying production mechanism.

Instead, further information should be provided in the future by angular 
distributions and polarization observables, as reported at higher excess 
energies by the DISTO collaboration~\cite{bal99}:
The importance of $K$ exchange for $\Lambda$ production, which might already 
be expected from elementary amplitudes~\footnote{The total $\pi p \rightarrow 
K Y$ cross section is in the range of $1\,\mbox{mb}$ compared to 
$10\,\mbox{mb}$ for $K^+ p$  scattering (see fig.~\ref{kpkp_kk_SsPsPp}a).} has 
been supported experimentally by data on the normal spin transfer coefficient 
$\mbox{D}_{NN}$ in exclusive data at an excess energy of $\mbox{Q} = 
430\,\mbox{MeV}$.
$\mbox{D}_{NN}$ is found to be large and the $\Lambda$ polarization correlated 
with the beam polarization is oriented opposite to the beam spin for forward 
$\Lambda$ production.
Based on the model described in~\cite{lag91}, calculations by Laget show 
$\mbox{D}_{NN}$ to discriminate between the $\pi$ and $K$ exchange mechanism 
from a different spin coupling at the vertices between the polarized proton in 
the initial and the hyperon in the final state~\footnote{For a more general 
discussion on the interpretation of spin transfer measurements 
see~\cite{vig92}.}. 
While from angular momentum and parity conservation $\pi$ exchange leads to 
$\mbox{D}_{NN} = +1$ for forward production, kaon emission results in a spin 
flip ($\mbox{D}_{NN} = -1$) and consequently, the latter is clearly favoured 
by the experimental result~\cite{bal99}~\footnote{The DISTO results show an 
opposite sign compared to inclusive $\mbox{D}_{NN}$ high energy ($\mbox{Q} = 
16.9\,\mbox{GeV}$) data~\cite{bra97}, which have been accounted for within 
quark model calculations~\cite{lia97}.}.
Recently, the negative sign of the spin transfer coefficient, i.e.\ the 
dominance of $K$ exchange for $\Lambda$ production, has been confirmed for two 
different beam momenta corresponding to excess energies of $319\,\mbox{MeV}$ 
and $200\,\mbox{MeV}$, respectively~\cite{mag01}.
However, unexpectedly large negative values are also seen for backward 
production, where the hyperon is preferentially associated with the 
unpolarized target proton --- an observation, which is so far not reproduced 
within the meson exchange picture.

The first close--to--threshold measurement of the $\Lambda$ recoil 
polarization at the COSY--TOF facility has shown a negative polarization for 
transverse momentum transfers $\mbox{p}_t \ge 
0.3\,\mbox{GeV/c}$~\cite{bilger217}~\footnote{A negative $\Lambda$ polarization 
is also observed for $\mbox{p}_t \ge 0.4\,\mbox{GeV/c}$ in proton--proton 
scattering at higher energies~\cite{lampolhigh}.}, which implies evidence for 
S-- and P--wave interference terms at the respective excess energy of 
$138\,\mbox{MeV}$ according to~\cite{kai99b}.

Differential observables, like the occupation in a Dalitz plot, or in 
projection invariant mass distributions of two--particle subsystems, might 
also provide important information to identify the relevant production 
mechanism.
A resonant production should influence for example the $\Lambda K^+$ invariant 
mass distribution significantly at excess energies in the range of 
$100\,\mbox{MeV}$, as calculated within the resonance model approach 
of~\cite{tsu97,tsu99} by Sibirtsev et al.~\cite{sib98}~\footnote{The 
calculations on differential distributions have recently been extended to the 
reaction $pp \rightarrow p K^+ \Sigma^0$ in~\cite{sib99b}.}.
First data has already been published~\cite{bilger217} and higher statistics 
distributions at several excess energies will be available from data taken at 
the COSY--TOF facility in the near future.

$\Lambda$ production on a neutron target, i.e.\ via the reactions $pn 
\rightarrow p(n) K^0(K^+) \Lambda$, might turn out to be sensitive on the 
relative strength of $\pi$, $\rho$ and $\eta$ exchange, as it has been 
discussed in case of $\eta$ production~\cite{fae97,calen2667}.\vspace{1ex}

An important aspect in the theoretical interpretation of data close to the 
production threshold is the interplay between the type of hyperon--nucleon 
interaction employed and the result of the model calculation.
As already discussed by Laget~\cite{lag91}, with experimental observables 
being significantly influenced by the strong final state interaction at low 
transverse momenta, exclusive close--to--threshold data should be expected to 
eventually constrain strongly $YN$ low--energy scattering amplitudes.
Within a meson--exchange approach, i.e.\ considering direct kaon emission as 
well as $\pi$ and $K$ exchange (figs.~\ref{pky_graph}a--b) and using an 
off--shell $T$ matrix description for the hyperon--nucleon 
interaction the sensitivity on the microscopic hyperon--nucleon potential used 
has been studied in detail by Kelkar and Jain~\cite{kel00} for $\Lambda$ 
production in comparison to the inclusive SATURNE data from~\cite{sie94}.
The authors conclude, that --- in order to separate different 
channels contributing at the $\Sigma N$ threshold --- \hspace{1ex} i) 
exclusive data is needed (which shall be discussed below) to constrain 
hyperon--nucleon interaction models and that \hspace{1ex} ii) an improved 
knowledge on the hyperon--nucleon interaction is prerequisite to further 
detailed investigations on the relevant production mechanisms.
We shall use these conclusions as guideline for the discussions during the 
next paragraphs.\vspace{1ex}

Due to the short hyperon lifetimes direct $Y N$ scattering experiments are 
difficult to perform and, consequently, low--energy scattering parameters are 
still rather unknown experimentally.
In the case of the $\Lambda$--proton system, experimental results have only 
been available at excess energies above $3.8\,\mbox{MeV}$~\cite{ale68,sec68}.
In contrast, a study in close--to--threshold production via the strong final 
state interaction covers an excess energy range down to threshold, otherwise 
inaccessible for elastic scattering experiments.

Assuming the energy dependence of the reaction $pp \rightarrow p K^+ \Lambda$ 
close--to--threshold being determined by eq.~\eqref{pklcross_eq}, i.e.\ by 
three--body phase--space modified by the strong $p \Lambda$ FSI and the Coulomb 
repulsion in the $p K^+$ subsystem, constraints on the spin--averaged low 
energy $\Lambda p$ scattering parameters have been extracted in~\cite{bal98b} 
from the differential occupation of Dalitz plot distributions obtained from 
measurements at the COSY--11 facility~\cite{bal98}.
However, in the analysis scattering length and effective range appeared
strongly correlated, as inherent to the parametrization used in
eqs.~\eqref{fsipl} and \eqref{pklcross_eq}. Thus, only a combined fit using the 
complementary information of low--energy $\Lambda p$ elastic scattering 
experiments~\cite{ale68,sec68} constrained by the Dalitz plot analysis allowed 
to derive separate spin--averaged values~\cite{bal98b}.
The result --- new spin averaged values of $\Lambda p$ S--wave scattering 
length and effective range $(\hat{a}_{p\Lambda},\hat{r}_{p\Lambda}) = 
(-2.0\,\mbox{fm},1.0\,\mbox{fm})$ --- is shown in figure~\ref{pl_param} 
together with other experimental and theoretical values, where the extracted
value of the $\Lambda p$ scattering length is clearly fixed by the old data.

Although the FSI approach to low--energy $\Lambda$--proton scattering 
proves to be a powerful tool, it should be noted that the above values 
from~\cite{bal98b} have been derived assuming a cross section ratio according 
to the number of possible magnetic spin quantum numbers, i.e.\ a ratio of three 
to one between the triplet and singlet $\Lambda p$ final states, which was 
confirmed in the early bubble chamber experiments.
Consequently, the analysis implies that the underlying reaction mechanism in 
proton--proton scattering does not favour either the singlet or triplet state, 
different to the assumed ratio of spin states.
Future threshold production experiments employing both polarized beam 
and target should dispel this concern and allow to extract singlet 
and triplet hyperon--nucleon scattering parameters separately.\vspace{1ex}

\begin{figure}[H]
\begin{center}
\epsfig{file=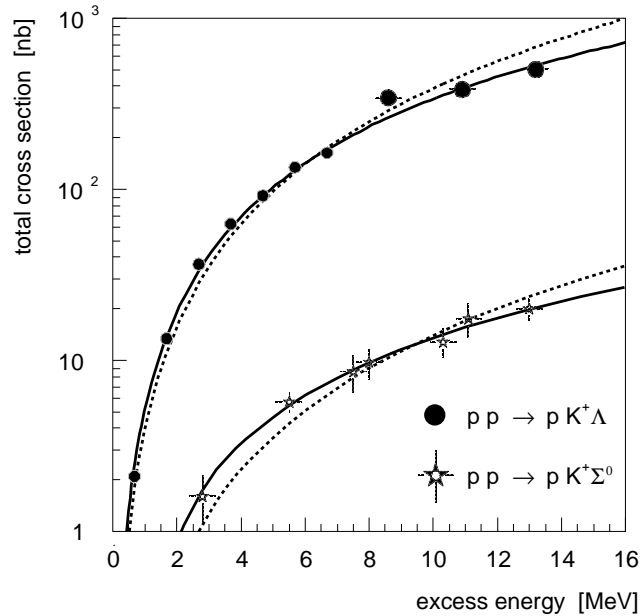,scale=0.62}
\end{center}
\caption{\label{pkls_cross} Total cross sections of the reactions $pp 
\rightarrow p K^+ \Lambda / \Sigma^0$~\cite{bal98,sewerin682}. 
In comparison, fits of the energy dependence for both hyperon production 
channels using a pure phase--space calculation (dotted lines) and phase--space 
modified by the proton--hyperon FSI (solid lines, eq.~\eqref{faldtwilkin}
\cite{fae97,faldt209,faldt2067}) are shown. }
\end{figure}

Eventually referring to the first conclusion of Kelkar and Jain~\cite{kel00} 
cited above, further constraints on the present understanding of 
close--to--threshold hyperon production have arisen from data on the reaction 
$pp \rightarrow p K^+ \Sigma^0$ taken at the COSY--11 
installation~\cite{sewerin682}:
At equal excess energies --- up to $13\,\mbox{MeV}$ with respect to the 
hyperon production thresholds --- the total cross sections have been 
determined for both $\Lambda$~\cite{bal98,sewerin682} and 
$\Sigma^0$~\cite{sewerin682} production.
A fit to the data including the hyperon--nucleon FSI is shown in 
figure~\ref{pkls_cross} and has been performed using the parametrization of 
the energy dependence developed by F\"aldt and Wilkin~\cite{fae97,faldt209,faldt2067} 
as given by eq.~\eqref{faldtwilkin}, yet neglecting the weak variation of the 
cross section due to the variation of the flux factor F.
It should be noted that Gasparyan~\cite{GAS_DIPLO} concludes that the
proton--hyperon FSI is less important for the $\Sigma N$ rather than for the  
$\Lambda N$ channel. A similar result gives the preliminary analysis
of new data taken at COSY--11 for both reaction channels up to excess energies 
of Q~=~60~MeV~\cite{kow02}.

The most striking feature of the data is the observed $\Sigma^0$ 
suppression in the energy range close--to--threshold with 
\begin{equation}
\label{rlamsig_eq}
{\cal{R}}_{\Lambda/\Sigma} =
 \frac{\sigma\left(pp \rightarrow p K^+ \Lambda\right)}
      {\sigma\left(pp \rightarrow p K^+ \Sigma^0\right)} = 
 28^{\,+6}_{\,-9},
\end{equation}
while at excess energies $\ge 300\,\mbox{MeV}$ this ratio is known to be about 
2.5~\cite{fla84}.
Considering only $\pi$ exchange, data on $\pi$ induced hyperon production via 
$\pi N \rightarrow K \Lambda\left(\Sigma^0\right)$ result in a ratio of 
${\cal{R}}_{\Lambda/\Sigma} \approx 0.9$~\cite{gas99}, clearly underestimating 
the experimental value of relation~\eqref{rlamsig_eq}.
Kaon exchange essentially relates the ratio ${\cal{R}}_{\Lambda/\Sigma}$ to 
the ratio of coupling constants squared $\mbox{g}^2_{N \Lambda K}/
\mbox{g}^2_{N \Sigma K}$.
Although there is some uncertainty in the literature on their 
values~\cite{hol89,reu94,mae89,dal82,mar81}, a $\Lambda/\Sigma^0$ production 
ratio of 27 follows from the suitable choice of the SU(6) 
prediction~\cite{dov84,swa63}, in good agreement with experiment. 
However, effects of final state interaction as well as the 
importance of $\pi$ exchange especially for $\Sigma^0$ production 
are completely neglected by this simple estimate.

Inclusive $K^+$ production data taken at the SPES~4 facility at SATURNE in 
proton--proton scattering at an excess energy of $\mbox{Q} = 252\,\mbox{MeV}$ 
with respect to the $p K^+ \Lambda$ threshold show enhancements in the 
invariant mass distribution at the $\Lambda p$ and $\Sigma N$ thresholds of 
similar magnitude~\cite{sie94}.
However, with only the kaon being detected, it is not clear whether the 
enhancement at the $\Sigma N$ threshold is due to $\Sigma$ production.
Qualitatively, a strong $\Sigma N \rightarrow \Lambda p$ final state 
conversion might account for both the inclusive results from SPES~4 as well as 
the $\Sigma^0$ depletion in the COSY--11 data.
In production experiments, evidence for $\Sigma N \rightarrow \Lambda p$
conversion effects is known experimentally from exclusive hyperon data
via $K^- d \rightarrow \pi^- \Lambda p$~\cite{tan69,cli68,tan73}
(for a theoretical interpretation see~\cite{kud71,dal83}).
In hypernuclear physics, the strong $\Sigma N \rightarrow \Lambda N$ mode
appears as the dominating decay channel of $\Sigma$--hypernuclei~\cite{ose90},
and the $\Sigma N$ interaction reflects in both shift and broadening of the
last X--rays observed from $\Sigma^-$ atoms~\cite{bat78}.
Theoretically, modern hyperon--nucleon interaction models show a significant
cusp effect for the $\Lambda N$ cross section at
the $\Sigma N$ threshold~\cite{hol89,mae89}.

In exploratory calculations performed within the framework of the 
J\"ulich meson exchange model~\cite{gas99}, taking into account both $\pi$ and 
$K$ exchange diagrams and rigorously including FSI effects in a coupled 
channel approach, a final state conversion is rather excluded as dominant 
origin of the observed $\Sigma^0$ suppression.
While $\Lambda$ production is found to be dominated by kaon exchange --- in 
agreement with the DISTO results at higher excess energies discussed above --- 
both $\pi$ and $K$ exchange turn out to contribute to the $\Sigma^0$ channel 
with similar strength. 
It is concluded by Gasparian et al.~\cite{gas99}, that a 
destructive interference of $\pi$ and $K$ exchange diagrams might explain the 
$\Sigma^0$ suppression close--to--threshold and a good agreement with the 
COSY--11 data is obtained after including the overall reduction $F_{ISI}$ in 
eq.~\eqref{pklcross_eq} due to the $pp$ interaction in the initial 
state~\cite{Batinic}.

An experimental study of $\Sigma$ production in different isospin 
configurations should provide a crucial test of the above interpretation:
For the reaction $pp \rightarrow n K^+ \Sigma^+$ an opposite interference 
pattern is found, i.e.\ the $n K^+ \Sigma^+$ channel is enhanced for a 
destructive pattern as compared to the constructive interference of $K$ and 
$\pi$ exchange. 
For the choice of a destructive interference favoured by the $p K^+ \Sigma^0$ 
channel, the corresponding ratio $\sigma\left(n K^+ \Sigma^+\right) / 
\sigma\left(p K^+ \Sigma^0\right) \approx 3$ is in good agreement with high 
energy data, whereas the alternative choice is rather difficult to reconcile 
with existing data at higher energies~\cite{gas99}.
In contrast, for the reaction $pp \rightarrow p K^0 \Sigma^+$ the same 
interference pattern occurs as for the $p K^+ \Sigma^0$ channel, resulting in 
a ratio $\sigma\left(p K^0 \Sigma^+\right)/\sigma\left(p K^+ \Sigma^0\right) 
\approx 3.3$ close--to--threshold for either choice~\cite{gas00}, while a ratio 
close to unity is found in the literature for higher energies~\cite{fla84}.

As stated by the authors of ref.~\cite{gas99}, contributions from direct 
production (fig.~\ref{pky_graph}a) as well as heavy meson exchanges 
(fig.~\ref{pky_graph}b) have been neglected so far in these calculations but 
might influence the $\Lambda/\Sigma^0$ production ratio 
(see also~\cite{tsu97,tsu99,kai99b}).
In fact, employing both a $\pi$ and $K$ exchange based meson exchange model 
without any interference of the amplitudes following~\cite{sib95} and the 
resonance model of~\cite{tsu97,tsu99} the close--to--threshold 
$\Lambda/\Sigma^0$ production ratio has been studied in~\cite{sib00b}.
Within a factor of two of the experimental error bars in~\cite{sewerin682} 
both models --- with parameters fixed at data from excess energies 
$\ge 1\,\mbox{GeV}$ --- are in reasonable agreement with the data, i.e.\ --- 
according to the authors --- the total cross section data close--to--threshold 
are not sensitive on the details of the model used~\cite{sib00b}~\footnote{It 
should be noted, that unlike~\cite{gas99} $\pi$ and $K$ exchange amplitudes 
are concluded by Sibirtsev et al.~\cite{sib00b} to contribute to both 
$\Lambda$ and $\Sigma^0$ production with similar magnitude. As mentioned 
in~\cite{sib00b}, an experimental determination of the kaon exchange 
contribution close--to--threshold --- e.g.\ from polarization observables as 
in~\cite{bal99} --- should be crucial to identify the dominant reaction 
mechanisms.}.

The latter study has been critically discussed in~\cite{shy01}, where the 
strangeness production is modeled in an effective Lagrangian approach 
following~\cite{shyammosel,shy99,eng96} and strangeness production proceeds 
via $\pi$, $\rho$, $\omega$ and $\sigma$ exchange exciting the nucleon 
S--wave resonance $N^*(1650)$ and the P--wave resonances $N^*(1710)$ and 
$N^*(1720)$.
For both the $\Lambda$ and $\Sigma^0$ production channel, $\pi$ exchange 
followed by an excitation of the $N^*(1650)$ is found to dominate the total 
cross section close--to--threshold (see also~\cite{fae97}), though an on--shell 
coupling of the $N^*(1650)$ to the $K^+ \Sigma^0$ channel is suppressed with 
the channel opening at $1686\,\mbox{MeV}/\mbox{c}^2$.
In contrast, the influence of the $N^*(1650)$ on $\Sigma^0$ production and 
possible interference effects of resonance contributions have been neglected 
in~\cite{sib00b}~\footnote{It has been remarked by Shyam et al.\ 
in~\cite{shy01} that the final state interaction employed in~\cite{sib00b} 
is ``at variance'' with~\cite{goldbergerwatson} and other approaches in the 
literature~\cite{shy99,dub86,moalem445}.}. 
At larger excess energies ($\mbox{Q} \ge 300\,\mbox{MeV}$) the $N^*(1710)$ is 
concluded in the work of Shyam et al.~\cite{shy01} to dominate both reaction 
channels, which would be expected when considering coupling constants 
only~\footnote{Accordingly, close--to--threshold $\Sigma^0$ production in the 
resonance model of~\cite{tsu99} proceeds preferrably via $\pi$ and $\eta$ 
exchange, the latter being enhanced for the $\Sigma^0$ compared to the 
$\Lambda$ channel due to the large branching ratio of the $N^*(1710)$ to the 
$N \eta$ channel. However, as pointed out in~\cite{sib99b} the $\eta$ exchange 
contribution is tainted with rather large uncertainties due to different 
values of the $\eta NN$ coupling constant extracted from experimental 
data (see~\cite{tia94} and section~\ref{Mwhs}).}.

It is worth noting, that the OBE calculations performed by 
Laget~\cite{lag91,lag01}, although not using the interference of pion and 
kaon exchange graphs explicitely and just choosing the relative sign to maximize the 
cross section, not only describe the recent close--to--threshold data on the 
$\Lambda/\Sigma^0$ ratio within a factor of two, as well as the polarization 
transfer results of the DISTO experiment~\cite{bal99}, but also give an 
accurate description of the $YN$ invariant mass distributions obtained in the 
inclusive measurements at SATURNE~\cite{sie94} cited above:
In particular, the FSI contribution from the direct $\Lambda p \rightarrow 
\Lambda p$ amplitude is found to significantly enhance the kaon spectrum just 
above the $\Lambda p$ threshold, while the strong $\Sigma N \rightarrow 
\Lambda p$ coupling ``induces only a small dip close to the $\Sigma N$ 
threshold'' in the $\pi$ and $K$ exchange amplitudes~\cite{lag91}.
On the other hand, it is this strong channel coupling and the resulting rapid 
variation of the ${}^3\mbox{S}_1$ $\Lambda N$ amplitude near the $\Sigma N$ 
threshold, which appears responsible for the sharp rise of the invariant mass 
distribution obtained at SPES~4 at the $\Sigma N$ threshold via the 
``kaon direct emission amplitude'', i.e.\ the coupling of one proton to 
a $K^+ Y$ pair and a following interaction with the second proton 
(fig.~\ref{pky_graph}a).
\begin{figure}[H]
\begin{center}
\epsfig{file=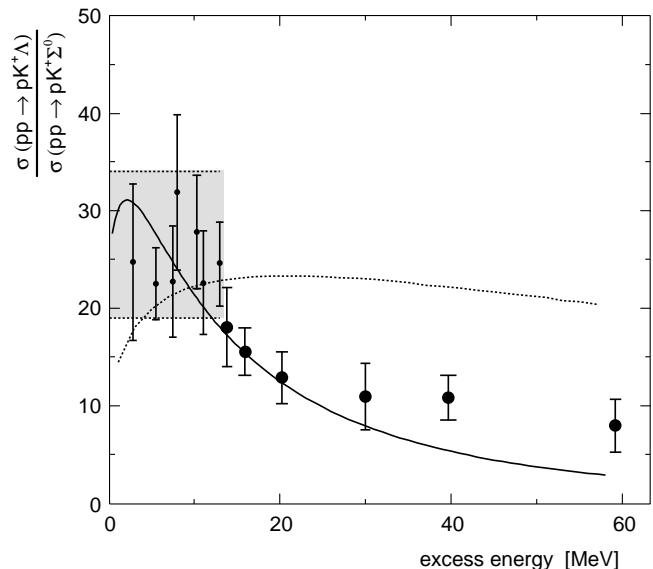,scale=0.43}
\end{center}

\vspace{-0.9cm}
\caption{\label{pkls_ratio} $\Lambda/\Sigma^0$ production ratio in 
proton--proton collisions as a function of the excess energy. Experimental 
data within the shaded area, which corresponds to the range described by 
relation~\eqref{rlamsig_eq}, are taken from~\cite{sewerin682}, data at higher 
excess energies from~\cite{kow02}. Calculations~\cite{gas01} are performed 
within the J\"ulich meson exchange model, assuming a destructive interference 
of $K$ and $\pi$ exchange and employing the microscopic $YN$ interaction 
models  Nijmegen~NSC89 (solid 
line~\cite{mae89}) and the new J\"ulich model (dotted line~\cite{hai01}), 
respectively.}
\end{figure}

Recently, measurements of the $\Lambda/\Sigma^0$ production ratio in 
proton--proton collisions have been extended up to excess energies of 
$\mbox{Q} = 60\,\mbox{MeV}$ at the COSY--11 installation~\cite{kow02}, in 
order to study the transition between the low--energy enhancement of the ratio 
reported in~\cite{sewerin682} and data at excess energies larger than 
$300\,\mbox{MeV}$.
In comparison to the experimental data, in figure~\ref{pkls_ratio} 
calculations are included obtained within the approach of Gasparian et 
al.~\cite{gas99} assuming a destructive interference of $\pi$ and $K$ exchange 
with different choices of the microscopic hyperon nucleon model~\cite{gas01}.

As it was already emphasized in~\cite{gas99}, the result depends on the 
details --- especially the off--shell properties --- of the microscopic 
hyperon--nucleon interaction employed, although the actual choice does not 
alter the general result in~\cite{gas99} of only the destructive interference 
of $K$ and $\pi$ exchange explaining the experimentally observed suppression 
of the $\Sigma^0$ signal close--to--threshold.
At the present stage of theoretical investigations, both the good agreement 
found for J\"ulich model~A~\cite{hol89} with the close--to--threshold result 
for ${\cal{R}}_{\Lambda/\Sigma}$ in relation~\ref{rlamsig_eq} and for the 
Nijmegen model (solid line in fig.~\ref{pkls_ratio}~\cite{mae89}) with the 
energy dependence of the cross section ratio in figure~\ref{pkls_ratio} 
should rather be regarded as accidental.
In the latter case an SU(2) breaking in the ${}^3\mbox{S}_1$ $\Sigma N$ 
channel had to be introduced~\cite{mae89}. Consequently, the relation between 
the $\Sigma^0 p$ amplitude with the $\Sigma^+ p$ and $\Sigma^- p$ channels 
becomes ambiguous. Only one of the choices leads to the good agreement with 
the data shown in figure~\ref{pkls_ratio} (solid line), whereas the other one 
results in a completely different prediction~\cite{haidpr}.

Calculations using the new J\"ulich model (dotted line in 
figure~\ref{pkls_ratio}~\cite{hai01}) do not reproduce the tendency of the 
experimental data.
It is suggested in~\cite{gas01} that neglecting the energy dependence of the
elementary amplitudes and assuming S--waves in the final state might 
no longer be justified beyond excess energies of $20\,\mbox{MeV}$.

Conclusions from figure~\ref{pkls_ratio}, as well as considerations by Kelkar 
and Jain~\cite{kel00} for the $pp \rightarrow p K^+ \Lambda$ reaction 
clearly demonstrate, that --- once the reaction 
mechanism for close--to--threshold hyperon production is understood --- 
exclusive hyperon production data should provide a strong constraint on the 
details of hyperon--nucleon interaction models.\vspace{1ex}

Finally, in future high resolution hyperon production experiments close to 
threshold may give further information on the possible existence of low--lying 
strange dibaryons.
Narrow, stable six--quark objects $D_s$ and $D_t$ with a $q^4 \otimes q^2$ 
cluster substructure and strangeness $\mbox{S} = -1$ coupling to P--wave 
nucleon--hyperon systems have been postulated from an extended MIT 
bag--model~\cite{aer85} below and above the $\Sigma N$ thresholds with widths 
in the range of $10\,\mbox{keV}$ to $10\,\mbox{MeV}$.
Experimental signatures consistent with a strange dibaryon hypothesis have 
been found both in strangeness transfer reactions $K^- d \rightarrow 
\pi^- Y N$~\cite{tan69,bra77,pie88} and in the inclusive $p p \rightarrow 
K^+ X$ data taken at SATURNE~\cite{sie94} at excess energies of 175 and 
$306\,\mbox{MeV}$ with respect to the $\Sigma^0 p$ threshold.
However, the statistical accuracy in these experiments turned out to be too 
small for unequivocal conclusions.
For a high resolution study of the inclusive $pp \rightarrow K^+ X$ reaction 
channel new experimental efforts are on their way at the BIG KARL spectrometer 
at COSY, with a resolution improved by a factor of five and a higher 
statistical accuracy compared to the SATURNE measurements~\cite{hin00}.

\section{Meson pair production }
\label{Mpp}

\subsection{Double pion production in NN scattering}
\label{DppiNNs}                           
Single and double pion production in elementary $\pi N$, $\gamma N$ and $NN$ 
reactions is an important source to gain information on nucleon--nucleon, 
nucleon--pion and pion--pion interactions. 
Apart from these fundamental aspects, the pion production on nucleons also 
enables the study of the properties of nucleonic resonances, which might be 
excited in the reaction processes. 
The reaction $\gamma N \rightarrow N\pi\pi$ for instance has been found to be 
influenced by the formation of the $N^*(1520)$ resonance~\cite{Gom94} with a 
subsequent decay into $\Delta \pi$. 
Therefore, studies on this reaction channel allow to gain information about 
the $N^*(1520) \rightarrow \Delta(1232) \pi$ decay amplitudes. 
Additionally, contributions of the $N^*(1440)$ resonance, followed by its 
decay $N^*(1440) \rightarrow 
N(\pi\pi)^{\mbox{\scriptsize T}=0}_{\mbox{\scriptsize S--wave}}$, play a 
non--negligible role in the near--threshold production of pion--pairs via the 
reaction channels $\gamma N \rightarrow N \pi\pi$ and $\pi N \rightarrow 
N \pi\pi$ \cite{Ose85}.\vspace{1ex}
 
In case of the pion--pair production in the nucleon--nucleon scattering 
similar contributions of nucleonic resonances are under discussion. 
Recently, detailed calculations on the two--pion production in 
nucleon--nucleon interactions have been performed~\cite{Alv98}, considering 
non--resonant terms as well as contributions from nucleonic resonances with 
subsequent decays into a nucleon and pions.
One of the main results of these studies is the prediction of a dominant 
effect of the $N^*(1440)$ Roper resonance, followed by its decay $N^*(1440) 
\rightarrow N(\pi\pi)^{\mbox{\scriptsize T}=0}_{\mbox{\scriptsize S--wave}}$, 
in the region of low excess energies. 
According to this, the absolute scale of the low energy total cross section of 
reaction channels, where the pions can be in an isospin zero state, e.g.\ $pp 
\rightarrow pp \pi^+\pi^-$ and $pp \rightarrow pp \pi^0\pi^0$, should be 
determined by the excitation of this resonance. 
In this region of excess energies contributions of non--resonant 
terms are discussed to be negligible for the description of the excitation 
function of both these reaction channels. 
However, at higher energies ($\mbox{Q} > 200\,\mbox{MeV}$) the excitation of 
$\Delta$ resonances has been evaluated to dominate the total cross sections.

A different situation is given in the case of the $pp \rightarrow 
pn \pi^+\pi^0$ reaction channel. 
Here contributions of the Roper resonance with the subsequent decay into 
$N(\pi\pi)^{\mbox{\scriptsize T}=0}_{\mbox{\scriptsize S--wave}}$ are 
forbidden by conservation laws and contributions of remaining $N^*$ production 
diagrams are suppressed at low excess energies. 
Since the non--resonant terms are also negligible in this case, the dominating 
processes are given by $\Delta$ resonance excitations.

It is worth noting that the discussed model calculations yield predictions for 
the excitation functions for different two--pion production reactions from 
threshold up to high excess energies ($\mbox{Q} \approx 300\,\mbox{MeV}$) and 
all are based on the same Feynman diagrams. 
However, especially in the close--to--threshold region the data situation for 
most double pion reaction channels is far from being optimal. 
Recently, three isospin--independent reaction channels on the two--pion 
production in proton--proton interactions, $pp \rightarrow pp \pi^+\pi^-$, 
$pp \rightarrow pp \pi^0\pi^0$ and $pp \rightarrow pn \pi^+\pi^0$, have been 
studied in high statistics precision measurements at the PROMICE/WASA facility 
at CELSIUS and resulted in the first total and differential cross sections in 
the near--threshold region~\cite{Joh00,brodowski}. 

\begin{figure}[H]
\hfill
\parbox{0.33\textwidth}
  {\epsfig{file=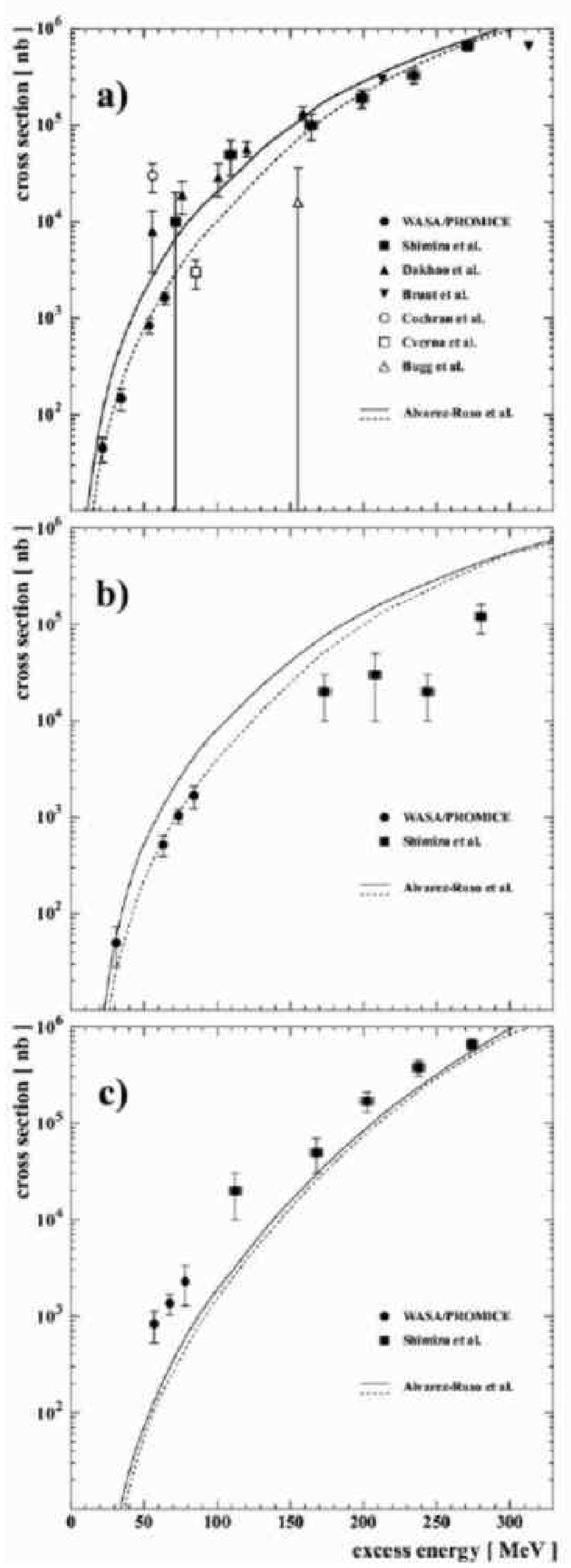,width=0.37\textwidth}} \hfill
\parbox{0.5\textwidth}
  {\caption{\label{twopiall}
  Total cross section data for the reactions:\newline
  a) $pp \rightarrow pp \pi^+\pi^-$, \newline
  b) $pp \rightarrow pp \pi^0\pi^0$ and \newline
  c) $pp \rightarrow pn \pi^+\pi^0$.  \newline
  The solid and the dashed curves represent model\newline 
  calculations from~\cite{Alv98}. \newline
  The data were taken from~\cite{Joh00,Coc72,Cve80,Bug64,Dak83,Shi82,Bru69,brodowski}.
  \protect\\
  \protect{\small 
    A version of the article including a figure of better quality can be found at the COSY-11 homepage:
   \protect\\
   $http://ikpe1101.ikp.kfa-juelich.de/cosy-11\protect\\
   /pub/List\_of\_Publications.html\#papers$}
 }}
\end{figure}

In fig.~\ref{twopiall}a the total cross sections for the $\pi^+\pi^-$ 
production in proton--proton scattering are presented up to an excess energy 
of $\mbox{Q} \approx 300\,\mbox{MeV}$. The solid and the dashed lines represent 
the above discussed predictions of ref.~\cite{Alv98} for this reaction channel 
and correspond to two possible solutions in the applied model. 
Obviously these calculations are in good agreement with the experimental data 
both close--to--threshold and in the region of higher excess energies, while the 
model seems to underestimate the data at intermediate energies. 
It should be noted that the data points from Dakhno et al.~\cite{Dak83} have 
been obtained using a deuterium filled bubble chamber and selecting events 
from the quasi--free two--pion production $pp(n) \rightarrow pp(n)\pi^+\pi^-$. 
However, in the analysis of the data the Fermi motion of the nucleons in the 
deuteron has been neglected. 
In later calculations an adequate consideration of the deuteron wave function 
has been found to shift the effective excess energy by $\sim 20\,\mbox{MeV}$ 
towards higher energies~\cite{Ste96}, corresponding to a shift of the data 
towards the region expected by the discussed model calculation.

Figure~\ref{twopiall}b displays the experimental situation for the reaction 
channel $pp \rightarrow pp\pi^0\pi^0$. 
The new data from PROMICE/WASA disfavour the solution of the model calculation 
indicated by the solid line but are in very good agreement with the prediction 
shown by the dashed line. 
However, the high energy data from ref.~\cite{Shi82} are in disagreement with 
both curves and are clearly overestimated.

Finally, the close--to--threshold total cross section data of the reaction 
channel $pp \rightarrow pn \pi^+\pi^0$ are presented in 
fig.~\ref{twopiall}c.\newline
Different to the previous situation now the data is underestimated by both 
solutions of the model calculations, especially in the near--threshold region. 
Since now a production via the $N^*(1440) \rightarrow 
N(\pi\pi)_{\mbox{\scriptsize S--wave}}^{\mbox{\scriptsize T}=0}$ is forbidden, 
this observation might indicate contributions from processes like $NN 
\rightarrow NN^* \rightarrow N \Delta \pi \rightarrow NN\pi\pi$ larger than 
predicted by theory.

It should be mentioned that contributions of higher partial waves and final 
state interactions have been neglected in the model calculations, indicated by 
the presented curves. 
However, the effect of the $pp$ FSI has been investigated and is reported to 
increase the close--to--threshold cross sections by nearly one order of 
magnitude, while at $\mbox{Q} \sim 120\,\mbox{MeV}$ this effect is reduced to 
an increase of a factor of $\sim 2$~\cite{Alv99}. 
According to this, considering the interaction of the final state baryons is 
expected to reduce the apparent discrepancy between the $\pi^+\pi^0$ total 
cross section data and the corresponding model calculations. 
Surprisingly, in case of the $\pi^+\pi^-$ and $\pi^0\pi^0$ pair production, 
the near--threshold data are described reasonable by these calculations 
neglecting the $pp$ FSI.\vspace{1ex}

Furthermore, signals for the contribution of higher partial waves are reported 
for the near--threshold data~\cite{Joh00}, which also have been neglected in 
the discussed model calculations. 
This observation might be interpreted as a signal of a heavy meson exchange 
($\sigma$, $\rho$) between the interacting protons.

\subsection{Double pion production in pd scattering} 
\label{Dppipds} 
It is well known and cited in the literature as the 
ABC--anomaly~\cite{ABC60,ABC63} that the missing mass enhancement at around 
$310\,\mbox{MeV}$ with a width of $50\,\mbox{MeV}$ observed in the inclusive 
measurements of the $pd \rightarrow {}^3He\,X^0$ reaction at a proton beam 
energy of $\mbox{T}_p \geq 745\,\mbox{MeV}$ (corresponding to an excess energy 
of $\mbox{Q} \geq 190\,\mbox{MeV}$ for the $\pi \pi$ production) is not excited 
in the ${}^3H\,X^+$ final state. 
Detailed investigations confirm that this observation is to be associated with 
the isospin--zero s--wave $\pi \pi$ double--pion state but additionally stress 
that the missing mass peak and its width vary with the beam 
momentum~\cite{BANAIGS73}. 
Further, since the isoscalar $\pi \pi$ scattering length is 
small~\cite{GRAYER74} the observed anomaly is supposed to be of kinematical 
origin and associated with the intermediate excitation of two $\Delta$ 
resonances~\cite{RISSER73}.\vspace{1ex}
\vspace{-1.8cm}
\begin{figure}[H]
\begin{center}
\epsfig{file=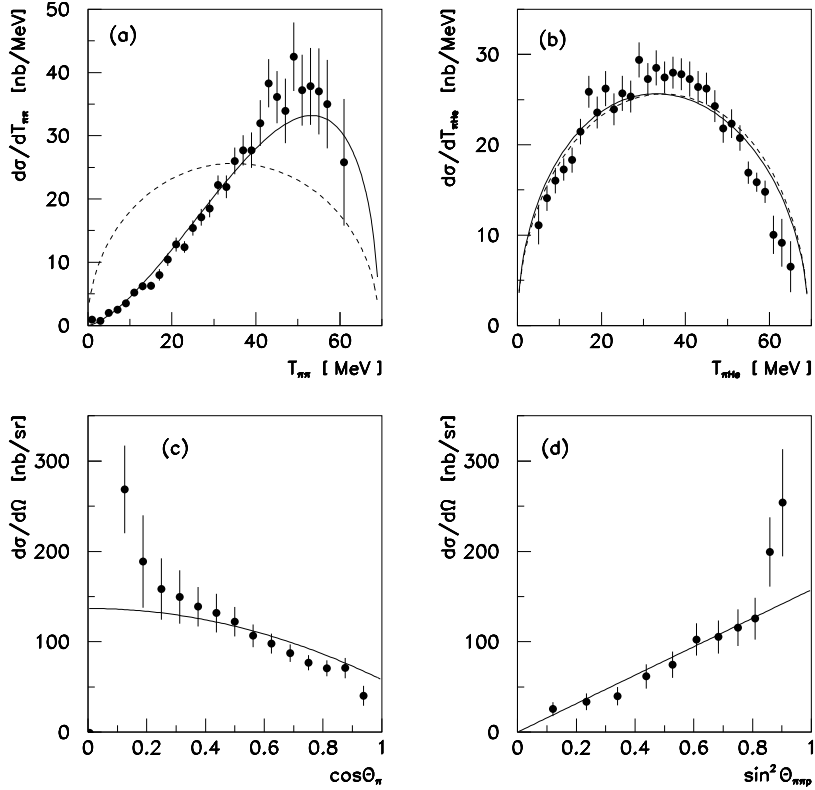,width=0.65\textwidth}
\end{center}
\vspace{-1.1cm}
\caption{\label{JAHN_MOMO_FIG03} Differential cross sections for the $pd 
\rightarrow {}^3He\, \pi^+ \pi^-$ reaction at $\mbox{T}_p = 546\,\mbox{MeV}$ 
as a function of four kinematical variables. Dashed curves represent 
phase--space predictions normalized to the data whereas the solid curves are 
calculations assuming the $\pi^+ \pi^-$ system to emerge in a relative 
p--wave, see~\cite{BELLEMANN99}.}
\end{figure}

Recently the MOMO 
({\bf{M}}onitor--{\bf{o}}f--{\bf{M}}esonic--{\bf{O}}bservables) collaboration 
at COSY investigated the $\pi\pi$ production closer to threshold in an 
exclusive measurement of the $pd \rightarrow {}^3He\,\pi^+\pi^-$ reaction at a 
beam energy of $\mbox{T}_p = 546\,\mbox{MeV}$ (excess energy $\mbox{Q} = 
70\,\mbox{MeV}$ for the $\pi \pi$ production), where the ABC enhancement is 
expected to show up near the centre of the available $\pi\pi$--invariant mass 
distribution~\cite{BELLEMANN99}.
\noindent

In figure~\ref{JAHN_MOMO_FIG03}a--d differential cross sections in terms of 
four kinematical variables are presented:\newline
a) The $\pi^+\pi^-$ excitation energy $\mbox{T}_{\pi\pi} = \mbox{m}_{\pi\pi} - 
2\mbox{m}_{\pi}$, where $\mbox{m}_{\pi\pi}$ is the two--pion invariant 
mass.\newline
b) The $\pi$--$\,{}^3He$ excitation energy.\newline
c) The angle ($\theta_{\pi}$) between the proton and one of the $\pi$'s in the 
overall centre--of--mass (c.m.) system.\newline
d) The angle ($\theta_{\pi \pi p}$) between the relative $\pi^+ \pi^-$ 
momentum and the beam axis in the overall c.m.\ system.\vspace{1ex}

The $\mbox{T}_{\pi\pi}$ distribution at this near--threshold data is pushed 
closer to the maximum values of excitation energy and is in contrast to the 
original ABC result~\cite{ABC60,ABC63} which showed an enhancement over 
phase--space at $\mbox{T}_{\pi\pi} \approx 30\,\mbox{MeV}$. 
The distribution of the $\pi$--$\,{}^3He$ excitation energy is rather 
consistent with the phase--space prediction. 
The significant anisotropy of $\theta_{\pi}$ confirms the importance of higher 
partial waves involved in the reaction process. 
The MOMO collaboration concludes~\cite{BELLEMANN99} that the agreement between 
the $\mbox{T}_{\pi\pi}$ data of fig.~\ref{JAHN_MOMO_FIG03}a with the solid 
line (phase--space multiplied by $\mbox{T}_{\pi\pi}$) and the fair linearity 
of $\sin^2 \theta_{\pi\pi p}$ indicates that the $\pi^+\pi^-$ system is 
essentially produced with an internal angular momentum $l =1$.

On the other hand, as outlined in section~\ref{DppiNNs},
the Valencia group~\cite{Alv98,ALV9810002} developed a model for the
two pion production resulting in a dynamical origin for the small invariant
masses of the low energy part, 
where the two pions can be in an isospin zero state. Here the reaction dynamics
is dominated by the intermediate excitation of the $N^*$(1440) Roper resonance 
followed by the $N^* \to \Delta \pi \to N (\pi \pi)^{T=0}_{s-wave}$ decay.

The experimental situation seems to be the following: s--wave $\pi\pi$ 
production has been observed in the $pd \rightarrow {}^3He\,\pi^+\pi^-$ 
reaction very close--to--threshold~\cite{BETKER96} and at high excess energies 
($\mbox{Q} \approx 200\,\mbox{MeV}$)~\cite{ABC60,ABC63} whereas at an 
intermediate excess energy range a p--wave $\pi^+\pi^-$ production process 
dominates. 
Similar results have been obtained in $np \rightarrow 
dX$~\cite{PLOUIN78,HOLLAS82} and in the $\pi^+ d \rightarrow \pi^+ \pi^+ n n\,
(\pi^+\pi^- pp)$~\cite{BONUTTI98,KERMANI98} reactions.  
\subsection{Double K--meson production in pd scattering} 
\label{Dkmpipds}
After the remarkable results of p--wave production in the $\pi^+\pi^-$ system 
the MOMO collaboration continued equivalent measurements for the $pd 
\rightarrow {}^3He\,K^+ K^-$ reaction at excess energies of $\mbox{Q} = 
56\,\mbox{MeV}$, $40\,\mbox{MeV}$ and $35\,\mbox{MeV}$.

\begin{figure}[H]
\hfill
\hspace{2.5cm}
\parbox{0.4\textwidth}
  {\epsfig{file=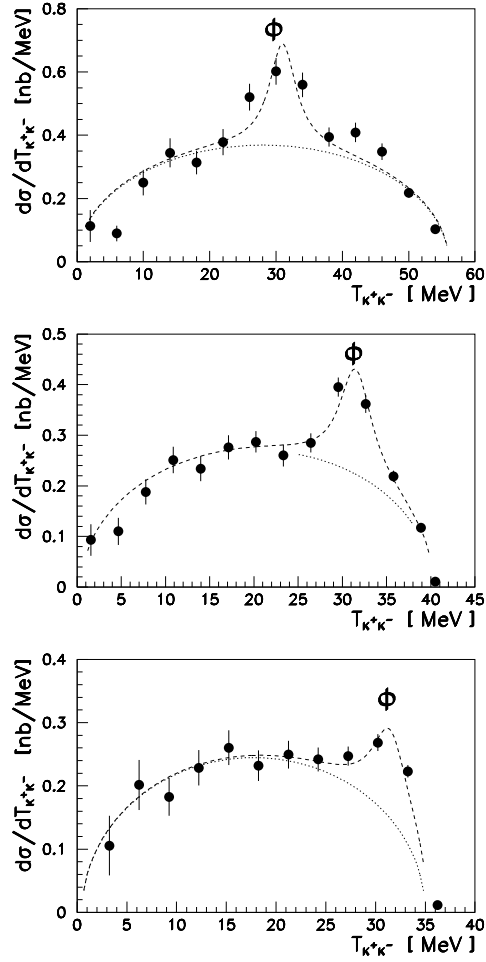,width=0.39\textwidth}} \hfill
\hspace{-1.5cm}
\parbox{0.25\textwidth}
  {\caption{\label{JAHN_KK_MOMO} $K^+ K^-$ invariant mass spectra from the 
  reaction $pd \rightarrow {}^3He\,K^+ K^-$ at $Q = 56\,\mbox{MeV}$, 
  $40\,\mbox{MeV}$ and $35\,\mbox{MeV}$ above threshold plotted in units of 
  the $K^+ K^-$ relative energy $\mbox{T}_{K^+ K^-}$, see~\cite{BELLEMANN01}.}}
\end{figure}

The $K^+ K^-$ invariant mass spectra in units of the $K^+ K^-$ relative energy 
$\mbox{T}_{K^+ K^-}$ are shown in figure~\ref{JAHN_KK_MOMO}. 
A clear signal of the $\phi$--meson production is seen above the continuous 
non--resonant $K^+ K^-$ spectrum. 
The dashed lines in all three parts of the figure correspond to simple 
phase--space calculations, where the Q--dependences of the cross sections 
($\sigma_{K^+ K^-}$ and $\sigma_{\phi}$) were taken to scale as 
$\sigma_{K^+ K^-} \propto \mbox{Q}^2$ for the three--body and 
$\sigma_{\phi} \propto \mbox{Q}^{1/2}$ for the two--body final state, 
respectively. 
The relative angular momentum between the two kaons was assumed to be $l = 0$.
No significant deviation of the data from these summed phase--space 
distributions is observed and thus it is interesting to note, that no 
indication of p--wave in the $K^+ K^-$ system is present in the mass spectra 
and angular distributions, in contrast to the above discussed data for the two 
pion production via the reaction $pd \rightarrow {}^3He\,\pi^+\pi^-$.
\vspace{-1.1cm}

\subsection{Strangeness dissociation into $\mbox{K}\overline{\mbox{K}}$--pairs}
\label{SdiKKp}
Measurements on the $K \overline{K}$ meson pair production are of 
considerable interest in the context of the continuing discussion on the 
nature of the scalar resonances $f_0(980)$ and $a_0(980)$, as already
discussed in section~\ref{Ssec}.

\begin{figure}[H]
\hfill
  { \centerline{\epsfig{file=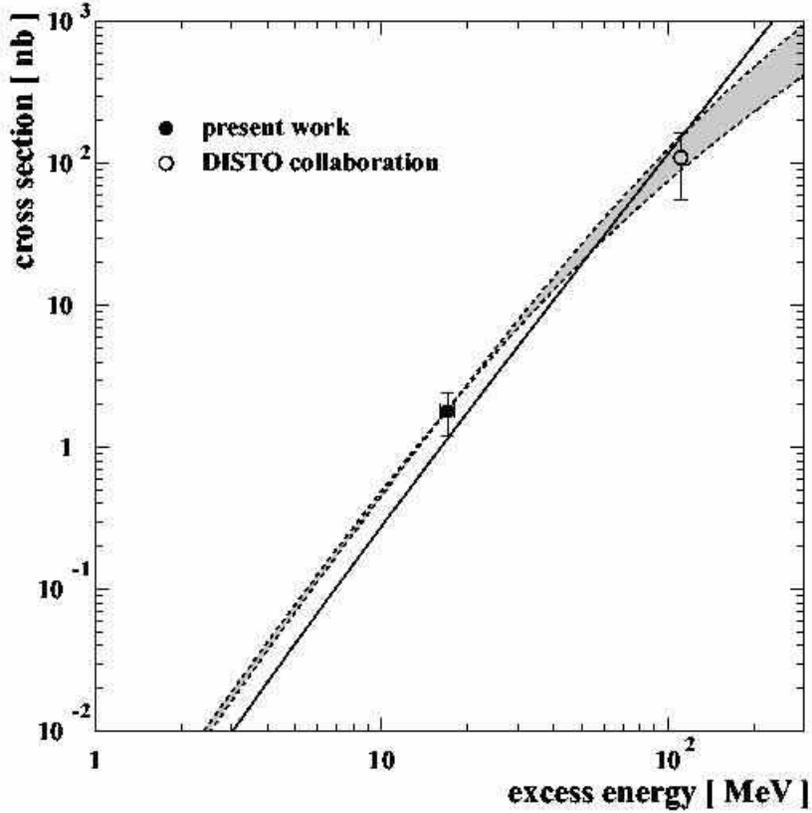,width=0.60\textwidth}}}
  {\caption{\label{kk} Total cross sections for the free $K^+K^-$ pair 
  production in proton--proton collisions \cite{quentmeier276,balestra7}. The 
  solid line indicates calculations on the basis of the non--resonant 
  four--body phase--space expectations fitted to the data points. The dashed 
  lines correspond to three--body phase--space for the resonant $K^+K^-$ 
  production via the $f_0$, normalized to the COSY--11 data point). The $pp$ 
  FSI and the Coulomb interaction have been taken into account. The effect of 
  the large uncertainty about the width of the $f_0(980)$ resonance ($\Gamma = 
  40 - 100\,\mbox{MeV}$~\cite{groom1}) is indicated by the dashed area.
\protect\\
  \protect{\small
    A version of the article including 
    a figure of better quality can be found at the COSY-11 homepage:
   \protect\\
   $http://ikpe1101.ikp.kfa-juelich.de/cosy-11/pub/List\_of\_Publications.html\#papers$}
}}
\end{figure}

Available close--to--threshold data~\cite{quentmeier276,balestra7} on the 
elementary antikaon production channel in proton--proton scattering, $pp 
\rightarrow pp K^+K^-$, are shown in figure~\ref{kk}.
In comparison, parametrizations of the energy dependence of the total cross 
section based on a four--body phase--space behaviour and for a resonant 
production via the $f_0$ are included.
Obviously, within the experimental error bars and neglecting the possible 
influence of higher partial waves~\footnote{Preliminary data by the DISTO 
collaboration~\cite{rit01} indicate that $K^+K^-$ production at an excess 
energy of $\mbox{Q} = 114\,\mbox{MeV}$ is still consistent with isotropic 
emission.}, the data is consistent with both a non--resonant and resonant 
production, as already concluded from figure~\ref{mm_kk_f0} in 
section~\ref{Ssec}, i.e. from the $K^+ K^-$ invariant mass distribution 
obtained at an excess energy of $\mbox{Q} = 
17\,\mbox{MeV}$~\cite{quentmeierphd}. 

Furthermore, exclusive $K^-$ production data are of special interest with 
respect to subthreshold kaon production experiments in nucleus--nucleus 
interactions, which are expected to probe the antikaon properties at high 
baryon density. 
Recent inclusive subthreshold measurements~\cite{lau99,sen01,oes01} resulted 
in comparable 
$K^+$ and $K^-$ yields at the same energy per nucleon below the production 
thresholds for the elementary reactions $pp \rightarrow K^{\pm}X$, 
which has been discussed as an indication for a possible restoration of chiral 
symmetry (as reviewed in~\cite{cas99}).

\begin{figure}[H]
\hfill
\parbox{0.6\textwidth}
  {\epsfig{file=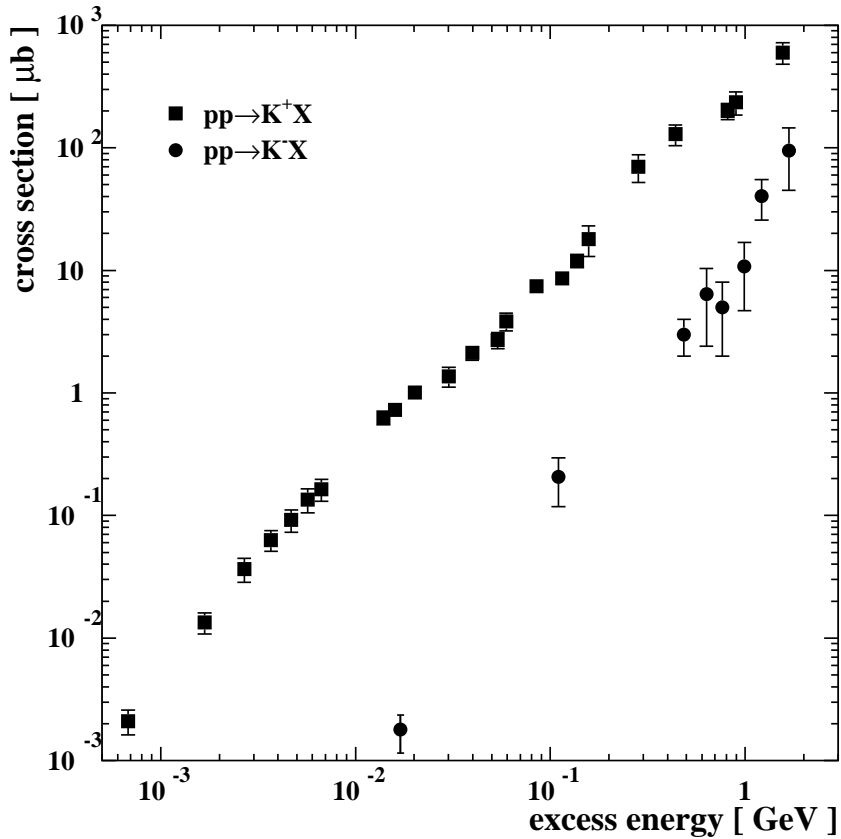,width=0.6\textwidth}} \hfill
\parbox{0.32\textwidth}
  {\caption{\label{kincl} World data on the inclusive total $K^+$ and $K^-$ 
  production cross sections in proton--proton collisions at excess energies 
  below $\mbox{Q} = 2\,\mbox{GeV}$. The data are taken from \cite{fla84,Lan88,
  sewerin682,bal99,kow02,mar01,hes01,quentmeier276,bal98,bilger217}.}}
\end{figure}

Available inclusive elementary $K^+$ and $K^-$ total cross sections in 
nucleon--nucleon interactions are shown in figure~\ref{kincl}.
In contrast to the results obtained in subthreshold measurements, in 
proton--proton scattering the inclusive $K^+$ cross section exceeds $K^-$ 
production by more than an order of magnitude in the near threshold region.
To explain this observation, different models~\cite{kap86,bro94,Waa96,
schaffner325,Lut98,ram00,koc94} consider kaons ($K^+,K^0$) and antikaons 
($K^-,\overline{K^0}$) to be subject of repulsive and attractive forces 
within the nuclear medium, respectively.
Recent theoretical studies using a chiral unitary approach 
to describe the $K^- N$ S--wave interaction allow for a 
self--consistent microscopic implementation of medium  
effects~\cite{Lut98,ram00}.
The $K^-$ self--energies obtained are consistent with kaonic atom
data, as shown in~\cite{bac00}.

\section{Meson production on light nuclei}
\label{Mpoln}
Near--threshold meson production experiments in nucleon--nucleon interactions 
might help to obtain fundamental information on the elementary production 
processes and final state interactions. 
Extending those measurements from the binary nucleon scattering to reactions 
with three or four participating nucleons, i.e.\ meson production in the 
proton--deuteron or deuteron--deuteron scattering, allows to study the role of 
further participating nucleons. 
This facilitates a gain in information on possible multi--step production 
processes, which are naturally absent in the nucleon--nucleon case.

\subsection{Dynamics in three and four nucleon systems}      
\label{Ditfns} 
Of considerable interest is the production of $\eta$ mesons in the $pd 
\rightarrow {}^3He\,\eta$ reaction. 
Studied at the SATURNE accelerator in the near threshold 
region~\cite{Ber88,May96} this channel is reported to expose remarkable 
features. 
In spite of the much higher momentum transfer, the observed threshold 
amplitude $f_{\eta}$ for the $\eta$ production, defined by the unpolarized 
centre--of--mass cross section
\begin{equation}
\label{hef}
\frac{d\sigma}{d\Omega} \, (pd \rightarrow {}^3He\,\eta) \;=\;
  \frac{\mbox{p}^*_{\eta}}{\mbox{p}^*_d}\;
  |f_{\eta} (pd \rightarrow {}^3He\,\eta)|^2,
\end{equation}
was found to be comparable to that for the $pd \rightarrow {}^3He\,\pi^0$ 
reaction at its respective threshold~\cite{Ber88}. 
In addition, while being consistent with s--wave production, 
$|f_{\eta}(\mbox{p}^*_\eta)|^2$ decreases by a factor of three from threshold 
up to a centre--of--mass momentum of emitted $\eta$ mesons of 
$\mbox{p}^*_{\eta} = 70\,\mbox{MeV/c}$ ($\widehat{=} \mbox{Q} \sim 
7\,\mbox{MeV}$).
 
This observed strong decrease in the near--threshold region can be addressed to 
a strong $\eta\;{}^3He$ final state interaction and is therefore of special 
interest in view of the existence of $\eta$--mesic nuclei. 
Although being the topic of several theoretical 
investigations~\cite{Hai86,Hai86b,rakityanskyR2043,Wyc95,Aba96,
wilkinR938,CHIANG738}, 
the possibility for the formation of quasi--bound ${}^3He\,\eta$ states is 
still an open question.
$~~$\\[-1.5cm] 
\begin{figure}[H]
\epsfig{file=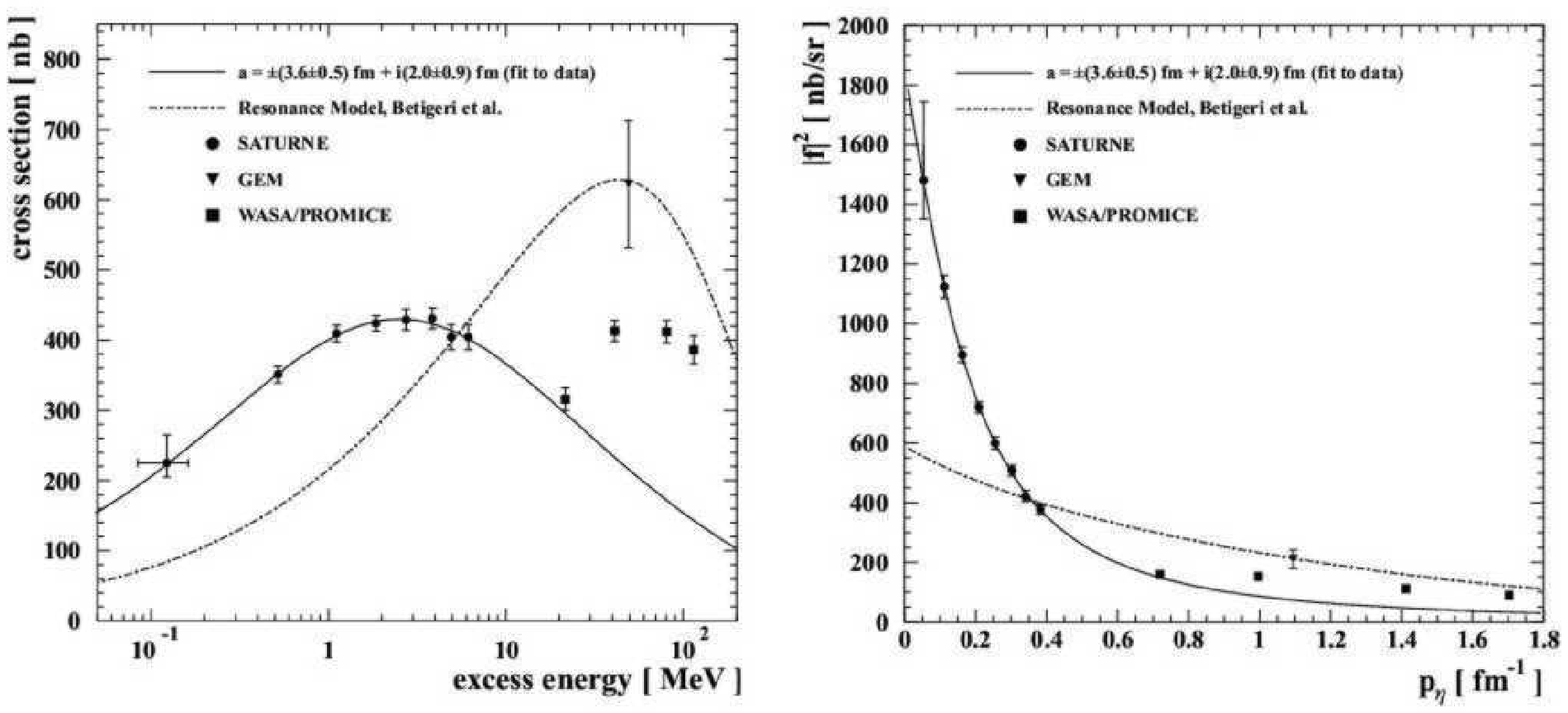,width=0.9\textwidth}

\vspace{-0.6cm}
\caption{\label{he3rev} Total cross section $\sigma$ and square of the 
production amplitude $f$ for the $pd \rightarrow {}^3He\,\eta$ reaction. The 
lines are explained in the text. The data are from 
references~\cite{May96,Bet00,Bil02} and show only the statistical uncertainties.
\protect\\
  \protect{\small
    A version of the article including a figure
    of better quality can be found at the COSY-11 homepage:
   \protect\\
   $http://ikpe1101.ikp.kfa-juelich.de/cosy-11/pub/List\_of\_Publications.html\#papers$}
}
\end{figure}

The experimental situation for the $ pd \rightarrow {}^3He\,\eta$ reaction is 
presented in fig.~\ref{he3rev}, displaying total cross section data from 
threshold up to an excess energy of $\mbox{Q} = 100\,\mbox{MeV}$.
 
\begin{figure}[H]
\begin{center}
\epsfig{file=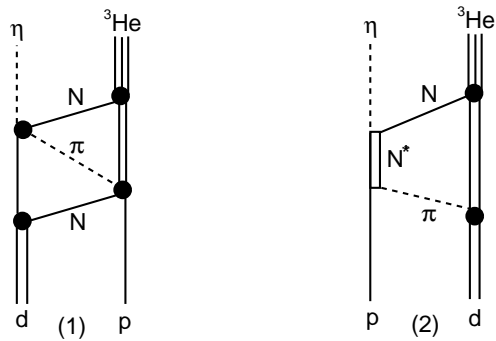,width=0.35\textwidth}
\end{center}

\vspace{-0.5cm}
\caption{\label{he3etafey} Production diagrams for the $pd \rightarrow 
{}^3He\,\eta$ reaction.}
\end{figure}

The SATURNE data~\cite{May96} (circles) expose an energy dependence, which 
clearly differs from pure phase--space expectations: 
According to naive two--body phase--space considerations, neglecting higher 
partial waves, an excitation function according to 
$\sigma\,\propto\,\mbox{Q}^{1/2}$ is expected. 
One promising ansatz to describe the reaction process was given by Kilian and 
Nann~\cite{Kil90}, taking into account both nucleons from the target and 
considering double--scattering diagrams. 
A sketch of such a double--scattering diagram is illustrated in the left part 
of fig.~\ref{he3etafey}.

\vspace{-0.4cm}

Calculations by F\"aldt and Wilkin \cite{Fal95} are in line with this approach 
and describe the $pd \rightarrow {}^3He\,\eta$ reaction as an initial $pp 
\rightarrow d \pi^+$ ($pn \rightarrow d \pi^0$) production followed by a final 
reaction according to $\pi^+ n \rightarrow \eta p$ ($\pi^0 p \rightarrow 
\eta p$).
These investigations lead to a prediction of the spin averaged threshold 
amplitude squared of $|f(\mbox{p}^*_{\eta} = 0)|^2 = 
0.69\,\mu\mbox{b}/\mbox{sr}$, underestimating the measured results by a factor 
of 2--3. 
Since none of the considered sub--reactions exposes a significant variation of 
the cross section in the relevant energy region, the observed energy 
dependence of the production amplitude $f_{\eta}$ has been interpreted to be a 
signal of a strong final state interaction of the $\eta$--nucleus system.

Introducing the ${}^3He\,\eta$ final state interaction by an s--wave 
scattering length formula

\begin{equation}
\label{he1}
f(\mbox{p}_\eta) \;=\; 
  \frac{f_B}{1 - i\;a_{\,{}^3He\,\eta}\;\mbox{p}^*_{\eta}}\,,
\end{equation}
with $f_{B}$  as the threshold amplitude,
the best description of the observed energy dependence of the SATURNE total 
cross section data is achieved using a scattering length of $a_{\,{}^3He\,\eta} 
= \pm (3.6 \pm 0.5)\,\mbox{fm} + i(2.0 \pm 0.9)\,\mbox{fm}$, as can be seen in 
fig.~\ref{he3rev} (solid line) with the absolute scale of the curve being 
scaled to fit the data. 
Such a large scattering length might be associated with the existence of a 
quasi--bound ${}^3He\,\eta$ state.  \\

In the right part of fig.~\ref{he3rev} the extracted production amplitude 
squared  $|f(\mbox{p}^*_{\eta})|^2$ is presented as function of the 
centre--of--mass momentum of the ejectiles. 
It should be noted that in the close vicinity of the production threshold and 
with a given cross section, the production amplitude squared 
$|f(\mbox{p}^*_{\eta})|^2$ depends crucially on the momentum 
$\mbox{p}^*_{\eta}$ and, therefore, on the exact masses of the ejectiles. 
Here a mass of the ${}^3He$--nucleus of $\mbox{m}_{\,{}^3He} = 
2.808392\,\mbox{GeV}/\mbox{c}^2$~\cite{Web99} and a meson mass of 
$\mbox{m}_{\eta} = 0.54730\,\mbox{GeV}/\mbox{c}^2$~\cite{groom1} have been 
applied.
While a constant amplitude would imply a behaviour according to pure s--wave 
phase--space considerations, the observed shape exposes a strong momentum 
dependence and can be reproduced well by the discussed formalism up to excess 
energies of $\mbox{Q} \approx 7\,\mbox{MeV}$, corresponding to 
$\mbox{p}^*_{\eta} \approx 0.35\,\mbox{fm}^{-1}$.\newline
However, recent measurements at higher excess energies from the GEM 
collaboration at COSY~\cite{Bet00} and from the WASA/PROMICE collaboration at 
CELSIUS~\cite{Bil02} clearly deviate from the predicted energy dependence of 
the cross section (fig.~\ref{he3rev}). 
Moreover, different to the SATURNE results, the centre--of--mass angular 
distribution of the emitted $\eta$ mesons strongly differs from an isotropic 
emission and exposes a peaking in forward direction. 
Motivated by results on the $\eta$ meson production in $\gamma p$ and $pp$ 
interactions, another approach to describe the production process has been 
evaluated~\cite{Bet00} where the $\eta$ meson creation is described by the 
excitation of the $\mbox{S}_{11}$ $N^*(1535)$ resonance as shown in 
figure~\ref{he3etafey} (right). 
Neglecting effects of other resonances, the production amplitude can be 
expressed by a Breit--Wigner distribution~\cite{Kru95}:
\begin{equation}
\label{bw1}
|f(\mbox{E})|^2 \;\sim\; 
 \left|\,
 \frac{\mbox{E}_R \Gamma_R}
   {(\mbox{E}_R^2 - \mbox{E}^2) - i\,\mbox{E}_R\Gamma(\mbox{E})} \,\right|^2\,,
\end{equation}
with $\mbox{E}_R$ as the mass of the $\mbox{S}_{11}$ resonance and $\Gamma_R$ 
as the corresponding width. 
Since the production threshold is located in the vicinity of the resonance 
mass $\mbox{E}_R$, the energy dependence of the width $\Gamma(\mbox{E})$ has 
to be considered:
\begin{equation}
\label{bw2}
\Gamma(\mbox{E}) = \Gamma_R 
  \left(\,b_{\eta}\,\frac{\mbox{p}^*_{\eta}}{\mbox{p}^*_{\eta R}} +  
          b_{\pi}\,\frac{\mbox{p}^*_{\pi}}{\mbox{p}^*_{\pi R}} + 
          b_{\pi \pi} \right).
 \end{equation}
$~~$\\
 
The parameters $b_i$ indicate the branching ratios for different decay modes 
of the resonance while the centre--of--mass momenta of the $\eta$ and $\pi$ 
mesons are represented by $\mbox{p}^*_i$. 
A corresponding calculation, scaled to fit the GEM data point, is given in 
figure~\ref{he3rev} by the dash--dotted line. 
Obviously, the prediction of this approach fails to describe the shape of the 
SATURNE data. 
This observation might indicate an influence of a strong ${}^3He\,\eta$ final state 
interaction only in the close vicinity of the production threshold. 
Alternatively, this effect might be caused by a transition between different
production processes.\\

Concluding, the $pd \rightarrow {}^3He\,\eta$ reaction appears to be an 
interesting and important channel to study both the meson--nucleus final 
state interaction as well as reaction processes involving three nucleons. 
However, the body of data is far from being optimal, especially in the 
near--threshold region. 
New data from PROMICE/WASA and COSY--11 is currently under evaluation and 
will improve the experimental situation.

\subsection{Quasi--bound states}
\label{Qbs}         
From the theoretical point of view the existence of mesic nuclei or 
quasi--bound meson--nucleus systems is not excluded, however, up to now a 
compelling experimental proof for the formation of such a state is still 
missing. 
It is obvious that the possibility to create such states crucially depends on 
both the properties of the meson and the nucleus as well as on the sign and 
the strength of the meson--nucleus interaction. 
Since the $\pi$--nucleon and $K^+$--nucleon interaction have been found to be 
repulsive, it is unlikely that they will form quasi--bound states. 
Contrary, there are evidences for an attractive $K^-$-nucleon interaction,
however, corresponding experiments would suffer from the low production cross 
sections for $K^-$ mesons. 
Furthermore, the Coulomb interaction is expected to screen corresponding 
physical observables and might lead to the formation of mesonic atoms bound by 
the Coulomb interactions.
Different to this, the $\eta$ meson is uncharged and the observation of an 
attractive $\eta$--nucleon interaction led to speculations concerning the 
existence of $\eta$--nuclear quasi--bound states. 
In the absence of $\eta$--meson beams such states bound by the strong 
interaction would offer a new possibility to study the $\eta$--nucleon 
interaction since the meson would be trapped for a relatively large time in 
nuclear medium. Predicted by Haider and Liu~\cite{Hai86,Hai86b}, detailed 
studies on the formation of such states have been the topic of recent 
investigations~\cite{rakityanskyR2043,Wyc95,Aba96,wilkinR938, CHIANG738,Li87}, 
resulting 
in different lower limits of the atomic number A for which bound states should 
exist. While e.g.\ in~\cite{Hai86b,Hai86} the lower mass number is estimated in 
the framework of an optical model approach to be $\mbox{A} = 12$, there are 
considerations that even for nuclei of lower masses ($d, t, {}^3He, {}^4He$) 
bound systems might exist~\cite{wilkinR938,rakityanskyR2043}, see also 
section~\ref{Mwhs}.\vspace{1ex}

\begin{figure}[H]
\hfill
\parbox{0.60\textwidth}
  {\epsfig{file=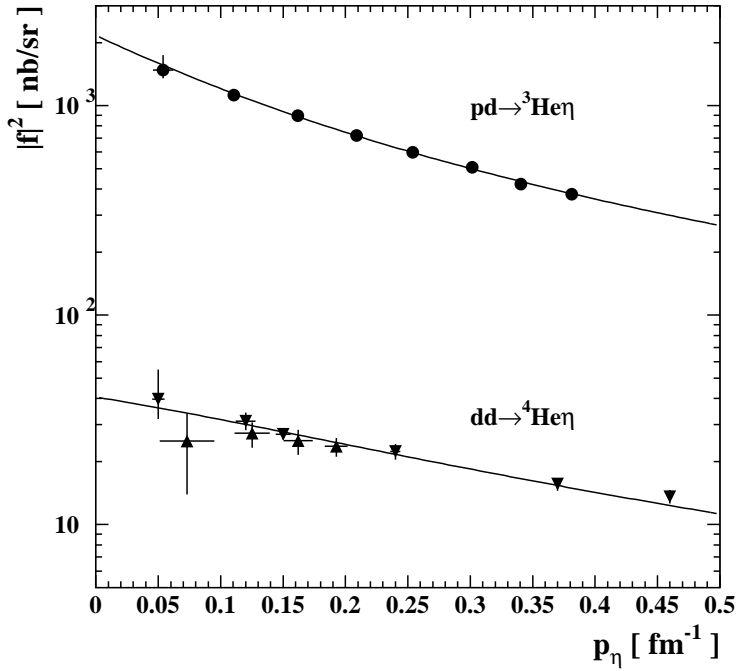,width=0.525\textwidth}} \hfill
\parbox{0.34\textwidth}
  {\caption{\label{he34} Averaged squared production amplitudes of the 
  reactions $pd \rightarrow {}^3He\,\eta$ and $dd \rightarrow {}^4He\,\eta$ as 
  function of the centre--of--mass $\eta$ momentum. The curves are discussed 
  in the text. The data are taken from~\cite{May96,Fra94,Wil97}.}}
\end{figure}

Experimental evidences for bound states have been found in the near--threshold 
production of $\eta$ mesons in the reaction channel $pd \rightarrow 
{}^3He\,\eta$. 
The unexpected large production amplitude as well as its rapid decrease with 
increasing energy (figure~\ref{he3rev}) is attributed to a strong s--wave 
final state interaction associated with a large $\eta$--$\,{}^3He$ scattering 
length~\cite{wilkinR938}.\newline
Such a large scattering length (see fit in figure~\ref{he3rev}) might imply 
the existence of a quasi--bound state. 
Assuming the ${}^3He\,\eta$ to form a quasi--bound state, it is therefore 
interesting to compare $pd \rightarrow {}^3He\,\eta$ production data with the 
ones of the reaction $dd \rightarrow {}^4He\,\eta$, where according to the 
higher mass number, the formation of a bound state should be even more 
probable. 
A comparison of corresponding near--threshold data is presented in 
figure~\ref{he34}.

A fit to the data using the scattering length formula~\eqref{he1} provides 
information on the scale of both the real and imaginary part of the scattering 
length of the meson--nucleus system but is unsensitive to the sign of the real 
part of $a_{He\,\eta}$. 
However, information on this sign can be obtained using a lowest--order 
optical potential~\cite{wilkinR938}:
\begin{equation}
\label{opt2}
2\,\mbox{m}_R\,V_{opt}(\mbox{r}) = 
  -4\pi\,\mbox{A}\,\rho(\mbox{r})\,a_{\,\eta\,N}
\end{equation}
with the $\eta$--nucleon reduced mass $\mbox{m}_R$, the mass number of the 
nucleus A and the $\eta$--nucleon scattering length $a_{\,\eta\,N}\,\approx\, 
(0.52 + i\,0.25)\,\mbox{fm}$.
Using Gaussian nuclear densities with rms radii of $\rho(\mbox{r}) = 
1.63\,\mbox{fm}$ ($1.9\,\mbox{fm}$) for the $\eta\,{}^3He$ ($\eta\,{}^4He$) 
system, meson--nucleus scattering lengths of
\begin{eqnarray}
\label{opt1}
a\,({}^3He\,\eta) \sim (-2.3 + i\,3.2)\,\mbox{fm} \\
a\,({}^4He\,\eta) \sim (-2.2 + i\,1.1)\,\mbox{fm}
\end{eqnarray}
with negative real parts for both systems have been obtained.

Fits to existing close--to--threshold data for both reaction channels using 
these scattering lengths and eq.~\eqref{he1} are presented in 
figure~\ref{he34}. 
Obviously, the data are reproduced well by this formalism.
These results imply the existence of quasi--bound meson--nucleus states in the 
systems $\eta\,{}^3He$ and $\eta\,{}^4He$, however, unambiguous evidence is 
still missing. 
Therefore, detailed investigations on both reaction channels as well as on 
reactions using heavier targets are highly recommended.

\subsection{Test of invariances}
\label{toi}  
Conservation laws restrict the quantum numbers for the production of mesons in 
the nucleon--nucleon interaction. 
Here especially the pion production is of considerable interest since it 
exhausts all inelasticity up to about $2\,\mbox{GeV/c}$ beam momentum in the 
nucleon--nucleon collision.\newline
As outlined in references~\cite{GMW,Rosenfeld,meyer54,niskanen2683}
the conservation of isospin allows the total cross section for pion production 
to be expressed as an incoherent sum of three independent partial cross 
sections given by: $\sigma_{11}$, $\sigma_{10}$ and $\sigma_{01}$, where the 
first (second) index indicates the isospin of the initial state (final state) 
two--nucleon system and $\sigma_{00}$ is ruled out due to isospin conservation 
arguments.\newline
The three cross sections can be determined via the reactions:
\begin{table}[H]
\begin{tabular}[l]{lclclclc}
pure $\sigma_{11}$ channels: & $\sigma(pp \rightarrow pp\pi^0)$, & 
  $\sigma(nn \rightarrow nn\pi^0)^*$  \\ 
pure $\sigma_{10}$ channels: & $\sigma(pp \rightarrow d\pi^+)$, & 
  $\sigma(np \rightarrow d\pi^0)$, & $\sigma(nn \rightarrow d\pi^-)^*$  \\
mixed $\sigma_{10} + \sigma_{11}$ channels: & 
  $\sigma(pp \rightarrow pn\pi^+)$, & $\sigma(nn \rightarrow np\pi^-)^*$  \\ 
mixed $\sigma_{10} + \sigma_{01}$ channels: & 
  $\sigma(np \rightarrow pn\pi^0)$, \\  
mixed $\sigma_{11} + \sigma_{01}$ channels: & 
  $\sigma(np \rightarrow pp\pi^-)$, & $\sigma(np \rightarrow nn\pi^+)$. 
\end{tabular}
\end{table}
Charge symmetry relates those reactions denoted by a $^*$ to their mirror 
systems given as the first one in the same row.

\begin{figure}[H]
\hfill
\parbox{0.65\textwidth}
  {\epsfig{file=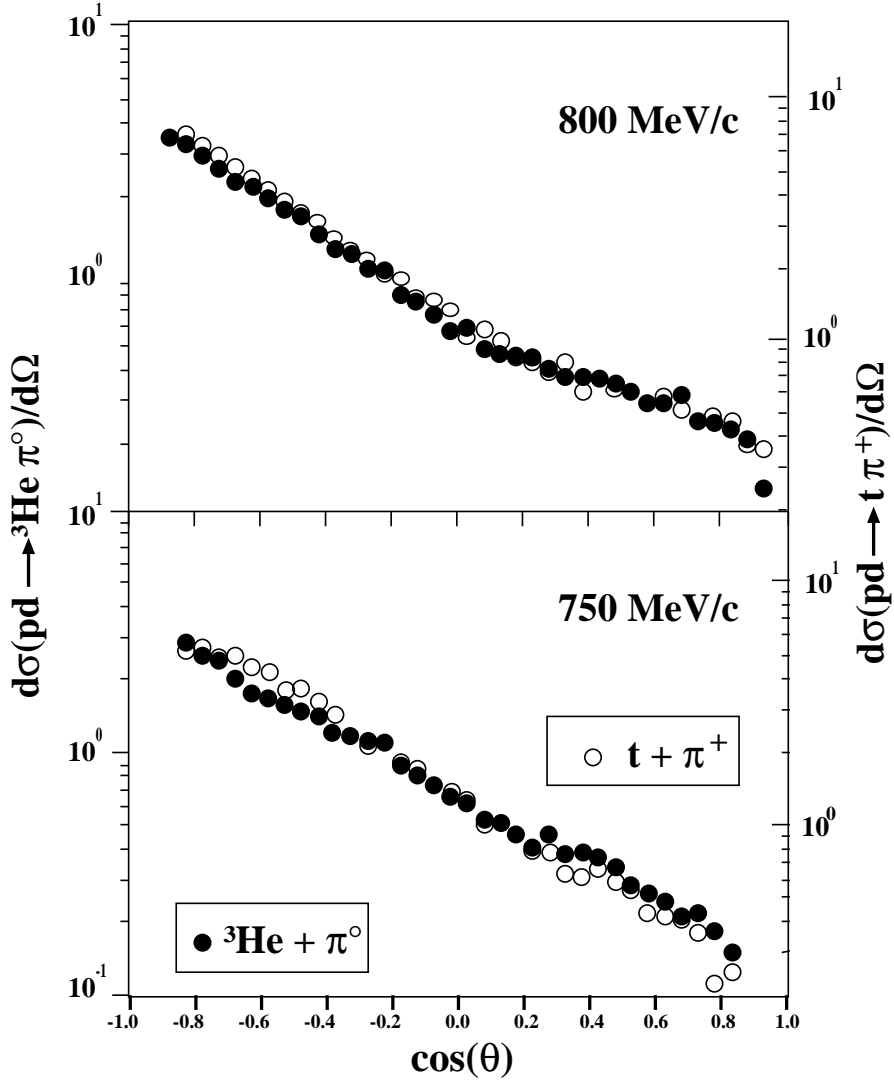,width=0.65\textwidth}} \hfill
\parbox{0.25\textwidth}
  {\caption{\label{Isospin_X_section} Differential cross sections for the two 
  isospin symmetric reactions $pd \rightarrow {}^3He\,\pi^0$ and $pd 
  \rightarrow {}^3H\,\pi^+$ at beam momenta of $800\,\mbox{MeV/c}$ and 
  $750\,\mbox{MeV/c}$, respectively. Data are from the GEM 
  collaboration~\cite{Betigeri}, the error bars are smaller than the symbols.}}
\end{figure}

The two reactions $pd \rightarrow {}^3H\,\pi^+$ and $pd 
\rightarrow {}^3He\,\pi^0$, which permit the study of isospin breaking effects
have been studied systematically~\cite{Betigeri}. 
Figure~\ref{Isospin_X_section} compares the differential cross sections for 
these two reactions at two proton momenta close to the production threshold.
At these momenta no significant differences in the differential cross sections
were observed. 
In fact, interpolating the cross sections to the same four--momentum--transfer 
yields an expected average ratio $R = \sigma(pd \rightarrow {}^3H\,\pi^+)/
\sigma(pd \rightarrow {}^3He\,\pi^0) \approx 2.0$ within the experimental 
uncertainties.
However, the COSY--GEM collaboration seems to have indications that a clear 
effect of isospin breaking is observed at a momentum of about $1.57\,\mbox{GeV/c}$, 
close to the $\eta$--production threshold~\cite{HAWRANEK}.

As pointed out in reference~\cite{Magiera,idee}, in context of QCD isospin and 
charge symmetry are broken due to both i) the mass difference between up and 
down quarks and ii) the electromagnetic interaction. 
The observation of mixing between mesons belonging to the same $SU(3)$ 
multiplets offers a unique possibility to study charge symmetry breaking 
effects. 
As suggested by Wilkin~\cite{C.Wilkin} 
a possible isospin symmetry breaking can manifest itself e.g.\ via external
$\pi^0$--$\eta$ mixing. 
Searching for such effects the authors of reference~\cite{Magiera} are 
investigating the cross section ratio for the $pd \rightarrow {}^3He\,\pi^0$ 
and $pd \rightarrow {}^3H\,\pi^+$ reactions in particular for large relative 
pion--proton angles and at beam energies of the corresponding $\eta$ 
production threshold. 
The idea is based on the assumption that --- besides the direct reaction --- 
an intermediately formed $\tilde{\eta}$ meson undergoes a transition to the 
final $\pi^0$--meson for the $pd \rightarrow {}^3He\,\pi^0$ reaction 
($pd \rightarrow {}^3He\,\tilde{\eta} \rightarrow {}^3He\,\pi^0$) with a yield 
governed by the strength of the mixing angle while such an intermediate 
state can not occur for the $pd \rightarrow {}^3H\,\pi^+$ case. 
The authors~\cite{Magiera} predict angular dependent cross section ratios as: 
$R_{\theta_{p \pi} = 180^\circ} = 2.4$ and $R_{\theta_{p \pi} = 0^\circ} = 
2.03$. 
The predicted symmetry breaking magnitude appears to be in reach of the 
experiments planned by the GEM collaboration at COSY.
  
Another direct evidence for the isospin invariance breaking would be the 
observation of the non--zero yield for the isospin forbidden $dd \rightarrow 
{}^4He\,\pi^0$ reaction.
The upgrade of the ANKE detection system at COSY will allow to search for the 
effect~\cite{ankephoton}.
\newpage

\section{Experimental facilities}
\label{Ef}
Studies of meson production at threshold require crucially high quality and 
precise quantitative knowledge of the accelerator beam quality especially
due to the strong cross section dependence on the excess energy. 
The rapid growth of the meson production cross section in the vicinity of the
kinematical threshold is evidently seen in figure~\ref{Dieter_Wasserfall}.
It has been shown in the previous sections that new precise data are available 
for different reaction channels. Here we present a very brief overview of
technical features of such experimental facilities as: accelerators, targets 
and detector arrangements where only a representative selection can be given.   

\subsection{Accelerators for hadronic physics at medium energies}
\label{Afhpame}                                     
Initiated by the rather high costs for experimenting with antiprotons and aiming
in a most efficient usage, storage rings have been developed. The possibility of
controlling the emittance of such beams by electron as well as stochastic cooling,
combined with some unique benefits of using internal targets lead to the
construction of medium energy cooler rings as the low energy antiproton ring
(LEAR) which might be regarded as the father for many facilities built.

\begin{figure}[H]         
        \parbox{0.6\textwidth}{\epsfig{file=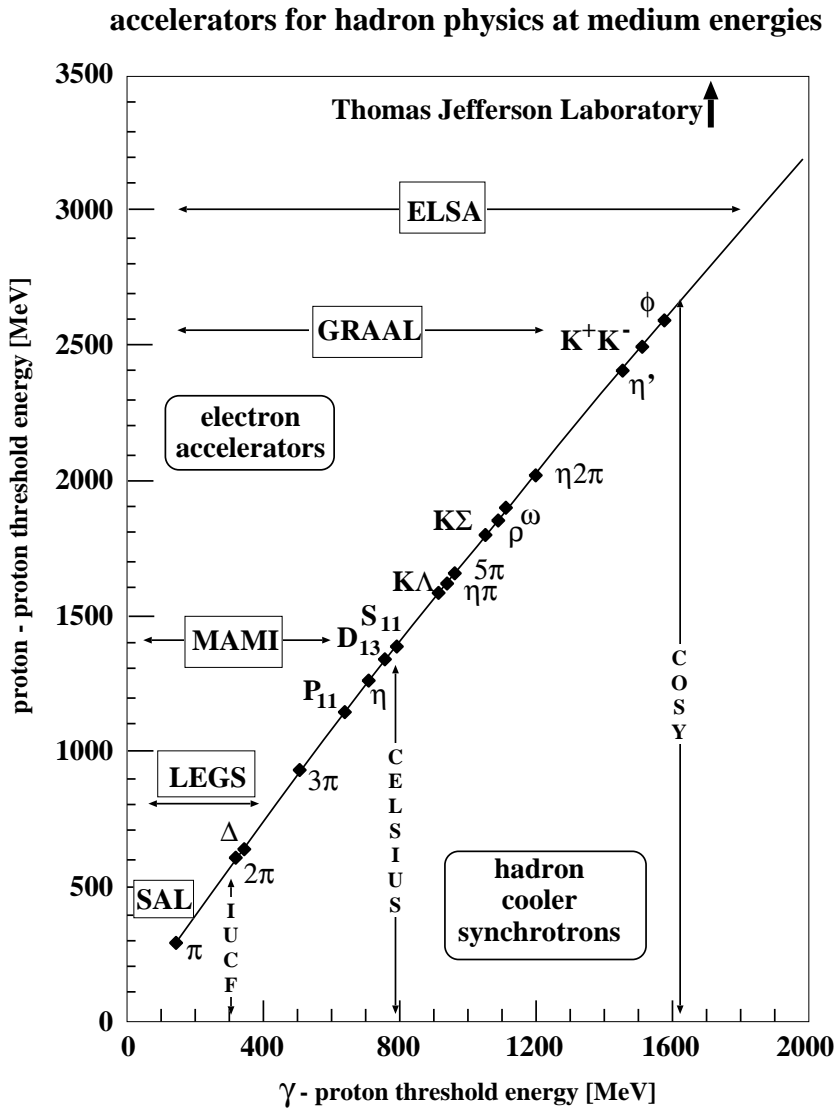,width=0.6\textwidth,angle=0}}
        \hfill
        \parbox{0.3\textwidth}{\vspace{-1cm}\caption{\small 
          Accelerators for medium energy physics.  
        \label{Beschleuniger}
        }}
      \end{figure}

The maximum available beam momentum limits the producible mass.
In figure~\ref{Beschleuniger} a comparison is presented between proton and
electron accelerators in the range of producing in the proton--proton and
$\gamma$--proton collision mesons up to the $\phi$--meson at 1.02 GeV/c$^2$.
This self--explaining figure demonstrates e.g. that the Uppsala accelerator
CELSIUS can reach beam momenta to produce the $\eta$--meson in the\\
proton--proton scattering which is about the same mass production range as that 
of MAMI via the $\gamma$--proton interaction.

Complementary to the investigations with charged hadronic beams the high 
precision secondary neutron beam of the TRIUMF facility~\cite{helmer588} 
permits the close--to--threshold meson production in neutron--proton collisions.\\

In Table~\ref{Table_Beschleugiger} some accelerators used for hadron physics at 
medium energies are listed, which are or were important for this field.\\

\begin{center}
Table~\ref{Table_Beschleugiger}$~~~$Hadron accelerators for hadron physics at medium 
energy, \\
LEAR and SATURNE stopped operation in 1997 and 1999, respectively. \\[0.3cm]

\begin{tabular}[l]{rcrcrcrcrc}
\hline
$~~$\\[-0.3cm]
accelerator  &typ            &momentum   &energy &beam        &cooling \\
             &               &MeV/c      &MeV                    \\
$~~$\\[-0.3cm]
\hline
\hline
$~~$\\[-0.3cm]
TRIUMF       &cyclotron      &1090       &500   &$n, \vec p$    & --\\
$~~$\\[-0.3cm]
IUCF         &synchrotron    &1090       &500   &A = $\vec 1$, 2, 3    &$e^-$ \\
$~~$\\[-0.3cm]
SIN--PSI     &cyclotron      &1220       &600   &$\vec p$    & --\\
$~~$\\[-0.3cm]
LAMPF   &linear accelerator  &1460       &800   &$\vec p$    & --\\
$~~$\\[-0.3cm]
LEAR         &synchrotron    &2000       &1270  &$\bar p$     &$e^-$ \\
$~~$\\[-0.3cm]
CELSIUS      &synchrotron    &2100       &1360  &$A = 1, 2, ...$    &$e^-$  \\
$~~$\\[-0.3cm]
COSY      &synchrotron       &3500       &2620  &$A = \vec 1, 2 $   &$~~~e^-$ plus stochasitc \\
$~~$\\[-0.3cm]
SATURNE      &synchrotron    &3770       &3000  &$A = 1 ... 40  $   & --  \\
$~~$\\[-0.3cm]
\hline
\hline
\label{Table_Beschleugiger}
\end{tabular} 
\end{center}

\subsection{Targets}      
\label{Targets}
   The choice of a certain target material for accelerator experiments as well as
   its chemical and mechanical properties is strongly correlated
   with the given experimental conditions and, certainly, with the
   reactions which should be investigated. Since in this article
   we concentrate on unpolarized meson production experiments off nucleons and
   deuterons, here we will restrict ourselves on corresponding
   targets.\\
   
   Proton and deuteron targets are commonly provided using
   hydrogen and deuterium as raw material. In the absence of free
   neutron targets deuterons can be used as substitute,
   considering the weakly bound deuterons as effective neutron
   targets and restricting on reactions with the bound protons acting
   only as spectator particles. For the sake of completeness we want to point out
   that in this case the Fermi motion of the nucleons has to be
   considered, especially in the regime of low excess energies.\\
  
   Near threshold production experiments are commonly connected
   with low cross sections in the order of picobarns or nanobarns.
   To obtain sufficient counting rates during experimental
   runs, high luminosities in the order of $\sim 10^{30}~$cm$^{-2}$s$^{-1}$
   are necessary and can be achieved using targets of high areal density
   and intense accelerator beams. Neglecting technical limitations like
   space charge effects, from the experimental point of view a
   combination of high intense ion beams and thin targets are
   preferable since in this case the quality of the cooled
   accelerator beam is nearly uninfluenced during the passage of
   the target, granting well defined kinematical conditions as
   well as single particle interactions.
   In practice one has to distinguish between internal and external
   beam experiments. In the latter case a well prepared and
   extracted accelerator beam is shot onto a target and beam
   particles not interacting with the target are lost. Therefore,
   liquid or solid state targets of sufficient high areal density
   have commonly to be used in order to obtain the required high
   luminosities. Contrary, in internal beam experiments carried out
   at storage rings the circulating beam particles pass the target
   material 10$^5$ to 10$^6$ times per second and projectiles not
   interacting with the target remain in the beam. Therefore, to
   achieve similar luminosities compared to external beam
   experiments only thin targets are necessary. Moreover, to preserve the
   quality of the circulating beam and to obtain life times of the
   accelerated beam up to hours, effective areal target densities
   below $\sim 10^{15}$ atoms/cm$^{2}$ are desirable and can be
   achieved using gas and cluster jets, pellets and thin fibers.
   Considering hydrogen or deuterium as raw material such targets
   can be realized in different ways.

    \subsubsection{Fiber Targets}
     Fiber targets with diameters of only a few micrometers
     are of special interest for storage ring
     experiments with the request of high target densities in
     combination with a point-like target and UHV
     conditions. Due to the
     relatively simple target installation, 4$\pi$ detection
     systems are possible. Using carbon for example as target
     material, fiber diameters down to 5$\mu$m are accessible,
     corresponding to $\sim 2\cdot 10^{14}$ atoms/cm$^2$
     \cite{Loz91} and allowing to compensate beam heating effects
     by beam cooling devices. Fiber targets for
     interactions off protons and deuterons naturally
     consist of chemical compounds like propylene such as the target
     for the internal
     beam experiment EDDA at COSY, where propylene fibers
     with diameters of $\sim 5\mu$m are used. However, apart
     from
     those advantages one has to keep in mind that the mass of such CH$_2$
     fibers is mostly given by ``carrier'' material carbon, which
     influences the
     performance of the data aquistion system and, moreover,
     causes a physical background in the data which has to be
     considered.

    \subsubsection{Liquid Targets}
     Liquid hydrogen and deuterium targets with their
     high volume densities ($\rho_{LH_2}~\sim$~0.07 g/cm$^3$)
     are reasonable for the use at external beam experiments.
     Depending on the experimental requirements large target volumes
     even up to several liters are accessible. However, precision
     experiments on the near threshold meson production force to
     use only small target volumes in the interaction region in order to
     minimize energy losses and
     effects of multiple scattering in the target material.
     An example of an experimental facility using a thin liquid hydrogen
     target
     is the COSY--TOF installation~\cite{Jaeckle94,Hassan97}. Here hydrogen is
     liquefyed using a commercial cryogenic cold head, providing a
     compact target volume of several millimeters in diameter. Since such
     targets naturally have to be encased in special vessels, the choice
     of appropriate entrance windows is of great importance to minimize the
     physical background. At COSY--TOF for example thin windows of
     only $\sim$~1~$\mu$m Mylar foil with a diameter of $\sim$~4~mm are
     routinely in use, granting good signal to background conditions.

    \subsubsection{Pellet Targets}
\begin{figure}[H]
        \vspace{0.4cm}
        \parbox{0.30\textwidth}{\epsfig{file=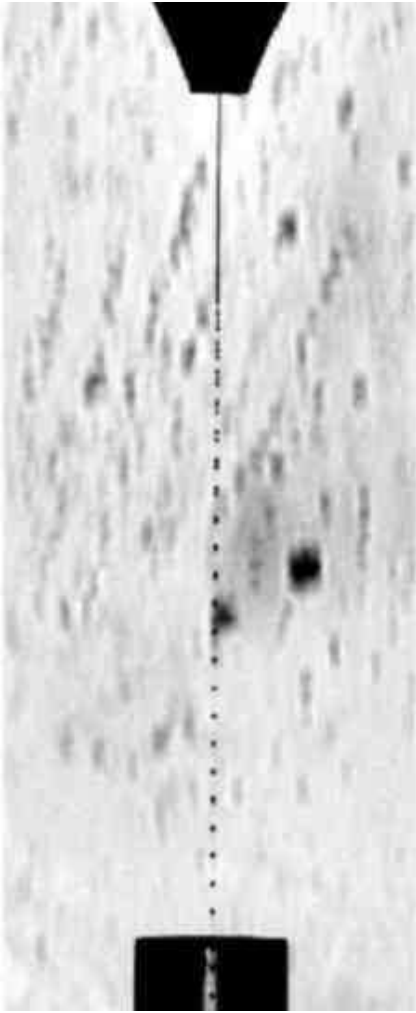,width=0.30\textwidth,angle=0}}
        \hfill
        \parbox{0.55\textwidth}{\vspace{-1cm}\caption{\small 
       Formation of deuterium pellets with diameters of $\sim$ 50 $\mu$m,
       the photograph was taken at the pellet target of the 
       CELSIUS/WASA detector at CELSIUS in Uppsala. 
        \label{pellets}
        }}
      \end{figure}

     Pellet target installations can be used to prepare streams of frozen
     gas micro-spheres as targets for internal beam experiments.
     Similar to fiber targets, pellet targets can be associated to the
     class of the solid state targets and allow to provide high
     areal densities in combination with point-like
     interaction regions, well suited for 4$\pi$ detector geometries.
     Moreover, contrary to fiber targets
     these installations allow to produce pure hydrogen or deuterium
     pellets without any carrier material, however, at the cost of
     an increased experimental effort and vacuum conditions of
     $\sim10^{-6}$ mbar in the scattering chamber. At the CELSIUS ring in
     Uppsala such a pellet target was build up for the WASA
     experiment \cite{Tro95,Eks96,EksT00}.
     The pellets are produced by pressing liquid hydrogen at triple point
     conditions (14 K, 72 mbar) through
     a glass nozzle of 20 $\mu$m diameter. By means of a
     piezo-electric transductor connected to the nozzle the stream
     of hydrogen is broken into uniformly sized micro-spheres
     (pellets) as shown in figure~\ref{pellets}. This device was designed to provide
     a stream of hydrogen pellets with diameters of 50 $\mu$m at a
     repetition rate of $\sim$ 68 kHz. While the pellets itself are
     locally very thick ($\sim10^{20}$ atoms/cm$^2$), the effective areal
     density
     in the scattering chamber is in the order of 10$^{16}$ atoms/cm$^2$
     due to the minimized geometrical beam-target overlap.

    \subsubsection{Gas Jet Targets}
     Supersonic gas jet beams can be used as targets for both internal
     and external beam experiments. Produced by feeding
     Laval-nozzles with gases like hydrogen or deuterium, high
     areal target densities ($\sim10^{16}$ atoms/cm$^2$) can easily be
     achieved close to the nozzle exit \cite{Spe90}. Similar to pellet
     targets, proton or deuteron targets without effects of windows or
     carrier materials can be provided by
     operating the installation with gases of high purity.
     However, since the volume density of a gas jet beam
     drastically decreases both in longitudinal and transversal
     direction, the nozzle has to be placed very close to
     the interaction point in the scattering chamber. Therefore,
     the advantage of high target densities of such installations
     is commonly connected with higher gas loads
     in the scattering chamber and a limited spatial resolution of the
     interaction point if no vertex detector is used. The great
     advantage of such targets is the possibility to use almost
     all gases as target material and to adjust the areal density
     by orders of magnitude by simply changing the gas input
     pressure.

    \subsubsection{Cluster Targets}
\begin{figure}[H]
        \vspace{0.4cm}
   \parbox{0.5\textwidth}{\epsfig{file=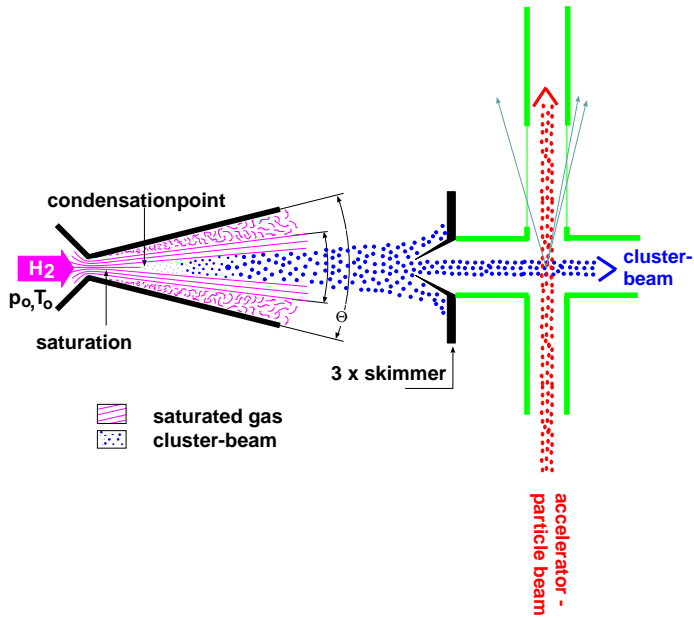,width=0.5\textwidth,angle=0}}
        \hfill
        \parbox{0.4\textwidth}{\vspace{-1cm}\caption{\small 
        Schematic drawing of the principle for the formation of clusters. 
        \label{cluster}
        }}
      \end{figure}

     Similar to conventional gas jet targets, cluster beams~\cite{DOMB} 
     can be produced by using fine nozzles with a convergent-divergent shape 
     (Laval-nozzle)  and, therefore,
     almost all gases can be used. Operated at appropriate low
     temperatures, a
     partially condensation of the used gas is possible and clusters
     with sizes of 10$^3$ to 10$^6$ can grow. Separated from the
     surrounding conventional gas beam by a set of collimators (see fig.~\ref{cluster}),
     cluster beams with a homogeneous volume density, a sharp
     boundary, a small angular divergence and well
     defined beam dimensions can be prepared for interaction with
     circulating accelerator beams~\cite{khoukaz275}. 
     Cluster beam
     installations are frequently used at cooler rings. Using Laval-nozzles
     with diameters of 11-16 $\mu$m, low nozzle temperatures of
     $\sim$25 K and gas input pressures of $\sim$18 bar,
     hydrogen and deuterium cluster
     beams with areal densities in the order of $\sim 10^{14}$
     atoms/cm$^2$ in combination with UHV conditions are accessible
     in the interaction region. Furthermore, similar to gas
     beam targets the areal density can easily be adjusted over orders of
     magnitude by changing the gas input pressure or the nozzle
     temperature~\cite{khoukaz275}.

\subsection{Typical detector arrangements}
\label{Tda}    
Special efforts are made for an optimal detector design to match the required
conditions for kinematically completeness and undisturbed determination of the
four--momentum vector of the ejectiles. A kinematically complete event is
characterized by knowing all four--momentum vectors of the participating
particles. The experimental information is limited by: i) statistical
fluctuations, ii) the precision for the determination of the 
components of the four--momentum vectors, iii) contributions from any kind of
background produced by means of the apparatus or due to physical reasons, 
iv) the efficiency of the detector covering not 100~$\%$ of the phase--space due
to size or construction limitations and finally v) the reconstruction 
efficiency due to unavoidable inadequate performance whatsoever. 
Here all collaborations have to meet the optimum between wishful thinking,
technical possibilities and affordable costs.
\subsubsection{Example of a 4$\pi$ internal--detector} 
The high luminosity and 4$\pi$ detector CELSIUS/WASA ({\bf{W}}ide {\bf{A}}ngle 
{\bf{S}}hower {\bf{A}}pparatus) facility~\cite{CALEN_96_NIM} is primarily 
\begin{figure}[H]
\begin{center}
\epsfig{file=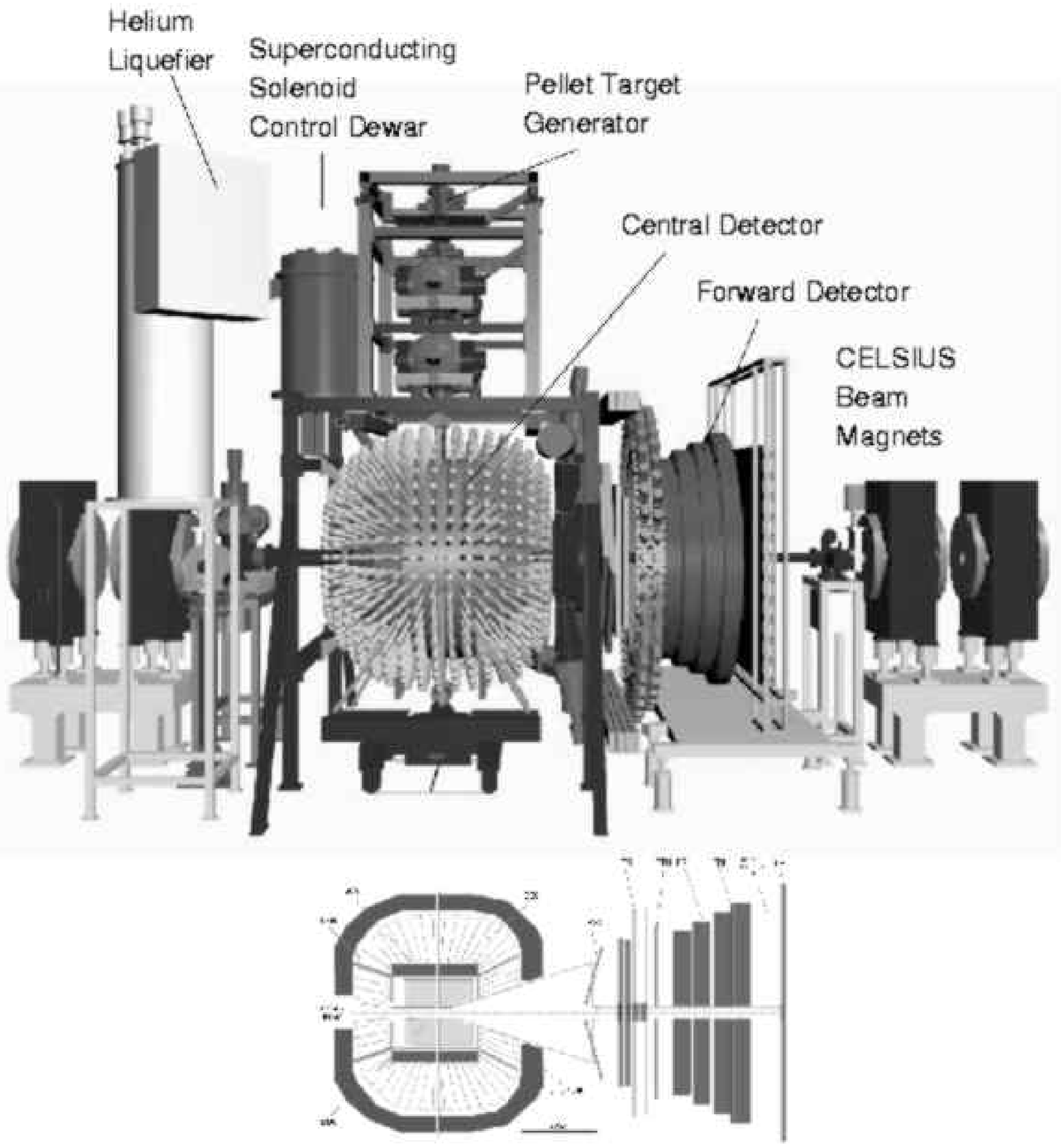,scale=0.65}
\vspace{-4cm}
\caption{Top: $~~~~~$CAD--view of the WASA detector facility \protect \\
$~~~~~~~~~~~~~~~~$Bottom: Cross section of the central and forward part of the 
WASA detector facility. 
  \protect\\
  \protect{\small 
    A version of the article including a photo of better quality can be found at the COSY-11 homepage:
   \protect\\
   $http://ikpe1101.ikp.kfa-juelich.de/cosy-11/pub/List\_of\_Publications.html\#papers$}
}
\label{Uppsala_WASA_detector}
\end{center}
\end{figure}
designed for both measuring rare decay processes of light mesons and
investigating meson production studies in the threshold region. The target and 
beam particles tend to be scattered at forward angles while the products of
the meson decay will be distributed rather isotropically. The detector is 
constructed following these constrains and in full operation an annual 
production rate of about $10 ^{~11} ~\eta$'s is expected in the $pp$ 
interaction.

The essential parts of the CELSIUS/WASA facility are: 
\begin{itemize}
\item
The pellet target with: 
\subitem
{\hspace{-0.65cm}}Pellets diameter and frequency of 25 -- 50~$\mu$m and a few kHz, respectively,\\
{\hspace{0.5cm}}where the beam diameter is about 3 mm,
$5 \times 10^{~15}$ atoms/cm$^2$. 
\item
The forward detector with:
\subitem
{\hspace{-0.65cm}}A segmented forward window counter (FWC) for fast timing in the first trigger level.\\
{\hspace{0.5cm}}A forward proportional chamber (FPC) out of straw chambers to measure the scattering angle.\\ 
{\hspace{0.5cm}}A forward trigger hodoscope (FTH) built out of three layers of circular scintillators used for:\\ 
{\hspace{0.5cm}}$~~~~~~$fast trigger decisions, rough track position determination and particle identification.\\
{\hspace{0.5cm}}A forward range hodoscope (FRH) as a plastic scintillator calorimeter for measurements of\\
{\hspace{0.5cm}}$~~~~~~$particle energy and identification.\\
{\hspace{0.5cm}}A forward range intermediate hodoscope (FRI) adding a position sensitivity to the FRH. 
\item
The zero degree spectrometer to filter out low rigidity reaction products 
employing HPGe-- \\ or telescopes of
Si--$\mu$-strip detectors, scintillation counters and CsI(Tl) crystals.
\item 
The central detector with:
\subitem
{\hspace{-0.65cm}}A scintillator electromagnetic calorimeter (SEC) consisting out of 1012 CsI(Na) \\
{\hspace{0.5cm}}$~~~~~~$between the superconducting solenoid (SCS) and the iron yoke. \\
{\hspace{0.5cm}}A mini drift chamber (MDC) with 1738 straw tubes for measuring tracks of charged
particles.
\end{itemize} 
The CELSIUS/WASA detector is completed and regular data taking has started in late 2001 on some $\pi$--production
reactions where the $\pi^0$'s are detected by their $\gamma \gamma$ decays. The tagging of the $\eta$--production
and the detection of the $\eta$'s by their $\gamma \gamma$ and 3$\pi^0$ decays
has been proven to work. In other words there is a high luminosity hadron 
facility ready for exciting physics. 

\subsubsection{Example of a centre--of--mass 4$\pi$ external--detector}
The COSY--TOF detection facility at the external beam line of COSY is a multi
purpose modular arrangement of segmented barrel and circular planar 
scintillation detectors for measuring the time-of-flight and the energy loss of
charged reaction products.

\begin{figure}[H]
\begin{center}
\epsfig{file=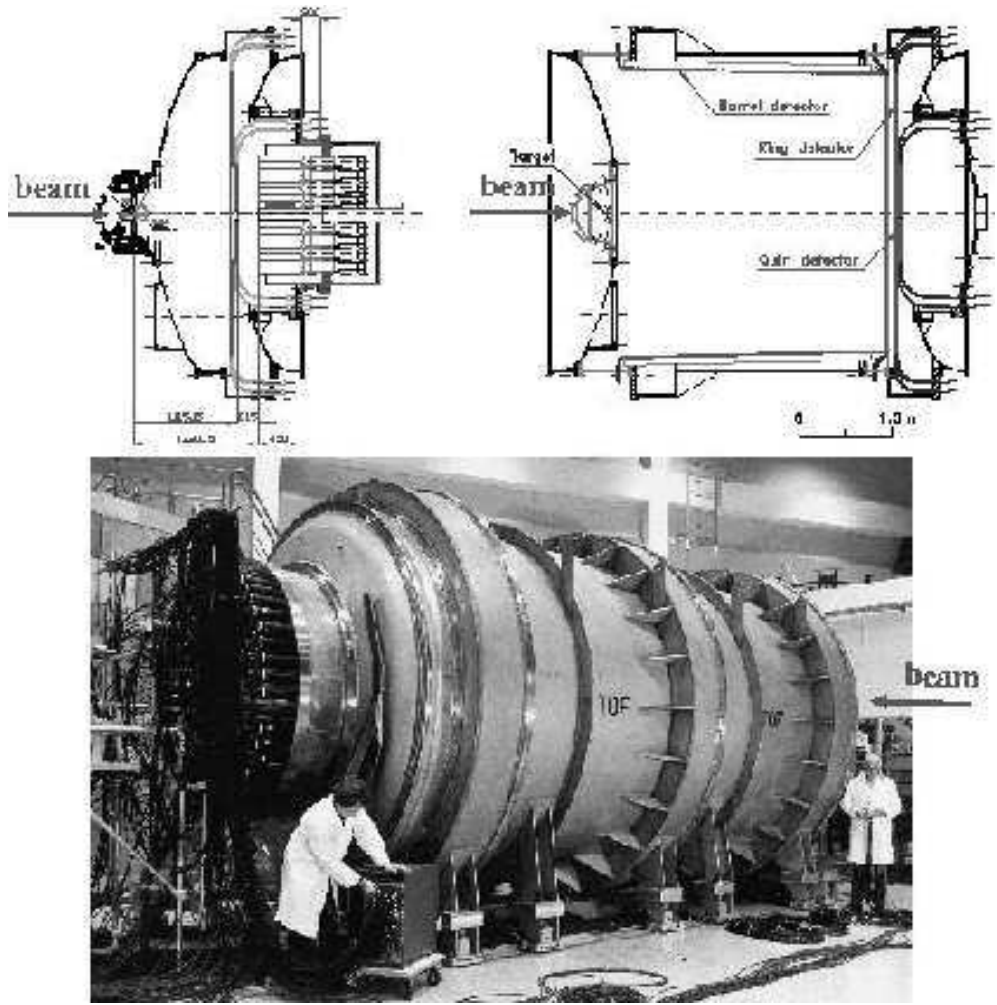,scale=0.67,angle=0}
\caption{Schematic view of the COSY--TOF detector facility in two
different set--ups, the short and the longer version. 
Bottom: A photo from the very long version of the COSY--TOF detector. 
  \protect\\
  \protect{\small 
    A version of the article including a photo of better quality can be found at the COSY-11 homepage:
   \protect\\
   $http://ikpe1101.ikp.kfa-juelich.de/cosy-11/pub/List\_of\_Publications.html\#papers$}
}
\label{TOF_ALL}
\end{center}
\end{figure}

In Fig.~\ref{TOF_ALL} two CAD-drawings of a short and 
an extended version as well as a photograph of the outer barrel in it's longest
assembly is shown. The modularity serves several purposes but essentially helps 
to adjust the system to the optimal dimensions according to the reaction under 
investigation. 

Three scintillators placed 1~m, 0.5~m and 0.1~m in front of the target are used
to veto a beam halo. Further the following components are employed:
\begin{itemize}
\item
A few mm$^3$ large liquid hydrogen/deuterium target with very thin
windows~\cite{Jaeckle94,Hassan97}
\item
Different granulated start counter systems optimized for the reaction type under
investigation.
\item
Up to three barrels equiped with granulated cylindrical detectors for timing, 
spatial resolution and energy loss determinations.
\item
Granulated three layer stop counters, divided into inner-- and outer circular scintillation detector
hodoscopes~\cite{Dahmen94}.
\item
A scintillation calorimeter for energy determination of charged and neutral
hadrons, placed behind the stop detector, see the short version of the facility,
left in fig.~\ref{TOF_ALL}.
\item
In preparation a straw drift chamber hodoscope, to be installed behind the start
detector systems. 
\end{itemize}
\vspace{-1cm}
 
\subsubsection{Example of a dedicated internal threshold detector}
\begin{figure}[H]
\begin{center}
\epsfig{file=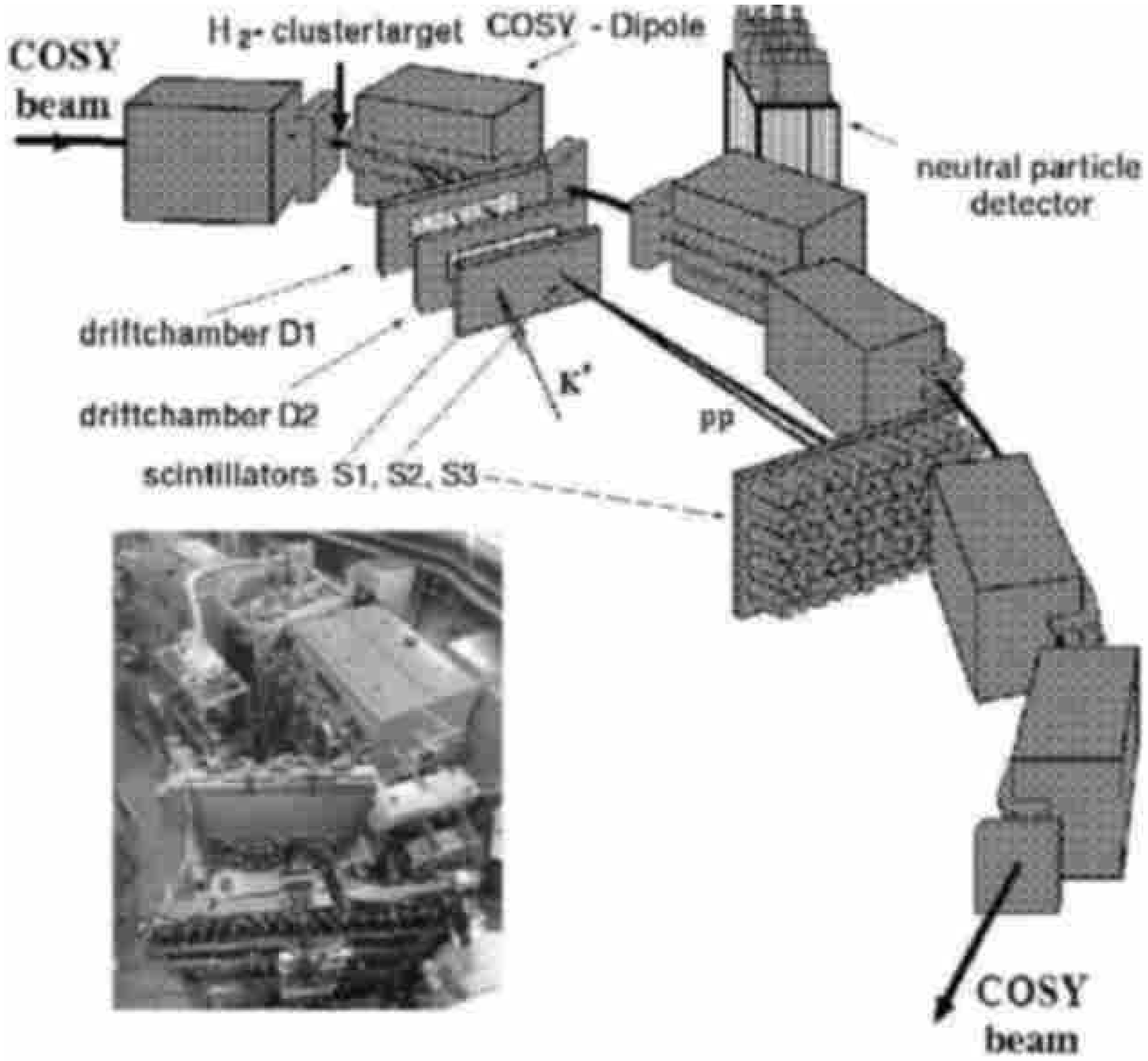,scale=0.8}
\caption{Top: $~~~~~$Top view of the COSY-11 zero degree facility. \protect \\
$~~~~~~~~~~~~~~~~$Bottom: A photographic view of the COSY--11 detector at 
the COSY ring. 
  \protect\\
  \protect{\small 
    A version of the article including a photo of better quality can be found at the COSY-11 homepage:
   \protect\\
   $http://ikpe1101.ikp.kfa-juelich.de/cosy-11/pub/List\_of\_Publications.html\#papers$}
}
\label{COSY-11_detector}
\end{center}
\end{figure}

The COSY--11 installation is an internal facility at COSY designed for threshold
meson production studies.
Close--to--threshold, due to the small transvers
momenta, all ejectiles are limited to a narrow forward cone in the laboratory 
system and therefore a high geometrical acceptance is attainable with a
comparatively small detector set--up.
The essential components of the COSY--11 installation are:
\begin{itemize}
\item
A cluster target located in front of a regular C--shaped COSY dipole producing
H$_2$ or D$_2$--clusters of up to 10$~^6$ molecules and densities of about 
10$~^{14}$ atoms/cm$^3$.
\item
An exit window (187 $\times$ 7.6 cm$^2$) of 30 $\mu$m Al and 2 $\times$ 150
$\mu$m carbon fibers to separate the ultra high beam line vacuum from
atmospheric pressure. It provides high acceptance compared to conventional
magnetic spectrometers especially in the horizontal plane.
\item
Two sets of drift chambers stacks (D1 and D2 in fig.~\ref{COSY-11_detector}) with 
six and eight planes, respectively and a position resolution of about 
200~$\mu$m in each plane, for particle track reconstruction and momentum
analysis through the magnetic field for positively charged ejectiles.
\end{itemize}
\vspace{-0.5cm}
 
\begin{itemize}
\item
Two scintillator hodoscopes (S1 and S2 in fig.~\ref{COSY-11_detector}) for fast
level trigger determination.
\item
A large area scintillation wall~\cite{Anton91} (2.2 $\times$ 1 m$^2$) S3 in 
fig.~\ref{COSY-11_detector} placed 9.1~m downstream, for fast trigger, 
time--of--flight and consecutively invariant mass determination.
\item
A detection system for negatively charged mesons consisting out of a
scintillator detector and an array of silicon pad detectors, both mounted inside
the dipole gap.
\item
A monitor system for elastically scattered particles, utilized for luminosity 
determination.
\item
In preparation for an extension: two wire chambers with a hexagonal drift 
structures
to be positioned at the side -- and at the front exit windows and a 
threshold \v Cerenkov counter behind the S1 hodoscope for pion background 
subtraction.  
\end{itemize}
 \section {Outlook}        
Storage rings with the flexibility of different projectiles, 
variable beam momenta, extremely good phase--space cooling and with the
unrivaled options of internal and external detection systems provide an 
excellent tool for investigations on meson production at threshold. 
Especially the high quality of relative and absolute energy resolutions 
available at modern accelerators are crucial for such studies.
  
In the present report, discussing the close--to--threshold meson production in
hadronic interactions, the natural restriction to investigations using only 
hadronic probes was made though interesting and informative results were
certainly obtained with electromagnetic tools as well.

The theory of strong interactions (QCD) has a convincing predictive power at
high energies. Perturbative expansions in the coupling constant no longer
converge for medium energy scales and the structure of QCD becomes increasingly
complicated.   
The threshold meson production in nucleon--nucleon scattering
offers valuable aspects for studying production, decay and interaction of
hadrons and investigating the reaction mechanisms, the dynamics of the system 
and appropriately chosen subsystems as well as the structure of hadrons
themselves. Such experiments probe the sensitivity of nucleon--nucleon
interactions at distances of $\le$~1~fm, while the long--range part of the
interaction is suppressed. Meson production in $pp \to ppX$ reactions 
(with $X~=~\pi^0, \eta, \omega, \phi$) 
is a process of strong inelasticity in proton--proton scattering.

\subsection{Polarized beams and/or targets}
\label{Pbaot}
In the present paper discussions utilizing polarized beam and/or 
targets are omitted. At the upgraded IUCF cooler ring a polarized proton 
beam in combination with a windowless polarized hydrogen target now 
permits measurements of all spin correlation coefficients for 
$\vec{p} \vec{p} \to NN\pi$ reactions. Very recently initial results for 
spin correlations were reported~\cite{Daehnick024003}. The goal of such
experiments is to qualify the energy dependence and the importance of 
higher partial waves by determining analyzing powers and spin correlations.
Here the hadron physics community certainly faces the challenge of new 
experimental techniques with necessary improvements of theoretical 
developments, since polarization observables are sensitive indicators of
reaction mechanisms, relative contribution of partial waves and transition
amplitudes.  

The production of $\eta$ mesons near threshold has been measured by 
different groups~\cite{bergdoltR2969,hibou41,calen39,smyrski182}. 
Very close--to--threshold only S--waves are supposed to contribute to the 
production process and in fact the S--wave production is large due to the 
presence of the N$^*$(1535)S$_{11}$ resonance which seems to act as an 
intermediate state.\\

Both TOF~\cite{TOFeta} and COSY--11~\cite{eta_menu} took recently rather high 
statistics data on the $\eta$ production at Q~$\approx$~15.5~MeV with fairly 
flat angular distributions as expected for an s--wave production. 
Still -- whether some data suggest or don't suggest -- differential cross 
section measurements might be too insensitive to reveal e.g. a small 
$Sd$--contribution. Data using a polarized beam,  however, might indicate 
effects arising from an interference term of the kind:
\begin{equation}
2 Re\{A_{Ss}A^*_{Sd}\} ~cos2 \theta^*_{\eta}
\end{equation}
$~~~$\\[-0.4cm]
where $A_{Ss}$ is the amplitude for the production of a final $pp$ pair with
angular momentum $L_{pp}~=~0$ and $l~=~0$ while for $A_{Sd}$
the $\eta$ meson has an angular momentum of $l_{\eta}~=~2$.\\
The importance of d--waves in $\eta$ production is well known from both 
$\pi^- p \to \eta n$~\cite{Deinet} and $\gamma p \to \eta p$~\cite{Kru95} 
reactions and the fact that this interference term is negative suggests that 
the $\eta$--production in proton--proton
scattering is governed mainly by $\rho$--exchange~\cite{calen190}.  
It is the aim of the COSY--11 group, followed by a recent feasibility 
study~\cite{Winter_Dipl}
to investigate the presence of this and other higher partial waves through the
study of the analyzing power $A_y$.

Though the asymmetry should also be studied with respect to the final 
proton directions, the s--d interference contributes significantly to the 
$\eta$ analyzing power defined as:
\begin{equation}
A_y := \frac{\mbox{Tr}(\mathcal{M}\,\sigma_y\,\mathcal{M}^\dagger)}
{\mbox{Tr}(\mathcal{M}\,\mathcal{M}^\dagger)}
\end{equation}
$~~~$\\[-0.40cm]
with the transition matrix element $\mathcal{M}$ and the Pauli matrix
$\sigma_y$.\\

To lowest order in the $\eta$ d--wave amplitude, the 
analyzing power is proportional to the imaginary part of the $s$--$d$ 
interference~\cite{COLINPRIVATE},
\begin{equation}
A_y \approx 2~ \frac{Im\{A_{Ss}A^*_{Sd}\}}{\arrowvert A_{Ss}\arrowvert ^2}~ 
\end{equation}
$~$\\[-0.40cm]
where $\phi^*_{\eta}~=~0$ corresponds to the plane of the beam 
(perpendicular to the polarization direction). On angular momentum grounds,
this signal is expected to vary like $(p^{cm}_{\eta})^2$, $i.e.$ roughly like Q.\\

The $P_{11}(1440)$ resonance could also be of importance for the $\eta$
production~\cite{LHP},
leading to p--waves in the final state. The interference between the $Ps$ and
$Pp$ amplitudes gives an analyzing power proportional to:
\begin{equation}
A_y \approx 2~ \frac{Im\{A_{Ps}A^*_{Pp}\}}{\arrowvert A_{Ss}\arrowvert ^2}~ 
\end{equation}
$~$\\[-0.40cm]
and such a signal has been observed in the $pp \to pp \pi^0$ reaction at low
energies~\cite{HOM}. The presence of the three $p$--waves in the interference
might suppress such a term close--to--threshold for $\eta$ production and no
unambiguous signal for them has been seen in the unpolarized differential cross
section. Nevertheless one must be prepared for their existence. The $Ss$--$Sd$
interference gives a maximal A$_y$ signal for $\theta^*_{\eta} \approx 45^0$ and
vanishes at $ 90^0$, whereas the $Ps$--$Pp$ signal should be largest at this
point.\\

Both the differential cross section and the analyzing power should be studied 
in order to detect effects associated with the onset of higher partial waves. 
It should be noted that the effects of incomplete acceptance are much less of a
problem for $A_y$ than for the $d\sigma / d\Omega$ measurement provided that 
the beam characteristics of the two polarization states are similar.  Finally, 
the results should help to settle the on--going discussion of whether the 
$\eta$ production is dominated by $\rho$~\cite{germond308,laget254,Santra,Gedalin}, 
$\omega$~\cite{vetter153} or $\eta$~\cite{Batinic} exchange.\\
 
\begin{figure}[H]
        \parbox{0.55\textwidth}{\epsfig{file=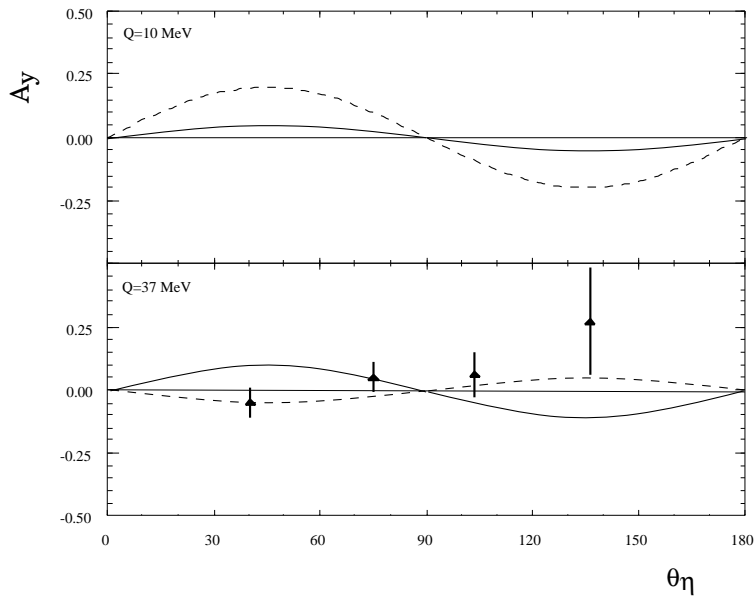,width=0.55\textwidth,angle=0}}
        \hfill
        \parbox{0.4\textwidth}{\vspace{-1cm}\caption{\small 
         Theoretical predictions for the analyzing power according 
         to~\cite{faldt427} (dotted line) and~\cite{nak01} (solid line) for 
         $Q=10$ and 37\,MeV. The solid triangles represent the measured 
         values~\cite{Winter_Dipl} at $Q=40\,$MeV. 
        \label{ayvergleich}
        }}
      \end{figure}

First data on the $\eta$ meson production with a polarized proton beam in 
the reaction have been taken at the internal experiment facility 
COSY-11~\cite{Winter_Dipl}. 
The measurement was performed with a proton beam momentum 
$\vec{P}_p~=2.096\,$GeV/c corresponding to an excess energy of $Q=40\,$MeV. \\
In Figure \ref{ayvergleich}  a comparison of the data (solid triangles) 
with theoretical predictions for $Q=37\,$MeV from F\"aldt and 
Wilkin~\cite{faldt427} (dotted line) and Nakayama \cite{nak01} (solid line)
is shown. In addition, the theoretical calculations for $Q=10\,$MeV are 
plotted, both based on the one meson exchange model. While 
F\"aldt and Wilkin predict a dominance of $\rho$ exchange (dashed line), 
Nakayama concludes a dominant $\pi$ and $\eta$ exchange (solid line). 
Though the data are consistent with $A_y=0$ within the error bars given, there 
seems to be a slight increase of $A_y$ towards backward scattering angles.
First results of the analyzing power -- deduced from the very first
feasibility study run -- suffer from statistics but tend to differ from both 
theoretical calculations, although the predictions of \cite{faldt427} show a 
quite good match. For a more precise statement further measurements are needed 
which could benefit from higher polarization now available at COSY. 

\subsection{Final state interaction}
\label{Fsinteraction}
An important physics aspect of the modern cooler ring facilities is the interaction 
of particles in the final state. A detailed study of the Dalitz plot of, e.g. the 
reaction $pp \to p K^+ \Lambda$ measured at COSY-11~\cite{bal98b}, made possible the
extraction of valuable information about the $\Lambda$--$N$ interaction. 

The $pn \to d \eta$ reaction, measured at TSL in
Uppsala~\cite{bilger64,calen2069,calen2667, calen2642}, limited significantly the
range of values for the $\eta$--$N$ scattering length. This quantity is the 
relevant input for calculations studying the existence of $\eta$--nucleus bound 
states. 

\subsection{Baryon resonances}
\label{Bresonanc}
Throughout the article the importance of an intermediate excitation of baryon 
resonances has been pointed out.\\
At excess energies of 1--2 GeV, baryon resonances begin to play an important 
role in nucleon--nucleon collisions. Most of our understanding about these 
resonances comes from pion and photon induced reactions
on the nucleon. Current quark models predict many more resonances than can be 
experimentally extracted from the data. A large program at both JLAB and ELSA 
focuses on this so-called missing resonance problem and the search for yet 
unknown baryonic states in electro-nuclear reactions. However, besides the 
possibility that resonances thus far simply went unnoticed because the quality 
of the data was insufficient, another explanation might be that these 
resonances couple only weakly to photons or pions. An accelerator like COSY,
built for nucleon--nucleon as well as nucleon--nucleus collisions will provide 
for the necessary experiments to fill that gap in the future. 

\subsection{Symmetries} 
\label{Symmetr}
The investigation of symmetries and their breaking has always led to 
significant insights into the inner workings of nature. Already a large 
program at COSY~\cite{Magiera,idee} is studying charge-symmetry breaking - an 
effect closely linked to the difference in the current masses of the light quarks.\\ 
Another interesting aspect of charge-symmetry breaking is its high sensitivity: 
Once, for instance, the $f_0$--$a_0$ mixing amplitude is isolated, it will provide 
us with insights about the nature of scalar resonances. Studies about
the structure of these resonances are under investigations at 
COSY~\cite{proceedingscosy11,Proposal,Ankeproposal}.\\
At the CELSIUS ring in Uppsala the CELSIUS/WASA detection system is in operation now
and exciting experiments especially on $\eta$--meson decay studies will start.

\subsection{Meson production in the three--nucleon system}
\label{Mpittns} 
Naively, one would think that near-threshold meson production in pd collisions 
can be fully understood in terms of a short-ranged $NN \to NN \pi$ process, with 
the extra nucleon acting as a spectator. However, backward $\pi$ production in the
$pd \to {}^3He \pi^0$ cannot be explained in terms of a spectator model. 
In the case of meson production in $pd$ collisions the number of 
participating amplitudes  is small as well, even though the transferred momentum is 
large. Hereby a unique environment is available for studying the interaction of
more than just two nucleons, e.g. three-nucleon forces. 
It is likely that eventually the research on three-nucleon
forces will focus on this process. Again, the role of polarization observables 
is important. 
 
\subsection{Apparative aspects}
\label{Aaspect}
For the experiments, the development of suitable experimental configurations, 
comprising polarized targets and associated detector systems that facilitate 
measurements of polarization observables is necessary. 

Present and future experiments become possible only through the outstanding 
performance of the cooler accelerators. With the new injector for COSY the situation 
will be further improved. It will be possible to fill the machine
up to the space-charge limit with polarized protons and deuterons, whereby 
electron-cooled beams at injection energy and stochastically-cooled beams 
at high energy of unprecedented brilliance will be available to the users.
Efficient correction schemes to preserve the polarization of protons 
during acceleration have already been established, while for deuterons no 
difficulties are anticipated, because no depolarizing resonances occur in the
energy range of COSY.  
New polarization experiments in the future will certainly call for further 
refinement of the spin-manipulation systems in COSY, e.g. longitudinally 
polarized internal and external beams, for which suitable
spin rotators have to be developed. 
 
\subsection{Epilogue} 
\label{epilog}
The above mentioned examples -- which will be the subject of a 
Summer School and Workshop on COSY Physics~\cite{school} --
demonstrate the richness of physics in meson production reactions. 
The drawback of this wealth comes at the expense that, apart from rare cases, 
it is difficult to extract a particular piece of information from the
data, e.g. both resonances and final state interactions modify the invariant 
mass plot in a coherent superposition.\\
Fortunately, the use of polarization in the initial state acts as a spin filter 
and different contributions to the interaction can be singled out, because they show 
up in the angular distributions of different spin combinations. It is important to 
stress, that as long as only the lowest partial waves contribute significantly 
to the process under study, the number of independent observables from experiments 
with polarized beam and target is sufficient to allow for a model independent 
partial wave analysis. 
\section*{Acknowledgements}
The authors would like to express their utmost graditude to the
numerous people who gave the various help needed in order to successfully
complete this article. \\
Especially we would like to thank R.~Czy{\.{z}}ykiewicz,
J.~Flood, J.~Haidenbauer, Ch.~Hanhart, T.~Sefzick and P.~Winter for both
scientific and technical support throughout the preparation of this manuscript.

\end{document}